\documentclass[letterpaper,11pt]{article}
\usepackage[margin=0.8in]{geometry}

\usepackage{graphicx}
\usepackage{algorithm}
\usepackage{algpseudocode} %
\usepackage{physics}

\usepackage{cmap} %
\usepackage[utf8]{inputenc}
\usepackage[english]{babel}
\usepackage[T1]{fontenc}
\usepackage{amsmath}
\usepackage{amssymb}
\usepackage{amsfonts}
\usepackage{bm}
\usepackage{bbm}
\usepackage{xcolor}
\definecolor{blueviolet}{rgb}{0.2, 0.2, 0.6}
\definecolor{webgreen}{rgb}{0,.5,0}
\definecolor{webbrown}{rgb}{.6,0,0}
\usepackage{setspace}
\usepackage[pdftex,
	bookmarks=false,
	colorlinks=true, %
	urlcolor=webbrown, 
	linkcolor=blueviolet, 
	citecolor=webgreen,
	pdfstartpage=1,
	pdfstartview={FitH},  %
	bookmarksopen=false
	]{hyperref}
\allowdisplaybreaks
\usepackage{tikz}
\usepackage{braket}
\usepackage[numbers,sort&compress]{natbib}
\usepackage{amsthm}
\usepackage{dsfont}
\usepackage{listings}
\usepackage[capitalize]{cleveref}
\usepackage{appendix}
\usepackage{booktabs}
\usepackage{authblk}
\usetikzlibrary{quantikz}
\usetikzlibrary{arrows.meta,calc,positioning,fit,backgrounds,decorations.pathreplacing,shapes.misc}
\definecolor{figblue}{HTML}{0072B2}
\definecolor{figorange}{HTML}{D55E00}
\definecolor{figgreen}{HTML}{009E73}
\definecolor{figpurple}{HTML}{AA4499}
\definecolor{figamber}{HTML}{E69F00}
\tikzset{
  figpanel/.style={draw=black!60, line width=0.5pt, rounded corners=4pt, inner sep=7pt},
  figbar/.style={figpanel, fill=black!4},
  figsub/.style={draw=black!50, line width=0.4pt, rounded corners=3pt, inner sep=5pt},
  figtitle/.style={font=\normalsize\bfseries},
  figstep/.style={font=\bfseries},
  fighead/.style={font=\normalsize\bfseries, text=figblue},
  figtext/.style={font=\normalsize, align=left},
  figarrow/.style={-{Stealth[length=4pt, width=3pt]}, line width=0.6pt},
  figarrowbig/.style={-{Stealth[length=6pt, width=5pt]}, line width=0.7pt},  %
}

\IfFileExists{titlesec.sty}{%
  \usepackage{titlesec}%
  \titleformat{\subsubsection}[block]
    {\centering\itshape}
    {\alph{subsubsection}.}
    {0.75em}
    {}%
  \renewcommand{\thesubsubsection}{\thesubsection.\alph{subsubsection}}%
}{%
  \PackageWarning{main_arxiv}{titlesec.sty not found; install with: tlmgr install titlesec. Subsubsections use alph numbering only.}%
  \renewcommand{\thesubsubsection}{\thesubsection.\alph{subsubsection}}%
}
  
\numberwithin{equation}{section}

\newtheorem{theorem}{Theorem}
\newtheorem{corollary}{Corollary}

\newtheorem{lemma}{Lemma}
\newtheorem{proposition}{Proposition}

\newtheorem{assumption}{Assumption}

\theoremstyle{definition}
\newtheorem{definition}{Definition}
\newtheorem{remark}{Remark}

\newcommand{\dlat}{d_{\mathrm{lat}}}  %

\newcommand{\vertiii}[1]{{\left\vert\kern-0.25ex\left\vert\kern-0.25ex\left\vert #1 \right\vert\kern-0.25ex\right\vert\kern-0.25ex\right\vert}}

\newcommand{\rom}[1]{\mathtt{\uppercase\expandafter{\romannumeral #1\relax}}}

\DeclareMathOperator*{\sign}{sign}

\renewcommand{\ketbra}[2]{\lvert #1 \rangle \! \langle #2 \rvert}

\renewcommand{\norm}[1]{\left\lVert#1\right\rVert}

\usepackage{xargs}
\usepackage[colorinlistoftodos,prependcaption,textsize=tiny]{todonotes}
\newcommandx{\lz}[2][1=]{\todo[linecolor=red,backgroundcolor=red!10,bordercolor=red,#1]{LZ: #2}}

\usepackage{mathtools}
\usepackage[normalem]{ulem}
\usepackage{wrapfig}
\usepackage{comment}
\usepackage{thmtools,thm-restate}
\usepackage[export]{adjustbox}
\usepackage{needspace}


\newtheorem*{theorem*}{Theorem}

\newtheorem*{task*}{Task}
\newtheorem*{proposition*}{Proposition}

\newcommand{\ee}{\end{equation}}

\usepackage{xcolor}

\def\:={\,\raisebox{0.85pt}{.}\hspace{-2.78pt}\raisebox{2.85pt}{.}\!\!=\,}
\def\=:{\,=\!\!\raisebox{0.85pt}{.}\hspace{-2.78pt}\raisebox{2.85pt}{.}\,}

\newtheoremstyle{mainbreak}{\topsep}{\topsep}{\itshape}{}{\bfseries}{.}{\newline}%
  {\thmname{#1}\thmnumber{ #2}\thmnote{ \normalfont\bfseries(#3)}}
\newtheoremstyle{mainbreakdef}{\topsep}{\topsep}{\normalfont}{}{\bfseries}{.}{\newline}%
  {\thmname{#1}\thmnumber{ #2}\thmnote{ \normalfont\bfseries(#3)}}
\newtheoremstyle{mainrun}{\topsep}{\topsep}{\itshape}{}{\bfseries}{.}{ }%
  {\thmname{#1}\thmnumber{ #2}\thmnote{ \normalfont\bfseries(#3)}}
\theoremstyle{mainbreak}
\newtheorem{theorembrk}[theorem]{Theorem}
\newtheorem{corollarybrk}[corollary]{Corollary}

\theoremstyle{mainbreakdef}

\theoremstyle{mainrun}
\newtheorem{theoremkey}[theorem]{Theorem}
\newtheorem{assumptionkey}[assumption]{Assumption}

\begin{document}
\allowdisplaybreaks

\title{Certified Discovery of Quantum Phase Diagrams}

\author[1]{Yuanxi Liu}
\author[1,2]{John Preskill}
\author[2,1]{Hsin-Yuan Huang}

\affil[1]{California Institute of Technology}
\affil[2]{Oratomic, Inc.}
\date{\today}

\maketitle

\begin{abstract}\normalsize
Can a machine provably discover new quantum phases of matter?
We show that the answer is yes under physically motivated assumptions.
For any smooth family of geometrically local Hamiltonians, we construct an efficient quantum algorithm that outputs the phase diagram to prescribed accuracy and provides a verifiable certificate for every transition it reports.
The algorithm takes as input only a classical description of the family together with unverified preparations of reference ground states; no order parameter, symmetry-breaking pattern, or training data are supplied. It  succeeds provided that at least one preparation is correct in each gapped region of parameter space.
The algorithm explores parameter space by quasi-adiabatically or adiabatically transporting each reference state, detecting the approach of a phase boundary by tapping the transported state with a frequency-filtered local probe: excitations created by the probe witness the closing of the energy gap.
We prove that the gap closing for continuous phase transitions can always be detected using the frequency-filtered gradient of the Hamiltonian.
The gap-detecting measurement is gentle, so when the ground state is unique a single reusable copy of each reference ground state supports the full exploration (for degenerate ground states, polynomially many copies suffice).
The quantum algorithm runs in polynomial time at inverse-polynomial gap resolution.
We also give a classical algorithm that runs in quasipolynomial time at constant gap resolution.
Numerical experiments demonstrate how our algorithm accurately determines the phase structure of several model quantum systems.
\end{abstract}

\pagenumbering{arabic} 
\setcounter{page}{1}

\addtocontents{toc}{\protect\setcounter{tocdepth}{0}}

\section{Introduction}
Quantum devices are beginning to prepare ground states and low-energy states of
models that nobody can solve: frustrated magnets, doped Hubbard-like models, and
engineered lattice gauge theories are already realized on programmable quantum
simulators~\cite{ebadi2021quantum, scholl2021quantum, semeghini2021probing,
shao2024antiferromagnetic, yang2020observation, zhou2022thermalization}, and
within a few years this will become routine. For such states, there is no answer key:
classical simulation cannot check them, and there is no textbook against which
to compare them. That absence is the whole point of building the machines.

The output we want from such a device is usually not the state itself but instead the
physical insight we can extract from it.
In particular, one would like a machine that \emph{provably discovers} phases of matter: given only a Hamiltonian family and a few unverified ground-state preparations, it outputs the phase diagram together with a certificate for every phase transition it reports. We call this task \emph{certified discovery of quantum phase diagrams}. Until now, whether such a machine is possible has remained open, and in the worst case it is not. Estimating a spectral gap is hard~\cite{ambainis2014physical, gharibian2019complexity}, as is locating a dynamical phase transition.~\cite{xu2026provable}). The gapped-versus-gapless question is undecidable in the thermodynamic limit~\cite{cubitt2015undecidability, bausch2020undecidability1d}, and the phase diagram of a one-parameter family can be uncomputable~\cite{bausch2021uncomputability}. Accordingly, quantum protocols for gap estimation ~\cite{cugini2025spectral, pages2026lowresource} come without worst-case guarantees.

What has existed instead are incomplete answers:
heuristic machine-learning and optimization-based methods classify or detect phases
without guarantees~\cite{carrasquilla2017machine, vannieuwenburg2017learning,
wang2016discovering, wetzel2017unsupervised, broecker2017machine, chng2017machine,
zhang2017quantum, rodrigueznieva2019identifying, greitemann2019probing,
carleo2019machine, che2025circuit, grossifonseca2025gradient, khosrojerdi2025unsupervised,
malyshev2026distance, huang2026detecting, zhou2026autonomous,
nys2026quantummatterexplorer}, discovering without certifying, and provable algorithms apply only to special
cases rather than generic geometrically local Hamiltonian families~\cite{knabe1988energy,rai2026hierarchy,lake2025exact,chaduteau2026agnostic}.
Additional rigorous results concern supervised learning~\cite{huang2022provably,wu2023quantum,jager2026provable}:
classical or quantum machine-learning algorithms provably \emph{classify}
phases after being trained on data in which phases are labeled. 
In that case, information about the phase structure is provided as an input; 
the algorithm does not discover new physics on its own.

\begin{figure}[t]
    \centering
    \includegraphics{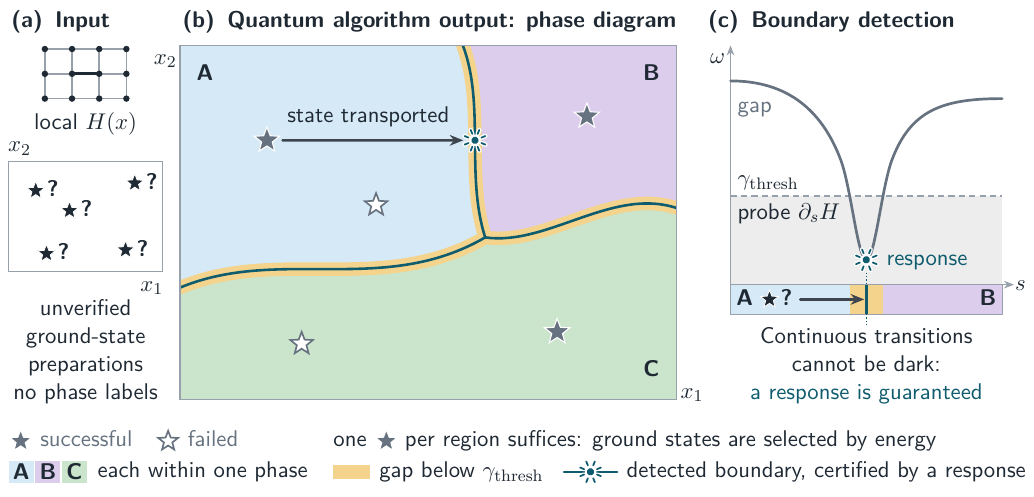}
    \caption{\textbf{Overview of certified discovery of quantum phase diagrams.}
    \textbf{(a)}~The input is a classical description of a geometrically
    local Hamiltonian family $H(x)=\sum_jh_j(x)$ and reference state
    preparations at a few points of the parameter space. The algorithm is
    not told which preparations succeed, and no phase labels are supplied.
    \textbf{(b)}~The output is the phase diagram to prescribed accuracy:
    points in the same tinted region are joined by chains of links whose endpoints are certified not to be in different
    phases, and each detected boundary is certified by a measured response
    and located to within the amber strip, where the gap is below the
    threshold $\gamma_{\mathrm{thresh}}$. The algorithm sweeps outward from
    every reference by quasi-adiabatic transport (one path is shown) and, where several
    transported states arrive at a point, selects among them by an energy
    test, so failed preparations do no harm: one successful preparation
    seeds discovery throughout its connected component of the resolved
    region. Gray stars mark the successful (filled) and failed (hollow)
    preparations, which the algorithm is not told.
    \textbf{(c)}~Gap closings are detected by tapping the transported
    state with a frequency-filtered local probe, the Hamiltonian gradient
    $\partial_sH$ along the path: a filtered response witnesses a gap
    below the threshold $\gamma_{\mathrm{thresh}}$, and for continuous
    transitions this response provably cannot vanish -- continuous
    transitions cannot be dark.}
    \label{fig:overview}
\end{figure}

In this work, we construct the missing machine: given a classical description of a local Hamiltonian family and unverified preparations of reference ground states, at least one of which succeeds in each gapped region, our efficient quantum algorithm provably outputs the phase diagram to prescribed accuracy, with no order parameter, symmetry pattern, or training labels supplied (Figure~\ref{fig:overview}).

This algorithm formalizes ideas that are familiar to physicists. In each gapped phase, a ground state is provided at a single point in parameter space -- perhaps an exact solution is available at that special point, or the ground state might be prepared accurately and efficiently by quantum simulation. Beginning at that point, the parameter space is explored systematically by quasi-adiabatically transporting the state, seeking points where the energy gap falls below a prescribed small value. The gap is detected by tapping the transported ground state with a local probe, filtered in frequency with cutoff $\Lambda$. Excitations created by the probe certify that the gap is less than $\Lambda$, hence approximately identifying the phase boundary where the energy gap vanishes in the thermodynamic limit. The gap-detecting measurement can be executed by an efficient quantum circuit; furthermore, it can be chosen to be very gentle, hardly perturbing the ground state at all, so that the same state can be used repeatedly by the algorithm. The runtime of the quantum algorithm is polynomial in the system size even if we choose to resolve the energy gap to inverse polynomial precision.

What happens if the closing of the energy gap is ``dark'' so that no local probe can detect the phase boundary? Fortunately, in the case of continuous transitions, this unfavorable situation can be excluded.\footnote{More precisely, a quantum phase transition cannot be dark unless the energy gap closes faster than inverse-polynomially as a function of system size.} If a continuous transition were dark, frequency-filtered quasi-adiabatic evolution across the phase boundary, which can be simulated by a constant-depth geometrically local quantum circuit, would preserve the ground state with high fidelity. The standard definition of quantum phases dictates that two quantum states connected by a constant-depth local unitary circuit are in the same phase~\cite{chen2010local}; therefore a genuine continuous phase transition cannot be dark. In fact, this argument guides our choice of local probe -- the (frequency-filtered) gradient of the Hamiltonian $H(x)$ with respect to the parameter $x$, the generator of quasi-adiabatic evolution, is guaranteed to detect the gap closing. 

In general, detecting a spectral gap is a very hard problem even for quantum computers~\cite{ambainis2014physical,gharibian2019complexity,cubitt2015undecidability,bausch2020undecidability1d,watson2021complexityapproximatingcriticalpoints}, so our quantum algorithm is surely foiled by dark transitions for some Hamiltonian families. Intuitively, quantum phase transitions might go undetected if two qualitatively different candidate ground states cross one another in energy at a surface in the parameter space, where the energy gap becomes exponentially small in the system size.%

A classical algorithm for phase learning, with the energy gap estimated to constant resolution, can be constructed following a similar strategy, in which quasi-adiabatic evolution and the response to a frequency-filtered probe are simulated in a finite-size patch of the system. Under a suitably strengthened assumption about the local visibility of the phase boundary, the runtime of this classical algorithm is quasipolynomial in the system size whenever the transport routes it follows have polylogarithmic length. 

In Section~\ref{sec:main} we provide a high-level description of the quantum and classical algorithms and our main theorems, and Section~\ref{sec:numerics-main} describes numerical simulations illustrating applications of the algorithms. Section~\ref{sec:discussion} contains some concluding remarks. Many technical details are in the appendices.

\section{Main results}
\label{sec:main}

\renewcommand{\thesubsubsection}{\thesubsection.\arabic{subsubsection}}
\IfFileExists{titlesec.sty}{%
  \titleformat{\subsubsection}[hang]
    {\normalfont\normalsize\bfseries}
    {\thesubsubsection}
    {0.75em}
    {}%
}{}

Let $H(x) = \sum_j h_j(x)$ be a family of geometrically local Hamiltonians on $n$ qubits, where each term $h_j$ acts on only a few neighboring qubits and varies smoothly as a function of the parameter $x \in \mathcal{M} = [-1,1]^m$ with $m = O(1)$. Our algorithm tracks how the spectral gap $\gamma(x)$ depends on $x$, seeking phase boundaries where $\gamma$ sinks to a small value that vanishes in the limit $n\to\infty$. When the ground state is unique, we define $\gamma(x)= E_1(x)-E_0(x)$, where $E_0$ is the ground-state energy and $E_1$ is the energy of the first excited state. If several ground states are nearly degenerate, we call this low-lying cluster the ground-state manifold $\mathcal G(x)$ and define $\gamma(x)$ as the energy difference separating it from the rest of the spectrum. We write $r(x)$ for its dimension (the ground-state degeneracy), $P_{\mathcal G}(x)$ for its projector, and $\rho_{\mathcal G}(x)=P_{\mathcal G}(x)/r(x)$ for the maximally mixed state on it (the maximally mixed ground state). This caveat is relevant for
broken-symmetry and topologically ordered phases, where splittings within the cluster are exponentially small in $n$, which should not be interpreted as signaling a phase boundary.

To draw a phase diagram, one needs to decide whether and where the spectral gap closes along paths in the parameter space. Therefore the core of our phase-learning algorithm is the certified detection of gap closings. In the worst case, this task is extremely hard. Estimating the spectral gap of a local Hamiltonian is $\mathsf{P}^{\mathsf{UQMA}[\log]}$-hard~\cite{ambainis2014physical,gharibian2019complexity}, and in the thermodynamic limit the gapped-versus-gapless question is undecidable for translation-invariant lattice systems in two dimensions~\cite{cubitt2015undecidability} or even in one dimension~\cite{bausch2020undecidability1d}. Locating a phase boundary to constant precision for a finite system is $\mathsf{P}^{\mathsf{QMA}_{\mathsf{EXP}}}$-complete~\cite{watson2021complexityapproximatingcriticalpoints}. No efficient algorithm can succeed unconditionally. But in all hard instances, properties of ground states change abruptly at the phase boundary. We will prove that quantum algorithms that efficiently detect gap closings can be constructed when the phase transition is continuous.\footnote{When we say that a phase transition is continuous, we mean that its finite-size gap closes no faster than an inverse polynomial in the system size $n$. In special cases, first-order transitions can also have inverse-polynomial minimum gaps rather than exponentially small ones~\cite{laumann2012quantum,campostrini2014finite}; our results apply to such first-order transitions as well.} %

Experimental physicists use a variety of heuristic methods to map phase diagrams: for example measuring observables such as order parameters, structure factors, or local correlations that behave differently in distinct phases~\cite{ebadi2021quantum, semeghini2021probing, keesling2019quantum, mazurenko2017cold, shao2024antiferromagnetic}. In general, though, it might not be easy to guess which operator signals a given transition. Our approach to detecting phase boundaries is inspired by physical experiments in which the investigator perturbs the ground state with a locally coupled low-frequency probe. Creation of low-energy excitations by a probe with frequency $\omega$ certifies an energy gap no larger than $\omega$.

To formalize this idea, we define the band-restricted spectral weight of
an operator $O$ with respect to the ground-state manifold,
\begin{equation}\label{eq-band-restricted-spectral-weight-main}
W^{O}_{[\gamma,\Lambda)}(x)
=\frac{1}{r(x)}
\sum_{\substack{0\le j<r(x)\le k\\ E_k(x)-E_j(x)<\Lambda}}
\bigl|\langle\psi_k(x)|O|\psi_j(x)\rangle\bigr|^2\,.
\end{equation}
It measures how strongly $O$ couples the maximally mixed ground state
$\rho_{\mathcal G}(x)$ to states above the manifold  with transition
energy at least $\gamma(x)$ but less than $\Lambda$; for a unique ground state it is the  %
sum of
squared transition matrix elements out of $|\psi_0(x)\rangle$. We probe
$\rho_{\mathcal G}(x)$ rather than a particular ground state because a
pure state can rotate within a degenerate manifold onto a state that a
given probe does not see, while $\rho_{\mathcal G}(x)$ is invariant under
such rotations. These rotations are harmless for transport in any case;
the only error that matters is leakage across the gap, and that is what
$W$ measures.

This quantity controls the transition rate in Fermi's golden rule and the
absorptive Kubo response if the system is probed by modulating $O$ with a
frequency spectrum uniformly distributed over a band from $\omega_0$ to
$\Lambda$, where $0<\omega_0<\gamma(x)$. Note, though, that our rigorous
analysis of the system's response to this probe does not require the use
of weak-coupling perturbation theory.

Non-negligible inelastic response to this probe certifies that the system has an energy gap less than $\Lambda$. But not detecting a response for a particular $O$ does not by itself ensure the gap to be greater than $\Lambda$. A different local probe could yield a different result. A small gap could even be invisible to all local probes because the spectral weight is negligible for any local $O$. Indeed, some discontinuous phase transitions are ``dark'' in this sense.
Theorem~\ref{thm:nondark-informal} states that continuous phase transitions obeying an inverse-polynomial finite-size gap condition cannot be dark.

To understand why continuous quantum phase transitions cannot be dark, begin by recalling the definition of a quantum phase. We fix a class $\mathcal{C}$ of geometrically local unitary circuits of constant depth $D = O(1)$ (for the definition of symmetry-protected phases we require the circuits to respect the symmetry), and we specify two functions of the system size
\begin{equation}
    \eta_P(n) > \eta_B(n) \ge 1/\mathrm{poly}(n).
\end{equation}
Two ground states $\rho_{\mathcal G}(x_A)$ and $\rho_{\mathcal G}(x_B)$ with $r(x_A)= r(x_B)$ are in the same phase if 
\begin{equation}\label{eq:same-phase}
    \operatorname{Tr}\bigl[P_{\mathcal G}(x_B)C\rho_{\mathcal G}(x_A)C^\dagger\bigr] \ge 1 - \eta_B(n) \, \text{ for some } C \in \mathcal{C},
\end{equation}
and are in different phases if 
\begin{equation}\label{eq:different-phase}
    \operatorname{Tr}\bigl[P_{\mathcal G}(x_B)C\rho_{\mathcal G}(x_A)C^\dagger\bigr] \le 1 - \eta_P(n) \, \text{ for all } C \in \mathcal{C}.
\end{equation}
Moreover, if $r_A \ne r_B$, then $\rho_{\mathcal G}(x_A)$ and $\rho_{\mathcal G}(x_B)$ are in different phases. 

Another essential ingredient in our arguments is quasi-adiabatic evolution. Let $H(s)$, $s \in [0,1]$, be a smooth parametrized path in the space of Hamiltonians. Suppose the gap of $H(s)$ stays open along the path, and let $F_\Delta(t)$ be a function of time whose Fourier transform has negligible support for $\Delta > 0$ (a \emph{filter function} with energy cutoff $\Delta$). The
\emph{quasi-adiabatic generator} $D_\Delta(s)$ and \emph{propagator} $U_\Delta(s)$
are~\cite{hastings2005quasiadiabatic, hastings2010locality,
bachmann2011automorphic}
\begin{equation}
D_\Delta(s) := i \int_{-\infty}^{\infty} F_\Delta(t)\, e^{itH(s)}\,
\partial_s H(s)\, e^{-itH(s)}\, dt,
\qquad
U_\Delta(s) := \mathcal{T}\exp\!\left(i\int_0^s D_\Delta(s')\,ds'\right),
\end{equation}
where $\mathcal{T}$ denotes time-ordering. The unitary
$U_\Delta := U_\Delta(1)$ is quasi-local and transports the ground state
along the path; on finite systems it is approximated by a
constant-depth geometrically local circuit. An $O(1)$-length gapped path in
parameter space therefore certifies that its endpoint ground states lie in
the same phase.  Conversely, if the endpoint ground states lie in different
phases, then along any $O(1)$-length path between them, the quasi-adiabatic
evolution $U_\Delta$ with constant cutoff $\Delta$ must incur at least
inverse-polynomial error:

\begin{equation}\label{eq-error-evolution-no-constant-depth}
\operatorname{Tr}\bigl[P_{\mathcal G}(x_B)U_\Delta\rho_{\mathcal G}(x_A)U_\Delta^\dagger\bigr]
\le 1-\eta_P(n)\,.
\end{equation}

One's naive expectation is that whether quasi-adiabatic continuation with
energy cutoff $\Delta$ successfully preserves the ground state along a
path in the parameter space hinges on whether $\Delta$ lies below the
true energy gap everywhere along the path. We prove a refined version of
this criterion: continuation is controlled by the spectral weight
$W^{\partial_sH}_{[\gamma,\Delta)}$ defined in
Eq.~\eqref{eq-band-restricted-spectral-weight-main}, where the operator
$O$ is the path-gradient operator $\partial_sH(s)$. Continuation succeeds
when this spectral weight is sufficiently small everywhere, even if the
cutoff lies above the gap. Stated contrapositively, %
if the gap is no smaller than 1/poly($n$) everywhere along the path,
then failure to preserve the ground state
as in Eq.~\eqref{eq-error-evolution-no-constant-depth} forces the
spectral weight of $\partial_sH(s)$ to be at least inverse-polynomially
large. 

\begin{theorembrk}[Continuous transitions are not dark; informal]
\label{thm:nondark-informal}
Let $H(s)=H(x(s))$, $s\in[0,1]$, be a smooth path of length $O(1)$ in parameter space whose endpoint ground states lie in different phases, and let $\gamma_{\min}$ be the smallest gap along the path. Fix a constant cutoff $\Delta=\Theta(1)$. Then the frequency-filtered path-gradient probe $\partial_sH$ elicits a response on a sizable portion of the path: there is a set $S_{\mathrm{high}}\subseteq[0,1]$ with
\begin{equation}\label{eq:nondark-main}
|S_{\mathrm{high}}|=\Omega\!\left(\frac{\gamma_{\min}\sqrt{\eta_P}}{n}\right),
\qquad
W^{\partial_sH}_{[\gamma,\Delta)}(x(s))=\Omega\!\left(\eta_P\,\gamma_{\min}^2\right)
\quad\text{for all }s\in S_{\mathrm{high}}.
\end{equation}
In particular, if $\gamma_{\min}\ge 1/\mathrm{poly}(n)$, a response with at least inverse-polynomial strength occurs on a fraction of the path at least inverse-polynomial in size.
The formal statement is Theorem~\ref{thm:cannot-be-dark}, proved in Appendix~\ref{sec:cannot-be-dark}.
\end{theorembrk}

Thus, quasi-adiabatic evolution plays two complementary roles in
our phase discovery procedure. When the gap is large, we can transport
the ground state across the parameter space efficiently, making only a
small error. Where the gap gets small, the path-gradient operator
$O=\partial_sH(s)$ can be employed as the probe that detects the gap
closing.

In the task of \emph{certified discovery of quantum phase diagrams}, the algorithm takes a list of points $x_1,\ldots,x_N$ in the parameter space $\mathcal{M}$, together with a constant threshold $\gamma_{\mathrm{thresh}}=\Theta(1)$, links some pairs of points, and sorts the points into clusters: two points share a cluster when a chain of links joins them. 
Its output comes with two guarantees. First, whenever two %
points are linked, their ground states are not in different phases.\footnote{Note that, because of the gap between $\eta_B$ and $\eta_P$ in the definition of a quantum phase, saying that two ground states ``are not in different phases'' is not equivalent to saying that the ground states are in the \emph{same} phase. Exhibiting a constant-depth local unitary circuit that maps ground state $A$ to a state that has overlap $1{-}\epsilon$ with ground state $B$, where $\eta_P > \epsilon > \eta_B$, shows that $A$ and $B$ are not in different phases but does not establish that $A$ and $B$ are in the same phase.} Second, any two points placed in different clusters are either in different phases, or cannot be joined by any path of bounded length in $\mathcal{M}$ without the gap dropping below $\gamma_{\mathrm{thresh}}$ somewhere along the way.\footnote{States in the same phase that are separated by gap closings within $\mathcal{M}$ do occur in some cases~\cite{jones2026quasiphases}, so the caveat is appropriate.} The outputs are guaranteed to be correct if the gap at each input point is at least $\gamma_{\mathrm{lb}}=1/\mathrm{poly}(n)$, which we call the \emph{resolution}; choosing a smaller $\gamma_{\mathrm{lb}}$ lets the algorithm resolve finer details of the phase diagram. Every gap closing the algorithm reports comes with a \emph{certificate}, a measured low-energy response to a local probe, 
which anyone with the same access to the supplied states can reproduce. Theorem~\ref{thm:nondark-informal} guarantees that such a response exists whenever the transition is continuous. 
The formal statement of this task is Definition~\ref{def:phase-clustering} (Appendix~\ref{sec:appendix-A}).

To solve this task, our phase learning algorithm requires as input a reference ground state that can be quasi-adiabatically transported to other parameter points; we therefore assume that it receives the outputs of ground-state preparations at a set of points $S$. In a real quantum device, the state preparations may fail, with no reliable flag of failure. We refer to each prepared state $C[x]$ as a \emph{reference state}, whether successful or not. The preparation is \emph{successful} when $\|C[x]-\rho_{\mathcal G}(x)\|_1\le\epsilon_{\mathrm{ref}}$ for a sufficiently small inverse-polynomial $\epsilon_{\mathrm{ref}}$.

\begin{assumptionkey}[Reference states]
\label{assumption:references-informal}
Consider the following region of parameter space:
\begin{equation}
\{x \in \mathcal{M} : \gamma(x) \ge \gamma_{\mathrm{lb}}\},
\end{equation}
where the resolution scale $\gamma_{\mathrm{lb}}$ is given as part of the
input. We call this the \emph{resolved region}. We assume the resolved region has at most polynomially many path-connected
components, and every path-connected component contains a point of $S$ where the ground-state preparation is successful. All other preparations may fail arbitrarily, and the algorithm is not told which succeeded.
\end{assumptionkey}

Some such assumption is unavoidable: if no preparation in a path-connected component succeeds, there is no valid state in that component to transport or probe, and the algorithm cannot notice, since a failed preparation may output any state and verifying that a state is a ground state is computationally hard in general. %

\begin{theoremkey}[Certified discovery of quantum phase diagrams; informal]
\label{thm:detection-informal}
Under Assumption~\ref{assumption:references-informal}, there is a quantum algorithm running in $\mathrm{poly}(n)$ time that, with probability at least $1-1/\mathrm{poly}(n)$, solves \emph{certified discovery of quantum phase diagrams}. When the ground state is unique, the algorithm uses a single copy of each supplied state, disturbing it by only an exponentially small amount; for degenerate ground states it uses polynomially many copies.
\end{theoremkey}

\noindent The formal statements are Assumption~\ref{assumption-reference-state} and Theorem~\ref{thm:detection} (Appendix~\ref{sec:local-detection}). The algorithm (Algorithm~\ref{alg:overall-dp}) is presented and the theorem is proved in Appendix~\ref{sec:detection}. 

The algorithm behind Theorem~\ref{thm:detection-informal} relies on non-local measurements which require a fault-tolerant quantum computer. Before turning to it, we first present two simpler algorithms. These require only spatially local measurements of the energy and of the response to the probe, and hence may be easier to execute on quantum platforms. However, to ensure that these simpler algorithms find the correct phase diagram we make assumptions that are not needed in the proof of Theorem~\ref{thm:detection-informal}.

To get started, we suppose that the Hamiltonian depends on just a single adjustable parameter.

\begin{theorembrk}[Certified discovery of quantum phase diagrams, one-parameter case; informal]
\label{thm:path-informal}
Under Assumption~\ref{assumption:references-informal}, there exists a quantum algorithm running in $\mathrm{poly}(n)$ time that, with probability at least $1-1/\mathrm{poly}(n)$, solves \emph{certified discovery of quantum phase diagrams} for $m=1$ with a \emph{relaxed guarantee}.
\end{theorembrk}

\noindent This one-parameter case is formalized as Theorem~\ref{thm:local-phase-diagram}. The algorithm (Algorithm~\ref{alg:overall}) is presented and the theorem is proved in Appendix~\ref{sec:local-detection}. The ``relaxed'' guarantee'' is stated in Remark~\ref{rem:phase-clustering-relaxation}; see the discussion below of Theorem 4, which generalizes Theorem 3, for further explanation.

Algorithm~\ref{alg:overall} uses quasi-adiabatic evolution with an energy cutoff $\Delta=\gamma_{\mathrm{lb}}/2$, which can be simulated in quantum polynomial time because $\gamma_{\mathrm{lb}}\ge 1/\mathrm{poly}(n)$. In one dimension, quasi-adiabatic transport starting from a successful reference state prepares accurate ground states at every point of the connected component in which the gap remains larger than $\gamma_{\mathrm{lb}}$. Since the algorithm does not know which reference is successful, it sweeps every input state in both directions, possibly transporting it through unresolved regions in which the gap is less than $\gamma_{\mathrm{lb}}$. When several states arrive at a point after being transported from different reference points, some of which may be failed preparations, the algorithm selects the correct ground state by locally estimating the energy of each state. 

Selecting the proper ground state requires no knowledge in advance about the ground-state energy or eigenstates. However, in the case where the ground state is degenerate, we do need to verify that the state is nearly maximally mixed on a subspace of known dimension. To address that nuance, Algorithm~\ref{alg:overall} includes a swap test that functions as a purity measurement. See the discussion below of Theorem~\ref{thm:diagram-informal} for an explanation of this test and why it is needed.

Once the proper ground state is selected, the spectral weight for the probe $O=\partial_sH$ supplies the signal for phase detection: on a resolved interval joining different phases, Theorem~\ref{thm:nondark-informal} forces a detectable response somewhere inside the interval.
To measure the spectral weight of Eq.~\eqref{eq-band-restricted-spectral-weight-main}, we must access the frequency-filtered response of the ground state to the probe. This can be achieved by quasi-adiabatic evolution with a suitable frequency cutoff. There is another method, though, which may be easier to implement in practice, and which also simplifies the theoretical analysis of the algorithm --  we may measure the probe's time-dependent correlation function and then apply \emph{classical} signal processing to the result. At a test point $x$, suppose we have measured the dissipative response $\chi_O(t;x)=i\operatorname{Tr}\bigl[\rho_{\mathcal G}(x)[O_x(t),O]\bigr]$, where $O_x(t)=e^{iH(x)t}Oe^{-iH(x)t}$. To isolate its low-frequency part we average against a real time kernel $\mathcal K$:
\begin{equation}\label{eq:filtered-response-main}
S^{O}_{\mathcal K}(x)
=\int_{-\infty}^{\infty}\mathcal K(t)\,\chi_O(t;x)\,dt
=\frac{1}{r(x)}\sum_{0\le j<r(x)\le k}
q(E_k(x)-E_j(x))\,
\bigl|\langle\psi_k(x)|O|\psi_j(x)\rangle\bigr|^2\,,
\end{equation}
where $q(\omega)=2\int\mathcal K(t)\sin(\omega t)\,dt$ is a smooth frequency window. Transitions within the ground-state manifold cancel because all its states have equal weight in $\rho_{\mathcal G}$, so a window retaining low frequencies isolates precisely the spectral weight of interest. Because $q$ is nonnegative on positive frequencies, the surviving terms cannot cancel one another, %
ensuring that any leakage from the ground state to low-lying excited states will be detected. Appendix~\ref{subsec:filtered-response-kernel} shows how $S^{O}_{\mathcal K}$ and $W^{O}_{[\gamma,\Lambda)}$ are related and how the kernel is chosen.

These measurements are local in space, but not in time. Each correlator $\chi_O(t;x)$ is obtained by a Hadamard test in which ancilla-controlled local Pauli operators are separated by ordinary evolution under $H(x)$ for a waiting time $t$, requiring neither ancilla control nor frequency filtering during the evolution.\footnote{With fresh copies of the reference state, these measurements need no ancilla (Lemma~\ref{lem:ancilla-free-response-correlator}, Appendix~\ref{sec:diabatic-signals}). Ref.~\cite{uhrich2017noninvasive} shows this for spin-$1/2$ systems and single-site observables; its noninvasive protocol uses a weak ancilla--system coupling.}
Smoothing the frequency window makes the time kernel decay rapidly, so that the integral defining $S^{O}_{\mathcal K}(x)$ can be estimated by a finite sum; polynomial evolution time and polynomially many measurements suffice for inverse-polynomial precision and frequency resolution. Were we to use frequency-filtered quantum dynamics to detect a gap closing, a sophisticated error analysis would be needed to ensure that our discovery algorithm is sufficiently accurate. Such complications are largely avoided if we rely on classical signal processing instead, greatly simplifying the proofs of our main results.

Both the response correlators and the mean energy used to select ground states (which is a sum of expectation values of the local terms $h_j(x)$) are observables at the test point, so the algorithm needs only expectation values in a ground state. Catalytic tomography~\cite{chen2024catalytictomography} measures such expectation values in the unique ground state of a known Hamiltonian while perturbing the state by only a negligible amount, so a single successful reference state can be used repeatedly to access all the input points in its cluster. For a degenerate manifold, we use a low-temperature Gibbs reference that approximates the maximally mixed ground state, and thermal shadow tomography~\cite{chen_gilyen_thermal_shadows} implements measurements that preserve it on average; the number of copies needed for $M$ normalized observables then grows only logarithmically in $M$. Appendix~\ref{sec:diabatic-signals} describes both implementations and bounds the required resources.

We now extend from discovering phase transitions in the one-parameter case to discovering a phase diagram in the case where the number of parameters is an arbitrary constant. Because the parameter space has constant dimension, the phase diagram can be learned by covering the parameter space with a grid and sorting the grid points into clusters. Though the algorithm acts on a discrete set of points, the resulting division into phases approximates the actual phase diagram, thanks to gap stability and Lipschitz continuity (Lemma~\ref{cor:gap-lipschitz}).\footnote{These are finite-size statements, with a Lipschitz constant polynomial in $n$; they do not require the gap to survive weak perturbations in the thermodynamic limit, which can fail for fine-tuned Hamiltonians~\cite{li2026stability}.} The difficulty is in the transport step. In one dimension there were only two directions to sweep, but in several dimensions there are exponentially many grid paths between a reference and an input point, so testing all paths to search for a gapped one might be too expensive.

There are two different ways to deal with this issue. One way is to sample grid paths at random, which works when gapped paths are not too rare and leads to Theorem~\ref{thm:diagram-informal} below.
Alternatively, one can search for a gapped path more systematically using dynamic programming, which 
leads to Theorem~\ref{thm:detection-informal}. The algorithm in Theorem~\ref{thm:diagram-informal} uses only spatially local measurements and is therefore less computationally demanding than the algorithm in Theorem~\ref{thm:detection-informal}, which calls for nonlocal measurements. On the other hand, to ensure that gapped paths are not too rare, Theorem~\ref{thm:diagram-informal} invokes an additional assumption that is not needed in Theorem~\ref{thm:detection-informal}.

The additional assumption concerns the geometry of the resolved regions of parameter space; namely, that each resolved region has a shape that remains stable as $n$ grows. Concretely, we require that each resolved point $x$ be reachable from a successful reference point $y$ by a path that stays a constant distance, independent of $n$, from the boundary of the connected component of the resolved region that contains $x$ and $y$. This assumption rules out bottlenecks in resolved regions that shrink to zero width as $n\to\infty$. Under this geometric assumption, a randomized procedure finds a gapped path from $y$ to $x$ with at least inverse-polynomial probability if such a path exists.
The precise geometric condition is Assumption~\ref{assumption:random-walk-navigation} (Appendix~\ref{sec:local-detection}). 

\begin{theorembrk}[Certified discovery of quantum phase diagrams, restricted case; informal]
\label{thm:diagram-informal}
Under Assumption~\ref{assumption:references-informal} and Assumption~\ref{assumption:random-walk-navigation}, there exists a quantum algorithm running in $\mathrm{poly}(n)$
time that, with probability at least $1-1/\mathrm{poly}(n)$, solves
certified discovery of quantum phase diagrams with a \emph{relaxed guarantee} (see Remark~\ref{rem:phase-clustering-relaxation}).

\end{theorembrk}

\noindent This result is Theorem~\ref{thm:local-phase-diagram-general}. The algorithm (Algorithm~\ref{alg:overall-general}) is presented and the  theorem is proved in Appendix~\ref{sec:local-detection}. It uses random walks to prepare ground states at points of the resolved region; the energy and the dissipative response are then estimated exactly as in the one-parameter case.

With each link, the algorithms of Theorems~\ref{thm:path-informal} and~\ref{thm:diagram-informal} report a path of bounded length between the two linked points (for $m=1$, the interval between them), tentatively asserting that these two points belong to the same phase. The ``relaxed'' guarantee states that these two points are sure not to be in different phases \emph{if} the gap is above $\gamma_{\mathrm{lb}}$ at every point along that reported path, not just at its two endpoints. Neither theorem rules out points being assigned to the same phase by the algorithm when they really belong to different phases, if the specified path crosses a region where the gap falls below the resolution $\gamma_{\mathrm{lb}}$, because a transition in an unresolved region might be unseen by local probes. We can improve the performance of the algorithm, at a polynomially larger cost, by choosing a smaller value of $\gamma_{\mathrm{lb}}$, hence separating phases that might otherwise be erroneously joined together. 

The algorithm could be simplified under a further assumption: that every prepared reference state, whether successful or not, is approximately maximally mixed on a subspace of known dimension. This maximally-mixed proviso would ensure that local energy measurements suffice for identifying the best candidate ground state at a test point $x$. Suppose that the reference at point $y$ is prepared correctly, with rank $r(y)$, and transported through the resolved region to test point $x$. The resulting state $\rho(x)$ is then sure to be nearly maximally mixed on the ground state manifold at $x$. Suppose another reference is prepared at point $z$, with known rank $r(z) = r(y)$, and also transported to $x$. In this case, though, the preparation may be faulty, or the path from $z$ to $x$ may veer away from resolved region. Because the transport from $z$ to $x$ is unitary, the resulting state $\rho'(x)$, though otherwise arbitrary, is sure to have rank $r'(x) = r(z)$. Therefore, if $\rho'(x)$ is not a good approximation to the maximally mixed state on the ground subspace at $x$, then it must have substantial support outside the ground subspace, and thus detectably larger mean energy than $\rho(x)$. Hence, just comparing the measured energies of $\rho(x)$ and $\rho'(x)$ suffices to flag $\rho'(x)$ as a failed ground state, indicating that $\rho(x)$ should be preferred. 

However, without the assurance that $r'(x)=r(x)$, local energy measurement alone would not suffice in general. It might be that $\rho'(x)$ has lower rank than $\rho(x)$, supported in a proper subspace of the actual ground space, in which case the mean energy alone provides no basis for rejecting this failed preparation. The maximally-mixed proviso excludes this unfavorable situation, empowering energy measurement to flag faulty putative ground states.

Algorithms~\ref{alg:overall} and \ref{alg:overall-general} perform the discovery tasks stated in Theorems~\ref{thm:path-informal} and \ref{thm:diagram-informal} without relying on any such proviso. Instead, for each reference state, these algorithms use a swap test on two copies to estimate the purity, which should be close to $1/r$ if the state is nearly maximally mixed on an $r$-dimensional subspace. At each resolved point, the correctly prepared ground state is selected by choosing the largest-rank state with nearly minimal energy. The discovery task can be achieved in quantum polynomial time with or without the extra assumption, but the algorithm is simpler if we accept the maximally-mixed proviso, because in that case the purity test is not needed.

To attain the stronger result Theorem~\ref{thm:detection-informal}, we retain Assumption~\ref{assumption:references-informal} but eschew Assumption~\ref{assumption:random-walk-navigation}. Algorithm~\ref{alg:overall-dp} finds gapped paths from successful reference points to other resolved points by a more sophisticated method than the random walk invoked in Theorem~\ref{thm:diagram-informal}. Assumption~\ref{assumption:references-informal} promises that such paths exist, but there are exponentially many possible grid paths, and paths might traverse dark gap closings which leave no locally detectable signature. To search for gapped paths systematically, we adopt the philosophy of dynamic programming. Rather than examining every route, one records at each point only the best way of reaching it found so far, and extends those records step by step, so that an exponential search collapses into polynomially many local updates. The dynamic program thereby certifies a connection to some successful preparation without knowing the path or the preparation in advance. The proof is given in Appendix~\ref{sec:detection}. 

Algorithm~\ref{alg:overall-dp} in Theorem~\ref{thm:detection-informal} flags phase boundaries by detecting leakage to low-energy excitations, but using a less direct and more robust method than Algorithm~\ref{alg:overall-general} in Theorem~\ref{thm:diagram-informal}. Given a successful state preparation at a point $x_0$, the dynamic program efficiently finds a path through the resolved region from $x_0$ to any other point $x$ in the same connected component as $x_0$; quasi-adiabatic transport with cutoff $\gamma_\text{lb}/2$ along that path thus prepares the ground state at $x$. Once the ground states have been established, one may address whether two neighboring points $x_1$ and $x_2$ in the connected component belong to the same phase by attempting quasi-adiabatic transport from $x_1$ to $x_2$ with a cutoff slightly below $\gamma_\text{thresh}$. Successful transport of the ground state under this low-depth unitary circuit certifies that $x_1$ and $x_2$ are not in different phases. Unsuccessful transport signals leakage to low-energy excitations, confirming a gap less than $\gamma_\text{thresh}$ and hence a phase boundary. The procedure is illustrated in Figure~\ref{fig:dp-exploration}.

Whether the transport from $x_1$ to $x_2$ successfully preserves the ground state is tested by measuring the normalized overlap of the transported state with the ground state prepared at $x_2$; a value close to 1 verifies successful ground-state transport. The dynamic program selects these ground states by measuring an approximate projector onto the ground state of $H(x)$. The measured quantity is \(\operatorname{Tr}[f_\nu(H(x))\sigma]\), where $\sigma$ is a candidate state and \(f_\nu\) is a smooth step function, chosen so that states in the ground-state manifold are accepted, and states well above the ground state are rejected. To prevent harmful accumulation of error during execution of the dynamic program, sharp energy resolution is required; hence the step function is chosen to reject states lying more than \(O(\gamma_{\mathrm{lb}})\) above the ground state. To attain this resolution, energy testing at point $x$ entails controlled evolution of the entire system under $H(x)$, or a block encoding of $H(x)$, for a time of order $1/\gamma_{\mathrm{lb}}$. Appendix~\ref{sec:detection} implements this evolution via a quantum singular value transformation, applying a polynomial approximation of the step function to the state before measuring.

Because Algorithm~\ref{alg:overall-dp} of Theorem~\ref{thm:detection-informal} calls for deep coherent circuits acting on the full system, a circuit-based fault-tolerant quantum computer would be needed to execute it. The spatially local measurements called for in Algorithm~\ref{alg:overall-general} of Theorem~\ref{thm:diagram-informal} are better suited for the capabilities of existing or near-term quantum simulators. In a controlled quantum simulator that realizes $H(x)$, a common preparation strategy is to start from a simple ground state and slowly vary the available Hamiltonian parameters, as in Rydberg-array experiments~\cite{bernien2017probing,ebadi2021quantum}. Such a parameter sweep, as in adiabatic state preparation, is ordinary time evolution under $H(x(t))$, with no quasi-adiabatic filter implemented. The quasi-adiabatic transport for state preparation used in Theorem~\ref{thm:diagram-informal} can be replaced by such a sweep on a controlled simulator, while gap detection still uses the frequency-filtered response. Local energy measurements could be invoked for state selection, and ancilla-free protocols could extract correlation functions, which can be classically filtered to detect phase boundaries. For a more detailed statement see Corollary~\ref{cor:adiabatic-phase-diagram}, proved in Appendix~\ref{subsec:experimental-implementations}. 

For the dynamic program of Algorithm~\ref{alg:overall-dp} as well, replacing quasi-adiabatic transport by adiabatic transport preserves the conclusion of Theorem~\ref{thm:detection-informal} under the same assumptions as before. A sufficiently smooth sweep of $x(t)$ with duration $\operatorname{polylog}(n)$ preserves gapped ground states, and hence can certify that neighboring resolved points do not lie in different phases. Theorem~\ref{thm:detection-informal} still requires fault-tolerant hardware, however, because the purity measurement and the global controlled ground-state-population test are needed to deliver its full guarantee.

Theorems~\ref{thm:detection-informal}-\ref{thm:diagram-informal} guarantee that phase boundaries can be detected where the gap falls below a specified constant value $\gamma_\mathrm{thresh}$. We chose $\gamma_\mathrm{thresh}$ to be a constant because quasi-adiabatic evolution with constant energy cutoff $\Delta$ can be simulated by a constant-depth quantum circuit. Therefore, if quantum phases are defined as in \eqref{eq:same-phase} and \eqref{eq:different-phase}, we are assured that the spectral weight is nonnegligible and the gap detectable at continuous transitions. But we might want to locate the phase boundary more precisely by choosing $\gamma_\mathrm{thresh}$ to be inverse polynomially small. In that case, we simulate evolution between neighboring grid points using a cutoff $\Delta <\gamma_\mathrm{thresh}$, and an outcome depending on $\Delta$ certifies a gap below $\gamma_\mathrm{thresh}$ just as in Theorem~\ref{thm:detection-informal}. However, quasi-adiabatic evolution with an inverse-polynomially small cutoff has polynomial depth rather than constant depth. Hence under our previous definition of a quantum phase, we can no longer guarantee that a phase boundary is necessarily detected. 

To address this issue, we may revise the definition of quantum phases by enlarging the class $\mathcal{C}$ of admissible quantum circuits to those with causal-cone radius (the distance over which the circuit propagates operator support) at most $c\,n^{1/\dlat}$, where $c$ is a positive constant and $\dlat$ is the spatial dimension. The exponent $1/\dlat$ is chosen advisedly; a circuit that transforms a trivial phase to a broken-symmetry or topologically ordered phase must have causal-cone radius comparable to the system diameter~\cite{haah2016invariant,bravyi2006lieb,Chen_2024}. Quasi-adiabatic evolution with energy cutoff $\Delta$ compiles into a circuit with causal-cone radius $\mathrm{polylog}(n)/\Delta$; therefore, the enlarged class $\mathcal{C}$ includes evolution with cutoff down to $n^{-1/\dlat}\,\mathrm{polylog}(n)$. An argument similar to the proof of Theorem~\ref{thm:detection-informal} then yields this corollary. 

\begin{corollarybrk}[Detection at an inverse-polynomial threshold]
\label{cor:detection-invariant-informal}
For phases that remain distinct under geometrically local circuits of
causal-cone radius $c\,n^{1/\dlat}$, the conclusion of
Theorem~\ref{thm:detection-informal} holds with
$\gamma_{\mathrm{thresh}} = \Omega\bigl(n^{-1/\dlat}\,\mathrm{polylog}(n)\bigr)$.
\end{corollarybrk}

Theorem~\ref{thm:diagram-informal} concerns a probabilistic quantum algorithm, but we can also construct a deterministic classical algorithm for phase diagram discovery, in which quasi-adiabatic evolution is simulated classically. In the quantum algorithm, we quasi-adiabatically transport the reference state to a grid point, where the state is tested for gap closure by measuring a local probe observable. Equivalently, the probe observable is evolved in the Heisenberg picture, and then measured in the reference state. In the classical algorithm, we choose $\gamma_\mathrm{thresh}= \Theta(1)$ and assume that the reference state's gap is larger than  $\gamma_\mathrm{thresh}$. The goal of the algorithm is to map the frontier where the gap falls to below $\gamma_\mathrm{thresh}$. For this purpose, quasi-adiabatic evolution with a constant energy cutoff suffices, which inflates the observable to a patch with radius $\textnormal{polylog}(n)$. To evaluate the expectation of this observable, a complete description of the reference state is not needed; rather the marginal states in patches of this size suffice. The evolved probe can be computed by exact diagonalization, and its expectation is determined by contracting with the reduced density operator on the patch (Appendix~\ref{sec:definitions-locality}--\ref{subsec:two-stage-truncation}). 

For the classical algorithm, an additional assumption is needed to ensure that the algorithm detects the phase boundary, which was not needed for the quantum algorithm. The argument in Theorem~\ref{thm:nondark-informal} that established the detectability of the boundary relied on faithful quasi-adiabatic transport of the ground state through regions where the gap is inverse-polynomially small, which requires an inverse-polynomially small cutoff and therefore cannot be efficiently simulated classically. Instead we assume \emph{local visibility} of the gap: near the threshold, some local observable %
couples the ground state to low-energy excitations with at least inverse-polynomial strength.
In terms of the spectral weight of Eq.~\eqref{eq-band-restricted-spectral-weight-main},
\begin{equation}\label{eq:local-visibility-main}
\max_{O \in \mathcal L}\;
W^{O}_{[\gamma(x),\,\eta\gamma_{\mathrm{thresh}})}(x)
\;\ge\; n^{-a}
\end{equation}
for constants $a > 0$ and $0<\eta<1$, whenever the gap $\gamma(x)$ is near the detection threshold, just below $\eta\gamma_{\mathrm{thresh}}$, as specified in Appendix~\ref{subsec:classical-main-results}. Here $\mathcal L$
ranges over normalized combinations of Pauli operators, each acting on
$O(1)$ neighboring qubits. The operator $O$ with nonnegligible spectral weight can be found efficiently by solving a principal-eigenvector problem. Putting these observations together, we find that under the local visibility assumption, the classical algorithm maps the phase boundary in quasipolynomial time.

\begin{theorembrk}[Quasipolynomial-time classical algorithm for %
quantum-phase-diagram discovery]
\label{thm:classical-classification-informal}
Let $\ket{\psi(x_\mathrm{ref})}$ be a reference ground state at
a point $x_{\mathrm{ref}}$ with $\gamma(x_{\mathrm{ref}})>\gamma_{\mathrm{thresh}}$, or the maximally mixed ground state
$\rho_{\mathcal G}(x_{\mathrm{ref}})$ if the ground state at $x_{\mathrm{ref}}$ is degenerate, and assume local visibility~\eqref{eq:local-visibility-main} holds near the detection threshold and the degeneracy is stable (Appendix~\ref{sec:classical-algorithm}). 
Assume that a classical description is available for the marginal state of $\ket{\psi(x_\mathrm{ref})}$ on balls of radius $\mathrm{polylog}(n)$. %
Then there is a deterministic classical algorithm
that covers the connected component of $\{x : \gamma(x) \ge \gamma_{\mathrm{thresh}}\}$ containing $x_{\mathrm{ref}}$, within path length $\mathrm{polylog}(n)$ of $x_{\mathrm{ref}}$, with runtime
\begin{equation}\label{eq:Tclass-main}
T_{\mathrm{class}}
 = \exp\!\left[ O\!\left(\left(\mathrm{polylog}\!\bigl(n^{4 + 2a}\bigr)\right)^{\dlat} \right) \right],
\end{equation}
where $a$ is the exponent in the local visibility
assumption~\eqref{eq:local-visibility-main} and $\dlat$ is the spatial
dimension.
\end{theorembrk}

\noindent This is Theorem~\ref{thm:classical-classification}. The classical algorithm (Algorithm~\ref{alg:classical}) is presented and the  theorem is proved in
Appendix~\ref{sec:classical-algorithm}.
If we wish to determine the phase boundary more precisely by choosing $\gamma_\text{thresh}= 1/\text{poly}(n)$, the classical algorithm is inefficient; the patch radius grows to poly$(n)$ size, and the cost of exact diagonalization becomes exponential.

\renewcommand{\thesubsubsection}{\thesubsection.\alph{subsubsection}}
\IfFileExists{titlesec.sty}{%
  \titleformat{\subsubsection}[block]
    {\centering\itshape}
    {\alph{subsubsection}.}
    {0.75em}
    {}%
}{}

\section{Numerical Simulations}
\label{sec:numerics-main}

To check the efficacy of our quantum algorithm, we simulated the phase diagram learning algorithm for a fixed number of parameters referenced in Theorem~\ref{thm:diagram-informal} (Algorithm~\ref{alg:overall-general}), including its subroutine (Algorithm~\ref{alg:gap_cert}), which certifies the spectral gap. Because our framework relies only on quasi-adiabatic continuation and operator locality, it is applicable to spin, bosonic, and fermionic lattice systems.
In particular, because our framework handles interacting fermions directly, the quantum Monte Carlo (QMC) sign problem is evaded.

We applied these algorithms to three lattice models: an interacting spinless fermion model with a quantum Monte Carlo (QMC) sign problem, the Kitaev honeycomb model, and the toric code in uniform fields. In each benchmark, we provide the algorithm with only two inputs.
The first input is the geometrically local Hamiltonian family $H(x)$, $x \in [-1,1]^m$, of Definition~\ref{definition: geometrically local hamiltonians}.
The second input is a single classically locally evaluable reference ground state $\rho_{\mathrm{ref}}$ at a gapped, exactly solvable point along with its gap $\gamma(x_{\mathrm{ref}})$.
We do not provide an order parameter, a symmetry-breaking pattern, or an expected universality class. 

Starting from the solvable point, Algorithm~\ref{alg:overall-general} explores the surrounding parameter space to map out the phase until the gapped path breaks down.
In each benchmark, the filtered-response test of Algorithm~\ref{alg:gap_cert} is applied at every grid vertex, detecting whether the spectral gap of Definition~\ref{definition: spectral gap}, the energy separation above the ground-state manifold, falls below the prescribed threshold value. When the ground space is degenerate, as for the twofold charge-density-wave ground space of the Fermi--Hubbard model and the fourfold ground spaces of the Kitaev model and the toric code on the torus, this is the gap above the whole manifold, not the splitting within it.

\begin{figure}[t]
    \centering
    \includegraphics[width=0.85\linewidth, trim=0 16 0 24, clip]{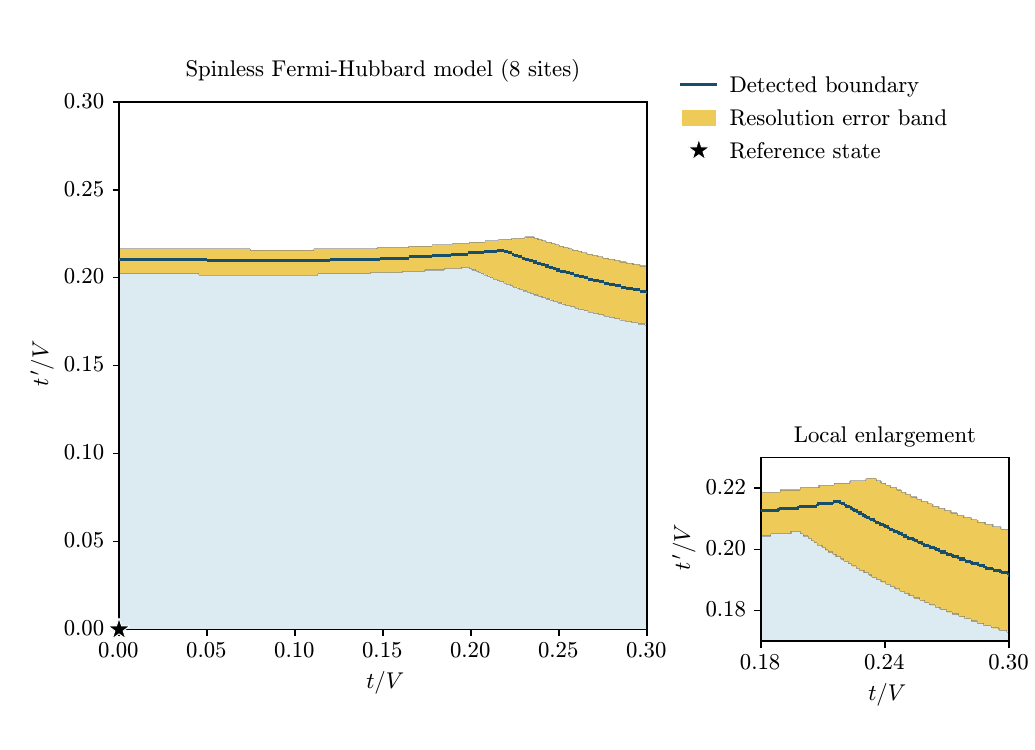}
    \caption{Classically computed phase diagram of the spinless Fermi--Hubbard model~\eqref{eq:fermion-hamiltonian-main} in the $(t/V, t'/V)$ plane.
The star marks the CDW reference state at $t = t' = 0$.
The blue region is certified to be connected to the reference state by a gapped path, with threshold $\gamma_{\mathrm{thresh}} = 1.1$ (in units of $V$) and margin $\epsilon_{\mathrm{thresh}} = 1/11$.
The solid line is the detected boundary.
The shaded band is the resolution error band, in which the spectral gap lies between $(1-\epsilon_{\mathrm{thresh}})\gamma_{\mathrm{thresh}}$ and $\gamma_{\mathrm{thresh}}$, and the certification may return either verdict.}
    \label{fig:hubbard-main}
\end{figure}

\begin{figure}[t]
    \centering
    \includegraphics[width=0.85\linewidth, trim=0 18 0 20, clip]{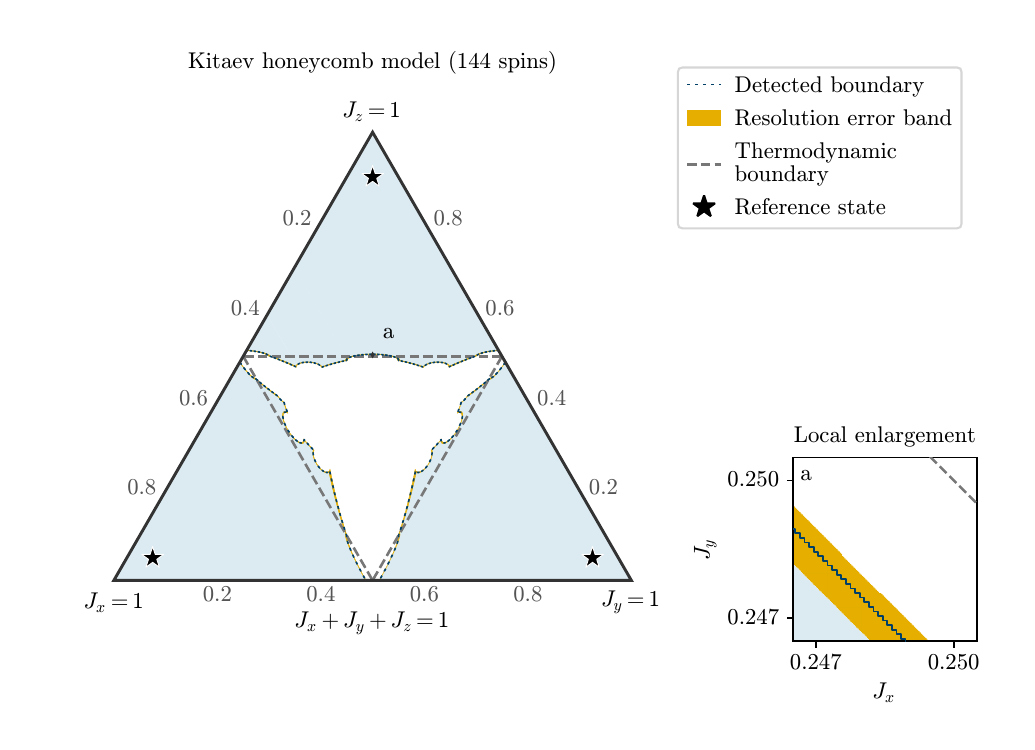}%
    \caption{Classically computed phase diagram of the Kitaev honeycomb model~\eqref{eq:kitaev-hamiltonian-main} on the simplex $J_x + J_y + J_z = 1$, for a periodic cluster of $n = 144$ spins.
The stars mark the three reference states, one in each gapped region $\mathcal{A}_x$, $\mathcal{A}_y$, $\mathcal{A}_z$.
The blue regions are certified to be connected to a reference state by a gapped path, with threshold $\gamma_{\mathrm{thresh}} = 0.11$ (in units where $J_x + J_y + J_z = 1$) and margin $\epsilon_{\mathrm{thresh}} = 1/11$.
The dotted line is the detected boundary.
The shaded band is the resolution error band, in which the spectral gap lies between $(1-\epsilon_{\mathrm{thresh}})\gamma_{\mathrm{thresh}}$ and $\gamma_{\mathrm{thresh}}$.
The dashed lines $J_\alpha = 1/2$ are the phase boundaries in the thermodynamic limit.
The inset enlarges the neighborhood of the point marked a on the line $J_z = 1/2$.}
    \label{fig:kitaev-main}
\end{figure}

\begin{figure}[t]
    \centering
    \includegraphics[width=0.9\linewidth, trim=0 22 0 24, clip]{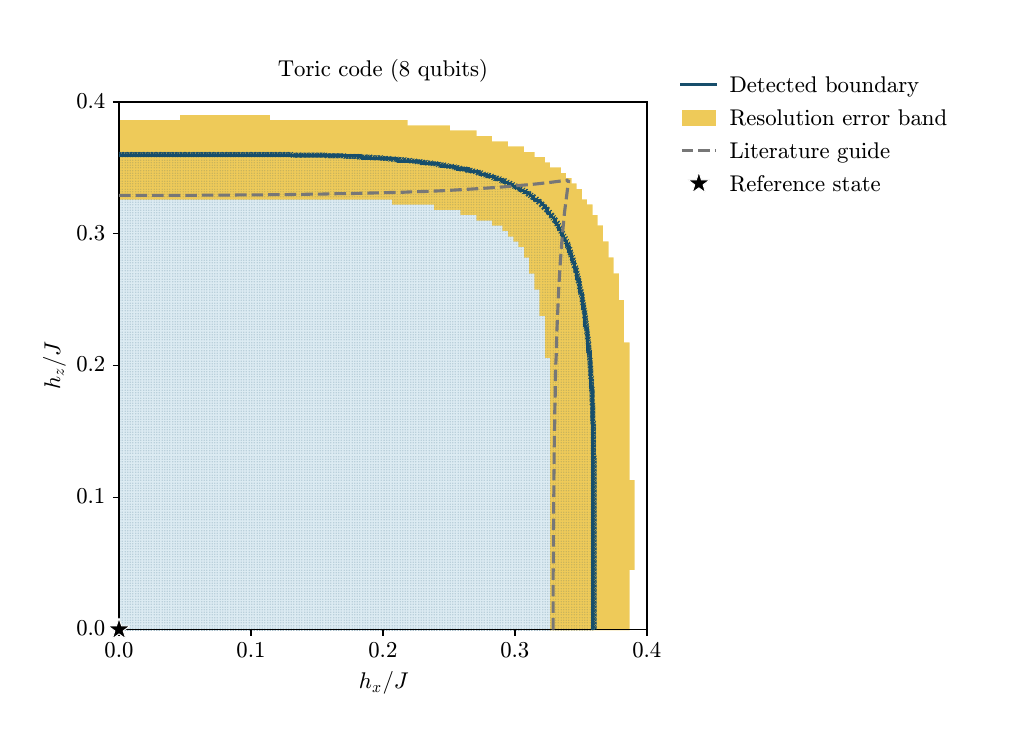}
    \caption{Classically computed phase diagram of the toric code in uniform fields, Eq.~\eqref{eq:toric-field-hamiltonian-main}, in the $(h_x/J, h_z/J)$ plane; $J$ is the coefficient of the stabilizer terms, set to one in Eq.~\eqref{eq:toric-field-hamiltonian-main}.
The star marks the stabilizer reference state at $h_x = h_z = 0$.
The blue region is certified as gapped-connected to the reference state, with threshold $\gamma_{\mathrm{thresh}} = 2.3485$ (in units of $J$) and margin $\epsilon_{\mathrm{thresh}} = 1/11$; the gap in this certification is the spectral gap above the fourfold ground space.
The solid line is the detected boundary.
The shaded band is the resolution error band, in which the spectral gap lies between $(1-\epsilon_{\mathrm{thresh}})\gamma_{\mathrm{thresh}}$ and $\gamma_{\mathrm{thresh}}$.
The dashed line is the phase boundary in the thermodynamic limit~\cite{WuDengProkofev2012}.}
    \label{fig:toric-main}
\end{figure}

Each of Figures~\ref{fig:hubbard-main}, \ref{fig:kitaev-main}, and \ref{fig:toric-main} shows the detected boundary of the certified region together with a resolution error band, the region of parameter space in which the spectral gap lies between $(1-\epsilon_{\mathrm{thresh}})\gamma_{\mathrm{thresh}}$ and $\gamma_{\mathrm{thresh}}$; the algorithm tests the gap at the threshold $\gamma_{\mathrm{thresh}}$ with a margin $\epsilon_{\mathrm{thresh}}$, so it need only locate the boundary to within this error bar. The resolution error band is computed independently of the algorithm, by exact diagonalization of the spectrum, whereas the detected boundary comes from simulating the algorithm itself: quasi-adiabatic transport of the reference state followed by the frequency-filtered response of the transported state to the probe. In all three benchmarks the detected boundary lies inside the band, as the guarantee of the algorithm requires. Appendix~\ref{app:numerics} provides further details of the lattice models.

\emph{Spinless Fermi--Hubbard model (fermion-sign regime).}
We study spinless fermions on an $N_x \times N_y$ square lattice (PBC; $N_x, N_y$ even) at half filling $N_f = N/2$, with the Hamiltonian
\begin{equation}\label{eq:fermion-hamiltonian-main}
H(t,t') = -t\sum_{\langle i,j\rangle}\!\bigl(c_i^\dagger c_j + \mathrm{h.c.}\bigr)
         - t'\sum_{\langle\!\langle i,j\rangle\!\rangle}\!\bigl(c_i^\dagger c_j + \mathrm{h.c.}\bigr)
         + V\sum_{\langle i,j\rangle} n_i n_j,
\end{equation}
where $n_i = c_i^\dagger c_i$, the symbols $\langle i,j\rangle$ and $\langle\!\langle i,j\rangle\!\rangle$ denote the nearest-neighbor and the next-nearest-neighbor (diagonal) bonds, and $V>0$. For $t' \neq 0$, the hopping is no longer bipartite, and QMC then develops a fermion sign problem~\cite{huffman2014solution,li2015solving}.
At $t = t' = 0$, the model is exactly solvable. At that point, the ground states are the two checkerboard charge-density-wave (CDW) product states, and the gap is $3V$.
We take this point as the reference.
Figure~\ref{fig:hubbard-main} shows the resulting phase diagram in the $(t,t')$ plane.
The blue region is the region certified as connected to the CDW reference by a gapped path. We expect this boundary to be a second-order transition between the CDW phase, which spontaneously breaks translation symmetry, and the gapless phase, which preserves it; see the analysis in Appendix~\ref{app:numerics}.

\emph{Kitaev honeycomb model (topological phase boundary).}
We next map the phase diagram of the Kitaev honeycomb model~\cite{kitaev2006anyons} with periodic boundary conditions,
\begin{equation}\label{eq:kitaev-hamiltonian-main}
H = -J_x \sum_{x\text{-links}} \sigma_j^x \sigma_k^x
    -J_y \sum_{y\text{-links}} \sigma_j^y \sigma_k^y
    -J_z \sum_{z\text{-links}} \sigma_j^z \sigma_k^z,
\end{equation}
with couplings restricted to the simplex $J_x + J_y + J_z = 1$ and $J_{x,y,z} \ge 0$, in periodic boundary condition.
The exact phase diagram consists of three gapped regions $\mathcal{A}_x, \mathcal{A}_y, \mathcal{A}_z$ near the corners, each with the same topological order as the toric code, as well as a gapless spin-liquid region $\mathcal{B}$ in the center.
The boundaries are the lines $J_\alpha = 1/2$, and the phase transition from $\mathcal{A}$ to $\mathcal{B}$ is topological.
We start from a gapped point in each of the three $\mathcal{A}$ regions, marked by stars in Fig.~\ref{fig:kitaev-main}.
At each reference point, $\rho_{\mathrm{ref}}$ is the equal mixture of the four vortex-free Wilson-loop ground states.
Figure~\ref{fig:kitaev-main} shows the certified regions for $n = 144$ spins.

For the system of 144 spins, the detected boundary exhibits undulations that also appear in the resolution error band. We compute this band independently by diagonalizing the quadratic Majorana Hamiltonians in the relevant flux and Wilson-loop sectors and reconstructing the physical low-energy spectrum entering Definition~\ref{definition: spectral gap}. The similar undulations in this independently computed band and the boundary detected by the filtered response support their interpretation as finite-size spectral effects. At finite system size, these contours need not coincide with the straight thermodynamic phase boundaries shown for comparison. Our certificate establishes that each accepted point is connected to a reference state by a path along which the finite-system spectral gap remains above $(1-\epsilon_{\mathrm{thresh}})\gamma_{\mathrm{thresh}}$.

\emph{Toric code in uniform fields (topological phase boundary).} Finally, we consider the toric code in uniform fields~\cite{Tupitsyn_2010}, with qubits residing on the edges of a periodic square lattice
\begin{equation}\label{eq:toric-field-hamiltonian-main}
H = -\sum_s A_s - \sum_p B_p - h_x \sum_j \sigma_j^x - h_z \sum_j \sigma_j^z,
\qquad
A_s = \!\!\prod_{j \in \mathrm{star}(s)}\!\! \sigma_j^x, \quad
B_p = \prod_{j \in \partial p} \sigma_j^z .
\end{equation}
Our reference state is a stabilizer ground state at the exactly solvable point $h_x = h_z = 0$, which lies in a gapped $\mathbb{Z}_2$ topologically ordered phase with a fourfold-degenerate ground space on the torus. Our algorithm certifies a topological phase surrounding the origin (Fig.~\ref{fig:toric-main}), in qualitative agreement with the phase diagram of the quantum model~\cite{WuDengProkofev2012}.

\section{Discussion}
\label{sec:discussion}

We have formulated efficient quantum algorithms that can discover quantum phase diagrams. These algorithms explore parameter space via quasi-adiabatic evolution of reusable ground states, and use  frequency-filtered local probes to detect phase boundaries where the energy gap becomes small; the detection is sure to succeed if the transition is continuous. Without further assumptions, discontinuous transitions might be ``dark'' and hence evade detection. We presented an algorithm that uses only local measurements and another that uses nonlocal measurements; the former is simpler to execute, but provides a weaker guarantee of correctness  than the latter. We also described a deterministic classical algorithm for phase discovery that runs in quasipolynomial time under an additional assumption. Our numerical simulations illustrate the effectiveness of the quantum algorithm at locating  symmetry-breaking and topological phase transitions in fermionic and spin lattice models. 

Our quantum algorithms estimate the expectation value of a frequency-filtered local probe on the putative ground state, with the filter applied in classical post-processing. Other methods that deliver this quantity could be used instead.
Ground-state filtering~\cite{LinTong2020} and Lindbladian cooling~\cite{DingChenLin2024} are potential alternative approaches to state preparation. For certain Hamiltonian families, the dissipative approach of Ref.~\cite{LiYangLin2026} extracts local observables from short-time dynamics without ever preparing the ground state. Applied to our probe, this might remove the need for a reference state altogether, while our guarantees for generic geometrically local Hamiltonians could still apply. On the practical side, more detailed resource estimates for the gap test, both on fault-tolerant hardware and on controlled quantum simulators where parameter sweeps replace filtered transport, would illuminate the prospects for near-term implementation of the discovery algorithm.

Three questions concern the scope of certified discovery itself. The first is whether phases can be discovered from copies of ground states alone, without a Hamiltonian~\cite{schuster2025hardness}. The task is well posed, because the definition of a gapped phase refers only to constant-depth circuits connecting the states, but a Hamiltonian-free procedure would need a new kind of certificate other than gap closings.

The second is whether the same guarantees can be obtained for mixed-state phases. Phases of mixed states can be defined by connectivity under local dissipative evolution rather than unitary circuits~\cite{coser2019classification,rakovszky2024defining,sang2024mixed,sang2025stability,ellison2025towards,bao2026mixed,fan2024diagnostics,ma2025topological,lessa2025strong}, with the spectral gap replaced by the gap of a local Lindbladian. Our construction, a local probe whose response is controlled by that gap and a transport step through parameter space, has a natural counterpart in this setting. Are continuous thermal transitions provably non-dark, and what plays the role of unverified reference preparations?

The third is whether other data beyond the location of phase boundaries can be certified. 
One would like to discover universal features of each phase or transition, such as new invariants under finite-depth local unitary circuits, with strong guarantees.

We expect progress on all these questions to follow the pattern of this work: a definition of the task that does not presuppose the answer, an efficiently measurable certificate, and a physical argument for why the certification cannot be fooled.

\vspace{0.5em}
\subsection*{Acknowledgments}
We thank Daniel Ranard, Biao Lian, Tao Zhang, Yongtao Zhan, and Alexei Kitaev for helpful discussions. We acknowledge support from the Institute for Quantum Information and Matter, an NSF Physics Frontiers Center (NSF Grant PHY-2317110), and from the U.S. Department of Energy, Office of Science,
National Quantum Information Science Research Centers, Quantum Systems Accelerator. YL and HH also acknowledge support from the Broadcom Innovation Fund.

\vspace{0.5em}
\noindent \textit{Competing interests.} JP and HH are shareholders and employees of Oratomic, Inc., which is developing fault-tolerant quantum computers.

\vspace{0.5em}
\noindent \textit{AI statement.} The main theorems and their proofs were developed without the use of AI. We then used a writing system based on large language models (LLMs), combining GPT, Claude, and our own fine-tuned LLM, to rewrite and polish the exposition of the appendices, clarify proof details, write out some supporting proofs from the authors' outlines, and suggest additional references. We also used LLMs, and for an earlier version of the appendices a structured verification system, to help check the proofs. The main text was written without AI assistance. The authors take full responsibility for the content.

\vspace{2.5em}
\appendix

\numberwithin{theorem}{section}
\numberwithin{corollary}{section}
\numberwithin{condition}{section}
\numberwithin{lemma}{section}
\numberwithin{proposition}{section}
\numberwithin{fact}{section}
\numberwithin{observation}{section}
\numberwithin{assumption}{section}
\numberwithin{example}{section}
\numberwithin{definition}{section}
\numberwithin{remark}{section}
\numberwithin{task}{section}
\makeatletter
\def\theHtheorem{\thesection.\arabic{theorem}}
\def\theHcorollary{\thesection.\arabic{corollary}}
\def\theHcondition{\thesection.\arabic{condition}}
\def\theHlemma{\thesection.\arabic{lemma}}
\def\theHproposition{\thesection.\arabic{proposition}}
\def\theHfact{\thesection.\arabic{fact}}
\def\theHobservation{\thesection.\arabic{observation}}
\def\theHassumption{\thesection.\arabic{assumption}}
\def\theHexample{\thesection.\arabic{example}}
\def\theHdefinition{\thesection.\arabic{definition}}
\def\theHremark{\thesection.\arabic{remark}}
\def\theHtask{\thesection.\arabic{task}}
\def\theHALG@line{\thealgorithm.\arabic{ALG@line}}
\makeatother

\addtocontents{toc}{\protect\setcounter{tocdepth}{2}}

\clearpage
\noindent 
\textit{\LARGE{}Appendices}

\tableofcontents

\relpenalty=9999 \binoppenalty=9999 %
\hyphenation{Dis-connected eigen-basis}
\newcommand\needlines[1]{\par\begingroup\dimen0=\pagegoal\advance\dimen0-\pagetotal\ifdim\dimen0<#1\baselineskip\newpage\fi\endgroup}

\section{Gapped Hamiltonians and Quantum Phases}\label{sec:appendix-A}

This appendix states the formal definitions behind the main text. It also collects the facts about spectral gaps and quasi-adiabatic evolution that the later appendices use.

\subsection{Systems, parameter space, and conventions}\label{sec:appA-setting}

\begin{definition}[Geometrically local Hamiltonian family]
\label{definition: geometrically local hamiltonians}
Consider a lattice of $n$ qubit sites in $\dlat$ spatial dimensions. The parameter space is the hypercube $\mathcal M = [-1,1]^m$ with $m = O(1)$. A geometrically local Hamiltonian family is a map $x \mapsto H(x)$ of the form
\begin{equation}\label{eq:hamiltonian-sum}
H(x) = \sum_{j=1}^{J} h_j(x),
\end{equation}
where $J = O(n)$ is the number of local terms. We require that each $h_j(x)$ acts on $O(1)$ sites contained in a ball of radius $O(1)$ and is analytic in $x$. Each lattice site must lie in the supports of at most $O(1)$ of the terms. For every unit vector $\widehat u \in \mathbb R^m$ we also require the bounds
\begin{equation}\label{eq:local-term-bounds}
\|h_j(x)\| \le 1,\qquad \|\partial_{\widehat u} h_j(x)\| \le 1,\qquad \|\partial^2_{\widehat u} h_j(x)\| \le 1,
\end{equation}
where $\partial_{\widehat u}$ is the derivative in the direction $\widehat u$.
\end{definition}

We use $\|\cdot\|$ to denote the operator norm (the largest singular value), $\|\cdot\|_1$ for the trace norm, $\|\cdot\|_F$ for the Frobenius (Hilbert--Schmidt) norm, and $\mathrm{Tr}$ for the trace. Distances between quantum states are measured in trace norm.

For an integrable function $\varphi$ of time, we define its Fourier transform as
\begin{equation}\label{eq:fourier-convention}
\widetilde{\varphi}(\omega) := \int_{-\infty}^{\infty} \varphi(t)\, e^{-i\omega t}\, dt.
\end{equation}

Asymptotic notations of $O(\cdot)$, $\Omega(\cdot)$, and $\Theta(\cdot)$ are with respect to the system size $n$. $\mathrm{poly}(n)$ is a fixed polynomial in $n$, and $\mathrm{polylog}(n)$ is a fixed polynomial in $\log n$. A quantity is negligible, written $\mathrm{negl}(n)$, if it is smaller than $n^{-k}$ for every fixed $k$ and all sufficiently large $n$.

\subsection{Paths in parameter space}\label{sec:appA-paths}

For a vector $v \in \mathbb{R}^m$, $|v|$ and $\|v\|$ both denote the Euclidean norm.
We use $\|x-y\|$ for the distance between two points $x, y$, and $|\dot{x}|$ for the speed of the paths defined next.
\begingroup\pretolerance=-1 \tolerance=1200 %
\begin{definition}[Path in parameter space]\label{def:path}
A path in the parameter space $\mathcal M$ is a piecewise-$C^1$ map $x(s) : [0,1] \to \mathcal M$, that is, a finite concatenation of $C^1$ maps. We write $\dot x(s) := \frac{d}{ds} x(s)$ for its velocity and $x_A := x(0)$, $x_B := x(1)$ for its endpoints. At the finitely many kink points, $\dot x(s)$ denotes either one-sided derivative. All displayed integrals and bounds are over the $C^1$ pieces. The length of the path is
\begin{equation}\label{eq:path-length}
L := \int_0^1 |\dot x(s)|\, ds .
\end{equation}
We require the path to be parametrized with bounded speed,
\begin{equation}\label{eq:path-speed}
|\dot x(s)| \le \zeta_s L \qquad \text{for all } s \in [0,1],
\end{equation}
where $\zeta_s \ge 1$ is a constant with $\zeta_s = O(1)$. Along the path we abbreviate $H(s) := H(x(s))$. We call
\begin{equation}\label{eq:path-derivative}
\partial_s H(s) = \dot x(s) \cdot \nabla_x H(x(s)) = |\dot x(s)|\; \widehat u(s) \cdot \nabla_x H(x(s)),\qquad \widehat u(s) := \frac{\dot x(s)}{|\dot x(s)|},
\end{equation}
the path-gradient operator. The second form holds wherever $\dot x(s) \ne 0$.
\end{definition}
\endgroup

Since the parameter space $\mathcal{M}$ is convex, for any $x, y \in \mathcal{M}$ the straight edge $x + \tau(y-x)$ with $\tau \in [0,1]$ lies in $\mathcal{M}$ and is a path.
Its length is $\norm{y-x}$ and its speed is constant, so $\zeta_s = 1$. We write
\begin{equation}\label{eq:straight-edge}
\begin{aligned}
H_{x \rightarrow y}(\tau) &:= H\left( x + \tau(y-x) \right),\qquad \widehat u_{xy} := \frac{y-x}{\norm{y-x}}, \\
\partial_\tau H_{x \rightarrow y}(\tau) &= \norm{y-x} \widehat u_{xy} \cdot \nabla_x H\left(x + \tau(y-x) \right),
\end{aligned}
\end{equation}
where $\widehat u_{xy} \cdot \nabla_x H$, the directional derivative along the edge, equals $\partial_\tau H_{x \rightarrow y}$ divided by the length of the edge.

\subsection{Spectrum, sectors, and spectral gaps}\label{subsec:sector-setup}

For $x \in \mathcal{M}$, $H(x)$ has eigenvalues $E_0(x) \le E_1(x) \le \cdots$ and orthonormal eigenstates $|\psi_k(x)\rangle$. With $\hbar = 1$, the transition frequency from level~$j$ to level~$k$ is
\begin{equation}\label{eq:transition-frequency}
    \omega_{kj}(x) := E_k(x) - E_j(x).
\end{equation}

We often track the $d$ lowest levels at fixed $d$, even where the dimension of the ground-state manifold changes.

\begin{definition}[Spectral sector]\label{def:spectral-sector}
Fix an integer $d \ge 1$.
The spectral sector of dimension $d$ at $x$ is the subspace spanned by the $d$ lowest eigenstates,
\begin{equation}\label{eq:sector}
\mathcal S_d(x) := \mathrm{span}\{|\psi_0(x)\rangle, \dots, |\psi_{d-1}(x)\rangle\}.
\end{equation}
Its projector is $P_d(x)$ and its sector state $\rho_d(x) := P_d(x)/d$ is the maximally mixed state on the sector.
The band gap of the sector is
\begin{equation}\label{eq:band-gap}
\gamma_d(x) := E_d(x) - E_{d-1}(x).
\end{equation}
The sector width is $E_{d-1}(x) - E_0(x)$.
When the context fixes $d$, we write $\mathcal S(x)$ and $P(x)$.
\end{definition}

We use sectors only at points $x$ where $\gamma_d(x) > 0$, so the sector is fixed by energy: $\mathcal{S}_d(x)$ is the span of all eigenstates with eigenvalue at most $E_{d-1}(x)$.
\begingroup\pretolerance=-1 \tolerance=2000
\begin{definition}[Ground-state manifold]
\label{definition: ground-state manifold}\label{definition: maximally mixed ground state}
We fix a tolerance $\delta_{\mathrm{gsm}} > 0$. Definition~\ref{def:phase-clustering} assumes it is smaller than the resolution. The ground-state degeneracy at $x$ is the number of levels within the tolerance of the ground-state energy,
\begin{equation}\label{eq:degeneracy-index}
r(x) := \#\{ j : E_j(x) - E_0(x) \le \delta_{\mathrm{gsm}} \}.
\end{equation}
The sector of dimension $r(x)$ is the ground-state manifold $\mathcal G(x) := \mathcal S_{r(x)}(x)$. We write $P_{\mathcal G}(x) := P_{r(x)}(x)$ for its projector. The maximally mixed ground state is
\begin{equation}\label{eq:gsm-state}
\rho_{\mathcal G}(x) := P_{\mathcal G}(x)/r(x).
\end{equation}
For $r(x) = 1$ it is the pure ground state $|\psi_0(x)\rangle\langle\psi_0(x)|$. By construction the width of the ground-state manifold is at most $\delta_{\mathrm{gsm}}$.
\end{definition}
\endgroup

\begin{figure}[t]
\centering
\definecolor{gapblue}{HTML}{0072B2}
\definecolor{thrgreen}{HTML}{009E73}
\tikzset{
  level/.style={line width=1.0pt, line cap=round},
  gaparrow/.style={gapblue, stealth-stealth, line width=0.7pt,
                   shorten >=1pt, shorten <=1pt},
}
\begin{tikzpicture}[xscale=0.60, yscale=1.05, every node/.style={font=\normalsize}]
\begin{scope}
  \draw[thrgreen, line width=0.9pt] (0,1.15) -- (10,1.15);
  \node[thrgreen!80!black, anchor=west] at (10.12,1.15) {$\delta_{\mathrm{gsm}}$};
  \draw[level] (0,0) -- (10,0);
  \node[anchor=east] at (-0.18,0) {$E_0(s)$};
  \draw[level] plot[domain=0:10, samples=60, smooth]
    (\x, {1.6 + sqrt(0.64 + 0.2023*(\x-5.5)^2)});
  \node[anchor=east] at (-0.18,4.2) {$E_1(s)$};
  \draw[gaparrow] (5.5,0.02) -- (5.5,2.38);
  \node[gapblue, anchor=west] at (5.65,1.75) {$\gamma_{\min}$};
  \node at (5,-0.45) {$r(s) = 1$};
  \draw[-stealth, line width=0.5pt] (-0.15,-0.95) -- (10.45,-0.95)
    node[anchor=west] {$s$};
  \foreach \x/\lab in {0/0, 10/1} {
    \draw[line width=0.5pt] (\x,-0.89) -- (\x,-1.01);
    \node[anchor=north] at (\x,-1.03) {\lab};
  }
  \node[anchor=west] at (-1.55,4.70) {\textbf{(a)}};
\end{scope}
\begin{scope}[shift={(15.0,0)}]
  \draw[thrgreen, line width=0.9pt] (0,1.15) -- (10,1.15);
  \node[thrgreen!80!black, anchor=west] at (10.12,1.15) {$\delta_{\mathrm{gsm}}$};
  \draw[level] (0,0) -- (10,0);
  \node[anchor=east] at (-0.18,0) {$E_0(s)$};
  \draw[level] plot[smooth, tension=0.7] coordinates
    {(0,0.44) (2.5,0.42) (5.5,0.40) (8,0.43) (10,0.45)};
  \node[anchor=east] at (-0.18,0.44) {$E_1(s)$};
  \draw[level] plot[smooth, tension=0.7] coordinates
    {(0,0.86) (2.5,0.84) (5.5,0.82) (8,0.85) (10,0.88)};
  \node[anchor=east] at (-0.18,0.86) {$E_2(s)$};
  \draw[level] plot[domain=0:10, samples=60, smooth]
    (\x, {1.6 + sqrt(0.64 + 0.2023*(\x-5.5)^2)});
  \node[anchor=east] at (-0.18,4.2) {$E_3(s)$};
  \draw[gaparrow] (5.5,0.84) -- (5.5,2.38);
  \node[gapblue, anchor=west] at (5.65,1.75) {$\gamma_{\min}$};
  \node at (5,-0.45) {$r(s) = 3$};
  \draw[-stealth, line width=0.5pt] (-0.15,-0.95) -- (10.45,-0.95)
    node[anchor=west] {$s$};
  \foreach \x/\lab in {0/0, 10/1} {
    \draw[line width=0.5pt] (\x,-0.89) -- (\x,-1.01);
    \node[anchor=north] at (\x,-1.03) {\lab};
  }
  \node[anchor=west] at (-1.55,4.70) {\textbf{(b)}};
\end{scope}
\end{tikzpicture}
\caption{When the ground-state degeneracy $r(s)$ is constant and $d = r(s)$,
the minimum band gap $\gamma_{\min}$ of Eq.~\eqref{eq:path-min-gap} is
the minimum spectral gap along the path (double arrows). Levels within
$\delta_{\mathrm{gsm}}$ of $E_0(s)$ belong to the ground-state manifold.
(a)~A unique ground state, $r(s) = 1$:
$\gamma_{\min} = \min_s\,\bigl(E_1(s) - E_0(s)\bigr)$.
(b)~A threefold degenerate ground-state manifold, $r(s) = 3$:
$\gamma_{\min} = \min_s\,\bigl(E_3(s) - E_2(s)\bigr)$.}
\label{fig:banded-path}
\end{figure}
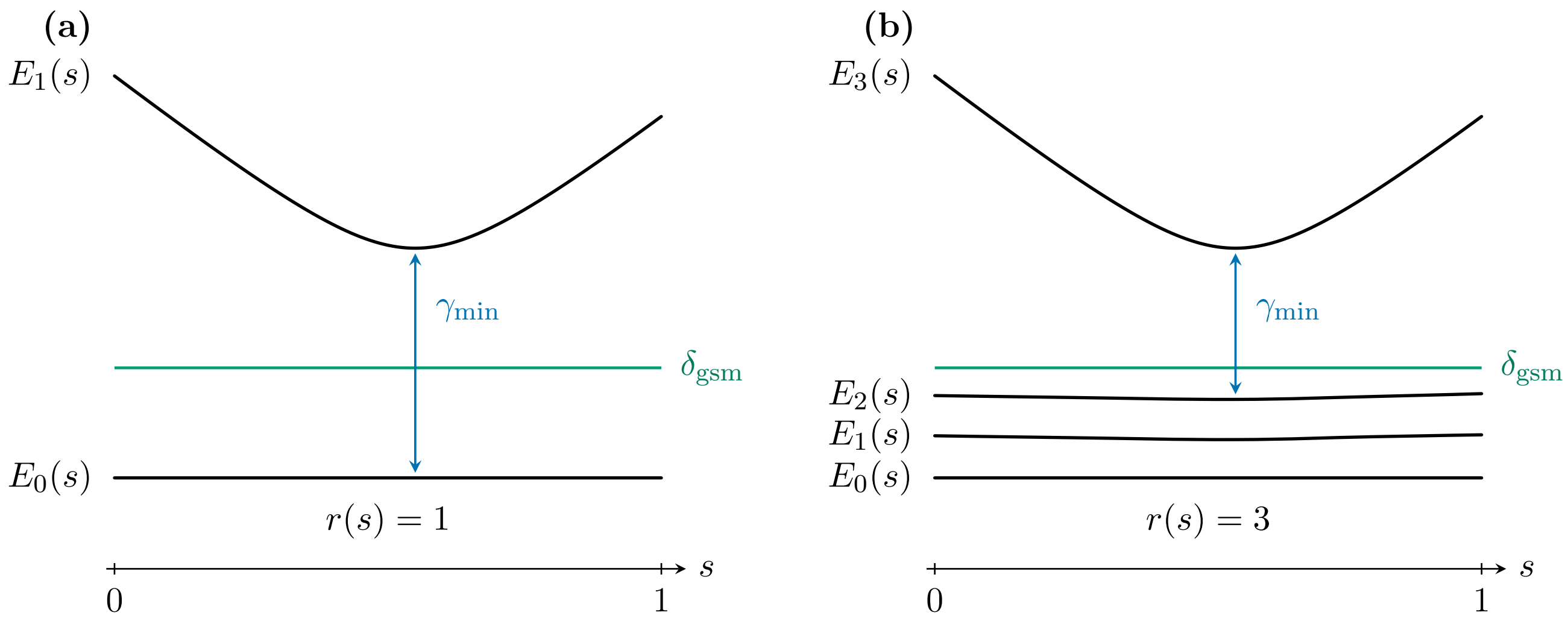

\begin{definition}[Spectral gap at a point and along a path]
\label{definition: spectral gap}
The spectral gap at $x$ is the band gap of the ground-state manifold,
\begin{equation}\label{eq:spectral-gap}
\gamma(x) := \gamma_{r(x)}(x) = E_{r(x)}(x) - E_{r(x)-1}(x).
\end{equation}
Along a path $x(s)$ (Definition~\ref{def:path}) we abbreviate $E_k(s)$, $|\psi_k(s)\rangle$, $\gamma_d(s)$, $r(s)$, $\mathcal S_d(s)$, and $P_d(s)$ for the same objects at $x(s)$. We write $r_A := r(x_A)$ and $r_B := r(x_B)$ for the ground-state degeneracies at the endpoints. For a fixed sector dimension $d$, the minimum band gap along the path is
\begin{equation}\label{eq:path-min-gap}
\gamma_{\min} := \min_{s \in [0,1]} \gamma_d(s).
\end{equation}
If the ground-state degeneracy equals $d$ along the path, $\gamma_{\min}$ is the minimum spectral gap (Figure~\ref{fig:banded-path}).
\end{definition}

\subsection{Elementary stability facts}\label{sec:gap-stability}

The learning algorithms in the later appendices replace continuous paths with paths on finite grids. Part (iii) of the next lemma bounds how quickly a band gap can change, so a grid path sufficiently close to a gapped path remains gapped.

\begin{lemma}[Spectral stability]\label{lem:weyl}\label{lem:gap-stability}\label{cor:gap-lipschitz}
Let $H$ and $K$ be two Hermitian matrices acting on the same finite-dimensional Hilbert space.
Let $E_0 \le E_1 \le \cdots$ and $E_0' \le E_1' \le \cdots$ be the eigenvalues of $H$ and $K$ in nondecreasing order.
\begin{enumerate}
\item[(i)]
Eigenvalues; Weyl's inequality. For each $k$,
\begin{equation}\label{eq:weyl-bound}
        |E_k' - E_k| \le \norm{K - H}.
    \end{equation}
\item[(ii)]
Band gaps. For each sector dimension $d \ge 1$, the band gaps $\gamma_d = E_d - E_{d-1}$ and $\gamma_d' = E_d' - E_{d-1}'$ satisfy
\begin{equation}\label{eq:gap-stability-bound}
        |\gamma_d' - \gamma_d| \le 2 \norm{K - H}.
    \end{equation}
\item[(iii)]
Lipschitz continuity in the parameter. Consider any $x, y \in \mathcal M$ and a family of Hamiltonians as in Definition~\ref{definition: geometrically local hamiltonians}. Then,
$\norm{H(x) - H(y)} \le J \norm{x - y}$. Hence, for each fixed sector dimension $d$,
\begin{equation}\label{eq:gap-lipschitz}
        |\gamma_d(x) - \gamma_d(y)| \le L_\gamma\, \|x - y\|,\qquad L_\gamma := 2J.
    \end{equation}
\end{enumerate}
\end{lemma}
\begin{proof}
(i) This is Weyl's perturbation theorem~\cite{Bha97:Matrix-Analysis}.
(ii) Each band gap is a difference of two eigenvalues with the same indices, so (i) applied twice gives the factor of two.
(iii) Integrating along the straight edge from $x$ to $y$ and using the term bounds of Definition~\ref{definition: geometrically local hamiltonians} gives
\begin{equation}
    \norm{H(x) - H(y)} \le \sum_{j=1}^{J} \norm{h_j(x) - h_j(y)} \le J \norm{x - y}.
\end{equation}
Then (ii) applies with $K = H(y)$.
\end{proof}

\subsection{Filter functions and the quasi-adiabatic propagator}\label{sec:appA-quasi-adiabatic}\label{sec:filter-function}

The quasi-adiabatic propagator is constructed from a filter function.

\begin{definition}[Filter function]\label{definition:filter-requirements}
Fix an energy scale $\Delta > 0$. A filter function at this energy scale is a function of time of the rescaled form
\begin{equation}\label{eq:filter-scaling}
    F_\Delta(t) := F_1(\Delta t),\qquad \text{so that}\qquad \widetilde F_\Delta(\omega) = \frac{1}{\Delta}\, \widetilde F_1\!\Bigl(\frac{\omega}{\Delta}\Bigr),
\end{equation}
where $F_1$ is a fixed integrable function that is odd and purely imaginary. The transform is defined by Eq.~\eqref{eq:fourier-convention}. It follows that $\widetilde F_\Delta$ is real, odd and continuous. We require the following properties.
\begin{enumerate}
\item
\emph{Fast decay.} The function $|F_\Delta(t)|$ decays faster than any polynomial in $|t|$, so the quasi-adiabatic generator in Definition~\ref{def:quasi-adiabatic-evolution} is quasi-local.
\item
\emph{Values above the cutoff.} $\widetilde F_\Delta(\omega) = -1/\omega$ for all $|\omega| \ge \Delta$. The quasi-adiabatic generator then agrees with the exact adiabatic generator on every transition of energy at least $\Delta$ (Lemma~\ref{lem:generator-matrix-elements}), so $\Delta$ is an energy cutoff.
\item
\emph{Value at zero.} $\widetilde F_\Delta(0) = 0$, which holds automatically since $\widetilde F_\Delta$ is odd and continuous.
\end{enumerate}
\end{definition}

Fix $\Delta > 0$. We use the filter from Bachmann, Michalakis, Nachtergaele and Sims~\cite{bachmann2011automorphic} with their energy scale $\gamma$ replaced by $\Delta$.
Their Fourier convention is that of Eq.~\eqref{eq:fourier-convention}.
We call this filter the BMNS filter.
It is the integral of a weight function whose Fourier transform vanishes for $|\omega| \ge \Delta$.
This band limit gives the filter its values above the cutoff.
The weight is
\begin{equation}\label{eq:bmns-weight}
w_\Delta(t) := c_\Delta \prod_{k=1}^{\infty} \Bigl( \frac{\sin a_k t}{a_k t} \Bigr)^2,\qquad a_k := \frac{a_1}{k \ln^2 k}\ \ (k \ge 2),\qquad \sum_{k=1}^{\infty} a_k = \frac{\Delta}{2},
\end{equation}
where $c_\Delta > 0$ is the normalization~\cite[Eq.~(2.1)]{bachmann2011automorphic}.
Here $a_1 > 0$ is chosen so that the $a_k$ sum to $\Delta/2$.
We derive the band limit from the partial products.
The Fourier transform of the first $N$ factors is $2\pi$ times the convolution of the triangular probability densities $(2a_k - |\omega|)_+/(4a_k^2)$ over $k \le N$, so it is even and non-negative and vanishes outside $[-\Delta, \Delta]$.
The partial products converge pointwise to $w_\Delta/c_\Delta$.
Since the first factor is integrable and dominates them, they converge in $L^1$.
In the limit, $\widetilde w_\Delta$ is even, non-negative and continuous, satisfies $\widetilde w_\Delta(0) = \int w_\Delta$, and obeys the band limit~\cite[text after Lemma~2.3]{bachmann2011automorphic},
\begin{equation}\label{eq:bmns-band-limit}
    \widetilde w_\Delta(\omega) = 0 \qquad \text{for } |\omega| \ge \Delta .
\end{equation}

\begin{lemma}[Properties of the BMNS weight function]\label{lemma:bmns-weight}
The infinite product in Eq.~\eqref{eq:bmns-weight} converges for every $t$ and defines an even, non-negative function $w_\Delta \in L^1(\mathbb R)$ with $0 \le w_\Delta(t) \le w_\Delta(0) = c_\Delta$. We can choose $c_\Delta$ so that $\int w_\Delta(t)\,dt = 1$. By changing scales, we have $w_\Delta(t) = \Delta w_1 (\Delta t)$, and hence $c_\Delta = \Delta c_1$. The constant satisfies $c_1 \leq 0.1031$. For every $t \geq e^{1 / \sqrt{2}} / \Delta$,
\begin{equation}\label{eq:bmns-decay}
    0 \le w_\Delta(t) \le 2(e\Delta)^2\, t \cdot \exp\!\Bigl( -\frac{2}{7}\, \frac{\Delta t}{\ln^2(\Delta t)} \Bigr).
\end{equation}
\end{lemma}
\begin{proof}
By~\cite[Lemma~2.3]{bachmann2011automorphic} with $\gamma = \Delta$, the product converges, and $w_\Delta$ is even, non-negative, integrable and normalizable and obeys the decay bound~\eqref{eq:bmns-decay}.\footnote{The decay bound holds, but the proof of~\cite[Lemma~2.3]{bachmann2011automorphic} needs corrections. It quotes $\gamma/(2\pi) < c_\gamma < \gamma/\pi$. The lower bound is false, since $c_1 \le 0.1031 < 1/(2\pi)$. In that proof, the step involving the factor $\ln^{4N}(N)$ fails for $N = 1, 2$, where $\ln N < 1$, but holds with $\max(1, \ln N)$ in place of $\ln N$. The derivation of the decay bound needs only $c_\gamma \le e^{-1/6}\gamma/\pi$. This bound on $c_\gamma$ also absorbs the factor $e^{1/(6N)}$ of Stirling's bound. In the cases $N = 0$ and $\gamma t < e$, which the proof does not treat, the right-hand side of Eq.~\eqref{eq:bmns-decay} exceeds $c_\gamma$.}
Every factor is at most one, so $w_\Delta(t) \le w_\Delta(0) = c_\Delta$.
Since $a_k \propto \Delta$, the product at scale $\Delta$ evaluated at $t$ equals the product at scale one evaluated at $\Delta t$. Normalizing gives $c_\Delta = \Delta c_1$ and $w_\Delta(t) = \Delta\,w_1(\Delta t)$.
For the bound on $c_1$, take $\Delta = 1$. The sum $\sum_{k \ge 2} 1/(k\ln^2 k) = 2.1097\ldots$ gives $a_1 = 1/(2 \cdot 3.1097\ldots) = 0.1608$, and $\Sigma_2 := \sum_k a_k^2 = 0.0566$.
For $|t| \le \sqrt{3/\Sigma_2}$, each factor lies in $[0, 1]$ and is at least $1 - a_k^2t^2/3$. The product is then at least $1 - t^2\Sigma_2/3$. Integrating over this interval gives $\int \prod_k (\sin a_k t/(a_k t))^2\,dt \ge \tfrac43\sqrt{3/\Sigma_2} = 9.707$.
Hence $c_1 \le \tfrac34\sqrt{\Sigma_2/3} \le 0.1031$.
\end{proof}

\begin{figure}[t]
\centering
\definecolor{gapblue}{HTML}{0072B2}
\begin{tikzpicture}
\begin{scope}
  \draw[->, line width=0.5pt] (-2.1,0) -- (2.2,0) node[anchor=west] {$\omega$};
  \draw[->, line width=0.5pt] (0,-0.12) -- (0,1.55);
  \draw[dashed, gray, line width=0.7pt] plot[domain=-2.05:2.05, samples=80]
    (\x, {exp(-1.4*\x*\x)});
  \draw[gapblue, line width=1.0pt]
    (-2.05,0) -- (-1,0) -- (-1,1) -- (1,1) -- (1,0) -- (2.05,0);
  \draw[line width=0.5pt] (-1,-0.05) -- (-1,0.05);
  \draw[line width=0.5pt] (1,-0.05) -- (1,0.05);
  \node[anchor=north] at (-1,-0.07) {$-a_k$};
  \node[anchor=north] at (1,-0.07) {$a_k$};
  \node[anchor=east] at (-1.08,1) {$\pi/a_k$};
  \node[gapblue, anchor=west] at (1.08,1.08) {$(\pi/a_k)\,\mathbf 1\{|\omega|\le a_k\}$};
  \node[gray, anchor=west] at (1.2,0.5) {Gaussian};
  \node[anchor=west] at (-2.1,2.2) {\textbf{(a)}};
\end{scope}
\begin{scope}[shift={(5.8,0)}]
  \draw[->, line width=0.5pt] (-2.1,0) -- (2.2,0) node[anchor=west] {$\omega$};
  \draw[->, line width=0.5pt] (0,-0.12) -- (0,1.55);
  \draw[gapblue, line width=1.0pt] (-2.05,0) -- (-1,0);
  \draw[gapblue, line width=1.0pt] (1,0) -- (2.05,0);
  \draw[gapblue, line width=1.0pt] plot[domain=-1:1, samples=100]
    (\x, {(1-\x*\x)^3});
  \draw[line width=0.5pt] (-1,-0.05) -- (-1,0.05);
  \draw[line width=0.5pt] (1,-0.05) -- (1,0.05);
  \node[anchor=north] at (-1,-0.07) {$-\Delta$};
  \node[anchor=north] at (1,-0.07) {$\Delta$};
  \draw[line width=0.5pt] (-0.05,1) -- (0.05,1);
  \node[anchor=south east] at (-0.08,1.03) {$1$};
  \node[gapblue, anchor=south west] at (0.12,1.03) {$\widetilde w_\Delta(\omega)$};
  \node[anchor=south] at (1.65,0.42) {exactly $0$};
  \draw[-stealth, line width=0.5pt] (1.65,0.36) -- (1.65,0.06);
  \node[anchor=east, align=right] at (-1.1,0.5) {smooth\\ tail};
  \node[anchor=west] at (-2.1,2.2) {\textbf{(b)}};
\end{scope}
\begin{scope}[shift={(11.7,0.7)}, yscale=0.5]
  \fill[gapblue!8] (-1,-2.0) rectangle (1,2.0);
  \draw[->, line width=0.5pt] (-2.7,0) -- (2.8,0) node[anchor=west] {$\omega$};
  \draw[->, line width=0.5pt] (0,-2.1) -- (0,2.2);
  \begin{scope}
    \clip (-2.7,-2.1) rectangle (2.8,2.2);
    \draw[dashed, gray, line width=0.7pt] plot[domain=0.30:2.65, samples=120]
      (\x, {-1/\x});
    \draw[dashed, gray, line width=0.7pt] plot[domain=-2.65:-0.30, samples=120]
      (\x, {-1/\x});
  \end{scope}
  \draw[gapblue, line width=1.0pt] plot[domain=-2.65:-1, samples=60]
    (\x, {-1/\x});
  \draw[gapblue, line width=1.0pt] plot[domain=-1:1, samples=100]
    (\x, {-3*\x + 3*\x*\x*\x - \x*\x*\x*\x*\x});
  \draw[gapblue, line width=1.0pt] plot[domain=1:2.65, samples=60]
    (\x, {-1/\x});
  \draw[line width=0.5pt] (-1,-0.1) -- (-1,0.1);
  \draw[line width=0.5pt] (1,-0.1) -- (1,0.1);
  \node[anchor=north] at (-1,-0.12) {$-\Delta$};
  \node[anchor=south] at (1,0.12) {$\Delta$};
  \node[gapblue, anchor=south west] at (-2.6,1.0) {$\widetilde F_\Delta(\omega)$};
  \node[gray, anchor=west] at (1.5,-1.3) {$-1/\omega$};
  \node[anchor=north east] at (-0.3,-1.35) {$\widetilde F_\Delta(0)=0$};
  \draw[-stealth, line width=0.5pt] (-0.4,-1.25) -- (-0.06,-0.12);
  \node[anchor=west] at (-2.7,3.0) {\textbf{(c)}};
\end{scope}
\end{tikzpicture}
\caption{The BMNS filter in frequency space (schematic profiles).
(a)~Each factor of the weight is the square of $\sin a_k t/(a_k t)$, whose Fourier transform $(\pi/a_k)\,\mathbf 1\{|\omega| \le a_k\}$
is a box of half-width $a_k$. A Gaussian (dashed) is shown for
contrast, since it is not band-limited. (b)~The weight
$\widetilde w_\Delta(\omega)$, a convolution of such boxes: even,
non-negative, $\widetilde w_\Delta(0) = \int w_\Delta = 1$, smooth,
and exactly zero for $|\omega| \ge \Delta = \sum_k 2a_k$. (c)~The
filter $\widetilde F_\Delta(\omega) = (\widetilde w_\Delta(\omega)-1)/\omega$,
Eq.~\eqref{eq:filter-fourier-identity}: odd, smooth through
$\omega = 0$ with $\widetilde F_\Delta(0) = 0$, and equal to
$-1/\omega$ (dashed) for $|\omega| \ge \Delta$, so that
$\widetilde F_\Delta + 1/\omega = \widetilde w_\Delta/\omega$ vanishes
outside the band (shaded).}
\label{fig:filter-shapes}
\end{figure}
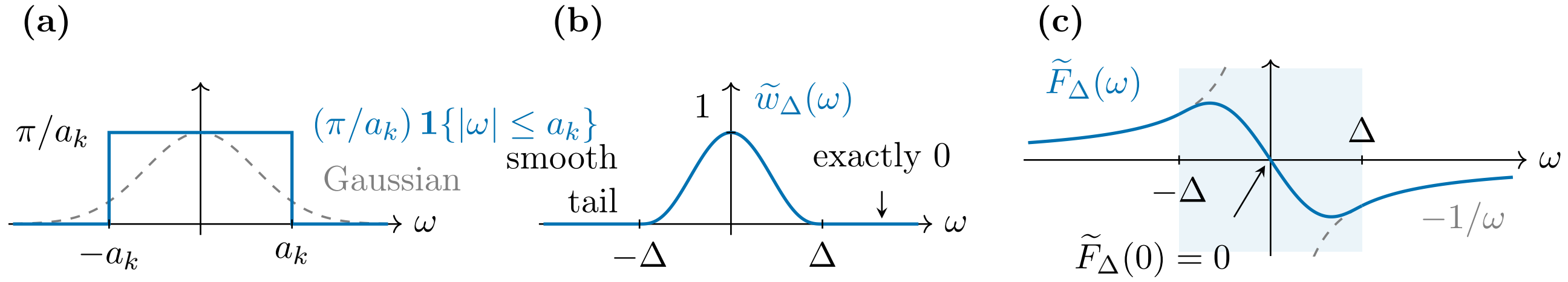

The filter is the integral of the weight,
\begin{equation}\label{eq:WDelta-def}
    F_\Delta(t) :=
    \begin{cases}
    \displaystyle -i\int_t^{\infty} w_\Delta(\xi)\,d\xi, & t > 0,\\[0.4em]
    0, & t = 0,\\[0.3em]
    \displaystyle \,i\int_{-\infty}^{t} w_\Delta(\xi)\,d\xi, & t < 0 ,
    \end{cases}
\end{equation}
so that, for $t \ne 0$, $iF_\Delta(t)$ is the odd integrated weight of~\cite[Eq.~(2.12)]{bachmann2011automorphic}.
Taking the Fourier transform of the distributional derivative $\partial_t\bigl(iF_\Delta(t)\bigr) = -w_\Delta(t) + \delta(t)$ yields $-\omega\,\widetilde F_\Delta(\omega) = 1 - \widetilde w_\Delta(\omega)$, i.e.,
\begin{equation}\label{eq:filter-fourier-identity}
    \widetilde F_\Delta(\omega) + \frac{1}{\omega} = \frac{\widetilde w_\Delta(\omega)}{\omega},\qquad \omega \ne 0,\qquad\text{and}\qquad \widetilde F_\Delta(\omega) + \frac{1}{\omega} = 0 \quad\text{for all } |\omega| \ge \Delta,
\end{equation}
where the second statement follows from the band limit~\eqref{eq:bmns-band-limit}. Figure~\ref{fig:filter-shapes} shows the boxes that build the weight, the weight and the filter in frequency space.

\begin{lemma}[BMNS filter: existence and properties]\label{lem:BMNS-filter}
The function $F_\Delta$ defined in Eq.~\eqref{eq:WDelta-def} is odd and purely imaginary. It scales as $F_\Delta(t) = F_1(\Delta t)$ and satisfies all requirements in Definition~\ref{definition:filter-requirements}.
Moreover, $0 \le \widetilde{w}_\Delta(\omega) \le 1$ holds for every $\omega$.  Consequently, Eq.~\eqref{eq:filter-fourier-identity} implies
\begin{equation}\label{eq:bmns-bracket-bound}
    \Bigl| \widetilde F_\Delta(\omega) + \frac{1}{\omega} \Bigr| \le \frac{1}{|\omega|}\qquad \text{for all } \omega \ne 0 .
\end{equation}
\end{lemma}
\begin{proof}
The filter $F_\Delta$ is odd, purely imaginary and decays sub-exponentially by~\cite[Lemma~2.6(i),(ii)]{bachmann2011automorphic}. The scaling $F_\Delta(t) = F_1(\Delta t)$ follows from $w_\Delta(t) = \Delta\, w_1(\Delta t)$ by $\xi \to \xi/\Delta$. Eq.~\eqref{eq:filter-fourier-identity} gives the values above the cutoff. The value at zero follows automatically (Definition~\ref{definition:filter-requirements}). Finally, $\widetilde w_\Delta \ge 0$ holds because the transforms of the partial products are non-negative and the products converge in $L^1$. With $|\widetilde w_\Delta(\omega)| \le \int w_\Delta = 1$, this implies Eq.~\eqref{eq:bmns-bracket-bound}.
\end{proof}

Explicit constants for the BMNS filter are needed because its total mass bounds the norm of the quasi-adiabatic generator and its tail controls the error from truncating the generator in time.

\begin{lemma}[Filter moments and tail]\label{lem:filter-constants}
The constant
\begin{equation}\label{eq:filter-constants}
    c_F := \Delta\!\int_{-\infty}^{\infty}\! |F_\Delta(t)|\,dt
\end{equation}
is finite and independent of $\Delta$. It is the first absolute moment of the weight at scale one,
\begin{equation}\label{eq:filter-constants-identities}
    c_F = \int_{-\infty}^{\infty} |u|\,w_1(u)\,du = 3.902 .
\end{equation}
Moreover, with $\mu_1 := 1/7$ and an absolute constant $C_{\mathrm{tail}} < 10^9$, for all $\xi \ge 0$,
\begin{equation}\label{eq:filter-tail-bound}
    \Delta \int_{|t| > \xi/\Delta} |F_\Delta(t)|\,dt
    \;\le\;
    C_{\mathrm{tail}}\, F_{\mathrm{ball}}(\xi),
    \qquad
    F_{\mathrm{ball}}(\xi) := \exp\!\left( -\mu_1\, \frac{\xi}{\log^2(e + \xi)} \right),
\end{equation}
and it holds with the same $C_{\mathrm{tail}}$ for every smaller value of $\mu_1 > 0$.
\end{lemma}
\begin{proof}
Off $t = 0$, $iF_\Delta$ is the function $W_\gamma$ of~\cite[Eq.~(6.35)]{nachtergaele2019quasi} at $\gamma = \Delta$.
Its weight $w_\gamma(t) = \gamma\,w(\gamma t)$ is our $w_\Delta$, and their $w$ is our $w_1$.
Because $w_1$ is even, the identity for $c_F$ is then $\Delta$ times~\cite[Eq.~(6.36)]{nachtergaele2019quasi}.
The moment is finite by the decay bound~\eqref{eq:bmns-decay}.
Its value comes from two independent quadratures of Eq.~\eqref{eq:bmns-weight}, which agree to $10^{-6}$.

For the tail bound, substituting $u = \Delta t$ gives
\begin{equation*}
    \Delta\int_{|t|>\xi/\Delta}|F_\Delta(t)|\,dt = 2\int_\xi^\infty |F_1(u)|\,du = 2\int_\xi^\infty (v - \xi)\,w_1(v)\,dv \le 2\int_\xi^\infty v\,w_1(v)\,dv.
\end{equation*}
At $\xi = 0$, this display also gives the identity for $c_F$, because $w_1$ is even.
For $\xi \ge 6000$ and $v \ge \xi$, we have $\sqrt v \ge \ln^2 v$, and $v/\ln^2 v$ is increasing. Then Eq.~\eqref{eq:bmns-decay} gives $w_1(v) \le 2e^2\,v\,\exp(-\tfrac17\,\xi/\ln^2\xi)\,\exp(-\tfrac17\sqrt v)$, whence
\begin{equation*}
    2\int_\xi^\infty v\,w_1(v)\,dv \le 4e^2 \exp\!\Bigl(-\frac{1}{7}\,\frac{\xi}{\ln^2 \xi}\Bigr) \int_0^\infty v^2 e^{-\sqrt v/7}\,dv = 8\cdot 5!\cdot 7^6\, e^2\, \exp\!\Bigl(-\frac{1}{7}\,\frac{\xi}{\ln^2 \xi}\Bigr),
\end{equation*}
because $\int_0^\infty v^2 e^{-\sqrt v/7}\,dv = 2\int_0^\infty s^5 e^{-s/7}\,ds = 2\cdot 5!\cdot 7^6$. Replacing $\ln^2\xi$ by $\ln^2(e+\xi)$ only weakens the bound.
For $\xi < 6000$, the left-hand side of Eq.~\eqref{eq:filter-tail-bound} is at most $c_F$, while $F_{\mathrm{ball}}(\xi) \ge F_{\mathrm{ball}}(6000)$. In both cases, the tail bound holds with $\mu_1 = 1/7$ and
\begin{equation*}
    C_{\mathrm{tail}} := \max\{8\cdot 5!\cdot 7^6 e^2,\; c_F/F_{\mathrm{ball}}(6000)\} = 8\cdot 5!\cdot 7^6 e^2 < 10^9.
\end{equation*}
Decreasing $\mu_1$ only increases $F_{\mathrm{ball}}$, so Eq.~\eqref{eq:filter-tail-bound} holds for any smaller $\mu_1$.
\end{proof}

\begin{definition}[Quasi-adiabatic generator and propagator]\label{def:quasi-adiabatic-evolution}
Let $x(s)$ be a path (Definition~\ref{def:path}), and $H(s) = H(x(s))$ be the Hamiltonian along this path.
Let $F_\Delta$ be a filter function with energy cutoff $\Delta > 0$.
We define the quasi-adiabatic generator
\begin{equation}\label{eq-quasi-adiabatic-generator}
    D_\Delta(s) := i \int_{-\infty}^{\infty} F_\Delta(t)\, e^{itH(s)}\, \partial_s H(s)\, e^{-itH(s)}\, dt,
\end{equation}
and the quasi-adiabatic propagator
\begin{equation}\label{eq-adiabatic-evolution}
\begin{gathered}
    U_\Delta[x](s) := \mathcal T \exp\!\Bigl( i \int_0^s D_\Delta(s')\, ds' \Bigr),
    \\
    \text{equivalently}\qquad
    \partial_s U_\Delta[x](s) = i\, D_\Delta(s)\, U_\Delta[x](s),\quad U_\Delta[x](0) = I,
\end{gathered}
\end{equation}
where $\mathcal T$ denotes time ordering (later generators to the left).
We denote by $U_\Delta[x] := U_\Delta[x](1)$ the propagator of the whole path. The bracket names the path.
If the path is fixed in the context, we write $U_\Delta(s)$ and $U_\Delta := U_\Delta(1)$.
For the straight edge between two points in Eq.~\eqref{eq:straight-edge}, $U_\Delta(x \to y)$ denotes the propagator of the whole edge.
We call $\Delta$ the energy cutoff of the evolution.
\end{definition}

Since $D_\Delta(s)\,ds$ depends only on $H(x(s))$ and $\partial_s H(s)\,ds$, the propagator $U_\Delta[x]$ is invariant under any orientation-preserving reparametrization.
The path $x$ followed by the path $y$ gives $U_\Delta[y]\, U_\Delta[x]$, and the reversed path $s \mapsto x(1-s)$ has generator $-D_\Delta(1-s)$ and propagator $U_\Delta[x]^\dagger$.

To compare $D_\Delta(s)$ with the exact adiabatic evolution, we write it in the eigenbasis of $H(s)$. In this basis, the condition $\widetilde F_\Delta(\omega) = -1/\omega$ for $|\omega| \ge \Delta$ (Definition~\ref{definition:filter-requirements}) makes $\Delta$ an energy cutoff. This comparison is used in Corollary~\ref{cor:gapped-path-same-phase} and Appendix~\ref{sec:cannot-be-dark}.

\begin{lemma}[Matrix elements of the generator]\label{lem:generator-matrix-elements}
Let $|\psi_k(s)\rangle$ and $E_k(s)$ be the eigenstates and eigenvalues of $H(s)$, and let $M_{kj}(s) := \langle \psi_k(s)| \partial_s H(s) |\psi_j(s)\rangle$. Then
\begin{equation}\label{eq:generator-matrix-elements}
    \langle \psi_k(s)| D_\Delta(s) |\psi_j(s)\rangle = -\,i\, M_{kj}(s)\, \widetilde F_\Delta(\omega_{kj}(s)),
\end{equation}
and $D_\Delta(s)$ is Hermitian, so that $U_\Delta[x](s)$ is unitary. Furthermore, $\langle \psi_k(s)| D_\Delta(s) |\psi_j(s)\rangle = i M_{kj}(s) / \omega_{kj}(s)$ whenever $|\omega_{kj}(s)| \ge \Delta$.
\end{lemma}
\begin{proof}
Inserting the eigen-decomposition of $H(s)$ into the definition of $D_\Delta(s)$ yields
\begin{equation}\label{eq:heisenberg-matrix-element}
    \langle \psi_k(s)| e^{itH(s)}\, \partial_s H(s)\, e^{-itH(s)} |\psi_j(s)\rangle = e^{i\omega_{kj}(s) t} M_{kj}(s).
\end{equation}
Therefore, the matrix element of $D_\Delta(s)$ is
\begin{align}
    i M_{kj}(s) \int_{-\infty}^{\infty} F_\Delta(t)\, e^{i\omega_{kj}(s) t}\, dt &= i M_{kj}(s)\, \widetilde F_\Delta(-\omega_{kj}(s)) \nonumber\\
    &= -i M_{kj}(s)\, \widetilde F_\Delta(\omega_{kj}(s)),
\end{align}
\looseness=-1 where we used that $\widetilde F_\Delta$ is odd. Conjugating this matrix element shows that $D_\Delta(s)$ is Hermitian. The matrix with entries $M_{kj}(s)$ is Hermitian, and $\widetilde F_\Delta$ is real and odd. The complex conjugate of $-i M_{kj}(s) \widetilde F_\Delta(\omega_{kj}(s))$ is therefore $i M_{jk}(s) \widetilde F_\Delta(\omega_{kj}(s)) = -i M_{jk}(s) \widetilde F_\Delta(\omega_{jk}(s))$, which is the matrix element with the two indices exchanged. The last claim follows from $\widetilde F_\Delta(\omega) = -1/\omega$ for $|\omega| \ge \Delta$.
\end{proof}

Exact adiabatic evolution transports every eigenstate error-free. Its generator has matrix element $iM_{kj}(s)/\omega_{kj}(s)$ between eigenstates of different energy.

\subsection{The band-restricted spectral weight}\label{sec:appA-spectral-weight}

\begin{definition}[Band-restricted spectral weight]
\label{def:band-restricted-local-spectral-weight}
Fix $x \in \mathcal M$, a sector dimension $d$ with $\gamma_d(x) > 0$ (Definition~\ref{def:spectral-sector}), a Hermitian operator $O$, and $\Lambda > 0$. The band-restricted spectral weight of $O$ sums the squared matrix elements of $O$ over the levels above the sector whose transition frequency is below $\Lambda$, and averages them over the $d$ sector states,
\begin{equation}\label{equation: local coupling window}
W^{O}_{[\gamma_d,\,\Lambda)}(x) := \frac{1}{d} \sum_{\substack{j < d \le k \\ \omega_{kj}(x) < \Lambda}} \bigl| \langle \psi_k(x) | O | \psi_j(x) \rangle \bigr|^2 .
\end{equation}
For the ground-state manifold, $d = r(x)$, we write $W^{O}_{[\gamma,\,\Lambda)}(x)$.
\end{definition}

\begin{figure}[t]
\centering
\tikzset{
  lvl/.style={black, line width=1.3pt, line cap=butt},
  lvlout/.style={black!35, line width=1.3pt, line cap=butt},
  elab/.style={figtext, anchor=east, inner sep=0pt},
  note/.style={figtext, text=black!55},
}%
\begin{tikzpicture}[x=1cm, y=1cm]
\fill[figblue!12] (1.0,1.70) rectangle (9.0,4.00);
\draw[figblue, line width=0.8pt] (0,4.00) -- (9.0,4.00);
\draw[black!60, line width=0.5pt, dash pattern=on 3pt off 2pt] (0,0.44) -- (9.0,0.44);
\draw[figarrow] (0,-0.35) -- (0,5.35);
\node[figtext, anchor=south, inner sep=2pt] at (0,5.35) {energy};
\foreach \y in {0,1.70} \draw[line width=0.5pt] (-0.07,\y) -- (0.07,\y);
\node[elab] at (-0.18,-0.015) {$E_0$};
\node[elab] at (-0.18,0.475) {$E_{d-1}$};
\node[elab] at (-0.18,1.70) {$E_{d}$};
\node[elab, text=figblue] at (-0.18,4.00) {$E_{d-1}+\Lambda$};
\foreach \y in {0,0.22,0.44} \draw[lvl] (1.0,\y) -- (3.6,\y);
\node[figtext, align=left, anchor=north west, inner sep=0pt] at (1.0,-0.28) {sector $\mathcal S(x)$: the $d$ lowest levels\\state $\rho_d(x)=P(x)/d$};
\draw[{Stealth[length=4pt, width=3pt]}-{Stealth[length=4pt, width=3pt]}, line width=0.5pt] (1.45,0.48) -- (1.45,1.66);
\node[figtext, anchor=west, inner sep=3pt] at (1.45,1.07) {$\gamma_d(x)$};
\node[note, font=\normalsize\itshape] at (6.2,1.07) {no levels};
\foreach \y in {1.70,2.15,2.55,3.05,3.45} \draw[lvl] (5.0,\y) -- (7.4,\y);
\node[figtext, anchor=west, inner sep=3pt] at (7.4,2.55) {$|\psi_k(x)\rangle$};
\foreach \y in {4.40,4.80} \draw[lvlout] (5.0,\y) -- (7.4,\y);
\node[note, anchor=west, inner sep=3pt] at (7.4,4.60) {above $E_{d-1}+\Lambda$: not counted};
\foreach \y/\w in {3.05/0.15, 2.15/0.25, 3.45/0.35, 2.55/0.55, 1.70/1.0} {
  \pgfmathsetmacro{\op}{50+50*\w}
  \pgfmathsetmacro{\lw}{0.5+1.3*\w}
  \draw[-{Stealth[length=5pt, width=4pt]}, figorange!\op, line width={\lw pt}]
    (3.62,0.22) .. controls (4.25,0.35*\y+0.25) and (4.45,\y) .. (4.94,\y);
}
\node[figtext, text=figorange, inner sep=0pt] at (3.94,2.35) {$O$};
\draw[line width=0.5pt, decorate, decoration={brace, amplitude=5pt}] (9.12,4.00) -- (9.12,1.70);
\node[figtext, anchor=west, align=left, inner sep=0pt] at (9.42,2.85) {$W^{O}_{[\gamma_d,\Lambda)}(x)$:\\[2pt]total weight\\in the window};
\draw[-{Stealth[length=5pt, width=4pt]}, figorange, line width=1.8pt] (7.4,-0.481) -- (8.15,-0.481);
\node[figtext, anchor=west, inner sep=3pt] at (8.18,-0.481) {large weight};
\draw[-{Stealth[length=5pt, width=4pt]}, figorange!57, line width=0.7pt] (7.4,-0.959) -- (8.15,-0.959);
\node[figtext, anchor=west, inner sep=3pt] at (8.18,-0.959) {small weight};
\end{tikzpicture}

\caption{The band-restricted spectral weight of Definition~\ref{def:band-restricted-local-spectral-weight}. Each arrow deposits the weight of the transitions from the sector into a level $|\psi_k(x)\rangle$ above it. The shaded window collects the transitions with frequency $\omega_{kj} < \Lambda$: it opens at the band gap, at the level $E_d(x)$, and, for the thin sector drawn, closes at $E_{d-1}(x)+\Lambda$.}
\label{fig:spectral-weight}
\end{figure}
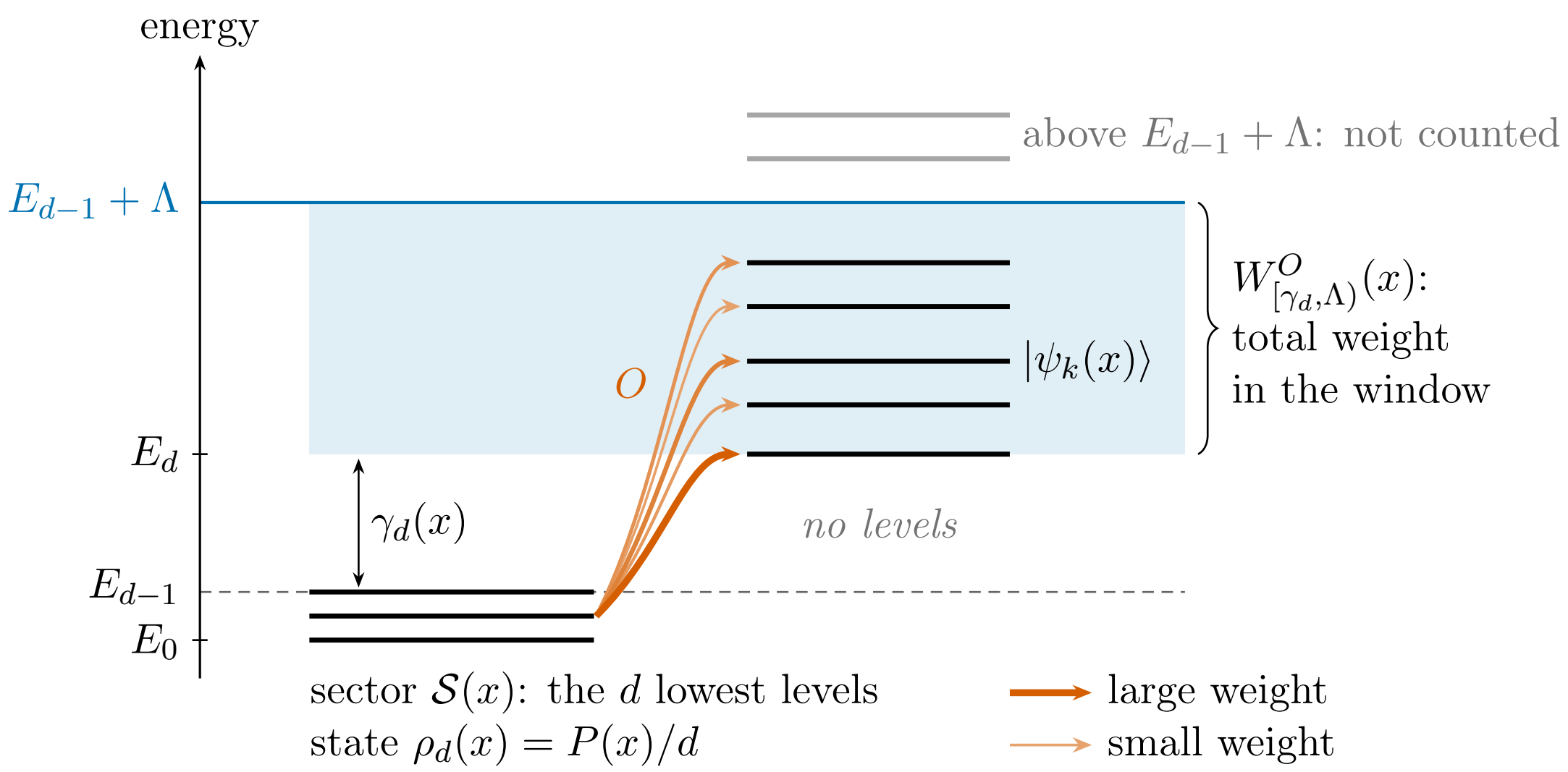

Each term in the sum satisfies $\omega_{kj}(x) \geq E_d(x) - E_{d-1}(x) = \gamma_d(x)$, so the spectral weight collects transitions in the window $[\gamma_d(x),\Lambda)$ (Figure~\ref{fig:spectral-weight}) and vanishes when $\Lambda \leq \gamma_d(x)$.
The band gap ensures no degenerate level straddles the boundary between the sector and the levels above it, so the spectral weight does not depend on the choice of eigenbasis within each degenerate level.
The spectral weight is quadratic in the operator,
\begin{equation}\label{eq:weight-scaling}
\begin{gathered}
    W^{cO}_{[\gamma_d,\,\Lambda)}(x) = c^2\, W^{O}_{[\gamma_d,\,\Lambda)}(x)\quad (c \in \mathbb R),
    \\
    \text{so that}\qquad W^{\partial_s H}_{[\gamma_d,\,\Lambda)}(x(s)) = |\dot x(s)|^2\; W^{\widehat u(s) \cdot \nabla_x H}_{[\gamma_d,\,\Lambda)}(x(s))
\end{gathered}
\end{equation}
by Eq.~\eqref{eq:path-derivative}.

Both experiments and the response measurements in our algorithms use local probes, so we restrict $O$ to the class $\mathcal{L}$ of normalized local operators with constant range $\ell = O(1)$,
\begin{equation}\label{eq:local-class}
    O = \sum_{P \in \mathcal{P}_\ell} z_P\, P,\qquad z_P \in \mathbb{R},\qquad \sum_{P \in \mathcal{P}_\ell} z_P^2 \leq 1,
\end{equation}
where $\mathcal{P}_\ell$ is the set of all Pauli strings that act nontrivially on at most $\ell$ sites in a ball of radius $\ell$.
Let $N_\mathcal{L} := |\mathcal{P}_\ell|$ denote the number of these Pauli strings, which equals the dimension of the span of $\mathcal{L}$ and is $O(n)$ since $\ell$ is constant.

\subsection{Finite-size quantum phases and certified discovery of quantum phase diagrams}\label{sec:finite-depth-qac}

Admissible circuits are defined so that they can contain the circuits that approximate the quasi-adiabatic propagator in Proposition~\ref{prop:fdlu-from-gapped-path} below.
These have depth $2^{\dlat}$ and gates on sets of polylogarithmic diameter.

\begin{definition}[Admissible circuits]\label{def:admissible-circuits}
Fix a radius function $R(n) = \mathrm{polylog}(n)$ and a depth bound $D = O(1)$ with $D \ge 2^{\dlat}$. We define a class $\mathcal C$ of admissible circuits. An admissible circuit is a quantum circuit of depth at most $D$ acting on the $n$ sites in which every gate acts on a set of diameter at most $R(n)$. In particular, an admissible circuit has a causal cone of radius at most $D\,R(n)$. For symmetry-protected phases, we additionally require that every gate respects the on-site symmetry protecting the phase. In that case, we also assume that every $h_j(x)$ respects it.
\end{definition}

\begin{definition}[Finite-size quantum phases]\label{def:finite-size-phase}
Fix $1/\mathrm{poly}(n) \le \eta_B(n) < \eta_P(n) < 1$. Let $x_A, x_B \in \mathcal M$ have ground-state degeneracies $r_A = r(x_A)$ and $r_B = r(x_B)$. If $r_A \ne r_B$, then $\rho_{\mathcal G}(x_A)$ and $\rho_{\mathcal G}(x_B)$ are in different phases. If $r_A = r_B = r$, we use the overlap $\operatorname{Tr}\bigl[P_{\mathcal G}(x_B)\,C\,\rho_{\mathcal G}(x_A)\,C^\dagger\bigr]$ of an admissible circuit $C \in \mathcal C$, the probability that the transported ground-state manifold of $x_A$ lands inside that of $x_B$.
\begin{enumerate}
\item[(i)] $\rho_{\mathcal G}(x_A)$ and $\rho_{\mathcal G}(x_B)$ are in the same phase if there exists $C \in \mathcal C$ with
\begin{equation}\label{eq:same-phase-appendix}
\operatorname{Tr}\bigl[P_{\mathcal G}(x_B)\,C\,\rho_{\mathcal G}(x_A)\,C^\dagger\bigr] \ge 1 - \eta_B(n).
\end{equation}
\item[(ii)] $\rho_{\mathcal G}(x_A)$ and $\rho_{\mathcal G}(x_B)$ are in different phases if for every $C \in \mathcal C$
\begin{equation}\label{eq:different-phase-appendix}
\operatorname{Tr}\bigl[P_{\mathcal G}(x_B)\,C\,\rho_{\mathcal G}(x_A)\,C^\dagger\bigr] \le 1 - \eta_P(n).
\end{equation}
\end{enumerate}
\end{definition}

Definition~\ref{def:finite-size-phase} does not depend on the order of the endpoints.
Since $\rho_{\mathcal G} = P_{\mathcal G}/r$ at both endpoints, cyclicity of the trace gives
\[
\operatorname{Tr}\bigl[P_{\mathcal G}(x_B)\,C\,\rho_{\mathcal G}(x_A)\,C^\dagger\bigr] = \operatorname{Tr}\bigl[P_{\mathcal G}(x_A)\,C^\dagger\,\rho_{\mathcal G}(x_B)\,C\bigr],
\]
the overlap of $C^\dagger$ with the endpoints exchanged.
The class $\mathcal{C}$ is closed under inversion, because $C^\dagger$ has the same depth and gate diameters as $C$ and inverse gates respect the protecting symmetry.

The phases are defined using admissible circuits in Definition~\ref{def:finite-size-phase}, but a path only gives the propagator $U_\Delta[x]$.
To prove Corollary~\ref{cor:gapped-path-same-phase}, we need to approximate $U_\Delta[x]$ by an admissible circuit.

\begin{proposition}[Circuit approximation of the quasi-adiabatic propagator~\cite{coser2019classification}]\label{prop:fdlu-from-gapped-path}
Let $x(s)$ be a path (Definition~\ref{def:path}) of length $L = O(1)$ and let $\Delta = \Theta(1)$ be a cutoff.
For every $k > 0$ and $\kappa > 0$, there exists a circuit $C$ of depth $2^{\dlat}$ and gate diameter $O(\log^{1 + \kappa} n)$ that approximates the quasi-adiabatic propagator for the entire path in operator norm,
\begin{equation}\label{eq:circuit-approximation}
    \bigl\| U_\Delta[x] - C \bigr\| \le n^{-k},
\end{equation}
where the constants depend on $k$, $\kappa$, $\Delta$, $\dlat$ and the filter.
If $R(n)$ is no smaller than the gate diameter, the circuit $C$ is admissible, $C \in \mathcal{C}$.
\end{proposition}

The proof follows Step 1.2 of Ref.~\cite{coser2019classification}.
The generator $D_\Delta(s)$ has one term for each local term of $\partial_s H(s)$.
First, by a Lieb--Robinson bound for each $H(s)$ and the sub-exponential decay of the BMNS filter, each term in $D_\Delta(s)$ is supported on a ball of radius $O(\log^{1 + \kappa} n / \Delta)$ up to error $n^{-k}$.
Second, the block decomposition of Ref.~\cite{coser2019classification} decomposes the time-ordered exponential into $2^{\dlat}$ layers of quasi-local gates, multiplied together.
Each lattice direction doubles the number of layers. If every $h_j(x)$ respects the protecting symmetry, so does every gate of this construction.

\begin{remark}[The circuit approximation applies to piecewise smooth paths]
\label{rem:piecewise-smooth}
Paths in Definition~\ref{def:path} may have finitely many kinks.
The construction of Ref.~\cite{coser2019classification} localizes each term in $D_\Delta(s)$ for a fixed $s$ and then applies the block decomposition to the time-ordered exponential.
Both steps require only that the local terms of $\partial_s H(s)$ be bounded and that $H(s)$ and $\partial_s H(s)$ be piecewise continuous in $s$, so Eq.~\eqref{eq:circuit-approximation} holds with the same constants.
\end{remark}

Proposition~\ref{prop:fdlu-from-gapped-path} does not require a gap along the path.
A gap is only needed to ensure that $U_\Delta$ maps the ground-state manifold of $x_A$ into that of $x_B$. The circuit then certifies the same phase.

\begin{corollary}[A gapped path certifies the same phase]\label{cor:gapped-path-same-phase}
Consider a path of length $L = O(1)$ from $x_A$ to $x_B$ with cutoff $\Delta = \Theta(1)$. Choose $k$ in Proposition~\ref{prop:fdlu-from-gapped-path} so that $2 n^{-k} \le \eta_B(n)$, which is possible because $\eta_B(n) \ge 1/\mathrm{poly}(n)$. Suppose the ground-state degeneracy is constant along the path, the spectral gap is $\gamma(x(s)) \ge \Delta$ for all $s \in [0,1]$, and $R(n)$ is at least the gate diameter of Proposition~\ref{prop:fdlu-from-gapped-path}. Then $\rho_{\mathcal G}(x_A)$ and $\rho_{\mathcal G}(x_B)$ are in the same phase.
\end{corollary}
\begin{proof}
The ground-state degeneracy is constant, so $r_A = r_B$.
Along the path, each transition that leaves the ground-state manifold has energy at least $\gamma(x(s)) \geq \Delta$.
By Lemma~\ref{lem:generator-matrix-elements}, on these transitions $D_\Delta(s)$ equals the exact adiabatic generator.
Then $U_\Delta$ carries the ground-state manifold of $x_A$ exactly onto that of $x_B$ (Corollary~\ref{cor:exact-transport}). In particular,
\begin{equation}\label{eq:exact-transport-overlap}
    \mathrm{Tr}\bigl[ P_{\mathcal G}(x_B)\, U_\Delta\, \rho_{\mathcal G}(x_A)\, U_\Delta^\dagger \bigr] = 1 .
\end{equation}
The circuit $C$ from Proposition~\ref{prop:fdlu-from-gapped-path} is admissible by the hypothesis on $R(n)$.
Replacing $U_\Delta$ by $C$ changes the overlap by at most $2 \lVert U_\Delta - C \rVert \leq 2 n^{-k} \leq \eta_B(n)$.
We used $\bigl| \mathrm{Tr}\bigl[ P ( U \rho U^\dagger - C \rho C^\dagger ) \bigr] \bigr| \le 2 \| U - C \|$ for any projector $P$, state $\rho$, and unitaries $U$, $C$~\cite{nielsen2002quantum}.
\end{proof}

The converse of Corollary~\ref{cor:gapped-path-same-phase} is not true in general.
Suppose that every path of length $O(1)$ from $x_A$ to $x_B$ inside $\mathcal{M}$ contains a point where the spectral gap closes.
Then it is still possible that $\rho_{\mathcal{G}}(x_A)$ and $\rho_{\mathcal{G}}(x_B)$ are in the same phase, because a gapped path may exist outside $\mathcal{M}$.
This caveat does not apply in the following setting.
Suppose that $\mathcal{M}$ is the space of all translation-invariant Hamiltonians with a fixed interaction range (a constant-dimensional space) and that $\mathcal{C}$ is restricted to translation-invariant circuits.
Then every separation of the parameter space into different phases is a genuine phase transition~\cite{essin2013classifying, cheng2016translational, barkeshli2019symmetry}.
In general, a gapped path joining two points in the same phase need not lie in the family.
Lemma~\ref{lem:same-phase-gapped-path} below shows that, when $\eta_B$ is small, every resolved pair in the same phase is joined by a path of Hamiltonians whose terms have polylogarithmic range and along which the band gap above the lowest $r$ levels stays open, where $r$ is the common ground-state degeneracy.
The maximally mixed ground state $\rho_{\mathcal G}(x)=P_{\mathcal G}(x)/r(x)$ determines the ground-state manifold, so the circuit condition of Definition~\ref{def:finite-size-phase} constrains ground-state manifolds and needs no choice of basis when $r(x)>1$.

\begin{lemma}[Same phase implies a gapped Hamiltonian path]\label{lem:same-phase-gapped-path}
Let $x_A,x_B\in\mathcal M$ have the same ground-state degeneracy $r$ and spectral gaps $\gamma(x_A),\gamma(x_B)\ge\gamma_{\mathrm{lb}}$. Suppose that an admissible circuit $C\in\mathcal C$ satisfies
\begin{equation}\label{eq:same-phase-path-overlap}
    \operatorname{Tr}\bigl[P_{\mathcal G}(x_B)\,C\,\rho_{\mathcal G}(x_A)\,C^\dagger\bigr]\ge1-\eta
    \qquad\text{with}\qquad
    8Jr\eta\le\gamma_{\mathrm{lb}}.
\end{equation}
Then $H(x_A)$ and $H(x_B)$ are joined by a continuous path of Hamiltonians, each a sum of at most $2J$ terms of norm at most one supported on balls of radius $O(D\,R(n))$, along which the band gap above the lowest $r$ levels is at least $\gamma_{\mathrm{lb}}/2$. If the protecting symmetry is unitary, every Hamiltonian on the path respects it.
\end{lemma}
\begin{proof}
The path has two segments. Write $C$ as at most $D$ layers of gates, where the gates of one layer act on disjoint sets of diameter at most $R(n)$. Each gate $g$ equals $e^{-iK_g}$ for the Hermitian operator $K_g$ with spectrum in $[-\pi,\pi)$ obtained from the spectral decomposition of $g$. This $K_g$ acts on the same set as $g$ and commutes with every operator that commutes with $g$. Replacing every gate $g$ by $e^{-itK_g}$ gives circuits $C(t)$, $t\in[0,1]$, with $C(0)=I$ and $C(1)=C$. On the first segment, $C(t)H(x_A)C(t)^\dagger$ is unitarily equivalent to $H(x_A)$, so its band gap above the lowest $r$ levels stays $\gamma(x_A)$, and each term $C(t)h_j(x_A)C(t)^\dagger$ is supported within distance $D\,R(n)$ of the support of $h_j(x_A)$.

The second segment is $H(t):=(1-t)H'+tH(x_B)$, $t\in[0,1]$, with $H':=CH(x_A)C^\dagger$. It is a sum of the $2J$ terms $(1-t)Ch_j(x_A)C^\dagger$ and $th_j(x_B)$, each of norm at most one. The projectors $Q:=CP_{\mathcal G}(x_A)C^\dagger$ and $P:=P_{\mathcal G}(x_B)$ have rank $r$ and project onto the lowest $r$ levels of $H'$ and $H(x_B)$. Since $\rho_{\mathcal G}(x_A)=P_{\mathcal G}(x_A)/r$ and $\operatorname{Tr}P=\operatorname{Tr}Q=r$, Eq.~\eqref{eq:same-phase-path-overlap} gives
\begin{equation*}
    \|(I-P)Q\|^2\le\operatorname{Tr}[(I-P)Q]\le r\eta,
    \qquad
    \|(I-Q)P\|^2\le\operatorname{Tr}[(I-Q)P]\le r\eta.
\end{equation*}
Let $E_0(\cdot)\le E_1(\cdot)\le\cdots$ denote the eigenvalues of a Hermitian matrix in increasing order. The spectrum of $H'$ lies in $[-J,J]$, and $H'$ commutes with $Q$. For a unit vector $\psi$ in the range of $P$, we therefore have $\langle\psi|H'|\psi\rangle\le E_{r-1}(H')+2J\|(I-Q)\psi\|^2\le E_{r-1}(H')+2Jr\eta$, and for a unit vector $\psi$ orthogonal to that range, $\langle\psi|H'|\psi\rangle\ge E_r(H')-2J\|Q\psi\|^2\ge E_r(H')-2Jr\eta$. The min-max principle with the range of $P$ then bounds $E_{r-1}(H(t))$ from above and $E_r(H(t))$ from below, and
\begin{equation*}
    E_r(H(t))-E_{r-1}(H(t))
    \ge
    (1-t)\bigl(\gamma(x_A)-4Jr\eta\bigr)+t\,\gamma(x_B)
    \ge
    \gamma_{\mathrm{lb}}-4Jr\eta
    \ge
    \frac{\gamma_{\mathrm{lb}}}{2}.
\end{equation*}
If the protecting symmetry is unitary, every gate and every $h_j(x)$ commute with it. Each $K_g$ is a polynomial in $g$, so every $K_g$, every $C(t)$ and every Hamiltonian on both segments commute with it as well.
\end{proof}

Lemma~\ref{lem:same-phase-gapped-path} needs small tolerances $\eta_B$ and $\eta_P$. It applies to every resolved pair in the same phase when $8Jr\eta_B\le\gamma_{\mathrm{lb}}$. Clause~(i) of Definition~\ref{def:phase-clustering} below guarantees only that linked resolved points are not in different phases, that is, they have the same ground-state degeneracy $r$ and some admissible circuit gives an overlap above $1-\eta_P$, so every link between resolved points has this interpretation when $8Jr\eta_P\le\gamma_{\mathrm{lb}}$. Since $J=O(n)$ and $\gamma_{\mathrm{lb}}\ge1/\mathrm{poly}(n)$, both bounds are inverse-polynomial whenever $r\le\mathrm{poly}(n)$, and so they are compatible with $\eta_B(n)\ge1/\mathrm{poly}(n)$ in Definition~\ref{def:finite-size-phase}. With overlap one, the same construction keeps the band gap at least $\min\{\gamma(x_A),\gamma(x_B)\}$ along the whole path.

\begin{definition}[Certified discovery of quantum phase diagrams]\label{def:phase-clustering}
\emergencystretch=1.5em
\emph{Input.} A threshold $\gamma_{\mathrm{thresh}} = \Theta(1)$, a resolution $1/\mathrm{poly}(n) \le \gamma_{\mathrm{lb}} \le \gamma_{\mathrm{thresh}}$, and points $x_1, \dots, x_N \in \mathcal{M}$ at which the phase diagram is to be learned.
Throughout, we assume that the tolerance from Definition~\ref{definition: ground-state manifold} satisfies $\delta_{\mathrm{gsm}} < \gamma_{\mathrm{lb}}$.

\emph{Output.} A symmetric relation called \emph{linked} on the input points $x_1, \dots, x_N \in \mathcal{M}$, and the partition of the points into \emph{clusters}, the connected components of the relation.

The output must satisfy the following requirements.
For every pair $x_i, x_j$ with $\gamma(x_i) \ge \gamma_{\mathrm{lb}}$ and $\gamma(x_j) \ge \gamma_{\mathrm{lb}}$,

\begin{enumerate}

\item[(i)]
 if $x_i$ and $x_j$ are linked, then $\rho_{\mathcal{G}}(x_i)$ and $\rho_{\mathcal{G}}(x_j)$ must not lie in different phases according to Definition~\ref{def:finite-size-phase};

\item[(ii)]
 if $x_i$ and $x_j$ are in different clusters, then either $\rho_{\mathcal{G}}(x_i)$ and $\rho_{\mathcal{G}}(x_j)$ lie in different phases, or for every path $x(s)$ of length $O(1)$ from $x_i$ to $x_j$ contained in $\mathcal{M}$, there exists a point $x(s)$ with $\gamma(x(s)) < \gamma_{\mathrm{thresh}}$.

\end{enumerate}
\end{definition}

Clause~(i) is weaker than requiring every two resolved points of a cluster to lie in the same phase.
A linked pair satisfying clause~(i) has equal ground-state degeneracies, and some admissible circuit $C$ connects the pair with overlap above $1-\eta_P$; this overlap may fall below the same-phase level $1-\eta_B$ from Definition~\ref{def:finite-size-phase}.
Moreover, clause~(i) applies to links, not to clusters.
Chaining these circuits connects any two points joined by a sequence of such links, but the depth and error can exceed their fixed bounds, so the relation of being connected by some admissible circuit is not transitive.

Clause~(ii) does not assert that different clusters are different phases: two gapped regimes in the same phase can be separated by gap closings along every path of length $O(1)$ in $\mathcal M$.
The quasi-phases of Ref.~\cite{jones2026quasiphases} are an example: a gapped path connects them in a larger parameter space, but no such path is built from small perturbations of their critical point.

\begin{remark}[Relaxed guarantee]
\label{rem:phase-clustering-relaxation}
In a weaker version of Definition~\ref{def:phase-clustering}, each link comes with a reported path $x(s)$ of length $O(1)$ in $\mathcal M$ between its endpoints. Clause~(i) then applies only when
\[
    \gamma(x(s))\geq\gamma_{\mathrm{lb}}
    \qquad\text{for all }s\in[0,1].
\]
All other requirements, including clause~(ii), are the same. The algorithm does not need to identify which reported paths satisfy this condition.
\end{remark}

\begin{definition}[Resolved region]\label{def:large-gap-component}
Fix a resolution $\gamma_{\mathrm{lb}}$ with $1/\mathrm{poly}(n) \le \gamma_{\mathrm{lb}}$ and $\delta_{\mathrm{gsm}} < \gamma_{\mathrm{lb}}$. A point $x \in \mathcal M$ with $\gamma(x) \ge \gamma_{\mathrm{lb}}$ is \emph{resolved}. The resolved points form the \emph{resolved region}. A \emph{connected component} $R$ of the resolved region means one of its path-connected components.
\end{definition}

The ground-state degeneracy is constant on each connected component of the resolved region.
Suppose a level enters or leaves the ground-state manifold along a path in the resolved region, which happens only where the level comes within $\delta_{\mathrm{gsm}}$ of the ground-state energy.
Consider nearby points of the path where the level lies outside the ground-state manifold.
Since the tolerance is smaller than the resolution, continuity keeps the level below $E_0 + \gamma_{\mathrm{lb}}$ there.
If $r$ is the ground-state degeneracy at such a point, the level lies at or above $E_r$.
The spectral gap there satisfies $\gamma = E_r - E_{r-1} \le E_r - E_0 < \gamma_{\mathrm{lb}}$.
Such a point is not resolved, a contradiction.
Hence each connected component $R$ of the resolved region carries a single ground-state degeneracy $r$, and $\gamma(x) = \gamma_r(x) \ge \gamma_{\mathrm{lb}}$ on $R$.

\section{Continuous Phase Transitions at Fixed Degeneracy Cannot Be Dark}
\label{sec:cannot-be-dark}

This appendix proves that continuous phase transitions at fixed degeneracy cannot be dark (Theorem~\ref{thm:cannot-be-dark}).
Consider a path of constant length whose endpoints have the same ground-state degeneracy $r$ and lie in different phases (Definition~\ref{def:finite-size-phase}).
Wherever the ground-state degeneracy is $r$, the band gap $\gamma_r$ of the lowest $r$ levels equals the spectral gap.
Suppose $\gamma_r$ is at least inverse-polynomial along the path.
Then the path-gradient operator $\partial_s H(s)$ has band-restricted spectral weight (Definition~\ref{def:band-restricted-local-spectral-weight}) at least inverse-polynomial below a constant cutoff.
This holds on a set of $s \in [0,1]$ of inverse-polynomial measure.
The proof has two steps.
Losing overlap under the quasi-adiabatic propagator requires spectral weight.
Different phases force this loss.
By definition, endpoints with different ground-state degeneracies lie in different phases, for example a unique ground state and a symmetry-broken or topologically ordered phase.
Appendices~\ref{sec:local-detection} and~\ref{sec:detection} handle that case by estimating the ground-state degeneracy.

Throughout, we work along a path $x : [0,1] \to \mathcal M$ from $x_A = x(0)$ to $x_B = x(1)$ with path length $L$ (Definition~\ref{def:path}).
The Hamiltonian $H(x)$ has $J$ local terms with directional derivatives of norm at most one (Definition~\ref{definition: geometrically local hamiltonians}).
The path has speed at most $\zeta_s L$, so $\| \partial_s H(s) \| \leq \zeta_s L J$.

\subsection{Lack of an admissible circuit implies non-negligible spectral weight}

The sector at $x$ is the span of the lowest $d$ levels (Appendix~\ref{subsec:sector-setup}), with projector $P(x)$, sector state $\rho_d(x)$, and band gap $\gamma_d$. We fix $d$ and write $\gamma_{\min}$ for the minimum of $\gamma_d$ along the path.

\begin{theorem}[Spectral weight of the path-gradient operator]\label{lem:path-gradient-spectral-weight}
Suppose the cutoff $\Delta > 0$ and the quasi-adiabatic propagator $U_\Delta$ along the path satisfy
\begin{equation}\label{eq:small-endpoint-overlap}
\mathrm{Tr}\bigl[P(x(1))\, U_\Delta\, \rho_d(x(0))\, U_\Delta^\dagger\bigr] \le 1 - \eta(n).
\end{equation}
For every number $0 \le C < 1$, there exists a subset $S_{\mathrm{high}} \subseteq [0,1]$ with measure
\begin{equation}\label{eq-S-bad-measure}
|S_{\mathrm{high}}| \ge \frac{(1-C)\,\gamma_{\min}\sqrt{\eta(n)}}{\zeta_s L J}
\end{equation}
such that for every $s \in S_{\mathrm{high}}$, the band-restricted spectral weight of the path-gradient operator (Definition~\ref{def:band-restricted-local-spectral-weight}) satisfies
\begin{equation}\label{eq:path-gradient-spectral-weight-lower-bound}
W^{\partial_s H}_{[\gamma_d,\,\Delta)}(s) \ge C^2\,\eta(n)\,\gamma_{\min}^2 .
\end{equation}
\end{theorem}
\begin{proof}
See Appendix~\ref{subsec:proof-complete-detection}.
\end{proof}

Different phases bound the overlap only for admissible circuits, not for the propagator $U_\Delta[x]$ (Definition~\ref{def:finite-size-phase}(ii)).
We therefore approximate $U_\Delta[x]$ by an admissible circuit $C$ at constant cutoff (Proposition~\ref{prop:fdlu-from-gapped-path}).
This approximation needs no gap along the path.

\begin{proposition}[Different phases bound the overlap after transport]\label{prop:phases-force-deficit}
Fix a cutoff $\Delta = \Theta(1)$ and a path $x(s)$ of length $L = O(1)$ with $r_A = r_B$, whose endpoints lie in different phases (Definition~\ref{def:finite-size-phase}(ii)) for an admissible class $\mathcal{C}$.
Proposition~\ref{prop:fdlu-from-gapped-path} gives a circuit that approximates $U_\Delta[x]$ to accuracy $\eta_P/8$.
Suppose that $R(n)$ is at least its gate diameter.
Then
\begin{equation}\label{eq:phases-force-deficit}
\mathrm{Tr}\bigl[ P_{\mathcal G}(x_B)\, U_\Delta[x]\, \rho_{\mathcal G}(x_A)\, U_\Delta[x]^\dagger \bigr] \le 1 - \tfrac{3}{4}\, \eta_P .
\end{equation}
\end{proposition}
\begin{proof}
Since $\eta_P \geq 1 / \mathrm{poly}(n)$, Proposition~\ref{prop:fdlu-from-gapped-path} gives a circuit $C \in \mathcal{C}$ with $\| U_{\Delta}[x] - C \| \leq \eta_P / 8$.
From Eq.~\eqref{eq:different-phase-appendix}, $\mathrm{Tr}[P_{\mathcal G}(x_B) C \rho_{\mathcal G}(x_A) C^\dagger]$ is at most $1 - \eta_P$.
Replacing $C$ with $U_\Delta[x]$ changes the overlap by at most $2\, \| U_\Delta[x] - C \| \le \eta_P/4$ (proof of Corollary~\ref{cor:gapped-path-same-phase}), which gives the claim.
\end{proof}

\begin{theorem}[Continuous phase transitions at fixed degeneracy cannot be dark]\label{thm:cannot-be-dark}
Fix a cutoff $\Delta = \Theta(1)$ and a path $x(s)$ that satisfy the hypotheses of Proposition~\ref{prop:phases-force-deficit}, with ground-state degeneracy $r$ at both endpoints.
Where $\dot x(s) \ne 0$, dividing the path-gradient operator by the speed $|\dot x(s)| \le \zeta_s L$ gives $\widehat u(s) \cdot \nabla_x H$ (Definition~\ref{def:path}).
Suppose the lowest $r$ levels have minimum band gap $\gamma_{\min} > 0$ (Eq.~\eqref{eq:path-min-gap} with $d = r$).
Then there is a set $S_{\mathrm{high}} \subseteq [0,1]$ of measure
\begin{equation*}
|S_{\mathrm{high}}| \ge \frac{\gamma_{\min}\sqrt{3\,\eta_P}}{4\, \zeta_s L J},
\end{equation*}
such that for every $s \in S_{\mathrm{high}}$
\begin{equation}\label{eq:cannot-be-dark-threshold}
W^{\widehat u(s) \cdot \nabla_x H}_{[\gamma_r,\,\Delta)}(x(s)) \ge \frac{3\,\eta_P\, \gamma_{\min}^2}{16\, (\zeta_s L)^2}.
\end{equation}
If $\gamma_{\min} \geq 1/\mathrm{poly}(n)$, both bounds are inverse polynomial in $n$.
\end{theorem}
\begin{proof}
See Appendix~\ref{subsec:proof-complete-detection}.
\end{proof}

\subsection{The diabatic coupling}\label{sec:gauge-transformation}

To prove Theorem~\ref{lem:path-gradient-spectral-weight}, we bound how fast $U_\Delta$ moves probability out of the sector.
We control this rate through the matrix elements $M_{kj}(s) = \langle \psi_k(s) | \partial_s H(s) | \psi_j(s) \rangle$ between the sector levels $j < d$ and the levels $k \ge d$ above them.
Their squared magnitudes, collected over transition frequencies $\omega_{kj}$ in the window $[\gamma_d, \Delta)$, form the band-restricted spectral weight of the path-gradient operator in Definition~\ref{def:band-restricted-local-spectral-weight},
\begin{equation}\label{eq:M_vcoup_relation}
W^{\partial_s H}_{[\gamma_d,\,\Delta)}(s) = \frac{1}{d} \sum_{\substack{j < d \le k \\ \omega_{kj}(s) < \Delta}} |M_{kj}(s)|^2 .
\end{equation}

\emph{Co-moving frame.}
As motivation only, we first derive the transport generator in the co-moving frame of the instantaneous eigenstates, wherever the eigenpairs can be chosen differentiable in $s$.
The rest of the argument uses a gauge-invariant form without this condition.
Where $\gamma_d(s) > 0$, no eigenvalue crosses the sector boundary.
Define
\begin{equation}\label{eq:Ugauge-def}
U_{\mathrm{gauge}}(s)
 := \sum_{j,j' < d}
   V^{\mathrm{sec}}_{jj'}(s)\,|\psi_j(s)\rangle\langle\psi_{j'}(0)|
 + \sum_{k,k'\ge d}
   V^{\mathrm{ex}}_{kk'}(s)\,|\psi_k(s)\rangle\langle\psi_{k'}(0)|,
\end{equation}
where $V^{\mathrm{sec}}$ and $V^{\mathrm{ex}}$ are unitary matrices, fixed below, acting within the sector and within its orthogonal complement, respectively.
The derivative of this frame defines
\begin{equation}\label{eq:Berry-connection-def}
A(s) := i\,\partial_s U_{\mathrm{gauge}}(s)\,U_{\mathrm{gauge}}^\dagger(s).
\end{equation}
Within each block $\bullet \in \{\mathrm{sec}, \mathrm{ex}\}$, its matrix elements are $i\langle\psi_k(s)|\partial_s\psi_{k'}(s)\rangle + i\bigl(\partial_s V^{\bullet}(s)\,V^{\bullet\dagger}(s)\bigr)_{kk'}$.
For $k, k'$ in the block $\bullet$, let $\widetilde A^{\bullet}_{kk'}(s) := i\langle\psi_k(s)|\partial_s\psi_{k'}(s)\rangle$.
We fix $V^{\mathrm{sec}}$ and $V^{\mathrm{ex}}$ by parallel transport:
\begin{equation}\label{eq:excited-gauge-fixing}
\partial_s V^{\bullet}(s) = i\,\widetilde A^{\bullet}(s)\,V^{\bullet}(s),
\qquad V^{\bullet}(0) = I,
\end{equation}
so that $A_{kk'}(s) = 0$ whenever $k$ and $k'$ lie in the same block.
For a quasi-degenerate sector, such as the ground-state manifold, $\widetilde A^{\mathrm{sec}}$ may be large, but this gauge choice cancels it exactly.
Across the blocks, $V^{\mathrm{sec}}$ and $V^{\mathrm{ex}}$ do not contribute, so the cross elements are $A_{kj}(s) = i\langle\psi_k(s)|\partial_s\psi_j(s)\rangle$.
Differentiating $H(s)|\psi_j(s)\rangle = E_j(s)|\psi_j(s)\rangle$ gives, for $k \ge d > j$,
\begin{equation}\label{eq:Berry-matrix-elements-beforefinal}
\langle\psi_k(s)|\partial_s\psi_j(s)\rangle
 = \frac{\langle\psi_k(s)|\partial_s H(s)|\psi_j(s)\rangle}
        {E_j(s) - E_k(s)}
 = -\frac{M_{kj}(s)}{\omega_{kj}(s)},
\end{equation}
so the transport generator is
\begin{equation}\label{eq:Berry-matrix-elements-final}
A_{kk'}(s) = 0 \ \ \text{($k, k'$ in the same block)},
\qquad
A_{kj}(s) = A_{jk}(s)^* = -i\,\frac{M_{kj}(s)}{\omega_{kj}(s)},
\qquad k \ge d > j.
\end{equation}

\emph{Gauge-invariant form.}
This gauge-fixed generator is block off-diagonal, with exactly these cross elements.
From here on, we define $A$ by its gauge-invariant form
\begin{equation}\label{eq:residual-def}
A(s) := i\,[\partial_s P(s),\, P(s)] ,
\end{equation}
the transport generator introduced by Kato~\cite{kato1950adiabatic}.
As a commutator with $P$, it also vanishes within each block.
Its cross elements are the same (Lemma~\ref{lem:projector-residual}(i) below).
Unlike the eigenbasis construction, it needs only the projector $P$.
Wherever $H$ is $C^1$ and $\gamma_d > 0$, the projector is $C^1$ as a contour integral of the resolvent of $H$.
There, $A$ is well defined and continuous, even at level crossings within either block.
Since $\partial_s P = -\,i\,[A, P]$ (Lemma~\ref{lem:projector-residual}(i)), the propagator would transport the sector exactly if its generator were $-A$.
The quasi-adiabatic generator $D_\Delta$ of Definition~\ref{def:quasi-adiabatic-evolution} differs from $-A$ by the effective generator
\begin{equation}
D_{\mathrm{eff}}(s) := A(s) + D_\Delta(s) .
\end{equation}
Probability leaves the sector only through its cross block $(1-P)\, D_{\mathrm{eff}}\, P$.

\begin{lemma}[The transport generator and the diabatic coupling]\label{lem:projector-residual}
Let $H(s)$ be $C^1$ and $\gamma_d(s) > 0$ on an interval.
We call $K_d := (1-P)\, D_{\mathrm{eff}}\, P$ the \emph{diabatic coupling}.
For every $s$ in the interval, and any orthonormal eigenbasis $\{|\psi_k(s)\rangle\}$ of $H(s)$, we have
\begin{enumerate}
\item[(i)]
\emph{Transport generator.}
$A$ is Hermitian, $\partial_s P = -\,i\,[A, P]$, and
\begin{equation}\label{eq:projector-derivative-elements}
\langle \psi_k | A | \psi_j \rangle = -\,i\,\frac{M_{kj}}{\omega_{kj}},
\qquad k \ge d > j .
\end{equation}
\item[(ii)]
\emph{Effective generator.}
Here $\widetilde F_\Delta$ is the Fourier transform of the filter.
The cross-block of $D_{\mathrm{eff}}$ satisfies
\begin{equation}\label{eq:projector-residual}
\langle \psi_k | K_d | \psi_j \rangle
= -\,i\,M_{kj}\,\Bigl[ \widetilde F_\Delta(\omega_{kj}) + \frac{1}{\omega_{kj}} \Bigr],
\qquad k \ge d > j,
\end{equation}
and
\begin{equation}\label{eq:coupling-defect}
\partial_s P - i\,[D_\Delta, P] = -\,i\,\bigl(K_d - K_d^\dagger\bigr) .
\end{equation}
\item[(iii)]
\emph{BMNS bound.}
For the BMNS filter of Bachmann, Michalakis, Nachtergaele and Sims, the elements in Eq.~\eqref{eq:projector-residual} are zero if $\omega_{kj} \ge \Delta$, and
\begin{equation}\label{eq:residual-bound}
\|K_d(s)\| \le \|K_d(s)\|_F \le \frac{\sqrt d}{\gamma_d(s)} \sqrt{W^{\partial_s H}_{[\gamma_d,\,\Delta)}(s)} .
\end{equation}
In particular, if $\Delta \le \gamma_d(s)$ on the interval, then $K_d \equiv 0$ on the interval and thus $\partial_s P = i\,[D_\Delta, P]$.
\end{enumerate}
\end{lemma}
\begin{proof}
(i) By its form, $A$ is Hermitian. Differentiating $P^2 = P$ shows that the derivative of a projector is block off-diagonal, $\partial_s P = P\, \partial_s P\, (1-P) + (1-P)\, \partial_s P\, P$~\cite[Eq.~(5)]{jansen2007bounds}.
Hence $-\,i\,[A, P] = [[\partial_s P, P], P] = \partial_s P\, P - 2 P\, \partial_s P\, P + P\, \partial_s P = \partial_s P$.
For $k \ge d > j$ we have $\langle \psi_k | P = 0$ and $P |\psi_j\rangle = |\psi_j\rangle$, so $\langle \psi_k | A | \psi_j \rangle = i \langle \psi_k | \partial_s P | \psi_j \rangle$.
Differentiating $H P = P H$ gives $\partial_s H\, P + H\, \partial_s P = \partial_s P\, H + P\, \partial_s H$, whose matrix element between $\langle \psi_k |$ and $|\psi_j\rangle$ is $M_{kj} + E_k \langle \psi_k | \partial_s P | \psi_j \rangle = E_j \langle \psi_k | \partial_s P | \psi_j \rangle$.
Since $\omega_{kj} \ge \gamma_d(s) > 0$, this gives Eq.~\eqref{eq:projector-derivative-elements}.

(ii) Lemma~\ref{lem:generator-matrix-elements} yields
$\langle \psi_k | D_\Delta | \psi_j \rangle = -\,i\,M_{kj}\,\widetilde F_\Delta(\omega_{kj})$ for all $k, j$.
For $k \ge d > j$, we have $\langle \psi_k | K_d | \psi_j \rangle = \langle \psi_k | D_{\mathrm{eff}} | \psi_j \rangle$.
Adding the matrix element of $A$ from (i) gives Eq.~\eqref{eq:projector-residual}.
By (i), we have $\partial_s P - i[D_\Delta, P] = -i[D_{\mathrm{eff}}, P]$.
Since $D_{\mathrm{eff}}$ is Hermitian (because $A$ is by (i) and $D_\Delta$ is by Lemma~\ref{lem:generator-matrix-elements}), we have $[D_{\mathrm{eff}}, P] = (1-P) D_{\mathrm{eff}} P - P D_{\mathrm{eff}} (1-P) = K_d - K_d^\dagger$.

(iii) Above the cutoff, $\widetilde F_\Delta(\omega) = -1/\omega$ (Eq.~\eqref{eq:filter-fourier-identity}).
The bracket in Eq.~\eqref{eq:projector-residual} therefore vanishes for $\omega_{kj} \ge \Delta$.
Below the cutoff, Eq.~\eqref{eq:bmns-bracket-bound} bounds its absolute value by $1/\omega_{kj} \le 1/\gamma_d(s)$.
Since $K_d = (1-P) K_d P$, its squared Frobenius norm is the sum of $|\langle \psi_k | K_d | \psi_j \rangle|^2$ over $j < d \le k$, which is therefore at most $\gamma_d(s)^{-2}$ times the sum of $|M_{kj}|^2$ over $j < d \le k$ with $\omega_{kj} < \Delta$.
By Eq.~\eqref{eq:M_vcoup_relation}, this bound equals $d\, W^{\partial_s H}_{[\gamma_d,\,\Delta)}(s)/\gamma_d(s)^2$.
The operator norm is at most the Frobenius norm.
When $\Delta \le \gamma_d$ on the interval, every $\omega_{kj}$ with $j < d \le k$ is at least $\Delta$, so $K_d \equiv 0$.
Then Eq.~\eqref{eq:coupling-defect} gives $\partial_s P = i[D_\Delta, P]$.
\end{proof}

\subsection{Error of quasi-adiabatic continuation from spectral weight}\label{subsec:order-one-evolution}

\begin{lemma}[Bounding the diabatic error by spectral weight]\label{lemma:sparse-coupling}\label{lemma:sparse-coupling-detailed}
Let $\Delta > 0$ be an energy cutoff, let $H(s)$ be piecewise $C^1$, and fix a sector dimension $d$.
\begin{enumerate}
\item[(a)] \emph{Pointwise bound.} At every $s$ where $H$ is $C^1$ and $\gamma_d(s) > 0$, the diabatic coupling of Lemma~\ref{lem:projector-residual} satisfies
\begin{equation}\label{eq:Deff-bound-via-coupling}
\bigl\| (1-P(s)) \bigl( \partial_s P(s) - i\,[D_\Delta(s), P(s)] \bigr) P(s) \bigr\|
= \|K_d(s)\|
\le \frac{\sqrt d}{\gamma_d(s)} \sqrt{W^{\partial_s H}_{[\gamma_d,\,\Delta)}(s)} .
\end{equation}
\item[(b)] \emph{Integrated bound.}
Suppose that $\gamma_d(s) > 0$ for every $s$.
Define the transported sector overlap
\begin{equation}\label{eq:transport-overlap-def}
T := \frac{1}{d}\, \mathrm{Tr}\bigl[ P_d(1)\, U_\Delta\, P_d(0)\, U_\Delta^\dagger \bigr].
\end{equation}
Then
\begin{equation}\label{eq:sqrt-transport-bound}
\sqrt{1 - T} \;\le\; \int_0^1 \frac{\sqrt{ W^{\partial_s H}_{[\gamma_d,\,\Delta)}(s) }}{\gamma_d(s)}\, ds .
\end{equation}
\end{enumerate}
\end{lemma}
\begin{proof}
Multiplying Eq.~\eqref{eq:coupling-defect} by $1-P$ on the left and by $P$ on the right gives $(1-P)\bigl(\partial_s P - i[D_\Delta, P]\bigr)P = -\,i\,K_d$.
Part (a) is therefore Eq.~\eqref{eq:residual-bound}, with $\|K_d\|$ expressed through $P$ and $D_\Delta$.
For part (b), we bound how fast the transported sector leaves the current sector.
Let $X(0)$ be an isometry from $\mathbb{C}^{d}$ onto the sector at $s = 0$, so that $X(0) X(0)^\dagger = P_d(0)$.
Let $X(s) := U_\Delta(s) X(0)$, so that $\|X(s)\| = 1$.
The part of $X(s)$ outside the current sector has size
\begin{equation*}
\beta(s) := \bigl\| \bigl(1 - P_d(s)\bigr) X(s) \bigr\|_F,
\qquad
\beta(0) = 0,
\qquad
\beta(1)^2 = \mathrm{Tr}\bigl[ \bigl(1 - P_d(1)\bigr) U_\Delta\, P_d(0)\, U_\Delta^\dagger \bigr] = d\,(1 - T).
\end{equation*}
Write $P = P_d(s)$.
On an interval on which $H$ is $C^1$, $\beta^2 = \mathrm{Tr}[X^\dagger (1-P) X]$ is $C^1$.
There, $\partial_s X = i D_\Delta X$ and Eq.~\eqref{eq:coupling-defect} give
\begin{equation*}
\partial_s \beta^2
= -\,\mathrm{Tr}\bigl[ X^\dagger \bigl(\partial_s P - i\,[D_\Delta, P]\bigr) X \bigr]
= i\,\mathrm{Tr}\bigl[ X^\dagger (K_d - K_d^\dagger)\, X \bigr]
= -2\,\mathrm{Im}\,\mathrm{Tr}\bigl[ X^\dagger (1-P)\, K_d\, P X \bigr],
\end{equation*}
where the last equality uses $K_d = (1-P) K_d P$.
The Cauchy--Schwarz inequality for the Frobenius inner product and $\|PX\| \le 1$ give $|\partial_s \beta^2| \le 2\, \|(1-P)X\|_F\, \|K_d\|_F\, \|PX\| \le 2\,\beta\, \|K_d\|_F$.
Hence $(\beta^2 + \epsilon^2)^{1/2}$ has derivative at most $\|K_d\|_F$ in absolute value for every $\epsilon > 0$.
Integrating between two points $s_1 < s_2$ of the interval and letting $\epsilon \to 0$ gives $\beta(s_2) \le \beta(s_1) + \int_{s_1}^{s_2} \|K_d(s)\|_F\, ds$.
At a kink of the path, $X$ and $P_d$ are continuous, so these bounds add up across it.
Combining the intervals and using Eq.~\eqref{eq:residual-bound} for $\|K_d(s)\|_F$ gives
\begin{equation*}
\sqrt{d\,(1-T)} = \beta(1) \le \beta(0) + \int_0^1 \|K_d(s)\|_F\, ds \le \sqrt{d} \int_0^1 \frac{\sqrt{ W^{\partial_s H}_{[\gamma_d,\,\Delta)}(s) }}{\gamma_d(s)}\, ds .
\end{equation*}
Dividing by $\sqrt{d}$ yields Eq.~\eqref{eq:sqrt-transport-bound}.
\end{proof}

\begin{corollary}[Exact transport below the gap]\label{cor:exact-transport}
Fix a sector dimension $d$ and suppose that $\Delta \leq \gamma_d(s)$ for all $s \in [0, 1]$.
Then $U_\Delta(1)$ carries $\mathcal{S}(0)$ onto $\mathcal{S}(1)$, i.e., $U_\Delta(1) P(0) U_\Delta(1)^\dagger = P(1)$.
\end{corollary}
\begin{proof}
On each $C^1$ piece of the path, Lemma~\ref{lem:projector-residual}(iii) gives $\partial_s P = i\,[D_\Delta, P]$, so $\partial_s \bigl( U_\Delta^\dagger P\, U_\Delta \bigr) = U_\Delta^\dagger \bigl( \partial_s P - i\,[D_\Delta, P] \bigr) U_\Delta = 0$.
Since $U_\Delta^\dagger P\, U_\Delta$ is continuous across the kinks, $U_\Delta(1)^\dagger P(1)\, U_\Delta(1) = P(0)$.
We thus recover the exact transport of~\cite[Proposition~2.4]{bachmann2011automorphic}, which has generator $D_\Delta$ by~\cite[Corollary~2.8]{bachmann2011automorphic}.
\end{proof}

\subsection{Proof of Theorems~\ref{lem:path-gradient-spectral-weight} and~\ref{thm:cannot-be-dark}}\label{subsec:proof-complete-detection}

\begin{proof}[Proof of Theorem~\ref{lem:path-gradient-spectral-weight}]
If $\gamma_{\min} = 0$, the empty set satisfies both bounds, so we assume $\gamma_{\min} > 0$.

Fix $0 \le C < 1$ and let
\begin{equation*}
S_{\mathrm{high}} := \Bigl\{ s \in [0,1] : W^{\partial_s H}_{[\gamma_d,\,\Delta)}(s) \ge C^2\,\eta(n)\,\gamma_{\min}^2 \Bigr\}.
\end{equation*}

\emph{Uniform bound.} For every $s \in [0,1]$,
\begin{equation}\label{eq:weight-uniform-upper-bound}
W^{\partial_s H}_{[\gamma_d,\,\Delta)}(s) \le \mathrm{Tr}\bigl[ \rho_d(s)\, (\partial_s H(s))^2 \bigr] \le \| \partial_s H(s) \|^2 \le (\zeta_s L J)^2 .
\end{equation}
The first inequality holds because dropping the restrictions $k \ge d$ and $\omega_{kj} < \Delta$ in Eq.~\eqref{eq:M_vcoup_relation} gives
\begin{equation*}
\frac{1}{d} \sum_{j < d} \sum_k |M_{kj}(s)|^2 = \frac{1}{d} \sum_{j < d} \langle \psi_j(s) | (\partial_s H(s))^2 | \psi_j(s) \rangle = \mathrm{Tr}\bigl[ \rho_d(s)\, (\partial_s H(s))^2 \bigr],
\end{equation*}
the middle inequality holds because $\rho_d(s)$ is a state, and the last is $\| \partial_s H(s) \| \leq \zeta_s L J$.

{\emergencystretch=2em \emph{Transport of the sector.} The overlap $T$ of Eq.~\eqref{eq:transport-overlap-def} equals $\mathrm{Tr}[P(1)\, U_\Delta\, \rho_d(0)\, U_\Delta^\dagger]$ because $\rho_d = P/d$.
Eq.~\eqref{eq:small-endpoint-overlap} therefore states $1 - T \ge \eta(n)$. Since $\gamma_d(s) \ge \gamma_{\min}$, Lemma~\ref{lemma:sparse-coupling}(b) gives
\begin{equation*}
\gamma_{\min}\sqrt{\eta(n)} \le \int_0^1 \sqrt{ W^{\partial_s H}_{[\gamma_d,\,\Delta)}(s) }\, ds .
\end{equation*}\par}

\emph{Measure bound.}
On the complement of $S_{\mathrm{high}}$, which has measure at most one, the integrand is at most $C\gamma_{\min}\sqrt{\eta(n)}$. Inside $S_{\mathrm{high}}$, it is at most $\zeta_s L J$ by the uniform bound.
We thus have
\begin{equation*}
\gamma_{\min}\sqrt{\eta(n)} \le C\,\gamma_{\min}\sqrt{\eta(n)} + \zeta_s L J\, |S_{\mathrm{high}}| ,
\end{equation*}
and rearranging gives Eq.~\eqref{eq-S-bad-measure}.
\end{proof}

\begin{proof}[Proof of Theorem~\ref{thm:cannot-be-dark}]
At both endpoints the ground-state degeneracy is $r$, so $P_{\mathcal G}(x_B) = P_r(x_B)$ and $\rho_{\mathcal G}(x_A) = \rho_r(x_A)$.
For the propagator $U_\Delta[x]$, Proposition~\ref{prop:phases-force-deficit} gives Eq.~\eqref{eq:small-endpoint-overlap} with $d = r$ and $\eta(n) = 3\eta_P/4$.
Theorem~\ref{lem:path-gradient-spectral-weight} with the number $C = 1/2$ gives a set $S_{\mathrm{high}}$ of measure at least $\gamma_{\min}\sqrt{\eta(n)}/(2\,\zeta_s L J) = \gamma_{\min}\sqrt{3\eta_P}/(4\,\zeta_s L J)$ on which $W^{\partial_s H}_{[\gamma_r,\,\Delta)}(s) \ge \eta(n)\gamma_{\min}^2/4 = 3\eta_P\gamma_{\min}^2/16$.
On $S_{\mathrm{high}}$ the weight is positive, so $\dot x(s) \ne 0$.
Because the weight is quadratic in the operator, Eq.~\eqref{eq:weight-scaling} and the speed bound $|\dot x(s)| \le \zeta_s L$ give $W^{\widehat u(s) \cdot \nabla_x H} = W^{\partial_s H}/|\dot x(s)|^2 \ge W^{\partial_s H}/(\zeta_s L)^2$, which is Eq.~\eqref{eq:cannot-be-dark-threshold}.
\end{proof}

\section{Measuring the Spectral Weight}\label{sec:diabatic-signals}

The previous appendix showed that continuous phase transitions at a fixed ground-state degeneracy are not dark: the path-gradient operator has non-negligible spectral weight. This appendix shows how to detect spectral weight efficiently on a quantum computer. The spectral weight is defined through the eigenstates of $H(x)$ and is nontrivial to measure directly. Instead, we measure a filtered response (Appendix~\ref{subsec:filtered-response-kernel}) defined through time evolution under $H(x)$. We write the Heisenberg evolution of an operator $Y$ under $H(x)$ for time~$t$ as
\begin{equation}\label{eq:heisenberg-picture}
    \tau^x_t(Y) := e^{i H(x) t} Y e^{-i H(x) t}.
\end{equation}
The main text writes $\tau^x_t(O)$ as $O_x(t)$.

\subsection{The frequency-filtered response}\label{subsec:filtered-response-kernel}\label{subsec-kernel-choice}

Fix $x \in \mathcal{M}$ and a sector dimension $d$ such that $\gamma_d(x) > 0$.
The sector state $\rho_d(x)$ is the maximally mixed state on the $d$ lowest levels (Appendix~\ref{subsec:sector-setup}).
The band gap $\gamma_d(x)$ is the energy gap above these levels.
For $d = r(x)$, we have $\rho_d(x) = \rho_{\mathcal G}(x)$ and $\gamma_d(x) = \gamma(x)$.
The dissipative response of a Hermitian probe $O$ in the sector state $\rho_d(x)$ is, in the notation of Eq.~\eqref{eq:heisenberg-picture},
\[
    \chi_O(t; x) := i\,\mathrm{Tr}\bigl[\rho_d(x)\, [\tau^x_t(O), O]\bigr].
\]
Expanding in the eigenbasis of $H(x)$, we have
\begin{equation}\label{eq:response-eigenbasis}
    \chi_O(t; x) = \frac{2}{d} \sum_{j \le d-1} \sum_{k \ge d} |\langle \psi_k(x) | O | \psi_j(x) \rangle|^2 \sin(\omega_{kj}(x) t).
\end{equation}
Only transitions out of the sector appear.
Even when the sector levels are split, terms with both levels in the sector cancel in pairs.
Such a term $(j, k)$ and its transpose $(k, j)$ share the same factor $1 / d$ (the sector state is maximally mixed) and the same squared matrix element ($O$ is Hermitian), but carry opposite signs ($\sin$ is odd).
The diagonal terms $j=k$ vanish as well.

Let $\mathcal{K}$ be a real integrable function of time, the \emph{kernel}, and $q$ be its sine transform,
\begin{equation}\label{eq:kernel-sine-transform}
    q(\omega) := 2 \int_{-\infty}^{\infty} \mathcal{K}(t) \sin(\omega t) dt,
\end{equation}
called the \emph{window}.
For odd $\mathcal{K}$, $q = 2 i \widetilde{\mathcal{K}}$ following the convention of Eq.~\eqref{eq:fourier-convention}.
The \emph{frequency-filtered response} of $O$ in state $\sigma$ is
\begin{equation}\label{eq:filtered-response-signal}
    S^O_{\mathcal{K}}(\sigma; x) := \int_{-\infty}^{\infty} \mathcal{K}(t) i \mathrm{Tr}\bigl[\sigma [\tau^x_t(O), O]\bigr] dt.
\end{equation}
We abbreviate $S^O_{\mathcal{K}}(\rho_d(x); x) = \int \mathcal{K}(t) \chi_O(t; x) dt$ as $S^O_{\mathcal{K}}(x)$.
Integrating Eq.~\eqref{eq:response-eigenbasis} against the kernel gives
\begin{equation}\label{eq:weighted-response}
    S^O_{\mathcal{K}}(x) = \frac{1}{d} \sum_{j \le d-1} \sum_{k \ge d} q(\omega_{kj}(x)) |\langle \psi_k(x) | O | \psi_j(x) \rangle|^2.
\end{equation}
With the odd sharp window of half-width $\Lambda$ in place of $q$, the sum above is exactly the spectral weight $W^O_{[\gamma_d, \Lambda)}(x)$.
This window and its kernel are
\begin{equation}\label{eq:sharp-window}
    q_{\mathrm{rect}}(\omega) := \sign(\omega) \mathbf{1}\{|\omega| < \Lambda\}, \qquad \text{kernel } \frac{1 - \cos(\Lambda t)}{2 \pi t}.
\end{equation}

To make the kernel decay fast, we smooth the window edges over a width $\delta$ with the BMNS weight $w_\delta$ of Eq.~\eqref{eq:bmns-weight} at scale $\delta$.
This weight is proportional to an infinite product whose $k$-th factor is $(\sin(a_k t)/(a_k t))^2$, where the numbers $a_k$ satisfy $\sum_k a_k = \delta / 2$.
By Lemma~\ref{lemma:bmns-weight}, $w_\delta$ is even, non-negative, and normalized, and satisfies $0 \le w_\delta(t) \le w_\delta(0) = \delta w_1(0)$.
Its Fourier transform satisfies $0 \le \widetilde w_\delta \le 1$ and, by Eq.~\eqref{eq:bmns-band-limit}, $\widetilde w_\delta(\omega) = 0$ whenever $|\omega| \ge \delta$.
Multiplying the sharp kernel by $w_\delta(t)/w_\delta(0)$ yields
\begin{equation}\label{eq:quantum-kernel}
    \mathcal{K}_{\Lambda, \delta}(t) := \frac{1 - \cos(\Lambda t)}{2 \pi t} \frac{w_\delta(t)}{w_\delta(0)}.
\end{equation}
This is equivalent to convolving the sharp window with the \emph{smoothing density} $\varphi_\delta(\omega) := \widetilde{w}_\delta(\omega) / (2 \pi w_\delta(0))$:
\begin{equation}\label{eq:quantum-window}
    q_{\Lambda, \delta} = q_{\mathrm{rect}} * \varphi_\delta.
\end{equation}
The density $\varphi_\delta$ is even, non-negative, supported on $[-\delta, \delta]$, and has unit mass, since $\int \widetilde{w}_\delta = 2 \pi w_\delta(0)$ by Fourier inversion at $t = 0$.

To obtain a closed form, define $\Phi_\delta(\nu) := \int_0^\nu \varphi_\delta(u) du$, which is odd and equals $1 / 2$ for $\nu \ge \delta$. Then Eq.~\eqref{eq:quantum-window} reads
\begin{equation}\label{eq:quantum-window-closed-form}
    q_{\Lambda, \delta}(\omega) = 2 \Phi_\delta(\omega) - \Phi_\delta(\omega + \Lambda) - \Phi_\delta(\omega - \Lambda).
\end{equation}

\begin{figure}[t]
\centering
\definecolor{gapblue}{HTML}{0072B2}
\begin{tikzpicture}[x=1cm, y=1.2cm]
  \fill[gapblue!8] (-0.6,-1.1) rectangle (0.6,1.1);
  \fill[gapblue!8] (3.4,0) rectangle (4.6,1.1);
  \fill[gapblue!8] (-4.6,-1.1) rectangle (-3.4,0);
  \draw[->, line width=0.5pt] (-5.6,0) -- (5.9,0) node[anchor=west] {$\omega$};
  \draw[->, line width=0.5pt] (0,-1.35) -- (0,1.5);
  \draw[gapblue, line width=1.0pt] (-5.4,0) -- (-4.6,0);
  \draw[gapblue, line width=1.0pt] plot[domain=-4.6:-3.4, samples=60]
    (\x, {-(6*((\x+4.6)/1.2)^5 - 15*((\x+4.6)/1.2)^4 + 10*((\x+4.6)/1.2)^3)});
  \draw[gapblue, line width=1.0pt] (-3.4,-1) -- (-0.6,-1);
  \draw[gapblue, line width=1.0pt] plot[domain=-0.6:0.6, samples=60]
    (\x, {2*(6*((\x+0.6)/1.2)^5 - 15*((\x+0.6)/1.2)^4 + 10*((\x+0.6)/1.2)^3) - 1});
  \draw[gapblue, line width=1.0pt] (0.6,1) -- (3.4,1);
  \draw[gapblue, line width=1.0pt] plot[domain=3.4:4.6, samples=60]
    (\x, {1 - (6*((\x-3.4)/1.2)^5 - 15*((\x-3.4)/1.2)^4 + 10*((\x-3.4)/1.2)^3)});
  \draw[gapblue, line width=1.0pt] (4.6,0) -- (5.4,0);
  \draw[dashed, gray, line width=0.7pt]
    (-5.4,0) -- (-4,0) -- (-4,-1) -- (0,-1) -- (0,1) -- (4,1) -- (4,0) -- (5.4,0);
  \draw[line width=0.5pt] (0.6,-0.05) -- (0.6,0.05);
  \node[anchor=north] at (0.6,-0.07) {$\delta$};
  \draw[line width=0.5pt] (3.4,-0.05) -- (3.4,0.05);
  \node[anchor=north] at (3.3,-0.07) {$\Lambda-\delta$};
  \draw[line width=0.5pt] (4.6,-0.05) -- (4.6,0.05);
  \node[anchor=north] at (4.75,-0.07) {$\Lambda+\delta$};
  \draw[line width=0.5pt] (-0.05,1) -- (0.05,1);
  \node[anchor=east] at (-0.08,1) {$1$};
  \draw[line width=0.5pt] (-0.05,-1) -- (0.05,-1);
  \node[anchor=west] at (0.08,-1) {$-1$};
  \node[gapblue, anchor=south] at (2,1.05) {$q_{\Lambda,\delta}(\omega)$};
  \node[gray, anchor=south] at (-4.15,0.3) {$q_{\mathrm{rect}}(\omega)$};
  \node[anchor=south] at (5.2,0.3) {exactly $0$};
\end{tikzpicture}
\caption{The window $q_{\Lambda,\delta}$ (solid) of
Eq.~\eqref{eq:quantum-window}, compared with the sharp window
$q_{\mathrm{rect}}$ (dashed). Both are odd. The shaded edges have width $2\delta$
(schematic profile).}
\label{fig:quantum-window}
\end{figure}
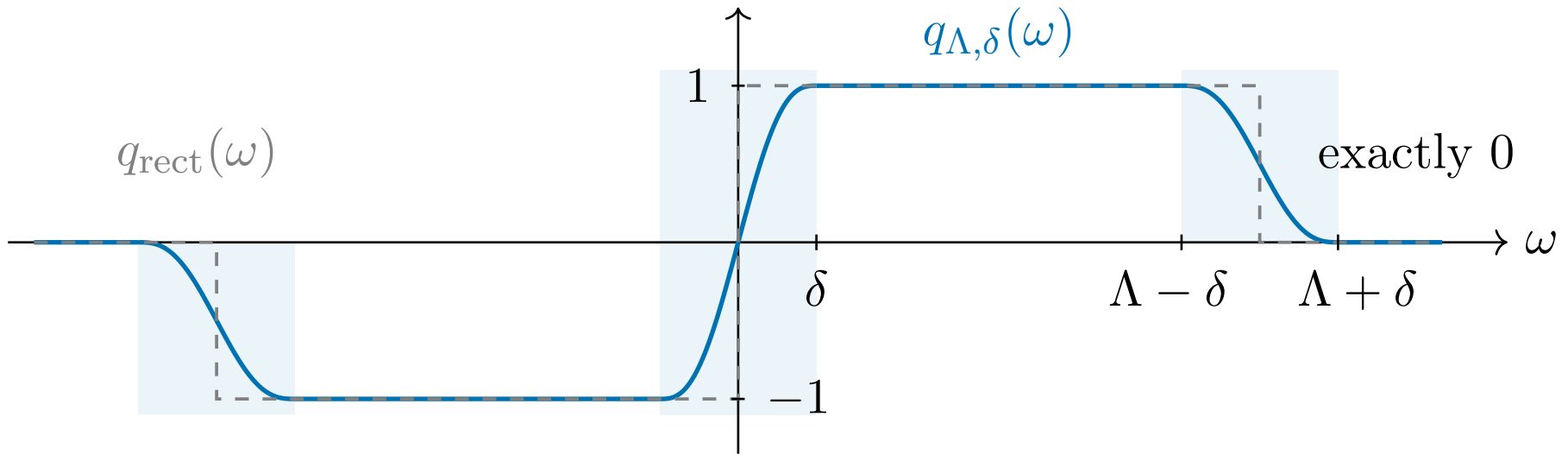

For $\delta \le \Lambda / 2$, the window is odd and for $\omega \ge 0$ satisfies
\begin{equation}\label{eq:quantum-window-shape}
    0 \le q_{\Lambda, \delta}(\omega) \le 1, \quad q_{\Lambda, \delta}(\omega) = 1 \ \text{ for } \delta \le \omega \le \Lambda - \delta, \quad q_{\Lambda, \delta}(\omega) = 0 \ \text{ for } \omega \ge \Lambda + \delta.
\end{equation}
Indeed, on $[\delta, \Lambda - \delta]$ the three $\Phi_\delta$ values in Eq.~\eqref{eq:quantum-window-closed-form} are $1 / 2$, $1 / 2$, and $-1 / 2$, while for $\omega \ge \Lambda + \delta$ all three are $1 / 2$.
On $[0, \delta]$ the window is $2 \Phi_\delta(\omega)$ and increases from $0$ to $1$. On $[\Lambda - \delta, \Lambda + \delta]$, it is $\frac{1}{2} - \Phi_\delta(\omega - \Lambda)$ and decreases from $1$ to $0$ through $q_{\Lambda, \delta}(\Lambda) = 1 / 2$.
The assumption $\delta \le \Lambda / 2$ ensures the flat top is non-empty and the two edge bands are disjoint (Figure~\ref{fig:quantum-window}).

\begin{lemma}[Bounds on the filtered response]\label{lem:quantum-kernel}
Let $0 < \delta \le \Lambda / 2$ and let $O$ be Hermitian. If the band gap satisfies $\gamma_d(x) \ge \delta$, then
\begin{equation}\label{eq:response-kernel-sandwich}
    W^O_{[\gamma_d, \Lambda - \delta)}(x) \le S^O_{\mathcal{K}_{\Lambda, \delta}}(x) \le W^O_{[\gamma_d, \Lambda + \delta)}(x).
\end{equation}
The upper bound holds for every $\gamma_d(x) > 0$, and so does the vanishing of the response
\begin{equation}\label{eq:response-kernel-null}
    S^O_{\mathcal{K}_{\Lambda, \delta}}(x) = 0 \qquad \text{whenever } \gamma_d(x) \ge \Lambda + \delta.
\end{equation}
\end{lemma}
\begin{proof}
All transition frequencies in Eq.~\eqref{eq:weighted-response} are positive and $\omega_{kj}(x) \ge \gamma_d(x)$.
By Eq.~\eqref{eq:quantum-window-shape}, each transition thus enters with a factor $q_{\Lambda, \delta}(\omega_{kj}(x))$ in $[0, 1]$.
This factor vanishes when $\omega_{kj}(x) \ge \Lambda + \delta$, which gives the upper bound for every $\gamma_d(x) > 0$.
When $\gamma_d(x) \ge \Lambda + \delta$, all factors vanish, so Eq.~\eqref{eq:response-kernel-null} holds.
For the lower bound, let $\gamma_d(x) \ge \delta$.
Any transition with $\omega_{kj}(x) < \Lambda - \delta$ then lies in $[\delta, \Lambda - \delta)$, where the factor equals one.
Keeping only these transitions and dropping the non-negative remainder yields the lower bound.
\end{proof}

For a time cutoff $T > 0$, the following integrals of a real-valued integrable kernel $\mathcal{K}$ control how much the filtered response varies and the cost of estimating it:
\begin{equation}\label{eq:kernel-moments-def}
\begin{gathered}
    K_0(\mathcal{K}) := \int_{-\infty}^\infty |\mathcal{K}(t)| dt, \qquad K_1(\mathcal{K}) := \int_{-\infty}^\infty |t| |\mathcal{K}(t)| dt,
    \\
    K_T := \int_{-T}^T |\mathcal{K}(t)| dt, \qquad K_{\mathrm{tail}}(T) := \int_{|t| > T} |\mathcal{K}(t)| dt,
\end{gathered}
\end{equation}
so that $K_T + K_{\mathrm{tail}}(T) = K_0(\mathcal{K})$.

\begin{lemma}[Decay, moments, and tail of the kernel]\label{lem:quantum-kernel-moments}
The kernel $\mathcal K_{\Lambda,\delta}$ is real and odd.
Let $0 < \delta \le \Lambda$. We have the following pointwise bound,
\begin{equation}\label{eq:quantum-kernel-pointwise}
|\mathcal K_{\Lambda,\delta}(t)| \le \frac{w_\delta(t)}{\pi\,|t|\,w_\delta(0)},
\end{equation}
and bounds on the first two moments:
\begin{equation}\label{eq:quantum-kernel-moments}
K_0(\mathcal K_{\Lambda,\delta}) \le \frac{2}{\pi}\ln\frac{\Lambda}{\delta} + C_{\mathcal K}, \qquad C_{\mathcal K} := \frac{1}{4\pi} + \frac{2}{\pi\,w_1(0)}, \qquad K_1(\mathcal K_{\Lambda,\delta}) = \frac{1}{2\pi\,w_1(0)\,\delta}.
\end{equation}
For every $T \ge e^{1/\sqrt 2}/\delta$, we have the tail bound
\begin{equation}\label{eq:quantum-kernel-tail}
\begin{aligned}
K_{\mathrm{tail}}(T) &\le \frac{4e^2}{\pi\,w_1(0)} \int_{\delta T}^{\infty} \exp\!\Bigl(-\frac{2}{7}\,\frac{u}{\ln^2 u}\Bigr)\,du \le C_{\mathcal K,\mathrm{tail}}\, \exp\!\Bigl(-\frac{1}{7}\,\frac{\delta T}{\ln^2(e+\delta T)}\Bigr), \\
C_{\mathcal K,\mathrm{tail}} &:= \frac{4e^2}{\pi\,w_1(0)} \int_{0}^{\infty} \exp\!\Bigl(-\frac{1}{7}\,\frac{u}{\ln^2(e+u)}\Bigr)\,du .
\end{aligned}
\end{equation}
\end{lemma}
\begin{proof}
Since $(1 - \cos(\Lambda t))/(2\pi t)$ is real and odd and $w_\delta$ is real, even, and non-negative, $\mathcal{K}_{\Lambda, \delta}$ is real and odd.
The bound $|1 - \cos(\Lambda t)| \le 2$ yields Eq.~\eqref{eq:quantum-kernel-pointwise}.

We bound the zeroth moment $K_0(\mathcal K_{\Lambda,\delta})$ by splitting the integral at $|t| = 1/\Lambda$: using $|1 - \cos(\Lambda t)| \le \Lambda^2 t^2 / 2$ and $0 \le w_\delta(t) \le w_\delta(0)$, we have
\[
|\mathcal{K}_{\Lambda, \delta}(t)| \le \frac{\Lambda^2 |t|}{4 \pi} \quad \text{for } |t| \le \frac{1}{\Lambda}, \quad |\mathcal{K}_{\Lambda, \delta}(t)| \le \frac{w_\delta(t)}{\pi |t| w_\delta(0)} \quad \text{for } |t| > \frac{1}{\Lambda}.
\]
Integrating the first inequality over $|t| \le 1 / \Lambda$ yields $1 / (4 \pi)$. For the second, we change variables to $u = \delta t$ and use $w_\delta(t) = \delta w_1(\delta t)$ to obtain
\[
\int_{|t| > 1 / \Lambda} \frac{w_\delta(t)}{\pi |t| w_\delta(0)} dt = \frac{2}{\pi} \int_{\delta / \Lambda}^\infty \frac{w_1(u)}{u w_1(0)} du \le \frac{2}{\pi} \left( \int_{\delta / \Lambda}^1 \frac{du}{u} + \frac{1}{w_1(0)} \int_1^\infty w_1(u) du \right) \le \frac{2}{\pi} \left( \ln \frac{\Lambda}{\delta} + \frac{1}{w_1(0)} \right),
\]
where the split at $u = 1$ uses $\delta \le \Lambda$, the first integral uses $w_1(u) \le w_1(0)$, and the second uses $\int w_1 = 1$.

For $K_1$, both factors in $|t| |\mathcal{K}_{\Lambda, \delta}(t)| = (1 - \cos(\Lambda t)) w_\delta(t) / (2 \pi w_\delta(0))$ are non-negative, so the integrand of $K_1$ can be written without an absolute value.
Moreover, $\int (1 - \cos(\Lambda t)) w_\delta(t) dt = 1 - \widetilde{w}_\delta(\Lambda) = 1$ since $w_\delta$ is even, $\int w_\delta = 1$, and $\widetilde{w}_\delta(\Lambda) = 0$ for $\Lambda \ge \delta$ by Eq.~\eqref{eq:bmns-band-limit}.
Hence $K_1(\mathcal{K}_{\Lambda, \delta}) = 1 / (2 \pi w_\delta(0)) = 1 / (2 \pi w_1(0) \delta)$.

For the tail, Eq.~\eqref{eq:quantum-kernel-pointwise} and the substitution $u = \delta t$ give $K_{\mathrm{tail}}(T) \le \frac{2}{\pi w_1(0)} \int_{\delta T}^\infty \frac{w_1(u)}{u} du$.
Using Eq.~\eqref{eq:bmns-decay}, $w_1(u) \le 2 e^2 u \exp(- \frac{2}{7} u / \ln^2 u)$ for $u \ge e^{1 / \sqrt{2}}$, we obtain the first inequality in Eq.~\eqref{eq:quantum-kernel-tail}.
The second inequality uses $\ln u \le \ln(e + u)$ for $u \ge 1$ and the fact that $u \mapsto u / \ln^2(e + u)$ is increasing for $u \ge 0$, as its derivative has the sign of $\ln(e + u) - 2 u / (e + u) \ge \ln 2$.
Therefore, for $u \ge \delta T$, we have $\exp(- \frac{2}{7} u / \ln^2 u) \le \exp(- \frac{1}{7} \delta T / \ln^2(e + \delta T)) \exp(- \frac{1}{7} u / \ln^2(e + u))$.
Integrating over $u \ge \delta T$ gives the second inequality.
The integral defining $C_{\mathcal{K}, \mathrm{tail}}$ converges because $u / \ln^2(e + u) \ge \sqrt{u}$ for $u \ge 6000$.
\end{proof}

\subsubsection{Lipschitz continuity and state robustness}\label{subsec:filtered-response-stability}

Our algorithms estimate the filtered response only at grid points and in approximate states.
The next lemma bounds how much it changes between nearby points and nearby states.

\begin{lemma}[Lipschitz continuity and state robustness of the filtered response]\label{lem:filtered-response-stability}
Suppose $\mathcal K$ is a real-valued integrable kernel with finite moments $K_0(\mathcal K)$ and $K_1(\mathcal K)$. Suppose $\mathcal N \subseteq \mathcal M$ is a convex set on which $\gamma_d(x) \ge g_* > 0$ for a fixed sector dimension $d$. Suppose $x \mapsto O(x)$ is a differentiable family of Hermitian observables. Suppose that for every $x \in \mathcal N$ and every unit direction $v \in \mathbb R^m$,
\[
\| \partial_v H(x) \| \le L_H, \qquad \| O(x) \| \le B_O, \qquad \| \partial_v O(x) \| \le L_O .
\]
Then for all $x, y \in \mathcal N$,
\begin{equation}\label{eq:general-response-lipschitz}
\bigl| S^{O(x)}_{\mathcal K}(x) - S^{O(y)}_{\mathcal K}(y) \bigr| \le L_{S,\mathcal K}\,\|x - y\|, \qquad L_{S,\mathcal K} := \frac{4 K_0(\mathcal K)\,B_O^2 L_H}{g_*} + 4 K_1(\mathcal K)\,B_O^2 L_H + 4 K_0(\mathcal K)\,B_O L_O .
\end{equation}
Moreover, for any Hermitian $O$ and any two states $\rho$, $\widetilde\rho$,
\begin{equation}\label{eq:response-trace-distance-robustness}
\bigl| S^{O}_{\mathcal K}(\widetilde\rho;x) - S^{O}_{\mathcal K}(\rho;x) \bigr| \le 2\,K_0(\mathcal K)\,\|O\|^2\,\|\widetilde\rho - \rho\|_1 .
\end{equation}
\end{lemma}

\begin{proof}
Fix a unit vector $v \in \mathbb{R}^m$ with directional derivative $\partial_v$.
We begin with the sector state $\rho_d(x) = P_d(x)/d$.
On $\mathcal{N}$, the sector projector $P_d(x)$ is a contour integral of the resolvent $(z - H(x))^{-1}$, along a contour that encloses the $d$ lowest eigenvalues of $H(x)$ and no others.
Because the gap $\gamma_d(x) > 0$ separates these eigenvalues from the higher ones, the representation is well defined, and it shows that $P_d(x)$ is smooth in $x$, even when levels cross within the sector or among the eigenvalues above it.
Differentiating under the integral sign gives
\[
P_d(x) = \frac{1}{2\pi i} \oint (z - H(x))^{-1} dz, \qquad \partial_v P_d(x) = \frac{1}{2\pi i} \oint (z - H(x))^{-1} \partial_v H(x) (z - H(x))^{-1} dz.
\]
As each matrix element with $j < d \le k$ has a single pole at $z = E_j(x)$, we have
\[
\langle \psi_k(x) | \partial_v P_d(x) | \psi_j(x) \rangle = \frac{1}{2\pi i} \oint \frac{\langle \psi_k(x) | \partial_v H(x) | \psi_j(x) \rangle}{(z - E_k(x)) (z - E_j(x))}\,dz = -\frac{\langle \psi_k(x) | \partial_v H(x) | \psi_j(x) \rangle}{\omega_{kj}(x)} \qquad (j < d \le k).
\]
Entries with $j, k < d$ have cancelling residues and entries with $j, k \ge d$ enclose no pole, so both vanish. Entries with $j \ge d > k$ yield the adjoint matrix element in the same way.
Hence, $\partial_v P_d(x) = X + X^\dagger$ where
\begin{equation}\label{eq:projector-derivative}
      X := -\sum_{j < d \le k} \frac{\langle \psi_k(x) | \partial_v H(x) | \psi_j(x) \rangle}{\omega_{kj}(x)}\; |\psi_k(x) \rangle \langle \psi_j(x)| .
\end{equation}
Since $X = X P_d(x)$, the matrix $X$ has rank at most $d$.
Its Frobenius norm obeys
\[
\| X \|_F^2 = \sum_{j < d \le k} \frac{|\bra{\psi_k(x)} \partial_v H(x) \ket{\psi_j(x)}|^2}{\omega_{kj}(x)^2} \le \frac{1}{\gamma_d(x)^2} \sum_{j < d} \bra{\psi_j(x)} (\partial_v H(x))^2 \ket{\psi_j(x)} \le \frac{d \| \partial_v H(x) \|^2}{\gamma_d(x)^2}.
\]
The first inequality uses $\omega_{kj}(x) \ge \gamma_d(x)$ for $j < d \le k$, and completeness of the eigenbasis to extend the sum over $k$ to all levels.
The second bounds each $\langle \psi_j(x) | (\partial_v H(x))^2 | \psi_j(x) \rangle$ by $\| \partial_v H(x) \|^2$.
Cauchy--Schwarz on the at most $d$ singular values of $X$ bounds its trace norm.
The triangle inequality then gives
\[
\| X \|_1 \le \sqrt{d} \| X \|_F \le \frac{d \| \partial_v H(x) \|}{\gamma_d(x)}, \qquad \| \partial_v \rho_d(x) \|_1 = \frac{1}{d} \| X + X^\dagger \|_1 \le \frac{2}{d} \| X \|_1 .
\]
Thus,
\begin{equation}\label{eq:sector-state-derivative}
\| \partial_v \rho_d(x) \|_1 \le \frac{2 \| \partial_v H(x) \|}{\gamma_d(x)} \le \frac{2 L_H}{g_*},
\end{equation}
because $\| \partial_v H(x) \| \le L_H$ and $\gamma_d(x) \ge g_*$.

Next, we bound the derivative of the time evolution with the product rule and Duhamel's formula $\partial_v e^{Y} = \int_0^1 e^{(1-u)Y}\,(\partial_v Y)\,e^{uY}\,du$.
With $Y = \pm i H(x) t$, its integrand is $\pm i t\,\partial_v H(x)$ between two unitaries, so
\[
\begin{aligned}
\partial_v \tau^x_t(O) &= (\partial_v e^{i H(x) t}) O e^{-i H(x) t} + e^{i H(x) t} O (\partial_v e^{-i H(x) t}),\\
\| \partial_v e^{\pm i H(x) t} \| &\le |t| \| \partial_v H(x) \| \le |t| L_H, \qquad \| \partial_v \tau^x_t(O) \| \le 2 |t| \| \partial_v H(x) \| \| O \| \le 2 |t| L_H \| O \| .
\end{aligned}
\]
The product rule then implies
\[
\begin{aligned}
\partial_v i \mathrm{Tr} \left[ \rho_d(x) \left[ \tau^x_t(O(x)), O(x) \right] \right] &= i \mathrm{Tr} \left[ \partial_v \rho_d(x) \left[ \tau^x_t(O(x)), O(x) \right] \right] + i \mathrm{Tr} \left[ \rho_d(x) \left[ \partial_v \tau^x_t(O(x)), O(x) \right] \right] \\
&\quad + i \mathrm{Tr} \left[ \rho_d(x) \left[ \tau^x_t(\partial_v O(x)), O(x) \right] \right] + i \mathrm{Tr} \left[ \rho_d(x) \left[ \tau^x_t(O(x)), \partial_v O(x) \right] \right],
\end{aligned}
\]
where $\partial_v \tau^x_t$ in the second term differentiates only the two time evolution operators, not $O$.

We use $| \mathrm{Tr} \left[ Y_1 \left[ Y_2, Y_3 \right] \right] | \le 2 \| Y_1 \|_1 \| Y_2 \| \| Y_3 \|$, $\mathrm{Tr}(\rho_d(x)) = 1$ and unitary invariance to bound
\[
\begin{aligned}
\left| i \mathrm{Tr} \left[ \partial_v \rho_d(x) \left[ \tau^x_t(O(x)), O(x) \right] \right] \right| &\le 2 \| \partial_v \rho_d(x) \|_1 \| O(x) \|^2 \le \frac{2 L_H}{g_*} \cdot 2 B_O^2 = \frac{4 B_O^2 L_H}{g_*}, \\
\left| i \mathrm{Tr} \left[ \rho_d(x) \left[ \partial_v \tau^x_t(O(x)), O(x) \right] \right] \right| &\le 2 \| \partial_v \tau^x_t(O(x)) \| \| O(x) \| \le 2 \cdot (2 |t| L_H B_O) \cdot B_O = 4 |t| B_O^2 L_H, \\
\left| i \mathrm{Tr} \left[ \rho_d(x) \left[ \tau^x_t(\partial_v O(x)), O(x) \right] \right] \right| &\le 2 \| \partial_v O(x) \| \| O(x) \| \le 2 B_O L_O, \\
\left| i \mathrm{Tr} \left[ \rho_d(x) \left[ \tau^x_t(O(x)), \partial_v O(x) \right] \right] \right| &\le 2 \| O(x) \| \| \partial_v O(x) \| \le 2 B_O L_O .
\end{aligned}
\]
Hence,
\begin{equation}\label{eq:response-derivative-bound}
\left| \partial_v i \mathrm{Tr} \left[ \rho_d(x) \left[ \tau^x_t(O(x)), O(x) \right] \right] \right| \le \frac{4 B_O^2 L_H}{g_*} + 4 |t| B_O^2 L_H + 4 B_O L_O .
\end{equation}
The derivative of the integrand is integrable against $| \mathcal{K}(t) |$, so differentiating under the integral sign gives $| \partial_v S^{O(x)}_{\mathcal{K}}(x) | \le L_{S, \mathcal{K}}$.
Integrating along the line segment from $x$ to $y$, which lies in $\mathcal{N}$ due to convexity of $\mathcal{N}$, gives Eq.~\eqref{eq:general-response-lipschitz}.

State robustness of $S^O_{\mathcal K}$ needs no derivatives, only linearity of the trace:
\[
i \mathrm{Tr} \left[ \widetilde{\rho} \left[ \tau^x_t(O), O \right] \right] - i \mathrm{Tr} \left[ \rho \left[ \tau^x_t(O), O \right] \right] = i \mathrm{Tr} \left[ (\widetilde{\rho} - \rho) \left[ \tau^x_t(O), O \right] \right] .
\]
H\"older's inequality and the commutator bound
$\|[\tau^x_t(O),O]\| \le 2\,\|\tau^x_t(O)\|\,\|O\| = 2\,\|O\|^2$
imply
\[
\left| i \mathrm{Tr} \left[ (\widetilde{\rho} - \rho) \left[ \tau^x_t(O), O \right] \right] \right| \le \| \widetilde{\rho} - \rho \|_1 \| \left[ \tau^x_t(O), O \right] \| \le 2 \| O \|^2 \| \widetilde{\rho} - \rho \|_1 .
\]
Integrating against $| \mathcal{K}(t) |$ yields Eq.~\eqref{eq:response-trace-distance-robustness}.
\end{proof}

For a geometrically local family, $H(x)$ has $J = O(n)$ local terms, each with parameter derivative at most one (Definition~\ref{definition: geometrically local hamiltonians}), so $L_H \le J$.
For the directional derivative $O(x) := \widehat u \cdot \nabla_x H(x)$ in a fixed unit direction $\widehat u$, we also have $\|O(x)\| \le J$.
The second directional derivatives of each $h_j$ are also at most one.
Polarization then gives
\begin{equation}\label{eq:probe-derivative-bounds}
\partial_{\widehat u}\partial_v h_j = \frac{1}{4}\bigl(((\widehat u + v)\cdot\nabla_x)^2 h_j - ((\widehat u - v)\cdot\nabla_x)^2 h_j\bigr), \qquad \|\partial_{\widehat u}\partial_v h_j\| \le \frac{1}{4}\bigl(|\widehat u+v|^2 + |\widehat u-v|^2\bigr) = 1 ,
\end{equation}
and hence $\|\partial_v O(x)\| \le J$.
Appendix~\ref{sec:local-detection} applies the lemma with $g_* = \gamma_{\mathrm{lb}}/2$.
Setting $B_O = L_O = L_H = J$, we then have
\begin{equation}\label{eq:gradient-probe-lipschitz}
L_{S,\mathcal K} \le \frac{8 J^3}{\gamma_{\mathrm{lb}}}\,K_0(\mathcal K) + 4 J^3\,K_1(\mathcal K) + 4 J^2\,K_0(\mathcal K) .
\end{equation}
For a fixed Hermitian observable $O$, the last term vanishes.
With $L_O = 0$ and $B_O = \|O\|$,
\begin{equation}\label{eq:fixed-probe-lipschitz}
\bigl| S^{O}_{\mathcal K}(x) - S^{O}_{\mathcal K}(y) \bigr| \le \|O\|^2\Bigl[\frac{8J}{\gamma_{\mathrm{lb}}}\,K_0(\mathcal K) + 4J\,K_1(\mathcal K)\Bigr]\,\|x - y\| ,
\end{equation}
which the classical algorithm of Appendix~\ref{sec:classical-algorithm} uses with its own value of $g_*$ in place of $\gamma_{\mathrm{lb}}/2$.
There, each probe $O \in \mathcal L$ combines the $N_{\mathcal L}$ Pauli strings $P \in \mathcal P_\ell$ with real coefficients $z_P$ whose squares sum to at most one (Eq.~\eqref{eq:local-class}).
Its operator norm can exceed one, but the triangle and Cauchy--Schwarz inequalities still give $\|O\| \le \sum_{P \in \mathcal P_\ell}|z_P| \le \sqrt{N_{\mathcal L}}\,\|z\|_2 \le \sqrt{N_{\mathcal L}}$.

For the kernel $\mathcal{K}_{\Lambda, \delta}$, Lemma~\ref{lem:quantum-kernel-moments} bounds $K_0$ and $K_1$.
With these bounds, Eq.~\eqref{eq:gradient-probe-lipschitz} shows that $L_{S, \mathcal{K}_{\Lambda, \delta}}$ is at most polynomial in $n$, $1/\delta$, and $1/\gamma_{\mathrm{lb}}$.

\subsection{Estimating the filtered response}\label{subsec:quantum-filtered-response}

\subsubsection{Measuring the response correlator for two Pauli strings}

We measure the filtered response in a reference state transported to $x$, not in $\rho_d(x)$.
Let $x_{\mathrm{ref}}$ be a point of $S$ with spectral gap $\geq \gamma_{\mathrm{lb}}$ whose preparation succeeds.
Its reference state $C[x_{\mathrm{ref}}]$ is then within $\epsilon_{\mathrm{ref}}$ of the maximally mixed ground state (Eq.~\eqref{eq:successful-preparation}).

To reach a grid point $x$, the local algorithms of Appendix~\ref{sec:local-detection} store a path from $x_{\mathrm{ref}}$ to $x$, with a cutoff for each edge.
The corresponding transport circuit $V$ is the product of edge circuits, each approximating the quasi-adiabatic propagator of one edge at its cutoff (Appendix~\ref{subsubsec:transport-edge}).
The \emph{candidate state} is
\begin{equation}\label{eq:response-candidate-state}
    \sigma_V := V\,C[x_{\mathrm{ref}}]\,V^\dagger.
\end{equation}

For any operator $A$, cyclicity of the trace gives the pull-back identity
\begin{equation}\label{eq:pullback-identity}
    \mathrm{Tr}\bigl[\sigma_V\,A\bigr] = \mathrm{Tr}\bigl[C[x_{\mathrm{ref}}]\,V^\dagger A\,V\bigr].
\end{equation}
Thus any expectation value in $\sigma_V$ can be measured on the reference state $C[x_{\mathrm{ref}}]$ after pulling the observable back through $V$.

Expanding the probe in Pauli strings turns the dissipative response into a weighted sum of the imaginary parts of correlators.
The response correlator for Hermitian unitaries $P$, $Q$ (Pauli strings) and waiting time $t$ is
\begin{equation}\label{eq:pauli-response-correlator}
    C_{P,Q}(t;x) := \mathrm{Tr}\bigl[\rho_d(x)\,\tau^{x}_{t}(P)\,Q\bigr].
\end{equation}

\begin{lemma}[Hadamard test for one response correlator]\label{lem:hadamard-response-correlator}
Prepare an ancilla in $|+\rangle$ and the system in a state $\rho$. Apply, in order,
\begin{enumerate}
\item $Q$, controlled by the ancilla being in $|1\rangle$;
\item $e^{-iH(x)t}$ to the system, without control;
\item $P$, controlled by the ancilla being in $|1\rangle$.
\end{enumerate}
Measure the ancilla in the $Y$ basis $\{|{+i}\rangle, |{-i}\rangle\}$ and assign $Y = +1$ to $|{+i}\rangle$ and $Y = -1$ to $|{-i}\rangle$. Then
\begin{equation}\label{eq:hadamard-response-expectation}
\mathbb E[Y] = \mathrm{Im}\,\mathrm{Tr}\bigl[\rho\,\tau^{x}_{t}(P)\,Q\bigr],
\end{equation}
which equals $\mathrm{Im}\,C_{P,Q}(t;x)$ for $\rho = \rho_d(x)$.
\end{lemma}
\begin{proof}
The circuit leaves the ancilla with $\langle\sigma^x\rangle+i\langle\sigma^y\rangle=\mathrm{Tr}[\rho\,\tau^{x}_{t}(P)\,Q]$.
For pure $\rho$, this is \cite[Eq.~(2) and Fig.~1]{pedernales2014efficient} with $n=2$, $|e\rangle=|0\rangle$, $|g\rangle=|1\rangle$, $\tilde U_c^0=Q$, $\tilde U_c^1=P$ and $U(t_1;0)=e^{-iH(x)t}$.
By linearity, it holds for every $\rho$.
Measuring in the $Y$ basis gives $\mathbb E[Y]=\langle\sigma^y\rangle$.
\end{proof}

The controlled operations can be replaced approximately by a weak ancilla--system coupling before the evolution, as in the noninvasive protocol of Ref.~\cite{uhrich2017noninvasive}.
For spin-$1/2$ systems and single-site observables, that protocol needs no ancilla at all.

That ancilla-free protocol~\cite[Section~VIII]{uhrich2017noninvasive} extends from single-site operators to Pauli strings, because it uses only $Q^2=I$.

\begin{lemma}[Ancilla-free protocol for one response correlator]\label{lem:ancilla-free-response-correlator}
\label{lemma:ancilla-free}
Prepare the system in a state $\rho$ and draw a sign $b\in\{+1,-1\}$ uniformly at random.
Apply $e^{ib\pi Q/4}$, then $e^{-iH(x)t}$, measure $P$ with outcome $a\in\{+1,-1\}$, and set $Y=-ab$.
Then
\begin{equation}\label{eq:ancilla-free-response-expectation}
\mathbb E[Y] = \mathrm{Im}\,\mathrm{Tr}\bigl[\rho\,\tau^{x}_{t}(P)\,Q\bigr],
\end{equation}
the same mean as in Lemma~\ref{lem:hadamard-response-correlator}.
\end{lemma}
\begin{proof}
Since $Q^2=I$, we have $e^{ib\pi Q/4}=(I+ibQ)/\sqrt2$, so for $A:=\tau^x_t(P)$,
\[
e^{-ib\pi Q/4}\,A\,e^{ib\pi Q/4}=\tfrac12\bigl(A+QAQ\bigr)+\tfrac{ib}{2}\,[A,Q].
\]
Since $\mathrm{Tr}\bigl[\rho\,[A,Q]\bigr]=2i\,\mathrm{Im}\,\mathrm{Tr}[\rho AQ]$, this gives $\mathbb E[a\mid b]=\tfrac12\mathrm{Tr}\bigl[\rho\,(A+QAQ)\bigr]-b\,\mathrm{Im}\,\mathrm{Tr}[\rho AQ]$.
Multiplying by $-b$ and averaging over $b$ removes the first term.
\end{proof}

\begin{figure}[t]
\centering
\begin{quantikz}[row sep=0.4cm, column sep=0.35cm]
\lstick{\textbf{(a)}} & & & & & & & & \lstick{\textbf{(b)}} & & & \\
\lstick{$\ket{+}$} & \ctrl{1} & \qw & \ctrl{1} & \meter{$Y$} & & & & & & & \\
\lstick{$\rho$} & \gate{Q} & \gate{e^{-iH(x)t}} & \gate{P} & \qw & & & & \lstick{$\rho$} & \gate{e^{ib\pi Q/4}} & \gate{e^{-iH(x)t}} & \meter{$P$}
\end{quantikz}
\caption{Two circuits whose estimates have mean $\mathrm{Im}\,\mathrm{Tr}[\rho\,\tau^x_t(P)\,Q]$. (a) Hadamard test (Lemma~\ref{lem:hadamard-response-correlator}): an ancilla in $\ket{+}$ controls the Pauli strings $Q$ and $P$ around the evolution and is measured in the $Y$ basis. (b) Ancilla-free protocol (Lemma~\ref{lem:ancilla-free-response-correlator}): a rotation $e^{ib\pi Q/4}$ with a uniformly random sign $b=\pm1$, the same evolution, and a measurement of $P$ with outcome $a$; the estimate is $-ab$.}
\label{fig:correlator-protocols}
\end{figure}
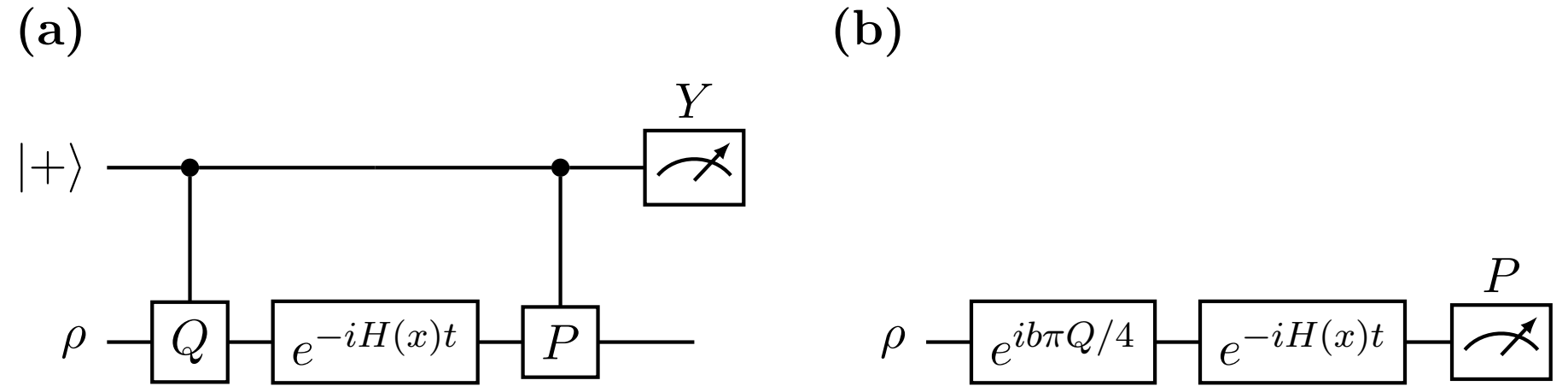

The rotation $e^{ib\pi Q/4}$ is a Clifford operation on the $O(1)$ sites of $Q$: a local pulse when $Q$ acts on a single site, and an entangling gate otherwise.
The Pauli $P$ is measured by single-site basis changes and readout, so this protocol needs no ancilla and no ancilla-controlled operation (Figure~\ref{fig:correlator-protocols} compares the two circuits).
The proofs of both lemmas hold for any unitary in place of $e^{-iH(x)t}$, so under approximate evolution the two protocols still have equal means.
The outcome $Y$ can therefore replace the Hadamard-test outcome in the fresh-copy method of Appendix~\ref{subsubsec:fresh-copy-adaptive}, with the same error.

We define the observable
\begin{equation}\label{eq:response-correlator-observable}
    A_{P,Q}(t;x) := \frac{1}{2i}\,\bigl[\tau^{x}_{t}(P),\,Q\bigr] .
\end{equation}
$A_{P,Q}(t;x)$ is Hermitian because $Q \, \tau^x_t(P) = (\tau^x_t(P) \, Q)^\dagger$. Since $Q \, \tau^x_t(P)$ and $\tau^x_t(P) \, Q$ are both products of unitaries, we have $\bigl\| A_{P,Q}(t;x) \bigr\| \le 1$. Moreover, $\mathrm{Tr}[\rho_d(x) \, Q \, \tau^x_t(P)] = C_{P,Q}(t;x)^*$, hence $\mathrm{Im}\,C_{P,Q}(t;x) = \mathrm{Tr}\bigl[\rho_d(x)\,A_{P,Q}(t;x)\bigr]$. 

In the candidate state $\sigma_V$, the imaginary part of the correlator becomes $c^{V}_{P,Q}(t;x) := \mathrm{Tr}\bigl[\sigma_V\,A_{P,Q}(t;x)\bigr]$.
By Eq.~\eqref{eq:pullback-identity},
\begin{equation}\label{eq:response-through-reference}
    c^{V}_{P,Q}(t;x) = \mathrm{Tr}\bigl[C[x_{\mathrm{ref}}]\,V^\dagger A_{P,Q}(t;x)\,V\bigr].
\end{equation}

Suppose we are given a unitary $\widetilde{U}_x(t)$ satisfying
\begin{equation}\label{eq:response-simulation-error}
    \bigl\| \widetilde U_x(t) - e^{-iH(x)t} \bigr\| \le \epsilon_{\mathrm{sim}} \qquad\text{for every } |t| \le T,
\end{equation}
and let $\widetilde{A}_{P,Q}(t;x) := \frac{1}{2i} \bigl(\widetilde{U}_x(t)^\dagger P \, \widetilde{U}_x(t) \, Q - Q \, \widetilde{U}_x(t)^\dagger P \, \widetilde{U}_x(t)\bigr)$ denote the observable in Eq.~\eqref{eq:response-correlator-observable} for which the Hamiltonian evolution is simulated by $\widetilde{U}_x(t)$. Since $\norm{\widetilde{U}_x(t)^\dagger P \, \widetilde{U}_x(t) \, Q - \tau^x_t(P) \, Q} \leq 2 \epsilon_{\mathrm{sim}}$ and the map $Y \mapsto (Y - Y^\dagger)/(2i)$ is norm non-increasing, we have
\begin{equation}\label{eq:correlator-observable-simulation-error}
    \bigl\| \widetilde A_{P,Q}(t;x) - A_{P,Q}(t;x) \bigr\| \le 2\epsilon_{\mathrm{sim}} .
\end{equation}
The tomography protocols of Appendix~\ref{subsubsec:reference-access-modes} take as input the pulled-back observable
\begin{equation}\label{eq:pulled-back-correlator-observable}
\widetilde B^{V}_{P,Q}(t;x) := V^\dagger\,\widetilde A_{P,Q}(t;x)\,V, \qquad \bigl\| \widetilde B^{V}_{P,Q}(t;x) \bigr\| \le 1 .
\end{equation}

These protocols assume access to each observable through a block encoding.
An efficient block encoding exists for $\widetilde{B}^V_{P,Q}(t;x)$, because $\widetilde{A}_{P,Q}(t;x)$ is a linear combination of two unitaries, $\widetilde{U}_x(t)^\dagger P \, \widetilde{U}_x(t) \, Q$ and its adjoint, with coefficients of modulus one half.
An ancilla qubit and controlled versions of both unitaries hence form a block encoding of $\widetilde{A}_{P,Q}(t;x)$ with normalization one.
A controlled version of $\widetilde U_x(t)^\dagger P\,\widetilde U_x(t)\,Q$ is $\widetilde U_x(t)^\dagger(\text{controlled-}P)\,\widetilde U_x(t)\,(\text{controlled-}Q)$, which controls only the two Pauli strings as in Lemma~\ref{lem:hadamard-response-correlator}.
Conjugating by the transport circuit $V$ and its adjoint $V^\dagger$, we obtain a block encoding of $\widetilde{B}^V_{P,Q}(t;x)$ with normalization one, in which $\widetilde U_x(t)^{\pm1}$ and $V$ are uncontrolled.

\subsubsection{Implementing one edge circuit}\label{subsubsec:transport-edge}

Each edge of a stored path has a stored cutoff $\Delta \ge \gamma_{\mathrm{lb}}/2$.
By Lemma~\ref{lemma:generator-discretization} below, this lower bound lets the transport circuit and its adjoint run efficiently.

For a grid edge from $y$ to $x$, we use the straight edge of Eq.~\eqref{eq:straight-edge}, along which the Hamiltonian is $H_{y \to x}(\tau) = H(y + \tau(x-y))$.
With the unit vector $\widehat u_{yx}$ from $y$ to $x$, the derivative along the edge is
\begin{equation}\label{eq:transport-edge-hamiltonian}
    \partial_\tau H_{y \to x}(\tau) = \|x-y\|\,\widehat u_{yx}\cdot\nabla_x H\bigl(y + \tau(x-y)\bigr), \qquad \tau \in [0,1],
\end{equation}
with derivative bound $\|\partial_\tau H_{y \to x}(\tau)\| \leq \|x-y\|\,J$, since $H$ is a sum of $J$ local terms each with parameter derivative at most one (Definition~\ref{definition: geometrically local hamiltonians}). Fix an energy cutoff $\Delta > 0$.
We use the BMNS filter function $F_{\Delta}$ of Lemma~\ref{lem:BMNS-filter}, which is odd, purely imaginary, and decays sub-exponentially.
With the edge as the path, Definition~\ref{def:quasi-adiabatic-evolution} gives the quasi-adiabatic generator at cutoff $\Delta$:
\begin{equation}\label{eq:transport-edge-generator}
    D_{\Delta}(\tau) = i\int_{-\infty}^{\infty} F_{\Delta}(u)\,e^{iuH_{y \to x}(\tau)}\,\partial_\tau H_{y \to x}(\tau)\,e^{-iuH_{y \to x}(\tau)}\,du.
\end{equation}
Its quasi-adiabatic propagator over the edge is $U_\Delta(y \to x)$.

\begin{lemma}[Edge circuit at cutoff $\Delta$]\label{lemma:generator-discretization}
Suppose that at every point $\tau \in [0,1]$ of the edge, we have block-encoding access to $H_{y \to x}(\tau)$ and $\partial_\tau H_{y \to x}(\tau)$, and Hamiltonian simulation access to $e^{-iuH_{y \to x}(\tau)}$. Then, for any energy cutoff $\Delta > 0$ and $\epsilon_U > 0$, we can implement a unitary $\widetilde U_{\Delta}(y \to x)$ that satisfies
\begin{equation}\label{eq:transport-edge-simulation-error}
\bigl\| \widetilde U_{\Delta}(y \to x) - U_{\Delta}(y \to x) \bigr\| \le \epsilon_U .
\end{equation}
The complexity of this circuit is polynomial if $\Delta \ge 1/\mathrm{poly}(n)$ and $\epsilon_U \ge 1/\mathrm{poly}(n)$.
\end{lemma}
\begin{proof}
We approximate the generator by a finite sum that can be block-encoded, and then simulate its evolution.
Abbreviate $F := F_{\Delta}$, $H(\tau) := H_{y \to x}(\tau)$, and $D(\tau) := D_{\Delta}(\tau)$.
Since $F$ is odd, Eq.~\eqref{eq:transport-edge-generator} becomes
\begin{equation}\label{eq:transport-generator-hermitian-form}
D(\tau) = i\int_0^\infty F(u)\Bigl( e^{iuH(\tau)}\,\partial_\tau H(\tau)\,e^{-iuH(\tau)} - e^{-iuH(\tau)}\,\partial_\tau H(\tau)\,e^{iuH(\tau)} \Bigr)\,du,
\end{equation}
and since $iF$ is real and the bracket is Hermitian, $D(\tau)$ is Hermitian.
First, we truncate at time $u_{\max}$.
Then, we approximate the integral on $[0, u_{\max}]$ using a positive quadrature rule with nodes $u_i$ and weights $\varpi_i$, for $i = 1, \dots, N_{\mathrm{quad}}$.
Finally, we replace each exact evolution $e^{-iu_i H(\tau)}$ by a unitary approximant $\widetilde U_i(\tau)$ satisfying $\bigl\| \widetilde U_i(\tau) - e^{-iu_i H(\tau)} \bigr\| \leq \epsilon_i^{\mathrm{evo}}$. The generator implemented in our algorithm is
\begin{equation}\label{eq:implemented-transport-generator}
    \widehat D(\tau) := i\sum_{i=1}^{N_{\mathrm{quad}}} \varpi_i\,F(u_i)\Bigl( \widetilde U_i(\tau)^\dagger\,\partial_\tau H(\tau)\,\widetilde U_i(\tau) - \widetilde U_i(\tau)\,\partial_\tau H(\tau)\,\widetilde U_i(\tau)^\dagger \Bigr).
\end{equation}
The error in the generator satisfies
\begin{equation}\label{eq:transport-generator-total-error}
    \int_0^1 \bigl\| D(\tau) - \widehat D(\tau) \bigr\|\,d\tau \leq \epsilon^{\mathrm{tail}} + \epsilon^{\mathrm{quad}} + \epsilon^{\mathrm{conj}},
\end{equation}
where
\begin{equation}\label{eq:transport-generator-tail-error}
    \epsilon^{\mathrm{tail}} \leq 2\int_0^1 \|\partial_\tau H(\tau)\|\,d\tau \int_{u_{\max}}^{\infty}|F(u)|\,du,
\end{equation}
\begin{equation}\label{eq:transport-generator-conjugation-error}
    \epsilon^{\mathrm{conj}} \leq 4\int_0^1 \|\partial_\tau H(\tau)\|\,d\tau \sum_{i=1}^{N_{\mathrm{quad}}} \varpi_i\,|F(u_i)|\,\epsilon_i^{\mathrm{evo}},
\end{equation}
and $\epsilon^{\mathrm{quad}}$ is the quadrature error in the truncated integral.
The bound on $\epsilon^{\mathrm{tail}}$ holds because the bracket in Eq.~\eqref{eq:transport-generator-hermitian-form} has norm at most $2\|\partial_\tau H(\tau)\|$.
By Eq.~\eqref{eq:filter-tail-bound} at cutoff $\Delta$, the filter mass beyond $u_{\max}$ obeys
\[
    \int_{u_{\max}}^{\infty}|F(u)|\,du \leq \frac{C_{\mathrm{tail}}}{\Delta}\,F_{\mathrm{ball}}(\Delta u_{\max}),
\]
and hence a truncation time $u_{\max} = \widetilde O\bigl(\Delta^{-1}\log(1/\epsilon_U)\bigr)$ suffices.
For $\epsilon^{\mathrm{quad}}$, the derivatives of the conjugated $\partial_\tau H(\tau)$ in Eq.~\eqref{eq:transport-generator-hermitian-form} satisfy
$\|\partial_u^k(e^{iuH(\tau)}\partial_\tau H(\tau)e^{-iuH(\tau)})\| \leq (2\|H(\tau)\|)^k\,\|\partial_\tau H(\tau)\|$.
A composite quadrature rule with node spacing inverse polynomial in $n$ can then achieve any inverse-polynomial $\epsilon^{\mathrm{quad}}$ with $N_{\mathrm{quad}} = \mathrm{poly}(n)$ nodes.
The bound on $\epsilon^{\mathrm{conj}}$ holds because replacing $e^{-iu_i H(\tau)}$ with $\widetilde U_i(\tau)$ changes each conjugate of $\partial_\tau H(\tau)$ in Eq.~\eqref{eq:implemented-transport-generator} by at most $2\epsilon_i^{\mathrm{evo}}\|\partial_\tau H(\tau)\|$, and each node has two conjugates.

Let $U(\tau)$ and $\widehat U(\tau)$ be the solutions to $\partial_\tau U = iDU$ and $\partial_\tau \widehat U = i\widehat D\widehat U$, respectively, with $U(0) = \widehat U(0) = I$.
The Duhamel identity for two time-dependent evolutions~\cite{childs2019theory} ensures that
\begin{equation}\label{eq:transport-duhamel-step}
    \bigl\| U(1) - \widehat U(1) \bigr\| \leq \int_0^1 \bigl\| D(\tau) - \widehat D(\tau) \bigr\|\,d\tau.
\end{equation}
The implemented generator of Eq.~\eqref{eq:implemented-transport-generator} is a finite linear combination of block-encoded conjugates of $\partial_\tau H(\tau)$.
Thus it can be block-encoded by the standard product and linear-combination constructions.
Given this block encoding, standard time-dependent Hamiltonian simulation implements $\widehat U(1)$ to any chosen error $\epsilon^{\mathrm{td}}$~\cite{berry2015simulating,berry2020timedependent}.
Its cost grows with the norm of the generator.
Bounding the integrand in Eq.~\eqref{eq:transport-edge-generator} gives
$\|D(\tau)\| \le \|\partial_\tau H(\tau)\|\int_{-\infty}^{\infty}|F(u)|\,du = c_F\,\|\partial_\tau H(\tau)\|/\Delta$.
By Eq.~\eqref{eq:filter-constants}, the constant $c_F$ equals $\Delta$ times the total mass of the filter and is independent of $\Delta$.
The implemented generator $\widehat D(\tau)$ obeys the same bound up to the truncation, quadrature, and conjugation errors already accounted for.
We choose the truncation time, number of quadrature nodes, and simulation precisions polynomially so that
\begin{equation}\label{eq:transport-error-budget}
    \epsilon^{\mathrm{tail}} + \epsilon^{\mathrm{quad}} + \epsilon^{\mathrm{conj}} + \epsilon^{\mathrm{td}} \leq \epsilon_U,
\end{equation}
which proves Eq.~\eqref{eq:transport-edge-simulation-error}.
\end{proof}

\subsubsection{Fresh-copy estimation and adaptive measurements}\label{subsubsec:fresh-copy-adaptive}

By default, each sample consumes fresh copies of the reference state.
Our algorithms choose later measurements from earlier outcomes.
The next lemma controls such adaptive measurements, with fresh copies and with one reused pure copy.

\begin{lemma}[Error bounds for adaptive measurements]\label{lem:adaptive-confidence}
\begin{enumerate}
\item[(a)] (Fresh copies for each estimate.) Let $M$ estimates be performed in sequence.
Each estimate uses fresh copies of the reference state.
Its observable and circuit may depend on the classical record of the earlier estimates.
Given that record, it fails with probability at most $\delta/M$. Then all $M$ estimates succeed, except with probability at most $\delta$.
\item[(b)] (Reusing a pure reference state.) Let $\psi = |\psi\rangle\langle\psi|$ be a pure state.
Let a quantum instrument applied to $\psi$ return the classical outcome $z$ with probability $p_z$ and the conditional state $\rho_z$, with marginal disturbance $\bigl\|\sum_z p_z\rho_z - \psi\bigr\|_1 \le d_c$. Then
\begin{equation}\label{eq:reuse-hybrid}
\sum_z p_z\,\|\rho_z - \psi\|_1
\;\le\; 2\sqrt{1 - \bra{\psi}\bigl(\textstyle\sum_z p_z\rho_z\bigr)\ket{\psi}}
\;\le\; \sqrt{2 d_c}\,.
\end{equation}
Consequently, suppose that $M$ instruments are applied in sequence to a single copy.
Each may depend on the earlier classical outcomes.
Each satisfies the marginal disturbance bound $d_c$ on the ideal input $\psi$.
In the ideal sequence, every instrument acts on a copy reset to $\psi$.
Then the joint classical--quantum output of the sequence is within $M\sqrt{2d_c}$ in trace norm of the output of the ideal sequence.
In particular, the total variation distance between the two classical transcripts is at most $M\sqrt{2d_c}/2$.
The final reference state, averaged over the record, is within $M\sqrt{2d_c}$ of $\psi$ in trace norm.
\end{enumerate}
\end{lemma}
\begin{proof}
For (a), let $F_j$ denote the failure event of the $j$-th estimate. We have
  \[
    \Pr\Bigl[\bigcup_{j=1}^M F_j\Bigr]\leq \sum_{j=1}^M\mathbb{E}\Bigl[\Pr\Bigl(F_j\Big|\text{classical record prior to step $j$}\Bigr)\Bigr]\leq M \cdot \frac{\delta}{M}=\delta\,.
  \]

For (b), the Fuchs--van de Graaf inequality for a pure state~\cite{watrous2018book} gives $\norm{\rho-\psi}_1\leq2\sqrt{1-\bra\psi\rho\ket\psi}$ for every state $\rho$.
Applying it to each $\rho_z$ and using the concavity of the square root (Jensen's inequality) gives the first inequality in Eq.~\eqref{eq:reuse-hybrid}.
For the second inequality, we use the variational characterization of the trace distance for the projector $\psi$ to obtain
  \[
    1-\bra{\psi}\sum_z p_z\rho_z\ket{\psi}=\mathrm{Tr}\Bigl[\psi\left(\psi-\sum_z p_z\rho_z\right)\Bigr]\leq\frac{1}{2}\norm{\psi-\sum_z p_z\rho_z}_1\leq\frac{d_c}{2}\,.
  \]
  Thus, $2\sqrt{1-\bra{\psi}\sum_z p_z\rho_z\ket{\psi}}\leq2\sqrt{d_c/2}=\sqrt{2d_c}$.
The reset version of the instrument returns $z$ with probability $p_z$ and resets the copy to $\psi$.
The left-hand side of Eq.~\eqref{eq:reuse-hybrid} is the distance in trace norm between the classical--quantum output $\sum_z p_z\ketbra{z}{z}\otimes\rho_z$ of the instrument and the output $\sum_z p_z\ketbra{z}{z}\otimes\psi$ of its reset version.
We replace the $M$ instruments by their reset versions one at a time, starting with the first.
After the first $k-1$ replacements, the input to the $k$-th instrument is $\psi$, so replacing it costs at most $\sqrt{2d_c}$.
The later adaptive instruments act as a single channel on the classical--quantum state, and channels are contractive in trace norm~\cite{nielsen2002quantum}.
Summing over the $M$ replacements gives $M\sqrt{2d_c}$.
The transcript bound follows because the total variation distance of the transcripts is at most half the distance in trace norm between the classical--quantum states.
The bound on the reference state follows by tracing out the record, which is contractive.
\end{proof}

\subsubsection{Protocols using fewer reference copies}\label{subsubsec:reference-access-modes}

\begin{lemma}[Catalytic ground-state tomography~\cite{chen2024catalytictomography}]\label{lemma:catalytic-tomography}
  Let $H_{\mathrm{ref}}$ be a known Hamiltonian with a unique ground state $|\psi_{\mathrm{ref}}\rangle$ separated from the rest of the spectrum by a gap $\gamma_{\mathrm{ref}} > 0$. Let $B$ be a Hermitian observable with $\|B\| \le 1$, and let $\epsilon_{\mathrm{cat}}, \delta_{\mathrm{cat}} \in (0,1)$. Given controlled Hamiltonian evolution $e^{\pm iH_{\mathrm{ref}}t}$ and a block encoding of $B$, there is a protocol which, from exactly one copy of $|\psi_{\mathrm{ref}}\rangle$, outputs an estimate $\widetilde b$ satisfying
  \begin{equation}\label{eq:catalytic-estimate}
      \Pr\Bigl[ \bigl| \widetilde b - \langle\psi_{\mathrm{ref}}|B|\psi_{\mathrm{ref}}\rangle \bigr| \le \epsilon_{\mathrm{cat}} \Bigr] \ge 1 - \delta_{\mathrm{cat}},
  \end{equation}
  and returns the copy in a state $\widetilde\rho_{\mathrm{ref}}$, averaged over the outcome of the protocol, satisfying
  \begin{equation}\label{eq:catalytic-disturbance}
      \bigl\| \widetilde\rho_{\mathrm{ref}} - |\psi_{\mathrm{ref}}\rangle\langle\psi_{\mathrm{ref}}| \bigr\|_1 \le \delta_{\mathrm{cat}} .
  \end{equation}
  The total controlled-evolution time is
  \begin{equation*}
      O\bigl(\log(1/\delta_{\mathrm{cat}})\,[\log(1/\delta_{\mathrm{cat}}) + \log(1/\epsilon_{\mathrm{cat}})]/(\gamma_{\mathrm{ref}}\,\epsilon_{\mathrm{cat}})\bigr),
  \end{equation*}
  and the protocol makes $O\bigl(\log(1/\delta_{\mathrm{cat}})/\epsilon_{\mathrm{cat}}\bigr)$ calls to the block encoding of $B$.
\end{lemma}

Lemma~\ref{lemma:catalytic-tomography} is Algorithm~III.1 in Ref.~\cite{chen2024catalytictomography}.
In this appendix, we use it when the reference state $C[x_{\mathrm{ref}}]$ $=|\psi_{\mathrm{ref}}\rangle\langle\psi_{\mathrm{ref}}|$ is the unique ground state of a known Hamiltonian $H_{\mathrm{ref}}$, which need not be $H(x_{\mathrm{ref}})$.
The protocol requires a known lower bound $\gamma_{\mathrm{ref}}$ on the gap of $H_{\mathrm{ref}}$, and this bound need only be inverse polynomial.

We reuse this single copy throughout the algorithm.
Let $N_{\mathrm{meas}}$ be the total number of catalytic calls to Lemma~\ref{lemma:catalytic-tomography} in the algorithm.
These calls are adaptive: each observable $B$ may depend on all earlier outcomes.
By Eq.~\eqref{eq:catalytic-disturbance}, each call disturbs the ideal copy $\psi_{\mathrm{ref}} := |\psi_{\mathrm{ref}}\rangle\langle\psi_{\mathrm{ref}}|$ by at most $\delta_{\mathrm{cat}}$ on average.
We therefore compare the actual run with an ideal run in which each call acts on $\psi_{\mathrm{ref}}$.
By Lemma~\ref{lem:adaptive-confidence}(b) with $d_c = \delta_{\mathrm{cat}}$ and $M = N_{\mathrm{meas}}$, the transcripts from the two runs are within $N_{\mathrm{meas}}\sqrt{2\delta_{\mathrm{cat}}}/2$ in total variation distance.
The output copy is within $N_{\mathrm{meas}}\sqrt{2\delta_{\mathrm{cat}}}$ of $\psi_{\mathrm{ref}}$ in trace norm.
Let $\widetilde{B}$ be the observable passed to the catalytic protocol in a call.
In the ideal run, the call outputs a value within $\epsilon_{\mathrm{cat}}$ of $\mathrm{Tr}[\psi_{\mathrm{ref}}\,\widetilde{B}]$, except with probability $\delta_{\mathrm{cat}}$.
Since $\widetilde{B}$ uses the simulated evolution, Eq.~\eqref{eq:correlator-observable-simulation-error} puts its expectation value within $2\epsilon_{\mathrm{sim}}$ of the exact pulled-back value.
Truncating each catalytic estimate $\widetilde{b}$ to the interval $[-1,1]$ only reduces the error.
Therefore, in the actual run, except with probability at most
\begin{equation}\label{eq:catalytic-correlator-error}
    N_{\mathrm{meas}}\,\delta_{\mathrm{cat}} + \frac{N_{\mathrm{meas}}\sqrt{2\delta_{\mathrm{cat}}}}{2},
    \qquad\text{every catalytic estimate has error at most}\qquad
    \epsilon_{\mathrm{cat}} + 2\epsilon_{\mathrm{sim}} .
\end{equation}
Since $N_{\mathrm{meas}} = \mathrm{poly}(n)$ and the cost of each catalytic call grows only polylogarithmically in $1/\delta_{\mathrm{cat}}$, we may choose an exponentially small
\begin{equation}\label{eq:catalytic-disturbance-choice}
    \delta_{\mathrm{cat}} \le \frac{\min\{\delta_{\mathrm{tot}}^2,\; e^{-2cn}\}}{8N_{\mathrm{meas}}^2}
\end{equation}
for any constant $c > 0$, where $\delta_{\mathrm{tot}}$ is the total failure budget.
Then $N_{\mathrm{meas}}\sqrt{2\delta_{\mathrm{cat}}} \le \min\{\delta_{\mathrm{tot}}, e^{-cn}\}/2$ and $N_{\mathrm{meas}}\delta_{\mathrm{cat}} \le \delta_{\mathrm{tot}}/8$.
Hence the total failure probability of all catalytic estimates is at most $3\delta_{\mathrm{tot}}/8$, and the cumulative disturbance of the copy is at most $e^{-cn}/2$.

\begin{lemma}[Thermal shadow tomography of a Gibbs state~\cite{chen_gilyen_thermal_shadows}]\label{lem:thermal-shadow-tomography}
  Let $\rho_\beta := e^{-\beta H}/\mathrm{Tr}[e^{-\beta H}]$ be the Gibbs state of the known Hamiltonian $H$. Let $A_1, \ldots, A_M$ be Hermitian observables with $\|A_j\| \le 1$ for all $j$. Suppose that independent copies of $\rho_\beta$, block encodings of $A_j$, and controlled Hamiltonian evolution $e^{\pm iHt}$ are available. Then for any $\epsilon_{\mathrm{cat}}, \delta_{\mathrm{cat}} \in (0,1)$, there exists a quantum protocol that outputs estimates $\widehat a_1, \ldots, \widehat a_M$ such that
\begin{equation*}
      \Pr\Bigl[ \max_{1 \le j \le M} \bigl| \widehat a_j - \mathrm{Tr}[\rho_\beta A_j] \bigr| \le \epsilon_{\mathrm{cat}} \Bigr] \ge 1 - \delta_{\mathrm{cat}}.
  \end{equation*}
The protocol consumes $O\bigl(\log(M/\delta_{\mathrm{cat}})/\epsilon_{\mathrm{cat}}^2\bigr)$ copies of $\rho_\beta$, queries the $A_j$ block encodings $\widetilde O(M/\epsilon_{\mathrm{cat}}^2)$ times, and runs controlled Hamiltonian evolution for a total time $\widetilde O(\beta M/\epsilon_{\mathrm{cat}}^2)$.
\end{lemma}

For a degenerate ground-state manifold of small enough width, the reference state can be a Gibbs state at a suitable temperature (Appendix~\ref{subsubsec:gibbs-window}).
Copies within a small enough inverse-polynomial trace distance of this Gibbs state also suffice, as the proof of Corollary~\ref{cor:reference-copy-count}(b) shows.
We apply this lemma with $H = H(x_{\mathrm{ref}})$ to batches of pulled-back observables.
A batch is fixed before any of its observables is measured, and between batches we adapt as in Lemma~\ref{lem:adaptive-confidence}(a).
The $A_j$ are the $\widetilde B^{V}_{P,Q}(t;x)$, block-encoded as described below Eq.~\eqref{eq:pulled-back-correlator-observable}.

\subsubsection{From correlator measurements to the filtered response}\label{subsubsec:total-estimation-error}

Consider first a Pauli expansion of the Hermitian probe $O$,
\begin{equation}\label{eq:probe-pauli-decomposition}
O = \sum_{P \in \mathcal P_O} z_P\,P, \qquad z_P \in \mathbb R, \qquad \|z\|_1 := \sum_{P \in \mathcal P_O}|z_P|,
\end{equation}
where $\mathcal P_O$ is a set of Pauli strings, and the coefficients $z_P$ can be efficiently computed.
For the directional derivative $O = \widehat u\cdot\nabla_x H(x)$ along a unit direction $\widehat u$, each term $\partial_{\widehat u} h_j(x)$ is supported on $O(1)$ sites and thus has an expansion in $O(1)$ Pauli strings, with coefficients at most $\|\partial_{\widehat u} h_j(x)\| \leq 1$ (Definition~\ref{definition: geometrically local hamiltonians}).
Then $|\mathcal P_O| = O(J) = O(n)$ and $\|z\|_1 = O(J) = O(n)$.

Using the correlators of Eq.~\eqref{eq:pauli-response-correlator}, the dissipative response expands as
\begin{equation}\label{eq:response-from-pauli-correlators}
\chi_O(t;x) = -2\sum_{P,Q \in \mathcal P_O} z_P z_Q\,\mathrm{Im}\,C_{P,Q}(t;x),
\end{equation}
whose derivation uses only Hermiticity.
The same identity therefore holds with any state in place of $\rho_d(x)$, such as the candidate state $\sigma_V$ of Eq.~\eqref{eq:response-candidate-state}.

We estimate the filtered response by averaging random samples over waiting times and pairs of Pauli strings.
We fix a time cutoff $T > 0$ and define the sampling density
\begin{equation}\label{eq:response-sampling-distributions}
p_T(t) := \frac{|\mathcal K(t)|}{K_T} \quad\text{on } [-T,T] .
\end{equation}
Next, we define the truncated response of the candidate state $\sigma_V$:
\begin{equation}\label{eq:candidate-truncated-response}
S^{O}_{\mathcal K,T}(\sigma_V;x) := \int_{-T}^{T} \mathcal K(t)\; i\,\mathrm{Tr}\bigl[\sigma_V\,[\tau^{x}_{t}(O),\,O]\bigr]\,dt .
\end{equation}
An ideal sample is formed as follows: draw $t \in [-T,T]$ from $p_T$, draw $P,Q$ independently, each with probability $|z_P|/\|z\|_1$, and compute the value
\begin{equation}\label{eq:ideal-response-sample}
X := -2K_T\,\|z\|_1^2\;\sign\bigl(\mathcal K(t)\,z_P z_Q\bigr)\; c^{V}_{P,Q}(t;x) .
\end{equation}
Applying Eq.~\eqref{eq:response-from-pauli-correlators} to $\sigma_V$ gives
\begin{equation}\label{eq:unbiased-candidate-response}
\mathbb E[X] = S^{O}_{\mathcal K,T}(\sigma_V;x),
\end{equation}
and, since $|c^{V}_{P,Q}(t;x)| \le 1$, every sample obeys $|X| \le 2K_T\,\|z\|_1^2$.

In the implementation, the kernel $\mathcal K_{\Lambda,\delta}$ involves the infinite product $w_\delta$ of Eq.~\eqref{eq:quantum-kernel}, so the waiting time $t$ is sampled from an implemented distribution $\widetilde p_T$, and the normalization used is a computed $\widetilde K_T$. These two approximations have precisions $\|\widetilde p_T - p_T\|_1 \le \eta_{\mathrm{samp}}$ and $|\widetilde K_T - K_T| \le \eta_K\,K_T$. The sample actually used is then
\begin{equation}\label{eq:actual-response-sample}
\widetilde X := -2\widetilde K_T\,\|z\|_1^2\;\sign\bigl(\mathcal K(t)\,z_P z_Q\bigr)\; \widetilde c^{V}_{P,Q}(t;x),
\end{equation}
where $\widetilde c^{V}_{P,Q}(t;x) \in [-1,1]$ is a correlator value supplied by one of three estimation methods: fresh copies, catalytic tomography, or thermal shadow tomography.
With fresh copies, it is a single-shot outcome $Y$.
Finally, the estimator is the empirical mean of these samples,
\begin{equation}\label{eq:filtered-response-estimator}
\widehat S^{O}_{\mathcal K,T}(\sigma_V;x) := \frac{1}{N_{\mathrm{samp}}}\sum_{i=1}^{N_{\mathrm{samp}}} \widetilde X_i .
\end{equation}

Each estimation method comes with an error $\epsilon_{\mathrm{corr}}$ and a failure probability $\delta_{\mathrm{mode}}$,
\begin{equation}\label{eq:access-mode-errors}
(\epsilon_{\mathrm{corr}},\,\delta_{\mathrm{mode}}) =
\begin{cases}
(2\epsilon_{\mathrm{sim}},\; 0) & \text{fresh copies},\\
(\epsilon_{\mathrm{cat}}+2\epsilon_{\mathrm{sim}},\; N_{\mathrm{meas}}\delta_{\mathrm{cat}}+N_{\mathrm{meas}}\sqrt{2\delta_{\mathrm{cat}}}/2) & \text{catalytic tomography, Eq.~\eqref{eq:catalytic-correlator-error}},\\
(\epsilon_{\mathrm{cat}}+2\epsilon_{\mathrm{sim}},\; \delta_{\mathrm{cat}}) & \text{thermal shadow tomography}.
\end{cases}
\end{equation}
For fresh copies, $\epsilon_{\mathrm{corr}}$ bounds the bias of each single-shot outcome, and for the two tomography protocols it bounds the error of every correlator value outside the failure event.

\begin{theorem}[Efficient estimation of the filtered response]\label{thm:quantum-filtered-response}\label{theorem:total-estimation-error}
Let $\mathcal K$ be a real kernel with $K_T < \infty$ and $K_{\mathrm{tail}}(T) \to 0$.
Our algorithms use the kernel $\mathcal K_{\Lambda,\delta}$ of Eq.~\eqref{eq:quantum-kernel}.
Suppose that
\begin{enumerate}
\item $0 < \delta \le \Lambda/2$, $\delta \ge 1/\mathrm{poly}(n)$, and $\Lambda \le \mathrm{poly}(n)$, where $\delta$ is the smoothing width of this kernel;
\item the probe $O$ has an efficiently computable Pauli decomposition, Eq.~\eqref{eq:probe-pauli-decomposition}, with $\|z\|_1 = \mathrm{poly}(n)$;
\item for every point of the family, in particular for $x$ and $x_{\mathrm{ref}}$, the Hamiltonian evolution $e^{-iH(x)t}$ can be simulated, without control, to precision $\epsilon_{\mathrm{sim}}$ in time $\mathrm{poly}(n, |t|, \log(1/\epsilon_{\mathrm{sim}}))$. For catalytic and thermal shadow tomography, controlled evolution under the known Hamiltonian of the reference state is also efficient;
\item the transport circuit $V$ contains at most $\mathrm{poly}(n)$ edge circuits, each at its stored cutoff and implemented as in Lemma~\ref{lemma:generator-discretization};
\item the waiting-time distribution and the normalization are implemented to precision $\eta_{\mathrm{samp}}$ and $\eta_K$ as above.
\end{enumerate}
Fix a time cutoff $T > 0$, let
\begin{equation}\label{eq:number-response-samples}
N_{\mathrm{samp}} \ge \frac{8K_T^2\,\|z\|_1^4}{\epsilon_{\mathrm{stat}}^2}\,\log\frac{2}{\delta_{\mathrm{fail}}},
\end{equation}
and let the correlator values be supplied by fresh copies, catalytic tomography, or thermal shadow tomography, with $\epsilon_{\mathrm{corr}}$ and $\delta_{\mathrm{mode}}$ given by Eq.~\eqref{eq:access-mode-errors}. Then, except with probability at most $\delta_{\mathrm{fail}} + \delta_{\mathrm{mode}}$, the estimator of Eq.~\eqref{eq:filtered-response-estimator} satisfies
\begin{equation}\label{eq:total-response-estimation-error}
\bigl| \widehat S^{O}_{\mathcal K,T}(\sigma_V;x) - S^{O}_{\mathcal K}(\sigma_V;x) \bigr| \le \epsilon_{\mathrm{stat}} + 2\|z\|_1^2\,K_{\mathrm{tail}}(T) + 2K_T\,\|z\|_1^2\,\bigl( \epsilon_{\mathrm{corr}} + \eta_{\mathrm{samp}} + \eta_K \bigr),
\end{equation}
where $S^{O}_{\mathcal K}(\sigma_V;x)$ is the filtered response of Eq.~\eqref{eq:filtered-response-signal} in the candidate state itself.
Suppose that $\gamma_{\mathrm{lb}} \ge 1/\mathrm{poly}(n)$ and that all precisions are inverse polynomial.
Suppose also that the kernel has $K_T = \mathrm{poly}(n)$ and an inverse-polynomial $K_{\mathrm{tail}}(T)$ at some $T = \mathrm{poly}(n)$.
Lemma~\ref{lem:quantum-kernel-moments} guarantees this for the kernel $\mathcal K_{\Lambda,\delta}$.
For catalytic tomography, suppose that $\gamma_{\mathrm{ref}}\ge1/\mathrm{poly}(n)$.
For thermal shadow tomography, suppose that the Gibbs reference state has inverse temperature $\beta\le\mathrm{poly}(n)$.
Then every estimation method runs in $\mathrm{poly}(n)$ time.
The methods use $N_{\mathrm{samp}}$ fresh copies, a single catalytic copy, or $O(\log(N_{\mathrm{samp}}/\delta_{\mathrm{cat}})/\epsilon_{\mathrm{cat}}^2)$ Gibbs copies per estimate.
\end{theorem}

\begin{proof}
For any state $\sigma$ we have $|i\,\mathrm{Tr}[\sigma [\tau_t^x(O), O]]| \le \|[\tau_t^x(O), O]\| \le 2\|O\|^2 \le 2\|z\|_1^2$. Integrating this bound against $|\mathcal K(t)|$ over $|t| > T$ gives
\begin{equation}\label{eq:truncation-term}
\bigl| S^{O}_{\mathcal K}(\sigma_V;x) - S^{O}_{\mathcal K,T}(\sigma_V;x) \bigr| \le 2\|z\|_1^2\,K_{\mathrm{tail}}(T).
\end{equation}

For the sampling, all classical draws $(t_i, P_i, Q_i)$ for $i = 1, \dots, N_{\mathrm{samp}}$ are made before any measurement, with $t_i$ drawn from $\widetilde p_T$. Their distribution therefore does not depend on any measurement outcome.
We compare the actual samples $\widetilde X_i$ with samples $X_i$ that differ from them only in the correlator value and the normalization.
For fresh copies, $X_i$ is not the ideal sample: it is the sample of Eq.~\eqref{eq:actual-response-sample} formed from the single-shot outcome $Y_i$, but with the exact normalization $K_T$. For the two tomography protocols, $X_i$ is the ideal sample of Eq.~\eqref{eq:ideal-response-sample}, formed with the exact correlator $c^{V}_{P_i,Q_i}(t_i;x)$.
In both cases the $X_i$ are mutually independent, and $|X_i| \le 2K_T\|z\|_1^2$.

The mean of $X_1$ is close to the truncated response.
Drawing $t_i$ from $\widetilde p_T$ instead of $p_T$ shifts the mean in Eq.~\eqref{eq:unbiased-candidate-response} by at most $2K_T\|z\|_1^2\,\eta_{\mathrm{samp}}$, since $|X_1| \le 2K_T\|z\|_1^2$ and $\|\widetilde p_T - p_T\|_1 \le \eta_{\mathrm{samp}}$.
In addition, for fresh copies, $X_i$ is formed from an outcome of the simulated evolution.
Applying Lemma~\ref{lem:hadamard-response-correlator} to $\sigma_V$ with $\widetilde U_x(t_i)$ in place of $e^{-iH(x)t_i}$ gives $\mathbb E[Y_i \mid t_i, P_i, Q_i] = \mathrm{Tr}[\sigma_V \widetilde A_{P_i,Q_i}(t_i;x)]$, which is within $2\epsilon_{\mathrm{sim}}$ of $c^{V}_{P_i,Q_i}(t_i;x)$ by Eq.~\eqref{eq:correlator-observable-simulation-error}.
Hence $\bigl| \mathbb E[X_1] - S^{O}_{\mathcal K,T}(\sigma_V;x) \bigr| \le 2K_T\|z\|_1^2\,(\eta_{\mathrm{samp}} + 2\epsilon_{\mathrm{sim}})$ for fresh copies, and the same bound without the $2\epsilon_{\mathrm{sim}}$ holds for the two tomography protocols.
By Hoeffding's inequality, the number of samples in Eq.~\eqref{eq:number-response-samples} ensures $\bigl| \frac{1}{N_{\mathrm{samp}}}\sum_i X_i - \mathbb E[X_1] \bigr| \le \epsilon_{\mathrm{stat}}$, except with probability $\delta_{\mathrm{fail}}$.

It remains to bound the readout error $|\widetilde X_i - X_i|$ on the event that the estimation method succeeds.
For every estimation method, the computed normalization contributes at most $|\widetilde K_T - K_T|\cdot 2\|z\|_1^2 \le 2K_T\|z\|_1^2\,\eta_K$, because the supplied value $\widetilde c^{V}_{P_i,Q_i}(t_i;x)$ lies in $[-1,1]$.
For fresh copies, $\widetilde X_i$ and $X_i$ are formed from the same outcome $Y_i$, so this is the whole readout error and $\delta_{\mathrm{mode}} = 0$. The simulation error $\epsilon_{\mathrm{corr}} = 2\epsilon_{\mathrm{sim}}$ enters only through the mean shift above.

For catalytic tomography, we use the error bound of Eq.~\eqref{eq:catalytic-correlator-error}. Outside the failure probability stated there, every estimate is within $\epsilon_{\mathrm{cat}} + 2\epsilon_{\mathrm{sim}}$ of the exact pulled-back correlator. Hence $|\widetilde X_i - X_i| \le 2K_T\|z\|_1^2\,(\epsilon_{\mathrm{cat}} + 2\epsilon_{\mathrm{sim}} + \eta_K)$ for all $i$.

For thermal shadow tomography, we apply Lemma~\ref{lem:thermal-shadow-tomography} with $H = H(x_{\mathrm{ref}})$, $M = N_{\mathrm{samp}}$, and $A_i = \widetilde B^{V}_{P_i,Q_i}(t_i;x)$.
The hypotheses hold: each $A_i$ has norm at most one by Eq.~\eqref{eq:pulled-back-correlator-observable}, and the list of the $A_i$ is fixed by $V$ and by the classical draws, which are made before the batch measurement.
Hence, except with probability $\delta_{\mathrm{cat}}$, all $N_{\mathrm{samp}}$ estimates of $\mathrm{Tr}[C[x_{\mathrm{ref}}]\widetilde B^{V}_{P_i,Q_i}(t_i;x)]$ are simultaneously within $\epsilon_{\mathrm{cat}}$. The simulated evolution adds another $2\epsilon_{\mathrm{sim}}$ by Eq.~\eqref{eq:correlator-observable-simulation-error}, so the readout bound is the same as for catalytic tomography.

Averaging the readout bound over $i$ and adding the mean shift, the Hoeffding term, and the truncation bound of Eq.~\eqref{eq:truncation-term} yields Eq.~\eqref{eq:total-response-estimation-error}, with the $\epsilon_{\mathrm{corr}}$ and $\delta_{\mathrm{mode}}$ for each estimation method as derived above.

We now count the running time. Lemma~\ref{lem:quantum-kernel-moments} and hypothesis~1 give $K_T \le K_0(\mathcal{K}_{\Lambda,\delta}) \le \frac{2}{\pi}\ln\frac{\Lambda}{\delta} + C_{\mathcal{K}} = O(\log n)$, so $N_{\mathrm{samp}}$ is polynomial in $n$ whenever $\epsilon_{\mathrm{stat}}$ is inverse polynomial.
To make the truncation error at most a target $\epsilon$, Eq.~\eqref{eq:quantum-kernel-tail} lets us choose
\begin{equation} \label{eq:quantum-response-time-cutoff}
T = \frac{1}{\delta}\,\widetilde O\Bigl( \log\frac{\|z\|_1^2}{\epsilon} \Bigr).
\end{equation}

Each measurement circuit consists of the transport circuit $V$, two controlled Pauli strings, and uncontrolled simulations of the evolution for time $|t| \le T = \mathrm{poly}(n)$.
The transport circuit has polynomially many edge circuits, each of polynomial cost by Lemma~\ref{lemma:generator-discretization}. The fresh-copy Hadamard test uses one simulation of $e^{-iH(x)t}$, while the block encoding of $\widetilde B^{V}_{P,Q}(t;x)$ for the two tomography protocols uses $e^{-iH(x)t}$ and its inverse.

For catalytic tomography, each call adds a controlled evolution under $H_{\mathrm{ref}}$ for time
\[
O\Bigl(\frac{\log(1/\delta_{\mathrm{cat}})\,[\log(1/\delta_{\mathrm{cat}}) + \log(1/\epsilon_{\mathrm{cat}})]}{\gamma_{\mathrm{ref}}\,\epsilon_{\mathrm{cat}}}\Bigr).
\]
This is polynomial in $n$, because $\gamma_{\mathrm{ref}} \ge 1/\mathrm{poly}(n)$ and, with the choice of Eq.~\eqref{eq:catalytic-disturbance-choice}, $\log(1/\delta_{\mathrm{cat}}) = O(n) + O(\log N_{\mathrm{meas}})$.
In thermal shadow tomography, the number of block-encoding calls is $\widetilde{O}(N_{\mathrm{samp}}/\epsilon_{\mathrm{cat}}^2)$, and the total time of controlled evolution under $H(x_{\mathrm{ref}})$ is $\widetilde{O}(\beta N_{\mathrm{samp}}/\epsilon_{\mathrm{cat}}^2)$, which is polynomial when $\beta \le \mathrm{poly}(n)$.

We set $\epsilon_{\mathrm{stat}}$, $\|z\|_1^2 K_{\mathrm{tail}}(T)$, and the terms scaled by $K_T\|z\|_1^2$ to fixed fractions of the target precision. Since $K_T = \mathrm{poly}(n)$ and $\|z\|_1 = \mathrm{poly}(n)$, every precision needed is inverse polynomial in $n$.

Finally, hypothesis~5 can be met for the kernel $\mathcal K_{\Lambda,\delta}$ with inverse-polynomial $\eta_{\mathrm{samp}}$ and $\eta_K$. We truncate the infinite product $w_\delta$ after $N$ factors. Each omitted factor lies between $1 - a_k^2 t^2 / 3$ and $1$.
Since $a_k = a_1/(k \ln^2 k)$ for $k \ge 2$ by Eq.~\eqref{eq:bmns-weight}, we have $\sum_{k > N} a_k^2 \le a_1^2 / (N \ln^4 N)$.
Hence, if $a_1^2 t^2 / (N \ln^4 N) \le 3/2$, the truncation changes $w_\delta(t)$ by a relative amount of at most $2 a_1^2 t^2 / (3 N \ln^4 N)$. With $a_1 \approx 0.16\,\delta$ and $|t| \le T$, a polynomially large $N$ makes both $\eta_{\mathrm{samp}}$ and $\eta_K$ inverse polynomial.
\end{proof}

We call the estimation procedure from the previous theorem \textsc{EstimateSignal}; it estimates the filtered response in the candidate state $\sigma_V$.
When $\sigma_V$ is close to $\rho_d(x)$, this response is close to $S^{O}_{\mathcal K}(x)$ by Lemma~\ref{lem:filtered-response-stability}.

\subsubsection{Optimizing the probe over local observables}

The classical algorithm in Appendix~\ref{sec:classical-algorithm} requires the largest filtered response over the normalized local probes of the class $\mathcal L$ of Eq.~\eqref{eq:local-class}, which we write as
\begin{equation}\label{eq:local-probe-coefficients}
    O(z) := \sum_{P \in \mathcal{P}_\ell} z_P P, \qquad z \in \mathbb{R}^{N_\mathcal{L}}, \qquad \|z\|_2 \le 1.
\end{equation}
Define the \emph{response matrix} $M_{\mathcal{K}}(x) \in \mathbb{R}^{N_{\mathcal{L}} \times N_{\mathcal{L}}}$ entrywise, for Pauli strings $P, Q \in \mathcal{P}_\ell$, by
\begin{equation}\label{eq:filtered-response-matrix-time}
[M_\mathcal{K}(x)]_{P,Q} := \frac{1}{2} \int_{-\infty}^\infty \mathcal{K}(t) \; i \mathrm{Tr}\Bigl[\rho_d(x) \bigl([\tau^x_t(P), Q] + [\tau^x_t(Q), P]\bigr)\Bigr]\,dt,
\end{equation}
which is well-defined for any integrable kernel function $\mathcal{K}$, since the integrand is bounded in absolute value by $2 |\mathcal{K}(t)|$.

The classical algorithm can compute this matrix only approximately.
The following lemma gives the largest filtered response and shows that an approximate matrix suffices.

\begin{lemma}[Optimization of the filtered response over local probes]\label{lem:optimize-filtered-response}
Let $\mathcal K$ be a real integrable kernel whose window satisfies $q(\omega) \ge 0$ for every $\omega > 0$, as does $q_{\Lambda,\delta}$ for $\delta \le \Lambda/2$ by Eq.~\eqref{eq:quantum-window-shape}, and let $\gamma_d(x) > 0$. Then $M_{\mathcal K}(x)$ is real, symmetric, and positive semidefinite, and for every real $z$
\begin{equation}\label{eq:filtered-response-quadratic-form}
S^{O(z)}_{\mathcal K}(x) = z^{\mathsf T} M_{\mathcal K}(x)\, z .
\end{equation}
Consequently
\begin{equation}\label{eq:filtered-response-maximum}
\max_{\|z\|_2 \le 1} S^{O(z)}_{\mathcal K}(x) = \lambda_{\max}\bigl(M_{\mathcal K}(x)\bigr),
\end{equation}
and the maximum is attained at a unit principal eigenvector of $M_{\mathcal K}(x)$. More generally, let $\widehat M_{\mathcal K}(x)$ be a real symmetric matrix with
\begin{equation}\label{eq:filtered-response-matrix-error}
\bigl\| \widehat M_{\mathcal K}(x) - M_{\mathcal K}(x) \bigr\| \le \epsilon,
\end{equation}
and let $\widehat z$ be a unit principal eigenvector of $\widehat M_{\mathcal K}(x)$. Then the probe $\widehat O := \sum_{P \in \mathcal P_\ell} \widehat z_P\,P$ satisfies
\begin{equation}\label{eq:approximate-optimal-probe}
S^{\widehat O}_{\mathcal K}(x) \ge \lambda_{\max}\bigl(M_{\mathcal K}(x)\bigr) - 2\epsilon .
\end{equation}
\end{lemma}
\begin{proof}
\tolerance=2000 \emergencystretch=1.5em
As $[\tau_t^{x}(P), Q]$ is anti-Hermitian and $\rho_d(x)$ is a state, $\operatorname{Tr}\left[ \rho_d(x) [\tau_t^{x}(P), Q] \right]$ is purely imaginary, and thus each entry of $M_{\mathcal{K}}(x)$ in Eq.~\eqref{eq:filtered-response-matrix-time} is real.
The symmetrization in Eq.~\eqref{eq:filtered-response-matrix-time} ensures that $M_{\mathcal{K}}(x)$ is symmetric.
Since $z$ is real, $O(z)$ is Hermitian.
The integrand of Eq.~\eqref{eq:filtered-response-matrix-time} is bilinear in $(P, Q)$, so $z^{\mathsf T} M_{\mathcal K}(x)\, z = \int \mathcal K(t)\, i \operatorname{Tr}\bigl[\rho_d(x) [\tau^x_t(O(z)), O(z)]\bigr]\, dt$.
By the definition of the filtered response, this is Eq.~\eqref{eq:filtered-response-quadratic-form}.
By Eq.~\eqref{eq:weighted-response}, this quadratic form equals $\frac{1}{d} \sum_{j<d,\, k \ge d} q(\omega_{kj})\, |\langle \psi_k | O(z) | \psi_j \rangle|^2$.
All cross-sector frequencies $\omega_{kj}$ with $j < d$ and $k \ge d$ are at least $\gamma_d(x) > 0$ and $q \ge 0$ on $\omega > 0$, so the quadratic form is non-negative and $M_{\mathcal{K}}(x)$ is positive semidefinite.
The value of the maximum then follows from the Rayleigh principle~\cite{Bha97:Matrix-Analysis}.
Since $\lambda_{\max} \geq 0$, the maximum over the unit ball is attained on the unit sphere, at a unit principal eigenvector of $M_{\mathcal{K}}(x)$.
We use the bound on $M_{\mathcal{K}}(x) - \widehat{M}_{\mathcal{K}}(x)$ twice: first in the quadratic form, then in the largest eigenvalue by Weyl's inequality (Lemma~\ref{lem:weyl}):
\[
S^{\widehat O}_{\mathcal K}(x) = \widehat z^{\mathsf T} M_{\mathcal K}(x)\,\widehat z \ge \widehat z^{\mathsf T} \widehat M_{\mathcal K}(x)\,\widehat z - \epsilon = \lambda_{\max}\bigl(\widehat M_{\mathcal K}(x)\bigr) - \epsilon \ge \lambda_{\max}\bigl(M_{\mathcal K}(x)\bigr) - 2\epsilon .
\]
As $\widehat{M}_{\mathcal{K}}(x)$ is real symmetric, we can choose the principal eigenvector $\widehat{z}$ to be real. Thus $\widehat{O} \in \mathcal{L}$.
\end{proof}

\subsection{Reference states and mean-energy estimation}\label{subsec:reference-states}

Every estimate above is taken in a state transported from a reference state $C[x_{\mathrm{ref}}]$.
When the sector of the lowest $r$ levels has sufficiently small width, a Gibbs state at a suitable temperature is close to the sector state.
Gibbs states can therefore serve as the successful reference states that the local algorithms of Appendix~\ref{sec:local-detection} assume.
The mean energy of a candidate state can also be estimated efficiently.

\subsubsection{Gibbs states as reference states}\label{subsubsec:gibbs-window}

Fix a parameter point $x\in\mathcal M$ and a sector dimension $r$, with sector state $P_r(x)/r$.
Let $w:=E_{r-1}(x)-E_0(x)$ be the sector width, and let $\rho_\beta=e^{-\beta H(x)}/\operatorname{Tr}e^{-\beta H(x)}$ be the Gibbs state at inverse temperature $\beta\ge0$.
For a target accuracy $\epsilon$, let the temperature $1/\beta$ lie between $w/\epsilon$ and $\gamma_r(x)/\log(2^n/\epsilon)$, up to constants.
The Gibbs state is then close to $P_r(x)/r$: the upper limit makes the $2^n-r$ levels above the sector nearly empty, and the lower limit makes the lowest $r$ levels nearly equally populated. The two limits are compatible when $w$ is small compared with $\epsilon\,\gamma_r(x)/\log(2^n/\epsilon)$.

\begin{lemma}[Approximation by a Gibbs state]
\label{lem:gibbs-window}
\label{lem:gibbs-ground-state-approximation}
With $x$, $r$, $w$ and $\rho_\beta$ as above, if the band gap satisfies $\gamma_r(x)>0$, then for every $\beta\ge0$
\begin{equation}
    \Bigl\|\rho_\beta-\frac{P_r(x)}r\Bigr\|_1
    \;\le\;
    e^{\beta w}-1+\frac{2(2^n-r)}{r}\,e^{-\beta\gamma_r(x)}.
    \label{eq:gibbs-window-bound}
\end{equation}
Consequently, for any $0<\epsilon<1$, the conditions
\begin{equation}
    \beta w\le\log\Bigl(1+\frac\epsilon2\Bigr),
    \qquad
    \beta\gamma_r(x)\ge\max\Bigl\{0,\;\log\frac{4(2^n-r)}{r\,\epsilon}\Bigr\}
    \label{eq:gibbs-window}
\end{equation}
are sufficient for $\|\rho_\beta-P_r(x)/r\|_1\le\epsilon$.
\end{lemma}

\begin{proof}
Since $P_r$ commutes with $e^{-\beta H}$, we have $\rho_\beta=(1-q)\tau_r+q\tau_\perp$.
Here $q:=\operatorname{Tr}[(I-P_r(x))\rho_\beta]$ is the population above the sector, and the states $\tau_r$ and $\tau_\perp$ are supported on the sector and on its complement.
The sector partition function obeys $Z_r:=\sum_{j<r}e^{-\beta E_j}\ge re^{-\beta E_{r-1}}$, so the population above the sector is bounded by
\[
    q=\frac{\sum_{k\ge r}e^{-\beta E_k}}{\operatorname{Tr}e^{-\beta H(x)}}
    \le\frac{\sum_{k\ge r}e^{-\beta E_k}}{Z_r}
    \le\frac{(2^n-r)\,e^{-\beta E_r}}{r\,e^{-\beta E_{r-1}}}
    =\frac{2^n-r}{r}\,e^{-\beta\gamma_r(x)},
\]
since each of the $2^n-r$ excited terms is at most $e^{-\beta E_r}$.
Consider the sector populations $p_j=e^{-\beta E_j}/Z_r$. Since $|E_i-E_j|\le w$ for $i,j<r$, each $p_j$ is between $e^{-\beta w}/r$ and $e^{\beta w}/r$.
It follows that $|p_j-1/r|\le(e^{\beta w}-1)/r$ for each $j$, because $e^{\beta w}-1\ge1-e^{-\beta w}$.
Summing these $r$ terms gives $\|\tau_r-P_r/r\|_1\le e^{\beta w}-1$.
We can write $\rho_\beta-P_r/r=(1-q)(\tau_r-P_r/r)+q(\tau_\perp-P_r/r)$.
Then the triangle inequality and $\|\tau_\perp-P_r/r\|_1\le2$ give Eq.~\eqref{eq:gibbs-window-bound}.

In Eq.~\eqref{eq:gibbs-window}, the condition on $\beta w$ makes $e^{\beta w}-1$ at most $\epsilon/2$, and the condition on $\beta\gamma_r(x)$ makes the term with $e^{-\beta\gamma_r(x)}$ at most $\epsilon/2$.
\end{proof}

\subsubsection{Estimating the mean energy}

Fix $x\in\mathcal M$, write $H(x)=\sum_{j=1}^{J}h_j(x)$ as in Definition~\ref{definition: geometrically local hamiltonians}, and put $B_H:=\sum_{j=1}^{J}\|h_j(x)\|\le J$. Let $V$ be a transport circuit of polynomially many edge circuits, and let $\sigma_V=VC[x_{\mathrm{ref}}]V^\dagger$ be the candidate state of Eq.~\eqref{eq:response-candidate-state}. Its mean energy $\mathrm{Tr}[H(x)\sigma_V]$ can be estimated using fresh copies or via tomography.

With fresh copies, one run draws an index $j$ with probability $\|h_j(x)\|/B_H$, applies $V$ to a fresh copy of $C[x_{\mathrm{ref}}]$, measures $h_j(x)$ in its eigenbasis, and records $X:=B_H\,\lambda/\|h_j(x)\|$ for the observed eigenvalue $\lambda$. The estimate $\widehat E$ is the empirical mean of $X$ over $N$ independent runs.

With catalytic or thermal shadow tomography, each term with $h_j(x)\ne0$ is pulled back and normalized to $B_j:=V^\dagger h_j(x)V/\|h_j(x)\|$.
Its expectation $\mathrm{Tr}[C[x_{\mathrm{ref}}]B_j]$ in the reference state is estimated by some $\widehat b_j$. The estimate is then $\widehat E:=\sum_{j:\,h_j(x)\ne0}\|h_j(x)\|\,\widehat b_j$.

\begin{corollary}[Mean energy of a candidate state]\label{cor:mean-energy-estimate}
\tolerance=2000 \emergencystretch=1.5em
With $x$, $V$, $\sigma_V$, $B_H$ and the two estimates $\widehat E$ as above, let $\eta>0$ and $\delta\in(0,1)$.
\begin{enumerate}
\item[(a)] (Fresh copies.) Each run gives $X\in[-B_H,B_H]$ with $\mathbb E[X]=\mathrm{Tr}[H(x)\sigma_V]$. If
\begin{equation}\label{eq:mean-energy-samples}
N\ge\frac{2B_H^2}{\eta^2}\log\frac2\delta,
\end{equation}
then $|\widehat E-\mathrm{Tr}[H(x)\sigma_V]|\le\eta$, except with probability at most $\delta$.
\item[(b)] (Catalytic or thermal shadow tomography.) Each $B_j$ is Hermitian with $\|B_j\|\le1$ and has a polynomial-size block encoding with normalization one. Suppose that $|\widehat b_j-\mathrm{Tr}[C[x_{\mathrm{ref}}]B_j]|\le\epsilon_{\mathrm{cat}}$ for every $j$ with $h_j(x)\ne0$.
Suppose also that these estimates come either from Lemma~\ref{lem:thermal-shadow-tomography} with $M\le J$, applied to independent copies of the Gibbs state of $H(x_{\mathrm{ref}})$, or from at most $J$ calls of Lemma~\ref{lemma:catalytic-tomography} on a single copy of the reference state. The second source applies when the reference state is the unique ground state of a known Hamiltonian with a known inverse-polynomial lower bound on its gap. Then
\begin{equation}\label{eq:mean-energy-tomography}
\bigl|\widehat E-\mathrm{Tr}[H(x)\sigma_V]\bigr|\le B_H\,\epsilon_{\mathrm{cat}} ,
\end{equation}
so that $\epsilon_{\mathrm{cat}}\le\eta/B_H$ gives error at most $\eta$ on the success event of the tomography protocol.
\item[(c)] (State robustness.) For any two states $\sigma$ and $\sigma'$,
\begin{equation}\label{eq:mean-energy-robustness}
\bigl|\mathrm{Tr}[H(x)\sigma]-\mathrm{Tr}[H(x)\sigma']\bigr|\le B_H\,\|\sigma-\sigma'\|_1 .
\end{equation}
\end{enumerate}
For $\eta\ge1/\mathrm{poly}(n)$ and $\delta\ge e^{-\mathrm{poly}(n)}$, every estimation method runs in $\mathrm{poly}(n)$ time, with $\delta_{\mathrm{cat}}$ chosen as in Eq.~\eqref{eq:catalytic-disturbance-choice} for catalytic tomography and a Gibbs reference state with $\beta\le\mathrm{poly}(n)$ for thermal shadow tomography.
\end{corollary}
\begin{proof}
For (a), $|\lambda|\le\|h_j(x)\|$ implies $|X|\le B_H$. Conditioned on $j$,
\[
\mathbb E[X\mid j]=\frac{B_H}{\|h_j(x)\|}\,\mathrm{Tr}\bigl[\sigma_V\,h_j(x)\bigr],
\qquad
\mathbb E[X]=\sum_{j:\,h_j(x)\ne0}\frac{\|h_j(x)\|}{B_H}\cdot\frac{B_H}{\|h_j(x)\|}\,\mathrm{Tr}\bigl[\sigma_V\,h_j(x)\bigr]=\mathrm{Tr}\bigl[H(x)\sigma_V\bigr],
\]
since terms with $h_j(x)=0$ contribute nothing to $H(x)$.
Hoeffding's inequality~\cite[Theorem~2.2.6]{vershynin2018high} for $N$ independent samples with range $2B_H$ implies Eq.~\eqref{eq:mean-energy-samples}.

For (b), let $\widehat h_j:=h_j(x)/\|h_j(x)\|$ be a Hermitian contraction on $O(1)$ sites. Define
\[
W_j:=\sigma^z\otimes\widehat h_j+\sigma^x\otimes\sqrt{I-\widehat h_j^{\,2}}
\]
acting on one ancilla qubit and the sites of $h_j(x)$.
The operator $W_j$ is Hermitian, and
\[
W_j^2=I\otimes\bigl(\widehat h_j^{\,2}+I-\widehat h_j^{\,2}\bigr)=I,
\]
because $\widehat h_j$ commutes with the square root and $\sigma^z\sigma^x=-\sigma^x\sigma^z$ cancels the cross terms.
Thus $W_j$ is a constant-size unitary with $(\langle0|\otimes I)\,W_j\,(|0\rangle\otimes I)=\widehat h_j$. Conjugating by the transport circuit, $(I\otimes V^\dagger)\,W_j\,(I\otimes V)$ block-encodes $B_j$ with normalization factor one and polynomial size.
Eq.~\eqref{eq:pullback-identity} gives $\mathrm{Tr}[C[x_{\mathrm{ref}}]B_j]=\mathrm{Tr}[\sigma_V\widehat h_j]$, hence
\[
\Bigl|\sum_{j:\,h_j(x)\ne0}\|h_j(x)\|\,\widehat b_j-\mathrm{Tr}\bigl[H(x)\sigma_V\bigr]\Bigr|
\le\sum_{j:\,h_j(x)\ne0}\|h_j(x)\|\,\bigl|\widehat b_j-\mathrm{Tr}[C[x_{\mathrm{ref}}]B_j]\bigr|\le B_H\,\epsilon_{\mathrm{cat}} .
\]
Appendix~\ref{subsubsec:reference-access-modes} verifies the remaining hypotheses of the two tomography lemmas.
For catalytic tomography, these calls count toward $N_{\mathrm{meas}}$ in Eq.~\eqref{eq:catalytic-disturbance-choice}.

For (c), Eq.~\eqref{eq:mean-energy-robustness} follows from H\"older's inequality and $\|H(x)\|\le B_H$.
The running times follow from Eq.~\eqref{eq:mean-energy-samples} and the copy, call, and evolution time bounds in Lemmas~\ref{lem:thermal-shadow-tomography} and~\ref{lemma:catalytic-tomography} with $\epsilon_{\mathrm{cat}}=\eta/B_H$.
\end{proof}

\section{Learning the Phase Diagram with Local Measurements}\label{sec:local-detection}

This appendix proves that we can efficiently learn the phase diagram in the one-parameter and restricted cases of the main text (Theorems~\ref{thm:local-phase-diagram} and~\ref{thm:local-phase-diagram-general}). The restricted case allows a fixed number of parameters but needs the geometric condition of Assumption~\ref{assumption:random-walk-navigation}. The most general result needs no geometric condition (Theorem~\ref{thm:detection}). We state it here because it shares the setup below. Appendix~\ref{sec:detection} proves it.

The algorithms in this appendix use local measurements and, when needed, a purity test. In particular, they estimate the energy, which is a sum of local terms, and response correlators built from local Pauli operators. These estimates can be obtained from fresh copies of the reference state or, under the assumption that each supplied reference state is the unique ground state of a known Hamiltonian, from a single copy of each reference state. In the latter case, catalytic tomography allows each reference state to be reused with exponentially small total disturbance. The implementations of these measurement protocols are described in Section~\ref{sec:diabatic-signals}.

The purity test measures two copies of a reference state at once. With the additional proviso that every reference state, successful or not, is approximately maximally mixed on a subspace of known dimension (Remark~\ref{rem:local-measurements}), the purity test is unnecessary. Subsection~\ref{subsec:experimental-implementations} replaces quasi-adiabatic evolution with parameter sweeps. Under the maximally-mixed proviso, experiments need only parameter sweeps and spatially local measurements, not a fault-tolerant quantum computer. In contrast, the dynamic program of Appendix~\ref{sec:detection} needs coherent energy filters of the whole Hamiltonian, and hence a fault-tolerant quantum computer.

\subsection{Setup and the main theorem}
\label{subsec:setup-main-theorem}

A preparation is run at polynomially many parameter points.
These points form the set $S\subset\mathcal M$.
We call the points of $S$ \emph{reference points}.
For each $x\in S$, the algorithm receives independent copies of the preparation's output $C[x]$.
We call $C[x]$ the \emph{reference state} at $x$.

The preparation at $x$ is \emph{successful} if its reference state is close to the maximally mixed ground state:
\begin{equation}\label{eq:successful-preparation}
    \bigl\|C[x]-\rho_{\mathcal G}(x)\bigr\|_1
    \le
    \epsilon_{\mathrm{ref}} .
\end{equation}
The reference precision satisfies $\epsilon_{\mathrm{ref}}\ge1/\mathrm{poly}(n)$, so preparations need only inverse-polynomial accuracy.
The proofs set $\epsilon_{\mathrm{ref}}$ to a small inverse polynomial.
We assume that the ground-state degeneracy is at most $r_{\max}$.
The resolution $\gamma_{\mathrm{lb}}$ and the threshold $\gamma_{\mathrm{thresh}}=\Theta(1)$ are those of Definition~\ref{def:phase-clustering}.
The algorithms of this appendix and of Appendix~\ref{sec:detection} test the gap at $(1-\epsilon_{\mathrm{thresh}})\gamma_{\mathrm{thresh}}$, using a margin $1/\mathrm{poly}(n)\le\epsilon_{\mathrm{thresh}}\le1/2$.

If the reference state at $x$ is the unique ground state of a known Hamiltonian with a known inverse-polynomial lower bound on its gap, we may instead be given one reusable copy of it.
This Hamiltonian need not be $H(x)$.

\begin{assumption}[Reference states]
\label{assumption-reference-state}
\label{assumption-reference-state-degenerate}
The resolved region (Definition~\ref{def:large-gap-component}) has at most polynomially many connected components, and each of them
contains a reference point whose preparation succeeds in the sense
of Eq.~\eqref{eq:successful-preparation}. The remaining preparations may
fail arbitrarily, and the algorithm is not told which preparations
succeed. The ground-state manifold is spanned by the levels within $\delta_{\mathrm{gsm}}$
of the ground-state energy (Definition~\ref{definition: ground-state manifold}). The tolerance $\delta_{\mathrm{gsm}}$ satisfies
\begin{equation}
    \delta_{\mathrm{gsm}}
    \le\frac{\gamma_{\mathrm{lb}}}{\max\{4r_{\max},\,16\}} .
    \label{eq:1d-gsm-tolerance}
\end{equation}
\end{assumption}

\begin{theorem}[Detailed form of Theorem~\ref{thm:detection-informal}]
\label{thm:detection}
Under Assumption~\ref{assumption-reference-state}, there exists a quantum
algorithm (Algorithm~\ref{alg:overall-dp}) running in $\mathrm{poly}(n)$ time that, with
probability at least $1-1/\mathrm{poly}(n)$, learns the phase diagram
(Definition~\ref{def:phase-clustering}).
Suppose that every reference state, successful or not, is the unique
ground state of a known Hamiltonian with a known inverse-polynomial lower bound on its gap, and that controlled evolution under these Hamiltonians is efficient.
Then Algorithm~\ref{alg:overall-dp} uses a single
copy of each reference state and
perturbs it by only an amount exponentially small in $n$
(Corollary~\ref{cor:reference-copy-count}(a)).
Otherwise,
Algorithm~\ref{alg:overall-dp} uses polynomially many copies of
each reference state.
If every reference state is within a small enough inverse-polynomial trace distance of a Gibbs state of a known Hamiltonian, with polynomially bounded inverse temperature and efficient controlled evolution, then Corollary~\ref{cor:reference-copy-count}(b) gives a smaller copy count.
\end{theorem}
\begin{proof}
See Subsection~\ref{app:main-algorithm-proof}.
\end{proof}

\begin{corollary}[Learning the components that contain a successful reference point]
\label{cor:components-with-reference-point}
Suppose only some connected components in
Assumption~\ref{assumption-reference-state} contain a successful reference point. Then
Algorithm~\ref{alg:overall-dp}, run on the same $S$ and without being told which
components these are, learns the phase diagram
(Definition~\ref{def:phase-clustering}) on the input points that lie in
these components.
\end{corollary}
\begin{proof}
See Subsection~\ref{app:main-algorithm-proof}.
\end{proof}

\subsection{The filtered-response test}
\label{subsec:spectral-gap-certification}

The \emph{filtered-response test} takes a candidate state at a grid point $x$ and measures its filtered response to a Hermitian operator $O$.
The local algorithms link two input points only if this test detects no low-energy response at any grid point of a path between them.
For a rank $r$, we use the sector objects of Definition~\ref{def:spectral-sector} at $d=r$: the projector $P_r(x)$ onto the lowest $r$ levels, the sector state $\rho_r(x)=P_r(x)/r$, and the band gap $\gamma_r(x)$.

The test uses the response kernel $\mathcal K_{\Lambda,\delta}$ of Eq.~\eqref{eq:quantum-kernel}, whose window edge $\Lambda$ lies below $\gamma_{\mathrm{thresh}}$:
\begin{equation}
    \delta
    :=
    \min\Bigl\{
        \frac{\gamma_{\mathrm{lb}}}2,\
        \frac{\epsilon_{\mathrm{thresh}}\gamma_{\mathrm{thresh}}}4
    \Bigr\},
    \qquad
    \Lambda:=(1-\epsilon_{\mathrm{thresh}})\gamma_{\mathrm{thresh}}+\delta .
    \label{eq:kernel-parameters}
\end{equation}
Whenever $\gamma_r(x)\ge\gamma_{\mathrm{lb}}/2$, Lemma~\ref{lem:quantum-kernel} sandwiches the response of $\rho_r(x)$ between two spectral weights,
\begin{equation}
    W^{O}_{[\gamma_r,(1-\epsilon_{\mathrm{thresh}})\gamma_{\mathrm{thresh}})}(x)
    \ \le\
    S^{O}_{\mathcal K_{\Lambda,\delta}}(x)
    \ \le\
    W^{O}_{[\gamma_r,(1-\epsilon_{\mathrm{thresh}}/2)\gamma_{\mathrm{thresh}})}(x).
    \label{eq:response-sandwich}
\end{equation}
The lemma applies because $\delta\le\gamma_{\mathrm{lb}}/2$ and $\delta\le\gamma_{\mathrm{thresh}}/8<\Lambda/2$ (as $\epsilon_{\mathrm{thresh}}\le1/2$). The lemma's windows end at $\Lambda-\delta=(1-\epsilon_{\mathrm{thresh}})\gamma_{\mathrm{thresh}}$ and $\Lambda+\delta\le(1-\epsilon_{\mathrm{thresh}}/2)\gamma_{\mathrm{thresh}}$.

The test returns \textsc{Disconnected} if the estimated response is at least $\epsilon_{\mathrm{sig}}/2$, where the signal scale $\epsilon_{\mathrm{sig}}$ is defined in Eq.~\eqref{eq:witness-scale} for one parameter and in Subsubsection~\ref{subsubsec:general-algorithm} for a fixed number of parameters. We show that every resolved tested path between different phases has a point with response at least $\epsilon_{\mathrm{sig}}$.
In one dimension, the probe is the derivative of the Hamiltonian:
\begin{equation}
    O(x):=\partial_xH(x).
    \label{eq:edge-probe}
\end{equation}

\begin{algorithm}
\caption{The filtered-response test}
\label{alg:edge-test}
\label{alg:gap_cert}
\begin{algorithmic}[1]
\Require A vertex $x$ of the refined grid (Subsection~\ref{subsec:grid-choice}), a Hermitian probe $O$ (in one dimension,
  $O(x)$ of Eq.~\eqref{eq:edge-probe}), and a candidate state $\sigma$,
  prepared as in
  Theorem~\ref{thm:quantum-filtered-response}.
\Ensure An estimated response $\widehat S$ and a status
  $\in\{\textsc{Connected},\textsc{Disconnected}\}$.
\State Estimate the filtered response
  $S^{O}_{\mathcal K_{\Lambda,\delta}}(\sigma;x)$,
  Eq.~\eqref{eq:filtered-response-signal},
  to additive precision $\epsilon_{\mathrm{sig}}/8$, using the
  estimator \textsc{EstimateSignal} of
  Theorem~\ref{thm:quantum-filtered-response}; denote
  the result by $\widehat S$.
\If{$\widehat S\ge\epsilon_{\mathrm{sig}}/2$}
  \State \Return \textsc{Disconnected}.
\Else
  \State \Return \textsc{Connected}.
\EndIf
\end{algorithmic}
\end{algorithm}

\begin{lemma}[Correctness of Algorithm~\ref{alg:gap_cert}]
\label{lem:edge-test}
\label{lem:gap-cert-efficiency}
Suppose the
test is run at $x$ with probe $O$, and that its estimate
$\widehat S(x)$ is within $\epsilon_{\mathrm{sig}}/4$ of the
response of the exact sector state,
\begin{equation}
    \Bigl|
        \widehat S(x)
        -
        S^{O}_{\mathcal K_{\Lambda,\delta}}\bigl(\rho_r(x);x\bigr)
    \Bigr|
    \le\frac{\epsilon_{\mathrm{sig}}}4 .
    \label{eq:edge-test-hypothesis}
\end{equation}
If
\begin{equation}
    \gamma_r(x)
    \ge\Bigl(1-\frac{\epsilon_{\mathrm{thresh}}}2\Bigr)\gamma_{\mathrm{thresh}},
    \label{eq:edge-test-connected}
\end{equation}
then Algorithm~\ref{alg:edge-test} returns \textsc{Connected}.
\end{lemma}

\begin{proof}
Since $\Lambda+\delta=(1-\epsilon_{\mathrm{thresh}})\gamma_{\mathrm{thresh}}+2\delta\le(1-\epsilon_{\mathrm{thresh}}/2)\gamma_{\mathrm{thresh}}$, the hypothesis in Eq.~\eqref{eq:edge-test-connected} implies $\gamma_r(x)\ge\Lambda+\delta$.
Hence the exact response is zero (Eq.~\eqref{eq:response-kernel-null}).
Then Eq.~\eqref{eq:edge-test-hypothesis} yields $\widehat S(x)\le\epsilon_{\mathrm{sig}}/4<\epsilon_{\mathrm{sig}}/2$, so the output is \textsc{Connected}.
\end{proof}

The proof of Theorem~\ref{thm:local-phase-diagram} shows that the test returns \textsc{Disconnected} at some vertex of a resolved interval whose endpoints lie in different phases (Eqs.~\eqref{eq:proof-witness-weight}--\eqref{eq:proof-witness-vertex}).

\subsection{Ground-state evolution and state selection}
\label{subsec:energy-rank-estimation}
\label{subsec:local-measurements}

We index the reference points as $x_a^{\mathrm{ref}}$, $a=1,\ldots,N_{\mathrm{ref}}$.
A failed preparation may output any state.
To compare candidate states by energy, the local algorithms first test the purity of the reference states.
The \emph{purity test} estimates the purity $\operatorname{Tr}[C[x_a^{\mathrm{ref}}]^2]$ of each reference state to accuracy $\epsilon_p/2$, using a swap test on pairs of copies.
For each rank $q\ge2$, it admits the reference point at rank $q$ if the estimate is within $2\epsilon_{\mathrm{ref}}+\epsilon_p/2$ of $1/q$.
Every reference point is admitted at rank $1$.
The test is therefore needed only if $r_{\max}\ge2$.
A state $\sigma$ satisfies the \emph{purity condition} for rank $q$ if $q=1$ or $\left|\operatorname{Tr}[\sigma^2]-\frac1q\right|\le2\epsilon_{\mathrm{ref}}+\epsilon_p$.
A reference state admitted at rank $q$ satisfies it.
A successful reference point of ground-state degeneracy $r$ is admitted at rank $r$.
Indeed, $\|C[x_a^{\mathrm{ref}}]-\rho_{\mathcal G}(x_a^{\mathrm{ref}})\|_1\le\epsilon_{\mathrm{ref}}$ puts the purity of its reference state within $2\epsilon_{\mathrm{ref}}$ of $1/r$.

The purity condition limits the weight that a state can put on a few levels.
For any rank-$d$ projector $\Pi$ and state $\sigma$, the Cauchy--Schwarz inequality gives the \emph{counting bound}
\begin{equation}
\operatorname{Tr}[\Pi\sigma]
\le\sqrt{d\operatorname{Tr}[\sigma^2]}.
\label{eq:purity-counting}
\end{equation}
Hence, if $2\epsilon_{\mathrm{ref}}+\epsilon_p\le1/(2q^2)$, no $q-1$ levels can have more than $1-1/(4q)$ of the weight of a state satisfying the purity condition for rank $q$.
The admitted ranks and mean-energy estimates identify the maximally mixed ground state for the filtered-response test.

The algorithm need not know where the ground-state degeneracy changes. Quasi-adiabatic evolution using $U_{\gamma_{\mathrm{lb}}/2}$ is unitary and therefore preserves purity. Hence each candidate state arriving at $x$ satisfies the purity condition at every rank at which its reference point is admitted. We say that the candidate state is admitted at these ranks. A gapped path from a successful reference point of ground-state degeneracy $r$ produces a state within $\epsilon_{\mathrm{ref}}$ of the sector state $\rho_r(x)$ at $x$ (Corollary~\ref{cor:exact-transport} with $d=r$). If the path stays in the resolved region, $\rho_r(x)$ is the maximally mixed ground state. A failed preparation, or a path that crosses an unresolved region, may produce an incorrect candidate state.

Since the algorithm does not know which candidate states are correct, selecting the lowest mean energy is not enough. A candidate state admitted at a rank smaller than $r(x)$ may appear to have lower energy than the maximally mixed ground state $\rho_r(x)$, while no candidate state admitted at a rank larger than $r(x)$ can have near-minimal energy. Lemma~\ref{lem:energy-rank-estimate} shows that $r(x)$ is the largest rank with a candidate state of near-minimal energy, so the algorithm need not know where a preparation or the transport fails.

At a vertex $x$, suppose for now that each rank $q$ has at most one candidate path, and let $V_q[x]$ denote its transport circuit. The candidate state $\sigma_{V_q[x]}$ is meant to approximate the maximally mixed state on the lowest $q$ levels.
At $x$, the mean energy of a density matrix $\sigma$ is $\operatorname{Tr}[H(x)\sigma]$.
Measuring the local terms of $H(x)$ on copies of $\sigma$ estimates it (Corollary~\ref{cor:mean-energy-estimate}).
For the lemma below, assume that $\sigma_{V_q[x]}$ satisfies the purity condition for rank $q$. Let $\widehat E_q[x]$ be an estimate of its mean energy up to the energy tolerance $\eta_E$, $\left|\widehat E_q[x]-\operatorname{Tr}[H(x)\sigma_{V_q[x]}]\right| \le\eta_E$. The rank comparisons below run over the ranks $q\in\{1,\ldots,r_{\max}\}$ that have a candidate path.

\begin{lemma}[Determining the ground-state degeneracy from mean energies]
\label{lem:energy-rank-estimate}
Let $x$ have ground-state degeneracy $r\le r_{\max}$, with
\begin{equation}
    E_{r-1}(x)-E_0(x)\le\delta_{\mathrm{gsm}},
    \qquad
    E_r(x)-E_{r-1}(x)\ge\gamma_{\mathrm{lb}}.
    \label{eq:energy-rank-gap}
\end{equation}
Suppose $V_r[x]$ is available and its candidate state satisfies
\begin{equation}
    \operatorname{Tr}[H(x)\sigma_{V_r[x]}]
    \le \operatorname{Tr}[H(x)\rho_r(x)]+\epsilon_E .
    \label{eq:correct-rank-energy}
\end{equation}
If
\begin{equation}
    2\delta_{\mathrm{gsm}}+2\epsilon_E+4\eta_E
    <\frac{\gamma_{\mathrm{lb}}}{4r_{\max}},
    \qquad
    2\epsilon_{\mathrm{ref}}+\epsilon_p\le\frac1{2r_{\max}^2},
    \label{eq:energy-rank-precision}
\end{equation}
then
\begin{equation}
    \widehat r(x):=
    \max\left\{
        q:\widehat E_q[x]
        \le\min_j\widehat E_j[x]
           +\delta_{\mathrm{gsm}}+\epsilon_E+2\eta_E
    \right\}
    \label{eq:energy-rank-estimate}
\end{equation}
equals $r$.
\end{lemma}

\begin{proof}

Each candidate state is a density matrix with energy at least $E_0(x)$, hence $\min_j\widehat E_j[x]\ge E_0(x)-\eta_E$. For the correct rank,
\[
    \widehat E_r[x]
    \le \operatorname{Tr}[H(x)\rho_r(x)]+\epsilon_E+\eta_E
    \le E_0(x)+\delta_{\mathrm{gsm}}+\epsilon_E+\eta_E.
\]
Therefore, rank $r$ is in the comparison window in Eq.~\eqref{eq:energy-rank-estimate}.

Now let $q>r$, so that $q\ge2$. By assumption, $\sigma_{V_q[x]}$ satisfies the purity condition for rank $q$.
Since $q\le r_{\max}$, the second condition in Eq.~\eqref{eq:energy-rank-precision} gives $2\epsilon_{\mathrm{ref}}+\epsilon_p\le1/(2q^2)$.
The counting bound then limits the weight of the lowest $r\le q-1$ levels to at most $1-1/(4q)$.
All other levels have energy at least $E_r(x)\ge E_0(x)+\gamma_{\mathrm{lb}}$.
Hence
\begin{equation}
    \operatorname{Tr}[H(x)\sigma_{V_q[x]}]
    \ge E_0(x)+\frac{E_r(x)-E_0(x)}{4q}
    \ge E_0(x)+\frac{\gamma_{\mathrm{lb}}}{4r_{\max}}.
\label{eq:larger-rank-energy}
\end{equation}
With the bound on $\widehat E_r[x]$ above,
\[
\begin{aligned}
    \widehat E_q[x]-\min_j\widehat E_j[x]
    &\ge\widehat E_q[x]-\widehat E_r[x]\\
    &\ge\frac{\gamma_{\mathrm{lb}}}{4r_{\max}}
          -\delta_{\mathrm{gsm}}-\epsilon_E-2\eta_E\\
    &>\delta_{\mathrm{gsm}}+\epsilon_E+2\eta_E,
\end{aligned}
\]
with the strict inequality from Eq.~\eqref{eq:energy-rank-precision}. Hence no larger rank qualifies, and $\widehat r(x)=r$.
\end{proof}

When several candidate states share a rank, we compare ranks using the minimum of their estimated energies.
If all estimated energies are $\eta_E$-accurate, this minimum is an $\eta_E$-accurate estimate of the lowest energy at this rank. Lemma~\ref{lem:energy-rank-estimate} then applies.

Fix a rank $r$ with $\gamma_r(x)>0$. A candidate path's \emph{ideal state} is obtained from its reference state by error-free transport. We show that the lowest-energy candidate path of this rank has an ideal state close to $\rho_r(x)$. Suppose that each estimated energy is within $\eta_E$ of the candidate path's ideal mean energy. Suppose also that some candidate path has ideal mean energy at most $\operatorname{Tr}[H(x)\rho_r(x)]+\epsilon_E$. Let $\sigma$ denote the ideal state of the candidate path with the lowest estimated energy. Comparing its estimated energy with that of such a candidate path gives $\operatorname{Tr}[H(x)\sigma]-\operatorname{Tr}[H(x)\rho_r(x)] \le \epsilon_E+2\eta_E$.

Write $\lambda:=1-\operatorname{Tr}[P_r(x)\sigma]$ for the population of $\sigma$ outside the lowest $r$ levels.
The lowest $r$ levels have energy at least $E_0(x)$.
All other levels have energy at least $E_r(x)$.
Hence $\operatorname{Tr}[H(x)\sigma]\ge E_0(x)+\lambda\bigl(E_r(x)-E_0(x)\bigr)$.
Since $\operatorname{Tr}[H(x)\rho_r(x)]\le E_{r-1}(x)$ and $E_r(x)-E_0(x)\ge\gamma_r(x)$,
\begin{equation}
  \gamma_r(x)\,\lambda
  \le \operatorname{Tr}[H(x)\sigma]-\operatorname{Tr}[H(x)\rho_r(x)]+E_{r-1}(x)-E_0(x).
  \label{eq:1d-energy-leakage}
\end{equation}
Suppose now that $\sigma$ satisfies the purity condition for rank $r$.
As in the proof of Corollary~\ref{cor:large-gap}, the purity condition turns $\lambda$ into a trace distance:
\begin{equation}
  \bigl\|\sigma-\rho_r(x)\bigr\|_1
  \le2\sqrt{\lambda}+\sqrt{2\lambda+r(2\epsilon_{\mathrm{ref}}+\epsilon_p)}.
  \label{eq:1d-selected-state}
\end{equation}

The last term of Eq.~\eqref{eq:1d-energy-leakage}, the width of the lowest $r$ levels, cannot be omitted: a state with the purity of $\rho_r(x)$ can put extra weight on the lowest level and spend the energy saved on population above the gap.
To keep this population small when $r_{\max}\ge2$, Theorems~\ref{thm:local-phase-diagram} and~\ref{thm:local-phase-diagram-general} assume a smaller tolerance:
\begin{equation}
  \delta_{\mathrm{gsm}}
  \le\frac{\epsilon_{\mathrm{ref}}\gamma_{\mathrm{lb}}}{2}.
  \label{eq:gsm-width}
\end{equation}
The intended ground-state manifold must have width at most $\epsilon_{\mathrm{ref}}\gamma_{\mathrm{lb}}/2$.
Otherwise, lowering $\delta_{\mathrm{gsm}}$ to meet this condition removes some of its levels from the ground-state manifold and changes the ground-state degeneracy.
The condition is mild: discrete symmetry breaking and topological order split their ground-state manifolds by amounts exponentially small in the linear system size, and the Gibbs reference states of Subsection~\ref{subsec:resources} already require a smaller $\delta_{\mathrm{gsm}}$.

If $\|\sigma_{V_r[x]}-\rho_r(x)\|_1\le\epsilon$, then Corollary~\ref{cor:mean-energy-estimate}(c) gives the energy hypothesis Eq.~\eqref{eq:correct-rank-energy} with $\epsilon_E=J\epsilon$. Here the $J$ local terms of $H(x)$ have norms summing to $B_H\le J$.

\subsection{The parameter-space grid}
\label{subsec:grid-choice}
\label{subsubsec:grid-gap-requirements}

Measurements and linking are performed on a refined grid.
We begin with a uniform cubic grid of spacing $\delta x$ and refine it by inserting coordinate hyperplanes through all input points and reference points, so that these points become vertices.
The refined grid has edges parallel to the coordinate axes.

Clause~(ii) of Definition~\ref{def:phase-clustering} concerns paths of length $O(1)$. Let $L_{\max}=O(1)$ bound their length.
Consider a path $x:[0,1]\to\mathcal M$ between two vertices of the refined grid, of length
$L\leq L_{\max}$. We choose a piecewise-linear grid path
$\widetilde x:[0,1]\to\mathcal M$ with
$\widetilde x(0)=x(0)$ and $\widetilde x(1)=x(1)$, such that
\begin{equation}
    \sup_{s\in[0,1]}
    \left\|
        x(s)-\widetilde x(s)
    \right\|
    \leq h_{\mathrm{grid}},
    \qquad
    \operatorname{length}(\widetilde x)
    \le 3mL,
    \qquad
    h_{\mathrm{grid}}:=2\sqrt m\,\delta x.
    \label{eq:grid-approximation}
\end{equation}
Such a path exists for every length $L$. Sample $x$ at $N+1$ points of equal arclength spacing $L/N$, with $N=\max\{1,\lceil L/\delta x\rceil\}$. Round each sample down coordinatewise to the refined grid, and join consecutive rounded points by axis-parallel grid steps.

We can parametrize $\widetilde x$ such that its speed is at most $\zeta_sL$ for some constant $\zeta_s=O(1)$, as Eq.~\eqref{eq:path-speed} requires of every path.

The spacing must meet three requirements, each an upper bound on $h_{\mathrm{grid}}$.
By Lemma~\ref{cor:gap-lipschitz}(iii), the band gap $\gamma_d$ for any fixed sector dimension $d$ satisfies the Lipschitz bound
\begin{equation}
    \left|
        \gamma_d(x)-\gamma_d(y)
    \right|
    \leq
    L_\gamma\left\|x-y\right\|,
    \qquad
    L_\gamma=2J.
    \label{eq:grid-gap-lipschitz}
\end{equation}
This finite-size bound holds even if the gap does not survive weak perturbations in the thermodynamic limit, as can happen for fine-tuned Hamiltonians~\cite{li2026stability}.
The first requirement, for gaps along $\widetilde x(s)$ on the scale of $\gamma_{\mathrm{lb}}$, is
\begin{equation}
    L_\gamma h_{\mathrm{grid}}
    \leq
    \frac{\gamma_{\mathrm{lb}}}{4}.
    \label{eq:grid-large-gap}
\end{equation}
This ensures that any path with $\gamma_r\ge\gamma_{\mathrm{lb}}$ at rank $r$ has a grid approximation with $\gamma_r\ge3\gamma_{\mathrm{lb}}/4$.

The second requirement comes from the threshold $\gamma_{\mathrm{thresh}}$. Clause~(ii) of Definition~\ref{def:phase-clustering} concerns paths with $\gamma(x(s))\ge\gamma_{\mathrm{thresh}}$. Such a path stays in one connected component of the resolved region, so its ground-state degeneracy is a constant $r$ and its gap is $\gamma_r$. The grid spacing condition
\begin{equation}
    L_\gamma h_{\mathrm{grid}}
    \le\frac{\epsilon_{\mathrm{thresh}}\gamma_{\mathrm{thresh}}}2
    \label{eq:grid-threshold-condition}
\end{equation}
ensures that the grid approximation $\widetilde x(s)$ satisfies
\begin{equation}
    \gamma_r(\widetilde x(s))
    \ge\Bigl(1-\frac{\epsilon_{\mathrm{thresh}}}2\Bigr)\gamma_{\mathrm{thresh}}
    \qquad\text{for all }s\in[0,1].
    \label{eq:grid-threshold-gap}
\end{equation}

The third requirement is that a response detectable between two vertices remains detectable at a tested vertex.
Every point of an edge lies within $h_{\mathrm{grid}}$ of an endpoint, which is a vertex of the refined grid and is therefore tested.
The filtered response $S_{\mathcal K}^{O}(x)$ of Eq.~\eqref{eq:weighted-response} for an observable $O$ must therefore vary slowly.
Take as probe the directional derivative $O(x)=\widehat u\cdot\nabla_x H(x)$ along a fixed unit vector $\widehat{u}$.
If the band gap $\gamma_d$ is at least $\gamma_{\mathrm{lb}}/2$ between $x$ and $y$, Lemma~\ref{lem:filtered-response-stability} gives
\begin{equation}
    \left|
        S_{\mathcal K}^{O(x)}(x)
        -
        S_{\mathcal K}^{O(y)}(y)
    \right|
    \leq
    L_{S,\mathcal K}
    \left\|x-y\right\|.
    \label{eq:grid-signal-lipschitz}
\end{equation}

Fix the kernel $\mathcal K=\mathcal{K}_{\Lambda,\delta}$ of Eq.~\eqref{eq:kernel-parameters}, and let $\epsilon_{\mathrm{sig}}$ be the signal scale of the algorithm in use. We require the grid constraint
\begin{equation}
    L_{S,\mathcal{K}_{\Lambda,\delta}} h_{\mathrm{grid}}
    \le
    \frac{\epsilon_{\mathrm{sig}}}{4},
    \label{eq:proof-grid-condition-lipschitz}
\end{equation}
where $L_{S,\mathcal{K}_{\Lambda,\delta}}$ is the signal Lipschitz constant for the probe in Eq.~\eqref{eq:edge-probe}.

We choose $h_{\mathrm{grid}}$ to meet the three upper bounds and set $\delta x=h_{\mathrm{grid}}/(2\sqrt m)$. The spacing is then inverse-polynomial: if $\delta$ and $\gamma_{\mathrm{lb}}$ are at least $1/\operatorname{poly}(n)$, the Lipschitz constants $L_\gamma$ and $L_{S,\mathcal K}$ are polynomially bounded, and the right-hand sides of the three requirements are at least $1/\operatorname{poly}(n)$.

Because $m=O(1)$, the number of vertices is
\begin{equation}
    N_{\mathrm{grid}}
    =
    O\left(
        \left(\frac{1}{\delta x}\right)^m
    \right)
    +
    \operatorname{poly}(n)
    =
    \operatorname{poly}(n),
    \label{eq:number-grid-points}
\end{equation}
where the additive term counts the vertices added by the hyperplanes through the polynomially many input and reference points.

\subsection{Learning in the one-parameter case}
\label{subsec:reference-sweeps}
\label{sec:proof-theorem-continuous-phase}

With one parameter, a reference point and a grid point are joined by a single interval, so no search over grid paths is needed.

\begin{theorem}[One-dimensional relaxation of
Theorem~\ref{thm:detection}]
\label{thm:local-phase-diagram}
Let $m=1$. Suppose that Assumption~\ref{assumption-reference-state} holds and, if $r_{\max}\ge2$,
that the tolerance satisfies Eq.~\eqref{eq:gsm-width}.
Then
there exists a quantum algorithm (Algorithm~\ref{alg:overall})
running in $\mathrm{poly}(n)$ time that, with probability at least
$1-1/\mathrm{poly}(n)$, learns the phase diagram with the relaxed
guarantee of Remark~\ref{rem:phase-clustering-relaxation}.
The algorithm uses $U_{\gamma_{\mathrm{lb}}/2}$ for transport and
estimates local energies and
the filtered response of Appendix~\ref{sec:diabatic-signals}.
If $r_{\max}\ge2$, it also runs the purity test of Subsection~\ref{subsec:energy-rank-estimation}.

As in Theorem~\ref{thm:detection}, when the reference states are unique ground states, the algorithm uses a single copy
of each and perturbs it by only
an amount exponentially small in $n$. Otherwise, it uses polynomially many
copies of each reference state.
\end{theorem}

\begin{remark}[Local measurements only]
\label{rem:local-measurements}
The purity test is the only step of Algorithm~\ref{alg:overall} that measures two copies at once.
To avoid it and use only local measurements, which are easier on a quantum simulator, we can additionally assume that every reference state, successful or not, is approximately maximally mixed on a subspace of known dimension:
\begin{equation}
    \bigl\|C[x_a^{\mathrm{ref}}]-Q_a/r_a\bigr\|_1
    \le\epsilon_{\mathrm{ref}}
    \qquad
    \text{for some projector $Q_a$ of known rank $r_a\le r_{\max}$.}
    \label{eq:class-promise}
\end{equation}
We call this assumption the \emph{maximally-mixed proviso}.
The projector $Q_a$ need not project onto the ground-state manifold. For a successful preparation, $Q_a=P_{r_a}(x_a^{\mathrm{ref}})$.
The known ranks then replace the purity test: the algorithm uses each reference point only at its rank $r_a$.
Under this proviso, the theorem holds without Eq.~\eqref{eq:gsm-width}: the original tolerance of Eq.~\eqref{eq:1d-gsm-tolerance} suffices.
In its proof, the ideal state of a candidate path is now obtained from $Q_a/r_a$ and therefore has the form $Q/q$ for a rank-$q$ projector $Q$.
Such a state has purity exactly $1/q$, so Eq.~\eqref{eq:1d-selected-state} still applies to it.
For $q>r$, Ky Fan's maximum principle~\cite{Bha97:Matrix-Analysis} gives
\[
\begin{aligned}
    \operatorname{Tr}[H(x)Q]/q
    &\ge\frac1q\sum_{i=0}^{q-1}E_i(x)\\
    &\ge \operatorname{Tr}[H(x)\rho_r(x)]
       +\frac{q-r}{q}\bigl(E_r(x)-\operatorname{Tr}[H(x)\rho_r(x)]\bigr)\\
    &\ge \operatorname{Tr}[H(x)\rho_r(x)]
       +\frac{\gamma_{\mathrm{lb}}}{r_{\max}} .
\end{aligned}
\]
The last step uses $\operatorname{Tr}[H(x)\rho_r(x)]\le E_{r-1}(x)$ and $(q-r)/q\ge1/r_{\max}$.
With this bound in place of Eq.~\eqref{eq:larger-rank-energy}, Lemma~\ref{lem:energy-rank-estimate} needs only $\delta_{\mathrm{gsm}}+2\epsilon_E+4\eta_E<\gamma_{\mathrm{lb}}/r_{\max}$, which holds under Eq.~\eqref{eq:1d-gsm-tolerance} when $\epsilon_E=\eta_E\le\gamma_{\mathrm{lb}}/(16r_{\max})$.
A state $Q/r$ puts at most $1/r$ of its weight on each level, so compared with $\rho_r(x)$ it only moves weight from the lowest $r$ levels to levels at or above $E_r(x)$.
Hence Eq.~\eqref{eq:1d-energy-leakage} holds without its last term.
\end{remark}

\subsubsection{The algorithm}
\label{subsubsec:1d-algorithm}

For every reference point $a$ and grid point $x$, let $V_a[x] :=U_{\gamma_{\mathrm{lb}}/2}(x_a^{\mathrm{ref}}\to x)$ be the quasi-adiabatic propagator along the interval from the reference point to $x$ (Definition~\ref{def:quasi-adiabatic-evolution}). The family of these circuits $V_a[x]$ is the \emph{quasi-adiabatic transport} from reference point $a$. We construct it edge by edge to the left and right of the reference point, and store only its classical description. We write $\sigma_{V_a[x]}$ for the candidate state actually implemented, as in Appendix~\ref{sec:diabatic-signals}.

For each reference point $a$ and grid point $x$, we estimate the mean energy of $\sigma_{V_a[x]}$ from the local terms of $H(x)$. We also run the filtered-response test of Algorithm~\ref{alg:gap_cert} on it, with the probe $O(x)$ of Eq.~\eqref{eq:edge-probe}. We denote the two estimates by $\widehat E_a(x)$ and $\widehat S_a(x)$.
The signal scale is
\begin{equation}
    \epsilon_{\mathrm{sig}}
    :=
    \frac{\eta_P(n)^2\,\gamma_{\mathrm{lb}}^2}{64\,r_{\max}},
    \label{eq:witness-scale}
\end{equation}
where, for equal ground-state degeneracies, different phases have overlap at most $1-\eta_P(n)$ under every admissible circuit (Definition~\ref{def:finite-size-phase}(ii)). Here $\sqrt{r_{\max}}$ bounds the factor $\sqrt d$ in the pointwise bound of Lemma~\ref{lemma:sparse-coupling}(a) for every sector dimension $d=r\le r_{\max}$.

At each vertex and for each rank $q$, we retain the lowest estimated energy among the candidate states admitted at rank $q$:
\begin{equation}
    \widehat E_q[x]
    :=\min_{a\text{ admitted at rank }q}\widehat E_a(x),
    \qquad
    a_q(x)\in\operatorname*{arg\,min}_{a\text{ admitted at rank }q}
                    \widehat E_a(x).
    \label{eq:1d-rank-energies}
\end{equation}
Ties are broken by a fixed convention.
The rank is then selected as in Lemma~\ref{lem:energy-rank-estimate} with $\epsilon_E=\eta_E$:
\begin{equation}
    \widehat r(x)
    :=\max\left\{
        q:\widehat E_q[x]\leq
        \min_j\widehat E_j[x]+\delta_{\mathrm{gsm}}+3\eta_E
    \right\}.
    \label{eq:1d-rank-selection}
\end{equation}
The \emph{selected state} at $x$ is the candidate state of $V_{a_{\widehat r(x)}(x)}[x]$.

Given an unordered input pair, order its endpoints so that $x_i\leq x_j$. Call the pair \emph{linked} if, for some rank $r$,
\begin{equation}
    \widehat r(v)=r,
    \qquad
    \widehat S_{a_r(v)}(v)<\frac{\epsilon_{\mathrm{sig}}}{2}
    \qquad
    \text{at every grid point }v\in[x_i,x_j].
    \label{eq:1d-band-silence}
\end{equation}
The reported path (Remark~\ref{rem:phase-clustering-relaxation}) is $[x_i,x_j]$, of length at most $2$.

\begin{algorithm}[!htbp]
\caption{\textsc{Learning-Phase-Diagram} in the one-parameter case}
\label{alg:overall}
\label{alg:phase_mapper}
\scriptsize
\begin{algorithmic}[1]
\Require $H(x)$, $x\in[-1,1]$; input points; reference states
$C[x_a^{\mathrm{ref}}]$.
\Ensure The linked relation and its clusters, with the guarantee of
Theorem~\ref{thm:local-phase-diagram}.
\State Form the refined grid of Subsection~\ref{subsec:grid-choice}.
\State If $r_{\max}\ge2$, run the purity test on every reference state
(Subsection~\ref{subsec:energy-rank-estimation}).
\For{each reference point $a$}
    \State Transport $C[x_a^{\mathrm{ref}}]$ from $x_a^{\mathrm{ref}}$ to both endpoints of
    $[-1,1]$ with
    $U_{\gamma_{\mathrm{lb}}/2}$.
    \State At each grid point $x$, estimate the mean energy and
    run Algorithm~\ref{alg:gap_cert} with the probe
    $O(x)=\partial_xH(x)$.
\EndFor
\State At each grid point, keep the lowest-energy candidate state admitted at
each rank and determine $\widehat r(x)$ using
Eq.~\eqref{eq:1d-rank-selection}.
\State Link two input points if, for some rank $r$, every grid point
of the interval between them has selected rank $r$ and its rank-$r$
candidate state returns \textsc{Connected}
(Eq.~\eqref{eq:1d-band-silence}); store the interval as the reported path.
\State \Return The links and their clusters.
\end{algorithmic}
\end{algorithm}

\subsubsection{Correctness}

\begin{proof}[Proof of Theorem~\ref{thm:local-phase-diagram}]
We run Algorithm~\ref{alg:overall} on $\mathcal M=[-1,1]$.

\emph{Precision and state selection.} Let $\sigma_a(x)$ be the ideal state obtained from the reference state $C[x_a^{\mathrm{ref}}]$,
\begin{equation}
    \sigma_a(x)
    :=U_{\gamma_{\mathrm{lb}}/2}(x_a^{\mathrm{ref}}\to x)\,
      C[x_a^{\mathrm{ref}}]\,
      U_{\gamma_{\mathrm{lb}}/2}(x_a^{\mathrm{ref}}\to x)^\dagger.
    \label{eq:1d-sweep-state}
\end{equation}
Each edge circuit has operator-norm error at most $\epsilon_U$, and we set
\begin{equation}
    \epsilon_0
    :=\epsilon_{\mathrm{ref}}
      +2N_{\mathrm{grid}}\epsilon_U.
    \label{eq:1d-implementation-error}
\end{equation}
Each transport circuit has at most $N_{\mathrm{grid}}-1$ edge circuits.
Hence the implemented state differs from $\sigma_a(x)$ by at most $\epsilon_0$ in trace norm.
We require
\begin{equation}
    J\epsilon_0\leq\frac{\eta_E}{2},
    \label{eq:1d-energy-budget-split}
\end{equation}
and estimate each implemented mean energy to precision $\eta_E/2$.
Thus every estimated energy satisfies $\left|\widehat E_a(x)-\operatorname{Tr}[H(x)\sigma_a(x)]\right|\leq\eta_E$.
We further require Eq.~\eqref{eq:energy-rank-precision} with $\epsilon_E=\eta_E$,
\begin{equation}
    2\delta_{\mathrm{gsm}}+6\eta_E
    <\frac{\gamma_{\mathrm{lb}}}{4r_{\max}},
    \qquad
    2\epsilon_{\mathrm{ref}}+\epsilon_p\le\frac1{2r_{\max}^2},
    \label{eq:1d-rank-budget}
\end{equation}
and the state budget of the filtered-response test,
\begin{equation}
    2K_0(\mathcal K_{\Lambda,\delta})\,\|O\|^2
    \left(
        \epsilon_0+4\sqrt{\frac{4\eta_E}{\gamma_{\mathrm{lb}}}+(r_{\max}+1)(2\epsilon_{\mathrm{ref}}+\epsilon_p)}
    \right)
    \le\frac{\epsilon_{\mathrm{sig}}}8.
    \label{eq:state-budget}
\end{equation}
In Eq.~\eqref{eq:state-budget}, the parentheses bound the trace distance between the kept rank-$r$ candidate state and the exact sector state.
By Eq.~\eqref{eq:response-trace-distance-robustness}, the prefactor turns this distance into a response error.
Suppose that the ideal state of the kept rank-$r$ candidate path has population $\lambda\le4\eta_E/\gamma_{\mathrm{lb}}+2\epsilon_{\mathrm{ref}}$ outside the lowest $r$ levels.
Since $2\sqrt\lambda+\sqrt{2\lambda+u}\le4\sqrt{\lambda+u}$ for all $u\ge0$, Eq.~\eqref{eq:1d-selected-state} then bounds its distance from $\rho_r(x)$ by the square-root term of the parentheses.
Eq.~\eqref{eq:1d-energy-leakage} keeps the population below this bound for band gaps down to $3\gamma_{\mathrm{lb}}/4$ and widths of the lowest $r$ levels up to $\epsilon_{\mathrm{ref}}\gamma_{\mathrm{lb}}$; such gaps and widths occur for a fixed number of parameters.

If $r_{\max}=1$, Eq.~\eqref{eq:1d-gsm-tolerance} gives $2\delta_{\mathrm{gsm}}\le\gamma_{\mathrm{lb}}/8$.
If $r_{\max}\ge2$, Eq.~\eqref{eq:gsm-width} gives $2\delta_{\mathrm{gsm}}\le\epsilon_{\mathrm{ref}}\gamma_{\mathrm{lb}}$.
For $r_{\max}\ge2$, the second condition in Eq.~\eqref{eq:1d-rank-budget} gives $\epsilon_{\mathrm{ref}}\le1/(4r_{\max}^2)\le1/(8r_{\max})$.
In both cases, $2\delta_{\mathrm{gsm}}\le\gamma_{\mathrm{lb}}/(8r_{\max})$.
Hence, if $\eta_E\le\gamma_{\mathrm{lb}}/(64r_{\max})$, the second condition in Eq.~\eqref{eq:1d-rank-budget} implies the first.
Write $\vartheta:=\epsilon_{\mathrm{sig}}/(16K_0(\mathcal K_{\Lambda,\delta})\|O\|^2)$ for the largest trace distance that the state budget allows.
We choose
\[
\begin{aligned}
    \eta_E
    &=\min\Bigl\{\frac{\gamma_{\mathrm{lb}}\vartheta^2}{1024},\,\frac{\gamma_{\mathrm{lb}}}{64r_{\max}}\Bigr\},\\
    \epsilon_{\mathrm{ref}}=\epsilon_p
    &=\min\Bigl\{\frac{\vartheta^2}{768(r_{\max}+1)},\,\frac{\eta_E}{4J},\,\frac{\vartheta}{8},\,\frac{1}{6r_{\max}^2}\Bigr\},\\
    2N_{\mathrm{grid}}\epsilon_U
    &\le\min\Bigl\{\frac{\eta_E}{4J},\,\frac{\vartheta}{8}\Bigr\}.
\end{aligned}
\]
These choices satisfy Eqs.~\eqref{eq:1d-energy-budget-split}--\eqref{eq:state-budget}, and Eq.~\eqref{eq:gsm-width} refers to this $\epsilon_{\mathrm{ref}}$.

Take any resolved vertex $x$ of ground-state degeneracy $r$. By Assumption~\ref{assumption-reference-state}, its connected component of the resolved region contains a successful reference point of ground-state degeneracy $r$. The purity test admits this point at rank $r$. The entire interval from this point to $x$ is resolved. Hence its candidate path at $x$ has an ideal state within $\epsilon_{\mathrm{ref}}$ of $\rho_r(x)=\rho_{\mathcal G}(x)$ (Corollary~\ref{cor:exact-transport}). By Eq.~\eqref{eq:1d-energy-budget-split}, its mean energy exceeds $\operatorname{Tr}[H(x)\rho_r(x)]$ by at most $J\epsilon_{\mathrm{ref}}\le\eta_E$. Lemma~\ref{lem:energy-rank-estimate} yields $\widehat r(x)=r$ for the minimum energies. If $r=1$, the width $E_{r-1}(x)-E_0(x)$ is zero. If $r\ge2$, then $r_{\max}\ge2$, and Eq.~\eqref{eq:gsm-width} bounds the width by $\epsilon_{\mathrm{ref}}\gamma_{\mathrm{lb}}/2$. With $\epsilon_E=\eta_E$ and $\gamma_r(x)\ge\gamma_{\mathrm{lb}}$, Eq.~\eqref{eq:1d-energy-leakage} bounds the population of the ideal state of the selected candidate path outside the lowest $r$ levels by $3\eta_E/\gamma_{\mathrm{lb}}+\epsilon_{\mathrm{ref}}/2$. By the discussion after Eq.~\eqref{eq:state-budget}, the implemented selected state is therefore within distance
\[
  \epsilon_0+4\sqrt{\frac{4\eta_E}{\gamma_{\mathrm{lb}}}+(r_{\max}+1)(2\epsilon_{\mathrm{ref}}+\epsilon_p)}
\]
of $\rho_{\mathcal G}(x)$. This, together with the precision of response estimation and the state budget, yields
\begin{equation}
    \left|
       \widehat S_{a_r(x)}(x)
       -S^{\partial_xH}_{\mathcal K_{\Lambda,\delta}}
          \bigl(\rho_r(x);x\bigr)
    \right|
    \leq\frac{\epsilon_{\mathrm{sig}}}{4}
    \qquad
    \text{whenever }\gamma(x)\geq\gamma_{\mathrm{lb}}.
    \label{eq:1d-selected-response}
\end{equation}

\emph{The connected case.}
Consider an interval $[x_i,x_j]$ satisfying
\begin{equation}
    \gamma(x)\geq
    \gamma_{\mathrm{thresh}}
    \qquad (x\in[x_i,x_j]).
    \label{eq:proof-connected-promise}
\end{equation}
Along it the ground-state degeneracy is a constant $r$, which the algorithm estimates correctly at every vertex. By Eq.~\eqref{eq:response-sandwich}, the exact sector response vanishes, so by Lemma~\ref{lem:edge-test}, the test on the selected state outputs \textsc{Connected} at every vertex. Hence Eq.~\eqref{eq:1d-band-silence} holds, and the algorithm links the endpoints.

\emph{The disconnected case on a resolved interval.}
Now suppose $\gamma(x)\geq\gamma_{\mathrm{lb}}$ on $[x_i,x_j]$ and the two ground states at the endpoints are in different phases.
We show that the test returns \textsc{Disconnected} at some vertex.
The ground-state degeneracy is again constant on this interval, say $r$.
Parametrize the interval by $x(s)=x_i+s(x_j-x_i)$ for $s\in[0,1]$.
We first bound the overlap after quasi-adiabatic transport (Definition~\ref{def:quasi-adiabatic-evolution}) with cutoff $(1-\epsilon_{\mathrm{thresh}})\gamma_{\mathrm{thresh}}$ instead of $\gamma_{\mathrm{lb}}/2$, since the circuit approximation below needs a constant cutoff. Admissible circuits (Definition~\ref{def:admissible-circuits}) have gates of diameter at most $R(n)$. Take $R(n)$ at least the gate diameter of Proposition~\ref{prop:fdlu-from-gapped-path} at accuracy $\eta_P(n)/8$. For the interval path, this proposition gives an admissible circuit $C$ with $\left\|U_{(1-\epsilon_{\mathrm{thresh}})\gamma_{\mathrm{thresh}}}[x]-C\right\| \leq\frac{\eta_P(n)}8$. By Definition~\ref{def:finite-size-phase}, $\operatorname{Tr}\left[ P_r(x_j)C\rho_r(x_i)C^\dagger \right]\leq1-\eta_P(n)$. Replacing $C$ with $U_{(1-\epsilon_{\mathrm{thresh}})\gamma_{\mathrm{thresh}}}[x]$ changes this overlap by at most $\eta_P(n)/4$, so
\begin{equation}
    \operatorname{Tr}\bigl[P_r(x_j)\,U_{(1-\epsilon_{\mathrm{thresh}})\gamma_{\mathrm{thresh}}}[x]\,\rho_r(x_i)\,U_{(1-\epsilon_{\mathrm{thresh}})\gamma_{\mathrm{thresh}}}[x]^\dagger\bigr]
    \leq1-\frac{3\eta_P(n)}4.
    \label{eq:proof-transport-deficit}
\end{equation}

If the band gap stayed at least $(1-\epsilon_{\mathrm{thresh}})\gamma_{\mathrm{thresh}}$ on the interval, exact transport would contradict Eq.~\eqref{eq:proof-transport-deficit}. Quantitatively, Theorem~\ref{lem:path-gradient-spectral-weight} turns this overlap bound into a point $s_0$ with spectral weight at least $\eta_P(n)\gamma_{\mathrm{lb}}^2/8$, since the band gap stays at least $\gamma_{\mathrm{lb}}$. Here $\eta(n)=\eta_P(n)/2$, and the theorem's constant is $C=1/2$ (not the circuit $C$ above). As $\eta_P(n) \le 1 < 2r_{\max}$, this implies the weaker bound
\[
    W^{\partial_sH}_{[\gamma_r,(1-\epsilon_{\mathrm{thresh}})\gamma_{\mathrm{thresh}})}
        (x(s_0))
    \geq
    \frac{\eta_P(n)^2\gamma_{\mathrm{lb}}^2}{16r_{\max}}.
\]
Since $\partial_sH=(x_j-x_i)\partial_xH$ and the spectral weight is quadratic in the probe, $z:=x(s_0)$ satisfies
\begin{equation}
    W^{\partial_xH}_{[\gamma_r,(1-\epsilon_{\mathrm{thresh}})\gamma_{\mathrm{thresh}})}(z)
    \geq
    \frac{\eta_P(n)^2\gamma_{\mathrm{lb}}^2}{64r_{\max}}
    =\epsilon_{\mathrm{sig}}.
    \label{eq:proof-witness-weight}
\end{equation}

Pick a vertex $v\in[x_i,x_j]$ within $h_{\mathrm{grid}}$ of $z$. Eqs.~\eqref{eq:response-sandwich} and~\eqref{eq:proof-grid-condition-lipschitz} imply
\begin{equation}
\begin{aligned}
    S^{\partial_xH}_{\mathcal K_{\Lambda,\delta}}(\rho_r(v);v)
    &\geq
    S^{\partial_xH}_{\mathcal K_{\Lambda,\delta}}(\rho_r(z);z)
       -L_{S,\mathcal K_{\Lambda,\delta}}h_{\mathrm{grid}}\\
    &\geq\frac{3\epsilon_{\mathrm{sig}}}{4}.
\end{aligned}
    \label{eq:proof-witness-vertex}
\end{equation}
By Eq.~\eqref{eq:1d-selected-response}, $\widehat S_{a_r(v)}(v)\geq\epsilon_{\mathrm{sig}}/2$, so the filtered-response test at $v$ outputs \textsc{Disconnected}, and the interval is not linked.

\emph{The phase-diagram guarantee.}
For clause~(i) of Definition~\ref{def:phase-clustering}, relaxed as in Remark~\ref{rem:phase-clustering-relaxation}, consider a link whose reported path is resolved.
If its endpoints were in different phases, then by the disconnected case the test on the selected state at some vertex would return \textsc{Disconnected}, which contradicts Eq.~\eqref{eq:1d-band-silence}. For clause~(ii), suppose there is a continuous path from $x_i$ to $x_j$ with spectral gap at least $\gamma_{\mathrm{thresh}}$ at every point. In one dimension, the image of the path contains the interval between its endpoints. Hence the connected case links $x_i$ and $x_j$ directly, so they belong to the same cluster. The measurement counts, estimation methods, and allocation of failure probabilities are in Subsection~\ref{subsec:resources}.
\end{proof}

\subsection{Extension to a fixed number of parameters}
\label{subsec:local-higher-dimensional}
\label{sec:proof-theorem-general}

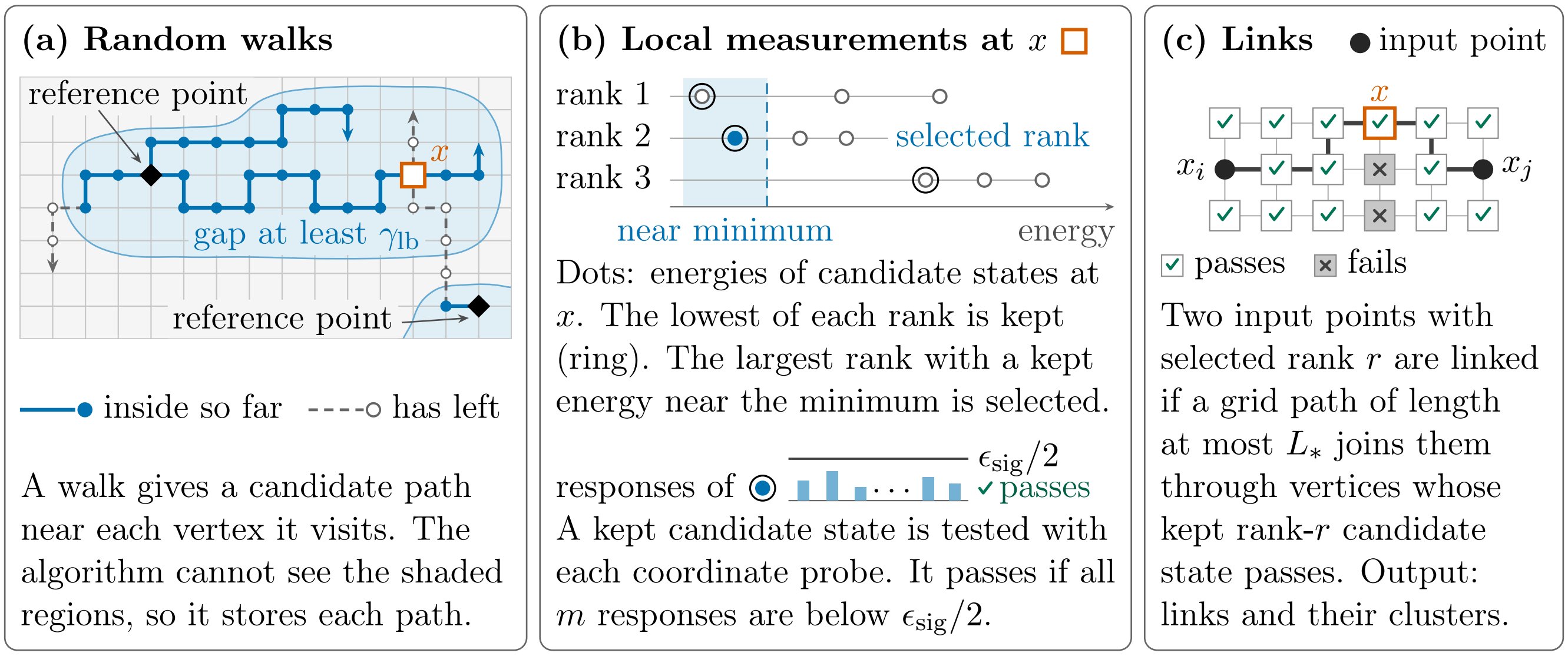
\begin{figure}[!t]
\centering
\tikzset{
  txt/.style={figtext, anchor=base west, inner sep=0pt},
  st/.style={figstep, anchor=base west, inner sep=0pt},
  cvert/.style={minimum size=0.34cm, inner sep=0pt, outer sep=0pt},
  ckey/.style={minimum size=0.24cm, inner sep=0pt, outer sep=0pt},
  inp/.style={circle, fill=black!85, inner sep=0pt, minimum size=6.5pt},
  tick/.pic={\draw[figgreen!75!black, line width=0.9pt, line cap=round, line join=round] (-0.07,0.015) -- (-0.025,-0.045) -- (0.075,0.085);},
  fail/.pic={\draw[black!75, line width=0.9pt, line cap=round] (-0.06,-0.06) -- (0.06,0.06) (-0.06,0.06) -- (0.06,-0.06);},
  walkin/.style={figblue, line width=1.2pt, line join=round},
  walkout/.style={black!60, line width=0.9pt, dash pattern=on 3pt off 2pt, line join=round},
  wtip/.style={-{Stealth[length=4.5pt, width=3.6pt]}},
  candin/.style={circle, fill=figblue, draw=figblue, line width=0.6pt, inner sep=0pt, minimum size=4.2pt},
  candout/.style={circle, fill=white, draw=black!60, line width=0.7pt, inner sep=0pt, minimum size=4.2pt},
  kept/.style={circle, draw=black, line width=0.6pt, inner sep=0pt, minimum size=8pt},
  xsq/.style={rectangle, draw=figorange, line width=1.1pt, fill=white, inner sep=0pt, outer sep=0pt, minimum size=0.26cm},
  mapin/.style={candin, minimum size=3pt},
  mapout/.style={candout, minimum size=3pt, line width=0.5pt},
  supplied/.style={rectangle, rotate=45, fill=black, inner sep=0pt, outer sep=0pt, minimum size=5.2pt},
}%
\begin{tikzpicture}[x=1cm, y=1cm]
\draw[figpanel] (0,0)     rectangle (5.86,-7.24);
\draw[figpanel] (6.01,0)  rectangle (12.65,-7.24);
\draw[figpanel] (12.80,0) rectangle (17.50,-7.24);
\node[st] at (0.17,-0.50) {(a) Random walks};
\begin{scope}
  \clip (0.17,-3.744) rectangle (5.69,-0.80);
  \fill[black!4] (0.17,-3.744) rectangle (5.69,-0.80);
  \begin{scope}[shift={(0.17,-3.744)}, x=0.368cm, y=0.368cm]
    \def\region{plot[smooth cycle, tension=0.7] coordinates {
        (1.3,4.6) (1.7,5.8) (3.2,6.35) (5.0,6.45) (6.4,6.6) (7.5,7.35) (9.0,7.65) (11.5,7.7)
        (13.4,7.55) (14.45,6.7) (14.7,5.2) (14.5,3.7) (13.7,2.75) (12.0,2.5) (9.5,2.45)
        (7.0,2.5) (5.0,2.5) (3.0,2.6) (1.75,3.3)}}
    \def\regiontwo{plot[smooth cycle, tension=0.7] coordinates {
        (11.9,-0.8) (12.0,0.6) (12.6,1.45) (13.6,1.65) (14.8,1.6) (15.8,1.2) (15.9,-0.8)}}
    \fill[figblue!12] \region;
    \fill[figblue!12] \regiontwo;
    \draw[black!22, line width=0.4pt] (0,0) grid[step=1] (15,8);
    \node[font=\normalsize, text=figblue, fill=figblue!12, inner sep=1.5pt] at (8.76,3.17) {gap at least $\gamma_{\mathrm{lb}}$};
    \begin{scope}[even odd rule]
      \clip (0,0) rectangle (15,8) \region;
      \node[font=\normalsize, fill=black!4, inner sep=1.5pt, anchor=base west] (sp) at (0.09,7.17) {reference point};
      \fill[black!4, nonzero rule] (0,8) rectangle ([yshift=-0.5pt]sp.north east) (0,8) rectangle ([xshift=0.5pt]sp.south west)
        ([xshift=-0.5pt]sp.south east) rectangle (6.98,8);%
    \end{scope}
    \node[font=\normalsize, fill=black!4, inner sep=1.5pt, anchor=base east] at (11.55,0.32) {reference point};
    \draw[figblue!60, line width=0.5pt] \region;
    \draw[figblue!60, line width=0.5pt] \regiontwo;
    \draw[black!22, line width=0.4pt] (0,0) rectangle (15,8);
    \draw[walkout, wtip] (13,1) -- (13,2) -- (13,3) -- (13,4) -- (12,4) -- (12,5) -- (12,6) -- (12,7);
    \draw[walkout, wtip] (2,4) -- (1,4) -- (1,3) -- (1,2);
    \draw[walkin] (14,1) -- (13,1);
    \draw[walkin] (4,5) -- (3,5) -- (2,5) -- (2,4);
    \draw[walkin, wtip] (4,5) -- (5,5) -- (5,4) -- (6,4) -- (7,4) -- (7,5) -- (8,5) -- (9,5) -- (9,4) -- (10,4)
      -- (11,4) -- (11,5) -- (12,5) -- (13,5) -- (14,5) -- (14,6);
    \draw[walkin, wtip] (4,5) -- (4,6) -- (5,6) -- (6,6) -- (7,6) -- (8,6) -- (8,7) -- (9,7) -- (10,7) -- (10,6);
    \foreach \p in {(13,2),(13,3),(13,4),(12,4),(12,6),(1,4),(1,3)} \node[mapout] at \p {};
    \foreach \p in {(5,5),(5,4),(6,4),(7,4),(7,5),(8,5),(9,5),(9,4),(10,4),(11,4),(11,5),(13,5),(14,5),(4,6),(5,6),
      (6,6),(7,6),(8,6),(8,7),(9,7),(10,7),(3,5),(2,5),(2,4),(13,1)} \node[mapin] at \p {};
    \node[supplied] at (4,5) {};
    \draw[black!70, line width=0.5pt, -{Stealth[length=3.5pt, width=2.8pt]}] (2.4,6.98) -- (3.7,5.36);
    \node[supplied] at (14,1) {};
    \draw[black!70, line width=0.5pt, -{Stealth[length=3.5pt, width=2.8pt]}] (11.65,0.5) -- (13.70,0.64);
    \node[xsq] at (12,5) {};
    \node[font=\normalsize, text=figorange, anchor=south west, inner sep=0pt] at (12.50,5.42) {$x$};
  \end{scope}
\end{scope}
\draw[walkin] (0.17,-4.54) -- (0.79,-4.54); \node[candin] at (0.90,-4.54) {};
\node[txt] at (1.10,-4.63) {inside so far};
\draw[walkout] (3.41,-4.54) -- (4.03,-4.54); \node[candout] at (4.14,-4.54) {};
\node[txt] at (4.34,-4.63) {has left};
\node[txt] at (0.17,-5.52) {A walk gives a candidate path};
\node[txt] at (0.17,-6.00) {near each vertex it visits. The};
\node[txt] at (0.17,-6.48) {algorithm cannot see the shaded};
\node[txt] at (0.17,-6.96) {regions, so it stores each path.};
\node[st] at (6.18,-0.50) {(b) Local measurements at $x$};
\node[xsq] at (12.01,-0.40) {};
\fill[figblue!12] (7.62,-2.26) rectangle (8.56,-0.82);
\draw[figblue, line width=0.5pt, dash pattern=on 3.5pt off 2.5pt] (8.56,-2.26) -- (8.56,-0.82);
\node[txt] at (6.18,-1.11) {rank 1};
\node[txt] at (6.18,-1.58) {rank 2};
\node[txt] at (6.18,-2.05) {rank 3};
\draw[black!35, line width=0.4pt] (7.47,-1.02) -- (10.50,-1.02);
\draw[black!35, line width=0.4pt] (7.47,-1.49) -- (9.92,-1.49);
\draw[black!35, line width=0.4pt] (7.47,-1.96) -- (11.65,-1.96);
\foreach \xx in {7.83,9.40,10.50} \node[candout] at (\xx,-1.02) {};
\node[candin] at (8.20,-1.49) {};
\foreach \xx in {8.93,9.45} \node[candout] at (\xx,-1.49) {};
\foreach \xx in {10.34,11.00,11.65} \node[candout] at (\xx,-1.96) {};
\node[kept] at (7.83,-1.02) {}; \node[kept] at (8.20,-1.49) {}; \node[kept] at (10.34,-1.96) {};
\node[txt, text=figblue] at (10.00,-1.58) {selected rank};
\draw[-{Stealth[length=3pt, width=2.5pt]}, black!60, line width=0.4pt] (7.47,-2.26) -- (12.46,-2.26);
\node[anchor=base, inner sep=0pt, font=\normalsize, text=figblue] at (8.09,-2.64) {near minimum};
\node[anchor=base east, inner sep=0pt, font=\normalsize, text=black!70] at (12.48,-2.64) {energy};
\node[txt] at (6.18,-3.12) {Dots: energies of candidate states at};
\node[txt] at (6.18,-3.60) {$x$. The lowest of each rank is kept};
\node[txt] at (6.18,-4.08) {(ring). The largest rank with a kept};
\node[txt] at (6.18,-4.56) {energy near the minimum is selected.};
\node[anchor=base west, inner sep=0pt, font=\normalsize] at (6.18,-5.52) {responses of};
\node[candin] at (8.51,-5.42) {}; \node[kept] at (8.51,-5.42) {};
\draw[black!60, line width=0.4pt] (8.80,-5.56) -- (10.82,-5.56);
\foreach \xx/\hh in {8.97/0.22, 9.29/0.33, 9.61/0.15, 10.37/0.26, 10.67/0.19} \fill[figblue!55] (\xx-0.065,-5.56) rectangle (\xx+0.065,-5.56+\hh);
\node[anchor=center, inner sep=0pt, font=\normalsize] at (9.99,-5.46) {$\cdots$};
\draw[black!75, line width=0.7pt] (8.80,-5.09) -- (10.82,-5.09);
\node[anchor=base west, inner sep=0pt, font=\normalsize] at (10.92,-5.173) {$\epsilon_{\mathrm{sig}}/2$};
\pic[scale=0.9] at (11.01,-5.435) {tick};
\node[anchor=base west, inner sep=0pt, font=\normalsize, text=figgreen!60!black] at (11.16,-5.52) {passes};
\node[txt] at (6.18,-6.00) {A kept candidate state is tested with};
\node[txt] at (6.18,-6.48) {each coordinate probe. It passes if all};
\node[txt] at (6.18,-6.96) {$m$ responses are below $\epsilon_{\mathrm{sig}}/2$.};
\node[st] at (12.97,-0.50) {(c) Links};
\node[inp] at (15.22,-0.42) {};
\node[anchor=base east, inner sep=0pt, font=\normalsize] at (17.33,-0.50) {input point};
\foreach \i/\xx in {1/13.70, 2/14.28, 3/14.86, 4/15.44, 5/16.02, 6/16.60}
  \foreach \j/\yy in {1/-1.32, 2/-1.84, 3/-2.36}
    \coordinate (g\i\j) at (\xx,\yy);
\foreach \j in {1,2,3} \draw[black!30, line width=0.4pt] (g1\j) -- (g6\j);
\foreach \i in {1,...,6} \draw[black!30, line width=0.4pt] (g\i1) -- (g\i3);
\draw[black!75, line width=1.4pt, line join=round] (g12) -- (g22) -- (g32) -- (g31) -- (g41) -- (g51) -- (g52) -- (g62);
\foreach \v in {g11,g21,g31,g51,g61,g22,g32,g52,g13,g23,g33,g53,g63} {
  \node[cvert, fill=white, draw=black!45, line width=0.4pt] at (\v) {}; \pic at (\v) {tick};}
\foreach \v in {g42,g43} {
  \node[cvert, fill=black!22, draw=black!45, line width=0.4pt] at (\v) {}; \pic at (\v) {fail};}
\node[cvert, fill=white, draw=figorange, line width=1.1pt] at (g41) {}; \pic at (g41) {tick};
\node[font=\normalsize, text=figorange, anchor=base, inner sep=0pt] at (15.44,-1.07) {$x$};
\node[inp] at (g12) {}; \node[inp] at (g62) {};
\node[font=\normalsize, anchor=east, inner sep=0pt] at ($(g12)+(-0.2,0)$) {$x_i$};
\node[font=\normalsize, anchor=west, inner sep=0pt] at ($(g62)+(0.2,0)$) {$x_j$};
\node[ckey, fill=white, draw=black!45, line width=0.4pt] at (13.11,-2.92) {}; \pic[scale=0.8] at (13.11,-2.92) {tick};
\node[txt] at (13.35,-3.00) {passes};
\node[ckey, fill=black!22, draw=black!45, line width=0.4pt] at (14.83,-2.92) {}; \pic[scale=0.8] at (14.83,-2.92) {fail};
\node[txt] at (15.07,-3.00) {fails};
\node[txt] at (12.97,-3.60) {Two input points with};
\node[txt] at (12.97,-4.08) {selected rank $r$ are linked};
\node[txt] at (12.97,-4.56) {if a grid path of length};
\node[txt] at (12.97,-5.04) {at most $L_*$ joins them};
\node[txt] at (12.97,-5.52) {through vertices whose};
\node[txt] at (12.97,-6.00) {kept rank-$r$ candidate};
\node[txt] at (12.97,-6.48) {state passes. Output:};
\node[txt] at (12.97,-6.96) {links and their clusters.};
\end{tikzpicture}
\caption{\textbf{Local algorithm for a fixed number of parameters}
(Algorithm~\ref{alg:overall-general}).
(a) From every reference point (two are shown), random walks transport its reference state with
$U_{\gamma_{\mathrm{lb}}/2}$. Shading marks the resolved region $\gamma\ge\gamma_{\mathrm{lb}}$. A candidate state
given before the walk leaves the connected component it started in (solid) is accurate if its reference point is successful.
(b) A candidate state is admitted at the ranks of its reference point. Energies are estimated from the local terms of $H(x)$.
Eq.~\eqref{eq:1d-rank-selection} then selects the rank. The dots include candidate states from walks not drawn in
(a). The filled dot is an accurate candidate state. The test is the filtered-response test of Algorithm~\ref{alg:gap_cert},
run with each of the $m$ coordinate probes $\partial_{x_k}H(x)$. A candidate state passes if every probe returns \textsc{Connected}.
(c) Here the length bound is $L_*=O(1)$. The kept rank-$r$ candidate states at both input points must pass too. If the reported path (thick
line) stays in the resolved region, the linked ground states are not in different phases
(Theorem~\ref{thm:local-phase-diagram-general} and Remark~\ref{rem:phase-clustering-relaxation}).}
\label{fig:local-fixed-dimension}
\end{figure}

In higher dimensions, the algorithm of Figure~\ref{fig:local-fixed-dimension} samples grid paths with random walks, since there are too many to try them all.
Under Assumption~\ref{assumption:random-walk-navigation}, polynomially many random walks find an accurate candidate state at every grid point near the resolved region, with high probability.
State selection (Subsection~\ref{subsec:energy-rank-estimation}) then chooses the desired state, as in one dimension.

\begin{assumption}[Geometry of the resolved region]
\label{assumption:random-walk-navigation}
There are constants $L_0,w_0>0$, independent of $n$,
with the following property.
For every point $x$ of a connected component
$R$ of the resolved region, there is a
successful reference point $x_{\mathrm{ref}}\in R$ and a path $\pi_x$ from
$x$ to $x_{\mathrm{ref}}$, contained in $R$, such that
\[
    \ell_x\leq L_0,
    \qquad
    \operatorname{dist}\!\left(\pi_x(s),\partial R\right)
    \geq
    \frac{w_0}{L_0}s-h_{\mathrm{w}}
    \quad (0\leq s\leq\ell_x).
\]
Here $\ell_x$ is the path length, $s$ is arclength measured
from $x$, and
$h_{\mathrm{w}}:=\gamma_{\mathrm{lb}}/(8L_\gamma)$ is the scale
of the random walks. The lower bound on the distance from $\partial R$ thus grows linearly in $s$ and is negative at $x$.
All lengths and distances are measured in physical
parameter-space units.
The paths need not be known to the algorithm.
\end{assumption}

\begin{theorem}[Constant-dimensional relaxation of
Theorem~\ref{thm:detection}]
\label{thm:local-phase-diagram-general}
Let $m=O(1)$ be fixed.
Under Assumptions~\ref{assumption-reference-state}
and~\ref{assumption:random-walk-navigation} and, if $r_{\max}\ge2$, the tolerance bound of
Eq.~\eqref{eq:gsm-width},
there exists a quantum algorithm (Algorithm~\ref{alg:overall-general}) running in $\mathrm{poly}(n)$
time that, with probability at least $1-1/\mathrm{poly}(n)$,
learns the phase diagram with the relaxed guarantee of
Remark~\ref{rem:phase-clustering-relaxation}. The algorithm uses
the transport and measurements of
Theorem~\ref{thm:local-phase-diagram} and has the same
guarantees on the copies of each reference state.
\end{theorem}

\begin{remark}[Local measurements only, fixed number of parameters]
\label{rem:local-measurements-general}
The maximally-mixed proviso of Remark~\ref{rem:local-measurements} applies here as well.
Under it, Algorithm~\ref{alg:overall-general} uses only local measurements.
The theorem then holds without Eq.~\eqref{eq:gsm-width}.
In its proof, the tightened grid condition of Subsubsection~\ref{subsubsec:random-walks} reduces to $L_\gamma h_{\mathrm{grid}}\le\gamma_{\mathrm{lb}}/8$.
\end{remark}

\subsubsection{Random walks on the grid}
\label{subsubsec:random-walks}

We want a candidate state at every vertex of the refined grid, including the input points.
We call these vertices \emph{targets}.
Each random walk starts at a reference point.
Walks move on the largest uniform cubic grid of spacing $\delta x_{\mathrm{w}}:=h_{\mathrm{w}}/(2\sqrt m)$ that fits in $\mathcal M$, centered in $\mathcal M$.
We call it the \emph{walk grid}.
Since $\gamma_{\mathrm{lb}}\ge1/\mathrm{poly}(n)$ and $L_\gamma=2J=O(n)$, the spacing $\delta x_{\mathrm{w}}$ is inverse-polynomial.
The targets and the reference points need not lie on this grid.
For any $y\in\mathcal M$, we therefore let $q(y)$ be the walk-grid point nearest to $y$.
If several walk-grid points are equally near, a fixed rule picks one.
A random walk from a reference point $x_{\mathrm{ref}}$ goes straight to $q(x_{\mathrm{ref}})$ and then takes random steps.
Each step is along one of the $2m$ coordinate directions, chosen uniformly at random.
If a step would leave $\mathcal M$, the walk stays where it is.
Between consecutive vertices, the path of a walk is the straight segment.

A candidate path to a target $x$ follows the walk until its first visit to $q(x)$ and then takes the straight segment from $q(x)$ to $x$.
One random walk thus gives a candidate path to every target $x$ whose $q(x)$ it visits.
Transporting the reference state $C[x_{\mathrm{ref}}]$ along a candidate path to $x$ gives a candidate state at $x$.
A walk may wander out of the resolved region.
Only some random walks therefore yield useful candidate paths (Figure~\ref{fig:local-fixed-dimension}(a)).

By Eq.~\eqref{eq:grid-approximation}, the grid approximation of a path in the resolved region can leave the region by up to $h_{\mathrm{grid}}$.
The proof of Theorem~\ref{thm:local-phase-diagram-general} therefore needs accurate candidate states at every vertex within $h_{\mathrm{grid}}$ of a connected component of the resolved region.
Fix such a component $R$ and a vertex $x$ of the refined grid within $h_{\mathrm{grid}}$ of $R$.
For a resolved vertex, $x\in R$.
Let $r$ be the ground-state degeneracy of $R$.
Call a candidate path to $x$ \emph{gapped} if $\gamma_r\ge3\gamma_{\mathrm{lb}}/4$ at every point of the path, end segments included. Transport along a gapped path is exact (Corollary~\ref{cor:exact-transport}).
So that $x$ lies within $h_{\mathrm{w}}$ of $R$, we strengthen Eq.~\eqref{eq:grid-large-gap} to the \emph{tightened grid condition} $L_\gamma h_{\mathrm{grid}}\le\gamma_{\mathrm{lb}}/8$.
Equivalently, $h_{\mathrm{grid}}\le h_{\mathrm{w}}$.
The candidate paths constructed below then run up to $2h_{\mathrm{w}}$ from $R$. Since $2L_\gamma h_{\mathrm{w}}=\gamma_{\mathrm{lb}}/4$, Eq.~\eqref{eq:grid-gap-lipschitz} gives $\gamma_r\ge3\gamma_{\mathrm{lb}}/4$ within $2h_{\mathrm{w}}$ of $R$. If $r_{\max}\ge2$, state selection near $R$ also needs the width of the lowest $r$ levels to stay at most $\epsilon_{\mathrm{ref}}\gamma_{\mathrm{lb}}$. The tightened grid condition then also includes $L_\gamma h_{\mathrm{grid}}\le\epsilon_{\mathrm{ref}}\gamma_{\mathrm{lb}}/2$. The walks do not need this finer scale. Write
\begin{equation}
    \kappa_x(x_{\mathrm{ref}})
    :=\Pr\bigl[\text{a random walk of $L_{\mathrm{walk}}$ steps from
    $x_{\mathrm{ref}}$ produces a gapped candidate path to $x$}\bigr]
    \label{eq:hit-probability}
\end{equation}
for the success probability of one random walk.
The next lemma fixes the number of steps $L_{\mathrm{walk}}$ and bounds this probability from below for some successful reference point $x_{\mathrm{ref}}\in R$.

\begin{lemma}[Geometry implies random-walk navigation]
\label{lem:geometric-navigation}
Fix $m=O(1)$ and impose the
tightened grid condition. Then
Assumption~\ref{assumption:random-walk-navigation} implies that there
are a number of steps $L_{\mathrm{walk}}$ and a probability $\kappa_0$ with
\[
    L_{\mathrm{walk}}=O((1+L_0/\delta x_{\mathrm{w}})^2)=\mathrm{poly}(n),
    \qquad
    \kappa_0\ge(1+L_0/\delta x_{\mathrm{w}})^{-O(1)}\ge1/\mathrm{poly}(n),
\]
such that every refined-grid point~$x$ within $h_{\mathrm{grid}}$ of a connected component $R$ of the resolved region
satisfies $\kappa_x(x_{\mathrm{ref}})\ge\kappa_0$ for some successful reference point $x_{\mathrm{ref}}\in R$.
\end{lemma}

\begin{lemma}[Independent random walks reach every target]
\label{lem:independent-trials}
Work in the setting of Lemma~\ref{lem:geometric-navigation}.
Fix a total failure probability $\delta_{\mathrm{tot}}$ and run
\begin{equation}
    T:=\left\lceil
        \kappa_0^{-1}
        \log\frac{2N_{\mathrm{grid}}^2}{\delta_{\mathrm{tot}}}
    \right\rceil
    \label{eq:trial-count}
\end{equation}
independent random walks of $L_{\mathrm{walk}}$ steps, starting from every reference point.
Then, with probability at least ${1-\delta_{\mathrm{tot}}/2}$, every refined-grid point within $h_{\mathrm{grid}}$ of a connected component $R$ of the resolved region receives a gapped candidate path from a random walk that starts at a successful reference point in $R$.
\end{lemma}

\begin{proof}
Fix such a point $x$, and let $x_{\mathrm{ref}}$ be the successful reference point given by Lemma~\ref{lem:geometric-navigation}. By that lemma, each random walk from $x_{\mathrm{ref}}$ independently produces a gapped candidate path to $x$ with probability at least $\kappa_0$. All $T$ random walks therefore fail with probability at most $(1-\kappa_0)^T\leq e^{-\kappa_0T} \leq \delta_{\mathrm{tot}}/(2N_{\mathrm{grid}}^2)$. Each component contains a reference point. Every reference point is a vertex of the refined grid. There are therefore at most $N_{\mathrm{grid}}N_{\mathrm{ref}}\leq N_{\mathrm{grid}}^2$ pairs of a vertex and a nearby component. A union bound over these pairs gives the claim.
\end{proof}

\subsubsection{The algorithm}
\label{subsubsec:general-algorithm}

The algorithm transports along random walks instead of intervals. From each reference point, run $T$ random walks of $L_{\mathrm{walk}}$ steps, with $L_{\mathrm{walk}}$ and $T$ as in Lemma~\ref{lem:independent-trials}. Implement the path of each random walk with $U_{\gamma_{\mathrm{lb}}/2}$ and store the circuit. A candidate path has at most $L_{\mathrm{walk}}+2$ segments. As in Eq.~\eqref{eq:1d-implementation-error}, the implemented state of each candidate path therefore lies within $\epsilon_{\mathrm{ref}} +2(L_{\mathrm{walk}}+2)\epsilon_U$ of its ideal state.
We impose the bound on the energy error, Eq.~\eqref{eq:1d-energy-budget-split}, with $\epsilon_{\mathrm{ref}} +2(L_{\mathrm{walk}}+2)\epsilon_U$ in place of $\epsilon_0$.
The same substitution applies to the state budget, Eq.~\eqref{eq:state-budget}.

At each vertex of the refined grid, we estimate the mean energy of every candidate state.
We also run the filtered-response test of Algorithm~\ref{alg:gap_cert} on every candidate state, once for each coordinate probe $\partial_{x_k}H(x)$, $k=1,\ldots,m$.
Let $L_*=O(L_{\max})$ bound the length of the grid approximations of Eq.~\eqref{eq:grid-approximation}. For example, $L_*=3mL_{\max}$ suffices. For reported paths of length at most $L_*$, we use the signal scale
\[
  \epsilon_{\mathrm{sig}}
  =
  \frac{\eta_P(n)^2\gamma_{\mathrm{lb}}^2}
       {16r_{\max}L_*^2},
\]
and impose the grid constraint, Eq.~\eqref{eq:proof-grid-condition-lipschitz}, and the state budget at this scale.
With $L_*=2$, as in one dimension, this equals the signal scale of Eq.~\eqref{eq:witness-scale}.
At this scale, we choose the precisions as in the proof of Theorem~\ref{thm:local-phase-diagram}, with $2(L_{\mathrm{walk}}+2)\epsilon_U$ in place of $2N_{\mathrm{grid}}\epsilon_U$.
Eq.~\eqref{eq:gsm-width} then refers to this $\epsilon_{\mathrm{ref}}$.

At each vertex that receives candidate states, we first keep the lowest-energy candidate state admitted at each rank, as in Eq.~\eqref{eq:1d-rank-energies}.
Then we determine the selected rank $\widehat r(x)$ by Eq.~\eqref{eq:1d-rank-selection}.
Lemma~\ref{lem:energy-rank-estimate} justifies this selection, and Figure~\ref{fig:local-fixed-dimension}(b) illustrates both steps.
We link two input points if both have selected rank $r$ and the refined grid contains a path between them, of length at most $L_*$, with the following property.
At every vertex of the path, all $m$ estimated responses of the kept rank-$r$ candidate state lie below $\epsilon_{\mathrm{sig}}/2$ (Figure~\ref{fig:local-fixed-dimension}(c)).
We store this path as the reported path of the link (Remark~\ref{rem:phase-clustering-relaxation}).
Unlike Eq.~\eqref{eq:1d-band-silence} in one dimension, this rule checks the selected rank only at the two input points, since the proof below needs it nowhere else.
The output is the links and their clusters; Algorithm~\ref{alg:overall-general} summarizes these steps.

\begin{algorithm}[!htbp]
\caption{\textsc{Learning-Phase-Diagram} for a fixed number of parameters}
\label{alg:overall-general}
\scriptsize
\begin{algorithmic}[1]
\Require $H(x)$, $x\in[-1,1]^m$; input points; reference states
$C[x_a^{\mathrm{ref}}]$; $L_{\mathrm{walk}}$ and $T$ from
Lemmas~\ref{lem:geometric-navigation}
and~\ref{lem:independent-trials}; length bound $L_*$.
\Ensure The linked relation and its clusters, with the guarantee of
Theorem~\ref{thm:local-phase-diagram-general}.
\State Form the refined grid of Subsection~\ref{subsec:grid-choice}.
\State If $r_{\max}\ge2$, run the purity test on every reference state
(Subsection~\ref{subsec:energy-rank-estimation}).
\For{each reference point $a$ and each of $T$ random walks}
    \State Run the walk for $L_{\mathrm{walk}}$ steps on the walk grid from
    $x_a^{\mathrm{ref}}$ (Subsubsection~\ref{subsubsec:random-walks}),
    transporting $C[x_a^{\mathrm{ref}}]$ with
    $U_{\gamma_{\mathrm{lb}}/2}$.
    \State At each vertex $x$ that receives a candidate state, estimate
    the mean energy and run Algorithm~\ref{alg:gap_cert} with each
    coordinate probe~$\partial_{x_k}H(x)$.
\EndFor
\State At each such vertex, keep the lowest-energy candidate state admitted at
each rank and determine $\widehat r(x)$ using
Eq.~\eqref{eq:1d-rank-selection}.
\State Link two input points with selected rank $r$ if they are
connected by a grid path of length at most $L_*$ along which the
kept rank-$r$ candidate state returns \textsc{Connected} for every coordinate
probe; store this path as the reported path.
\State \Return The links and their clusters.
\end{algorithmic}
\end{algorithm}

\subsubsection{Correctness}

\begin{proof}[Proof of Theorem~\ref{thm:local-phase-diagram-general}]
Run Algorithm~\ref{alg:overall-general}.

\emph{Coverage and state selection.}
We condition on the event of Lemma~\ref{lem:independent-trials} and on all estimates being within their accuracies.
By the allocation of failure probabilities in Subsection~\ref{subsec:resources}, both hold with probability at least $1-\delta_{\mathrm{tot}}$.

Let $x$ be a vertex within distance $h_{\mathrm{grid}}$ of a connected component $R$ of the resolved region.
Let $r$ be the ground-state degeneracy of $R$.
On the event of Lemma~\ref{lem:independent-trials}, $x$ gets a gapped candidate path from a successful reference point in $R$.
By the exact transport of Corollary~\ref{cor:exact-transport}, the ideal state of this candidate path is within $\epsilon_{\mathrm{ref}}$ of $\rho_r(x)$.
By the tightened grid condition, $\gamma_r(x)\ge3\gamma_{\mathrm{lb}}/4$.
At a point of $R$ within $h_{\mathrm{grid}}$ of $x$, the width $E_{r-1}-E_0$ is at most $\delta_{\mathrm{gsm}}$.
Each level is $J$-Lipschitz in the parameters (Lemma~\ref{lem:gap-stability}).
If $r\ge2$, the width at $x$ is therefore at most $\delta_{\mathrm{gsm}}+L_\gamma h_{\mathrm{grid}}$.
Eq.~\eqref{eq:gsm-width} and the tightened grid condition bound this by $\epsilon_{\mathrm{ref}}\gamma_{\mathrm{lb}}$.
With $\epsilon_{\mathrm{ref}}+2(L_{\mathrm{walk}}+2)\epsilon_U$ in place of $\epsilon_0$, the step \emph{Precision and state selection} of the proof of Theorem~\ref{thm:local-phase-diagram} now applies.
The state budget, Eq.~\eqref{eq:state-budget}, already covers band gaps as small as $3\gamma_{\mathrm{lb}}/4$ and widths up to $\epsilon_{\mathrm{ref}}\gamma_{\mathrm{lb}}$.
If $x$ is resolved, that step gives $\widehat r(x)=r$.
Whether or not $x$ is resolved, the $m$ estimated coordinate responses of its kept rank-$r$ candidate state are within $\epsilon_{\mathrm{sig}}/4$ of those of $\rho_r(x)$, as in Eq.~\eqref{eq:1d-selected-response}.

\emph{The connected case.}
Suppose two input points are connected by a path of length at most $L_{\max}$ with gap at least $\gamma_{\mathrm{thresh}}$ throughout.
The ground-state degeneracy along the path is then a constant $r$.
Since both input points are resolved, their selected rank is $r$.
The grid approximation of the path, Eq.~\eqref{eq:grid-approximation}, has length at most $L_*$.
By the Lipschitz bound on the gap, its rank-$r$ band gap is at least $(1-\epsilon_{\mathrm{thresh}}/2)\gamma_{\mathrm{thresh}}$, as in Eq.~\eqref{eq:grid-threshold-gap}.
The grid approximation also stays within distance $h_{\mathrm{grid}}$ of the connected component of the resolved region that contains the path.
At every vertex of the grid approximation, the estimated responses of the kept rank-$r$ candidate state are therefore $\epsilon_{\mathrm{sig}}/4$-accurate.
By Lemma~\ref{lem:edge-test}, each of these $m$ estimated responses is below $\epsilon_{\mathrm{sig}}/2$.
Hence the two input points are linked.
Input points in different clusters therefore cannot be connected by such a path, which proves clause (ii) of Definition~\ref{def:phase-clustering}.

\emph{The disconnected case.}
Consider a link whose reported path is resolved.
Let $r=\widehat r(x_i)$ be the selected rank of its endpoints.
The reported path lies in a single connected component of the resolved region.
Since the endpoints are resolved, its ground-state degeneracy is their selected rank $r$.
Suppose for contradiction that the endpoints lie in different phases.
Parametrize the reported path with constant speed on $[0,1]$.
The speed then equals the path length, which is at most $L_*$.
By Remark~\ref{rem:piecewise-smooth}, the circuit approximation of the quasi-adiabatic propagator holds for this piecewise smooth path.
As in the one-parameter proof, quasi-adiabatic transport along the reported path with cutoff $(1-\epsilon_{\mathrm{thresh}})\gamma_{\mathrm{thresh}}$ then fails, in the sense of Eq.~\eqref{eq:proof-transport-deficit}.
Theorem~\ref{lem:path-gradient-spectral-weight} turns this failure into a set of positive measure of points $s_0$ with
\[
    W^{\partial_sH}_{[\gamma_r,(1-\epsilon_{\mathrm{thresh}})\gamma_{\mathrm{thresh}})}(x(s_0))
    \geq
    \frac{\eta_P(n)^2\gamma_{\mathrm{lb}}^2}{16r_{\max}}.
\]
Because this set has positive measure and the path has only finitely many corners, we may take $s_0$ inside an edge, say along the $k$-th coordinate.
There $\partial_sH=\pm\ell\,\partial_{x_k}H$, where $\ell\le L_*$ is the speed.
The spectral weight is quadratic in the probe, so $z:=x(s_0)$ satisfies
\[
    W^{\partial_{x_k}H}_{[\gamma_r,(1-\epsilon_{\mathrm{thresh}})\gamma_{\mathrm{thresh}})}(z)
    \geq
    \frac{\eta_P(n)^2\gamma_{\mathrm{lb}}^2}{16r_{\max}L_*^2}
    =\epsilon_{\mathrm{sig}}.
\]
An endpoint $v$ of this edge is a tested vertex within $\delta x\le h_{\mathrm{grid}}$ of $z$.
As in Eq.~\eqref{eq:proof-witness-vertex}, the sandwich bound on the response and the grid constraint give $S^{\partial_{x_k}H}_{\mathcal K_{\Lambda,\delta}}(\rho_r(v);v)\ge3\epsilon_{\mathrm{sig}}/4$.
Since the kept rank-$r$ candidate state at $v$ has $\epsilon_{\mathrm{sig}}/4$-accurate responses, its estimated response is at least $\epsilon_{\mathrm{sig}}/2$, which contradicts the link and proves clause (i).

Subsection~\ref{subsec:resources} shows that the resources are polynomial and gives the guarantees on the copies of each reference state.
\end{proof}

\subsubsection{Proof of Lemma~\ref{lem:geometric-navigation}}

\begin{proof}[Proof of Lemma~\ref{lem:geometric-navigation}]
The walk can follow the path of Assumption~\ref{assumption:random-walk-navigation} backward from the reference point by crossing a chain of logarithmically many lattice cubes.
Each crossing succeeds with constant probability.

Fix $x$ and a point $p\in R$ with $\|x-p\|\le h_{\mathrm{grid}}$. For a resolved vertex, $p=x$.
Let $\pi_p:[0,\ell_p]\to R$ be the arclength-parametrized path from $p$ to $x_{\mathrm{ref}}$ given by this assumption.
By the tightened grid condition, $\|x-p\|\le h_{\mathrm{grid}}\le h_{\mathrm{w}}$.

\emph{Rounding paths to the walk grid.}
Sample the curve at intervals of arclength at most $\delta x_{\mathrm{w}}$, round each sample point by $q$ and join consecutive samples by the shortest paths on the walk grid in each coordinate.
Each join consists of at most $2m$ steps and remains in $\mathcal M$.
A portion of length $O(\delta x_{\mathrm{w}})$ is thus approximated by only $O(1)$ steps of the walk.
Each coordinate of a point on a join differs from that of the preceding sample point on $\pi_p$ by at most $\delta x_{\mathrm{w}}+\delta x_{\mathrm{w}}/2$.
Every point on a join is therefore within distance $(3/2)\sqrt m\,\delta x_{\mathrm{w}}<h_{\mathrm{w}}$ from that sample point.
The straight-line connections at the original endpoints are within distance $\sqrt m\,\delta x_{\mathrm{w}}/2$ from these endpoints.
The joins and connections therefore lie within $2h_{\mathrm{w}}$ of $R$.
Hence they satisfy $\gamma_r\geq3\gamma_{\mathrm{lb}}/4$.

\emph{The box estimate.}
Fix a large enough lattice cube.
Its \emph{central half} is the concentric subcube of half the side.
Fix also a subcube of the central half whose side is at least a fixed fraction of the cube's side.
The \emph{box estimate} concerns a walk from any vertex of the central half.
At a time proportional to the squared side length, it is in the subcube without having left the cube.
This happens with probability bounded below by a positive constant, uniformly in the starting vertex.

We prove the box estimate by diagonalizing the killed transition matrix of the walk. This matrix advances the walk but kills it the moment it leaves the cube. Translate the cube so that its interior vertices form $B=\{1,\ldots,d-1\}^m$, with the killing boundary at coordinates $0$ and $d$. The product sine basis diagonalizes this matrix, with eigenvalues $\lambda_{\boldsymbol k}$ and eigenvectors $\phi_{\boldsymbol k}$ given by
\[
    \lambda_{\boldsymbol k}
    =\frac1m\sum_{i=1}^m
        \cos\frac{\pi k_i}{d},
    \qquad
    \phi_{\boldsymbol k}(u)
    =\left(\frac2d\right)^{m/2}
      \prod_{i=1}^m\sin\frac{\pi k_i u_i}{d},
\]
where $1\leq k_i\leq d-1$.
In $t$ steps, the probability of going from $u$ to $v$ inside the cube is a sum over these modes.
We show that the principal mode dominates it.
The principal eigenvalue is $\lambda_{\boldsymbol1}=\cos(\pi/d)$.
For positive eigenvalues,
\[
    \frac{\lambda_{\boldsymbol k}}
         {\lambda_{\boldsymbol1}}
    \leq
    \exp\!\left[
        -\frac{2}{md^2}\sum_{i=1}^m(k_i^2-1)
    \right],
    \qquad
    \frac{|\phi_{\boldsymbol k}(u)
               \phi_{\boldsymbol k}(v)|}
         {\phi_{\boldsymbol1}(u)\phi_{\boldsymbol1}(v)}
    \leq\prod_{i=1}^m k_i^2.
\]
Both bounds follow from the identity
\[
    \cos\frac{\pi}{d}-\cos\frac{\pi k}{d}
    =
    2\sin\frac{(k+1)\pi}{2d}
     \sin\frac{(k-1)\pi}{2d}
    \geq\frac{2(k^2-1)}{d^2}
\]
together with the elementary inequality $|\sin(k\theta)|\leq k\sin\theta$ for $0\leq\theta\leq\pi$.

We organize the sum over modes by pairing $\boldsymbol k$ with $d\boldsymbol1-\boldsymbol k$, which accounts for opposite eigenvalues. When $t\equiv\sum_i(u_i-v_i)\pmod2$, the paired positive modes are doubled, while zero modes do not contribute for $t\geq1$. We choose $t=\lceil C d^2\rceil$ for a new, large enough constant $C$ depending only on $m$.
Summing the per-mode comparisons, the total contribution of all nonprincipal positive
modes, relative to the principal mode, is at most
\[
    \left(
        \sum_{k\geq1}k^2e^{-2C(k^2-1)/m}
    \right)^m-1<\frac12.
\]
For central $u,v$, the principal eigenvector product is at least a constant times $d^{-m}$, and $\lambda_{\boldsymbol1}^{t}$ is at least a positive constant.
Hence, if the parity is compatible, the probability of reaching a single vertex $v$ satisfies
\[
    \Pr_u[X_t=v,\ X_0,\ldots,X_t\in B]
    \geq c_m d^{-m}.
\]
For sufficiently large $d$, the subcube contains of order $d^m$ vertices of each parity.
Summing this bound over these vertices gives the box estimate.

\emph{Constants and the short-path case.} Choose a small constant $a>0$ depending only on $m$ and $w_0/L_0$, and a large constant $K$ independent of $n$.
If $\ell_p\leq K\delta x_{\mathrm{w}}$, rounding the curve to the walk grid gives an $O(1)$-step path from $q(x_{\mathrm{ref}})$ to $q(p)$.
The walk follows this path with positive constant probability.

\emph{The chain of cubes.} Now suppose $\ell_p>K\delta x_{\mathrm{w}}$.
The clearance of $\pi_p$ from $\partial R$ may shrink linearly toward $p$, so the cubes must shrink toward $p$ to stay inside $R$.
We set the scales
\[
    s_0=\ell_p,\qquad s_{j+1}=(1-a/16)s_j,
\]
until the first index with $s_N\leq K\delta x_{\mathrm{w}}$.
Let $B_j$ be the cube of walk-grid points centered at $q(\pi_p(s_j))$ with maximum norm radius
\[
    r_j=\left\lfloor\frac{a s_j}{\delta x_{\mathrm{w}}}\right\rfloor
\]
measured in grid steps, and let $A_j$ be its concentric subcube of radius $\lfloor r_j/8\rfloor$.

Every cube $B_j$ lies in $R$.
Indeed, the clearance at $\pi_p(s_j)$ is at least $(w_0/L_0)s_j-h_{\mathrm{w}}$, whereas the cube radius plus the rounding error is at most $\sqrt m\,(r_j+1/2)\delta x_{\mathrm{w}}$.
For the choice of $a$ and $K$ below, the second bound is the smaller. The centers of consecutive cubes are close,
\[
    \|q(\pi_p(s_j))-q(\pi_p(s_{j+1}))\|_\infty
    \leq a s_j/16+\delta x_{\mathrm{w}},
\]
and consecutive radii are comparable.
In particular, $A_j$ and $A_{j+1}$ lie in the central half of $B_j$, and the side of $A_{j+1}$ is at least a fixed fraction of that of $B_j$.
All of these properties hold simultaneously for $a\leq\min\{1,w_0/L_0\}/(100\sqrt m)$ and $K$ large enough to absorb rounding and clearance errors.
Finally, since $s_N>(1-a/16)K\delta x_{\mathrm{w}}$, every cube is large enough to apply the box estimate.

\emph{Chaining the crossings and completing the walk.}
The walk starts at $q(x_{\mathrm{ref}})\in A_0$.
Fix $j<N$. By the box estimate, a walk from any point of $A_j$ reaches $A_{j+1}$ after a time of order $r_j^2$ without ever leaving $B_j$.
This happens with probability bounded below by a positive constant.
By the Markov property, the crossings chain one after another.
All of them therefore succeed with probability at least this constant to the power $N$, where
\[
    N=O\!\left(
        \log\left(1+\frac{\ell_p}{\delta x_{\mathrm{w}}}\right)
    \right),
    \qquad
    \sum_{j<N}r_j^2
    =O\!\left(
        \left(\frac{\ell_p}{\delta x_{\mathrm{w}}}\right)^2
    \right).
\]

From any arrival point in $A_N$, take a coordinate path inside $A_N$ to its center, then follow the rounded approximation of $\pi_p([0,s_N])$ backward to $q(p)$. Since $s_N=O(\delta x_{\mathrm{w}})$ and $r_N=O(1)$, this adds only $O(1)$ steps and costs another constant probability factor. Finally, connect $q(p)$ to $q(x)$ by a coordinate-wise grid path. Since $\|q(p)-q(x)\|_\infty \le\|p-x\|+\sqrt m\,\delta x_{\mathrm{w}}\le h_{\mathrm{w}} +\sqrt m\,\delta x_{\mathrm{w}}$, this adds $O(1)$ steps and costs one more constant probability factor.
Every point of this connection is $2h_{\mathrm{w}}$-close to $p\in R$.

Hence each segment of the completed path, including the initial and final connections, is $2h_{\mathrm{w}}$-close to $R$.
Eq.~\eqref{eq:grid-gap-lipschitz} then gives $\gamma_r\geq3\gamma_{\mathrm{lb}}/4$ on the whole path.

The walk may visit $q(x)$ before it completes this path.
Its initial segment up to the first visit then still has $\gamma_r\geq3\gamma_{\mathrm{lb}}/4$.
The final connection to $x$ also stays valid.
Stopping at the first visit therefore preserves the guarantee.

Combining these bounds gives, uniformly over all targets, a number of steps $O((1+L_0/\delta x_{\mathrm{w}})^2)$ and a success probability at least ${(1+L_0/\delta x_{\mathrm{w}})^{-O(1)}}$.
\end{proof}

\subsection{Experimental implementations}
\label{subsec:experimental-implementations}

A \emph{parameter sweep} is ordinary time evolution under $H(x(t))$ as $x(t)$ moves slowly. We assume that the experiment can prepare the reference states of Subsection~\ref{subsec:setup-main-theorem} and run prescribed smooth parameter sweeps along the algorithms' grid paths and straight end segments, with turns and revisits. Each parameter sweep must have polynomial duration and inverse-polynomial implementation error.

Starting from $\rho_{\mathcal G}(x(0))$, we choose the schedule such that
\[
    \|\rho(t)-\rho_{\mathcal G}(x(t))\|_1\leq\epsilon_U
\]
until $x(t)$ first exits the connected component of the resolved region containing $x(0)$. Since the spectral gap is at least $\gamma_{\mathrm{lb}}$ before this exit, the adiabatic theorem for an isolated band~\cite[Theorem~3]{jansen2007bounds} gives such a schedule of polynomial duration. Errors in the reference state contribute an additional $\epsilon_{\mathrm{ref}}$. After this exit, the parameter sweep may continue without tracking the ground state, but must keep the same coherent dynamics.

The estimator of Theorem~\ref{thm:quantum-filtered-response} samples waiting times of both signs. Negative waiting times need backward evolution.
For a state that commutes with $H(x)$, non-negative waiting times suffice.

\begin{corollary}[Response estimation without backward evolution]\label{cor:forward-time-response}
Let $\sigma_V$ be the measured candidate state.
Let $\rho$ be a state that commutes with $H(x)$, such as $\rho_r(x)$.
Suppose that $\mathcal K$ is odd, as $\mathcal K_{\Lambda,\delta}$ is (Lemma~\ref{lem:quantum-kernel-moments}).
Assume the hypotheses of Theorem~\ref{thm:quantum-filtered-response}.
Its estimator draws each waiting time from the probability density $p_T(t)=|\mathcal K(t)|/K_T$ on $[-T,T]$.
Draw the waiting times instead from $p^{+}_T(t):=2|\mathcal K(t)|/K_T$ on $[0,T]$, and otherwise form the estimator of Eq.~\eqref{eq:filtered-response-estimator} unchanged.
Then the resulting estimate satisfies the error bound of Eq.~\eqref{eq:total-response-estimation-error}, with $S^{O}_{\mathcal K}(\rho;x)$ in place of $S^{O}_{\mathcal K}(\sigma_V;x)$ and with an additional error $2K_T\|O\|^2{\|\sigma_V-\rho\|_1}$.
As in that theorem, this holds except with probability at most $\delta_{\mathrm{fail}}+\delta_{\mathrm{mode}}$.
With fresh copies, all waiting-time evolutions $e^{-iH(x)t}$ then run forward in time.
\end{corollary}
\begin{proof}
Here $\tau^x_t(O)$ is $O$ evolved for time $t$ under $H(x)$.
Since $\rho$ commutes with $H(x)$, cyclicity of the trace gives $\mathrm{Tr}\bigl[\rho\,[\tau^x_{-t}(O),O]\bigr]=-\mathrm{Tr}\bigl[\rho\,[\tau^x_t(O),O]\bigr]$.
The dissipative response of $\rho$ is therefore odd in $t$.

As $\mathcal K$ is odd, $S^{O}_{\mathcal K}(\rho;x)=2\int_0^\infty\mathcal K(t)\,i\,\mathrm{Tr}\bigl[\rho\,[\tau^x_t(O),O]\bigr]\,dt$.
Draw $t$ from $p^{+}_T$ and form the ideal sample $X$ of Eq.~\eqref{eq:ideal-response-sample}.
As in Eq.~\eqref{eq:unbiased-candidate-response}, its mean is $2\int_0^T\mathcal K(t)\,i\,\mathrm{Tr}\bigl[\sigma_V[\tau^x_t(O),O]\bigr]\,dt$, and it still obeys $|X|\le2K_T\|z\|_1^2$.
Replacing $\sigma_V$ by $\rho$ in this mean costs at most $2K_T\|O\|^2\|\sigma_V-\rho\|_1$, because $\bigl|\mathrm{Tr}\bigl[(\sigma_V-\rho)[\tau^x_t(O),O]\bigr]\bigr|\le2\|O\|^2\|\sigma_V-\rho\|_1$ and $\int_0^T|\mathcal K(t)|\,dt=K_T/2$.
Since $|\mathcal K|$ is even, taking $|t|$ for $t$ drawn from the implemented distribution $\widetilde p_T$ samples $p^{+}_T$ to its precision $\eta_{\mathrm{samp}}$.
The truncation, sampling and readout errors are therefore bounded as in the proof of Theorem~\ref{thm:quantum-filtered-response}.
\end{proof}

The local algorithms compare their estimates with the filtered response of the sector state $\rho_r(x)$, which commutes with $H(x)$.
Since $K_T\le K_0(\mathcal K)$, the state budget, Eq.~\eqref{eq:state-budget}, covers the additional error.
With fresh copies and Lemma~\ref{lem:ancilla-free-response-correlator}, the response measurements therefore need neither an ancilla nor backward evolution.

The local algorithms transport candidate states with the quasi-adiabatic propagator $U_{\gamma_{\mathrm{lb}}/2}$, but parameter sweeps suffice, as the following corollary shows.

\begin{corollary}[Experimental implementations of
Theorems~\ref{thm:local-phase-diagram}
and~\ref{thm:local-phase-diagram-general}]
\label{cor:adiabatic-phase-diagram}
Suppose that the experiment can prepare the reference states
and run the parameter sweeps described above.
Under Assumption~\ref{assumption-reference-state}, the maximally-mixed proviso
of Eq.~\eqref{eq:class-promise}, and,
for $m>1$, Assumption~\ref{assumption:random-walk-navigation},
the same relaxed guarantee then follows from parameter
sweeps, local energy estimates, and filtered-response measurements.
The total evolution time and the number of copies of each
reference state are polynomial.

Alternatively, in one dimension, suppose that a polynomial-time
preparation outputs $\rho_{\mathcal G}(x)$ to the
required accuracy at every resolved grid point, with arbitrary
outputs elsewhere. Then the same guarantee follows from
filtered-response measurements on these states alone, without transport.
\end{corollary}

\begin{proof}[Proof of Corollary~\ref{cor:adiabatic-phase-diagram}]
\emph{Parameter sweeps in one dimension.}
In Algorithm~\ref{alg:overall}, replace quasi-adiabatic transport by parameter sweeps, which are also unitary and so preserve the form $Q/q$ that Remark~\ref{rem:local-measurements} uses.
At a resolved vertex $x$ with ground-state degeneracy $r$, take the monotone parameter sweep from a successful reference point in its connected component of the resolved region (Assumption~\ref{assumption-reference-state}).
This sweep stays in the component, so by the setup of this subsection its tracking error is at most $\epsilon_U$.
This adds to the error $\epsilon_0$ of Eq.~\eqref{eq:1d-implementation-error}.
Within $\epsilon_0$, $\epsilon_U$ also bounds the implementation error of the parameter sweep along each grid edge.
With $J(\epsilon_U+\epsilon_0)\leq\eta_E$, the resulting candidate state's energy exceeds $\operatorname{Tr}[H(x)\rho_r(x)]$ by at most $\epsilon_E=\eta_E$, as Eq.~\eqref{eq:1d-rank-selection} requires.
The rest of the proof of Theorem~\ref{thm:local-phase-diagram} carries over.
The maximally-mixed proviso changes only the steps listed in Remark~\ref{rem:local-measurements}.

\emph{Parameter sweeps in higher dimensions.}
In Algorithm~\ref{alg:overall-general}, replace $U_{\gamma_{\mathrm{lb}}/2}$ along the sampled paths, end segments included, by parameter sweeps.
On a gapped path (Subsubsection~\ref{subsubsec:random-walks}), the lowest $r$ levels form a band isolated by $\gamma_r\geq3\gamma_{\mathrm{lb}}/4$, even outside the resolved region.
The adiabatic theorem for an isolated band~\cite[Theorem~3]{jansen2007bounds} then bounds the band tracking error by $\epsilon_U$ after a polynomial duration.
Its hypotheses hold.
$H(x)$ has polynomially bounded first and second derivatives (Definition~\ref{definition: geometrically local hamiltonians}), and smoothing each corner over a length at most $\delta x$ gives polynomially bounded speed and acceleration.
The smoothed path stays within $\delta x<h_{\mathrm{grid}}$ of the gapped path.
By Eq.~\eqref{eq:grid-gap-lipschitz} and the tightened grid condition, the smoothed path therefore has $\gamma_r\geq5\gamma_{\mathrm{lb}}/8$, an inverse polynomial.
With $\epsilon_{\mathrm{ref}}+2(L_{\mathrm{walk}}+2)\epsilon_U$ in place of $\epsilon_0$ in the condition $J(\epsilon_U+\epsilon_0)\le\eta_E$, the selection, response, and linking arguments of Theorem~\ref{thm:local-phase-diagram-general} apply.
The maximally-mixed proviso again changes only the steps listed in Remark~\ref{rem:local-measurements}.

\emph{Evolution time and copies.} We estimate each observable by repeating the preparation, parameter sweep, and measurement in Appendix~\ref{sec:diabatic-signals}.
Estimating polynomially many bounded observables to inverse polynomial precision requires polynomially many repetitions, each taking polynomial time, which yields the claimed total evolution time and number of copies.

\emph{Preparation at every resolved grid point.}
In Algorithm~\ref{alg:overall}, omit quasi-adiabatic transport and state selection, and at each grid point run Algorithm~\ref{alg:gap_cert} on the state prepared at that point. Link two input points when all estimates in the interval between them are below $\epsilon_{\mathrm{sig}}/2$.
The preparation accuracy and the estimation error guarantee that Eq.~\eqref{eq:1d-selected-response} holds at every resolved vertex. Hence, the connected and disconnected cases of the proof of Theorem~\ref{thm:local-phase-diagram} apply, yielding clause~(i) on resolved intervals and clause~(ii). Neither case uses the state prepared at an unresolved vertex, and the resource bounds apply as before.
\end{proof}

\subsection{Resources}
\label{subsec:resources}

\subsubsection{Counts, correlator sampling, and precision}

In the one-parameter case, there are $N_{\mathrm{ref}}N_{\mathrm{grid}}$ candidate states, each with one energy estimate and one response estimate.
If $r_{\max}\ge2$, each reference state also has one purity estimate, used for all ranks.
A swap test returns $\pm1$ with mean equal to the purity.
Hoeffding's inequality then gives accuracy $\epsilon_p/2$ from $O(\epsilon_p^{-2}\log(N_{\mathrm{ref}}/\delta_{\mathrm{tot}}))$ pairs of copies.
For $N$ input points, testing all input intervals requires at most $O(N^2N_{\mathrm{grid}})$ comparisons.
Computing the clusters takes polynomial time.

For total failure probability $\delta_{\mathrm{tot}}$, Eq.~\eqref{eq:number-response-samples} allows
\begin{equation}
    N_{\mathrm{corr}}
    =
    O\left(
       \frac{K_T^2\,\|z\|_1^4}
            {\epsilon_{\mathrm{sig}}^2}
       \log\frac{2N_{\mathrm{ref}}N_{\mathrm{grid}}}
                {\delta_{\mathrm{tot}}}
    \right)
    \label{eq:1d-correlator-count}
\end{equation}
sampled correlators per response estimate.
Here $K_T$ is the truncated kernel norm, and $\|z\|_1$ is the Pauli one-norm of $O=\partial_xH(x)$, maximized over $x$.
For each reference state, the number of normalized observables to estimate is at most
\begin{equation}
    M=O\left(N_{\mathrm{grid}}
       \left(J+N_{\mathrm{corr}}\right)\right).
    \label{eq:1d-observable-count}
\end{equation}
These are the $J$ local energy terms and the sampled correlator observables at each grid point $x$, each pulled back by the transport circuit $V_a[x]$ for measurement on the reference state.
We choose their estimation precision to be
\begin{equation}
    \epsilon_*
    :=\min\left\{
       \frac{\eta_E}{2J},
       \frac{\epsilon_{\mathrm{sig}}}
       {64K_T\|z\|_1^2}
    \right\}.
    \label{eq:1d-tomography-precision}
\end{equation}
The first term ensures $\eta_E/2$-accurate mean energies, and the second is for estimating the filtered responses, whose time truncation, simulation, and sampling errors are controlled by Theorem~\ref{thm:quantum-filtered-response}.
All precisions are inverse-polynomial and all observables have polynomial-sized implementations.

\subsubsection{Fixed dimension}

There are $N_{\mathrm{ref}}T$ random walks of at most $L_{\mathrm{walk}}+2$ segments each. Each reference state has at most $M'=O\bigl(TN_{\mathrm{grid}}(J+mN_{\mathrm{corr}})\bigr)$ normalized observables. Here $N_{\mathrm{corr}}$ and $\epsilon_*$ are as in Eqs.~\eqref{eq:1d-correlator-count} and~\eqref{eq:1d-tomography-precision}, except that the signal scale $\epsilon_{\mathrm{sig}}$ is the rescaled one of Subsection~\ref{subsec:local-higher-dimensional} and the logarithm in Eq.~\eqref{eq:1d-correlator-count} becomes $\log(2N_{\mathrm{ref}}M'/\delta_{\mathrm{tot}})$, where any polynomial upper bound on $M'$ may be used. Since $T$, $L_{\mathrm{walk}}$, and $N_{\mathrm{grid}}$ are polynomial, all counts and durations remain polynomial and all precisions inverse-polynomial.

The analysis below of unique ground states applies with $M'$ instead of $M$.
Allocating $\delta_{\mathrm{tot}}/2$ to Lemma~\ref{lem:independent-trials} and $\delta_{\mathrm{tot}}/2$ to the estimates yields total failure probability $1/\mathrm{poly}(n)$.

\subsubsection{Unique ground states}

As in Theorem~\ref{thm:detection}, we also assume efficient controlled evolution under the known Hamiltonian of each reference state.
As in Corollary~\ref{cor:reference-copy-count}(a), reference states that are unique ground states make every resolved point non-degenerate.
We may therefore take $r_{\max}=1$ and skip the purity test.
Catalytic tomography (Lemma~\ref{lemma:catalytic-tomography}) then estimates the above observables using a single copy of each reference state. With $\delta_{\mathrm{cat}}$ chosen as in Eq.~\eqref{eq:catalytic-disturbance-choice} for a constant $c>0$ and $N_{\mathrm{meas}}=N_{\mathrm{ref}}M$, all estimates succeed except with probability at most $3\delta_{\mathrm{tot}}/8$, as computed after that equation. The total disturbance of the reference states is at most $e^{-cn}$. The lemma's cost is polylogarithmic in the inverse disturbance, so the overall runtime is polynomial.

\subsubsection{Gibbs reference states}

To make the Gibbs preparations at resolved points successful, we take $\beta_{\mathrm{ref}}$ and assume a bound on the tolerance $\delta_{\mathrm{gsm}}$:
\[
    \beta_{\mathrm{ref}}
    =
    \frac{1}{\gamma_{\mathrm{lb}}}
    \log\left(\frac{8\cdot 2^n}{\epsilon_{\mathrm{ref}}}\right),
    \qquad
    \delta_{\mathrm{gsm}}
    \le
    \gamma_{\mathrm{lb}}\,
    \frac{\log(1+\epsilon_{\mathrm{ref}}/4)}
         {\log(8\cdot 2^n/\epsilon_{\mathrm{ref}})} .
\]
By Lemma~\ref{lem:gibbs-ground-state-approximation} with $\epsilon=\epsilon_{\mathrm{ref}}/2$, the Gibbs state $\rho_{\beta_{\mathrm{ref}}}(x)$ of $H(x)$ at inverse temperature $\beta_{\mathrm{ref}}$ satisfies $\|\rho_{\beta_{\mathrm{ref}}}(x)-\rho_{\mathcal G}(x)\|_1 \le\epsilon_{\mathrm{ref}}/2$ at every $x$ with $\gamma(x)\ge\gamma_{\mathrm{lb}}$. Thus, by the triangle inequality, any $C[x]$ within $\epsilon_{\mathrm{ref}}/2$ of $\rho_{\beta_{\mathrm{ref}}}(x)$ in trace norm provides the successful preparations required by Assumption~\ref{assumption-reference-state}. The bound on $\delta_{\mathrm{gsm}}$ above implies Eq.~\eqref{eq:gsm-width}, because $\log(1+\epsilon_{\mathrm{ref}}/4)\le\epsilon_{\mathrm{ref}}/4$ and $\log(8\cdot2^n/\epsilon_{\mathrm{ref}})\ge\log16>1$.
Theorems~\ref{thm:local-phase-diagram} and~\ref{thm:local-phase-diagram-general} therefore apply to Gibbs reference states.
Gibbs states at unresolved points need not satisfy the maximally-mixed proviso.
Remark~\ref{rem:local-measurements} therefore does not cover them.

For every estimation method, a union bound over sampling and readout errors bounds the total failure probability by $\delta_{\mathrm{tot}}=1/\mathrm{poly}(n)$.
Gibbs reference states, other reference states, and the parameter-sweep implementation use independent sampling with polynomially many copies, as counted by Eqs.~\eqref{eq:number-response-samples} and~\eqref{eq:mean-energy-samples}.
\section{Learning the Phase Diagram with Dynamic Programming}
\label{sec:detection}

This appendix proves Theorem~\ref{thm:detection}. Our algorithm learns the phase diagram efficiently, with the full guarantee of Definition~\ref{def:phase-clustering}. Unlike the local algorithms with $r_{\max}\ge2$, it does not need the tolerance bound of Eq.~\eqref{eq:gsm-width}. Unlike Theorem~\ref{thm:local-phase-diagram-general}, it needs no geometric condition.
If only some connected components of the resolved region contain a successful reference point, the guarantee holds in those components (Corollary~\ref{cor:components-with-reference-point}).
In such a component, a gapped grid path joins a successful reference point to each grid point (Lemma~\ref{lem:gapped-grid-path}). Transporting the reference state along this path prepares the ground state there accurately.
But in several dimensions there are exponentially many grid paths, and we know neither which of them are gapped nor which preparations succeeded.

\subsection{Overview of the algorithm}
\label{app:overview}

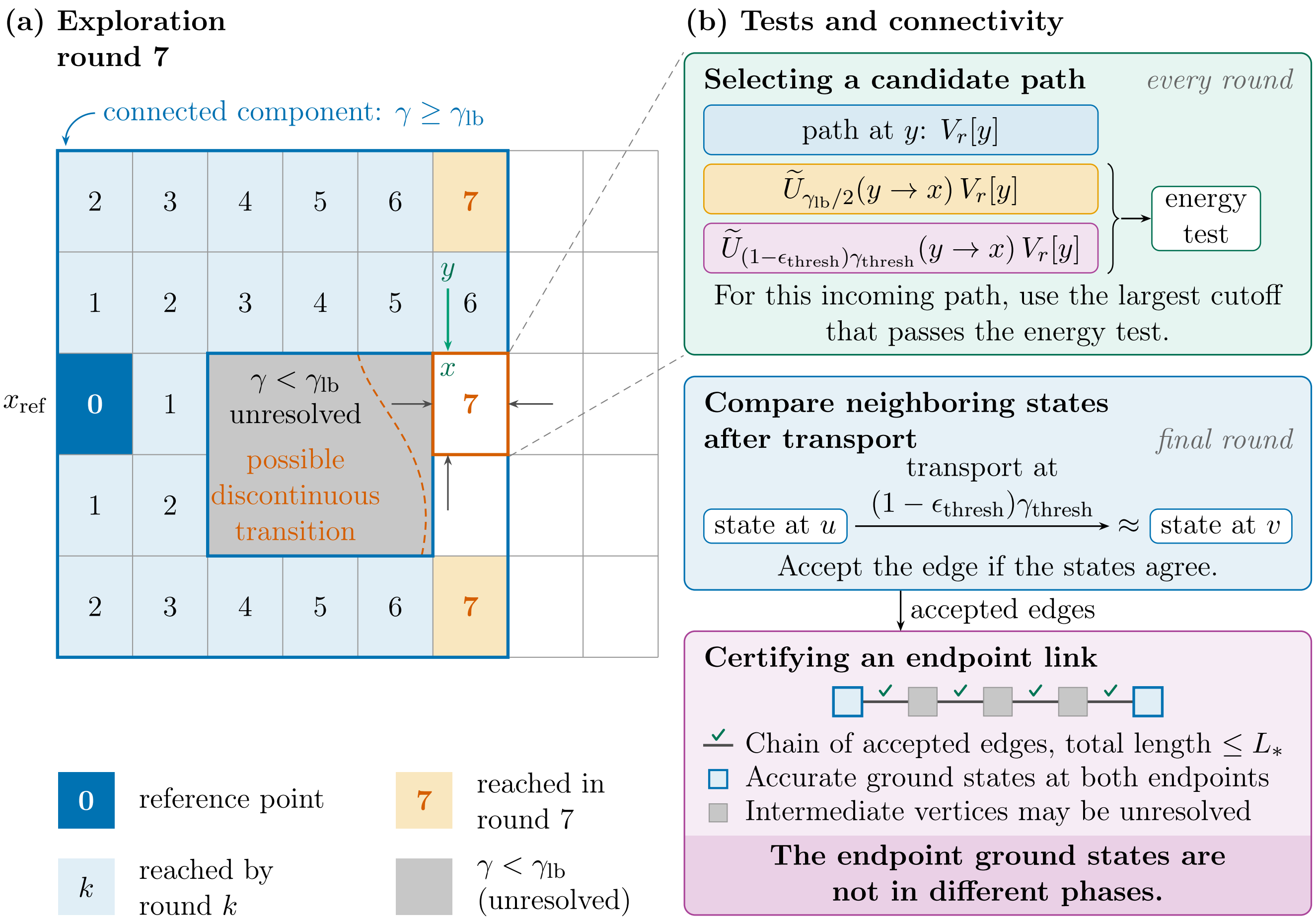
\begin{figure}[!t]
\centering
\tikzset{
  cell/.style={minimum width=1.0cm, minimum height=1.35cm, inner sep=0pt, anchor=center, font=\normalsize},
  cellnum/.style={cell, fill=figblue!12},
  cellnew/.style={cell, fill=figamber!25, text=figorange, font=\bfseries},
  gridline/.style={black!40, line width=0.4pt},
  zoom/.style={black!50, dash pattern=on 3pt off 2pt, line width=0.4pt},
  bbox/.style={rounded corners=4pt, line width=0.6pt, inner sep=0pt},
  pill/.style={rounded corners=3pt, line width=0.5pt, inner xsep=5pt, inner ysep=1.6pt, minimum width=5.25cm, minimum height=0.66cm, text height=10.7pt, text depth=3.9pt, font=\normalsize},
  spill/.style={rounded corners=3pt, line width=0.5pt, inner xsep=4pt, inner ysep=1pt, text height=8.2pt, text depth=2.7pt, font=\normalsize, fill=white, draw=figblue},
  lsq/.style={minimum size=0.75cm, inner sep=0pt, anchor=west, font=\normalsize},
  ulab/.style={font=\normalsize, inner sep=1pt},
  bline/.style={figtext, align=center, inner sep=0pt},
  stmt/.style={figtext, anchor=west, inner sep=0pt, text height=8.2pt, text depth=2.7pt},
  tagtxt/.style={font=\normalsize\itshape, text=black!60, anchor=base east, inner sep=0pt},
  spacer/.style={inner sep=0pt, outer sep=0pt, minimum height=0pt, minimum width=8.35cm},
  cvert/.style={minimum size=0.38cm, inner sep=0pt, outer sep=0pt},
  ckey/.style={minimum size=0.24cm, inner sep=0pt, outer sep=0pt},
  tick/.pic={\draw[figgreen!75!black, line width=0.9pt, line cap=round, line join=round] (-0.07,0.015) -- (-0.025,-0.045) -- (0.075,0.085);},
  gridxf/.style={yshift=0.1cm, xscale=1.0, yscale=1.35},
}%
\begin{tikzpicture}
\node[figtitle, anchor=north west, inner sep=0pt] at (8.35, 2.0) {(b) Tests and connectivity};
\node[spacer] (w1) at (12.525, 1.40) {};
\node[figstep, anchor=north west, inner sep=0pt] (h1) at (8.60, 1.20) {Selecting a candidate path};
\node[tagtxt] at (16.45,0 |- h1.base) {every round};
\node[pill, fill=figblue!15, draw=figblue, anchor=north west] (p1) at ($(h1.south west)+(0,-0.13)$) {path at $y$: $V_r[y]$};
\node[pill, fill=figamber!25, draw=figamber, below=0.11 of p1] (p2) {$\widetilde U_{\gamma_{\mathrm{lb}}/2}(y\to x)\,V_r[y]$};
\node[pill, fill=figpurple!15, draw=figpurple, below=0.11 of p2] (p3) {$\widetilde U_{(1-\epsilon_{\mathrm{thresh}})\gamma_{\mathrm{thresh}}}(y\to x)\,V_r[y]$};
\draw[decorate, decoration={brace, amplitude=4pt}, line width=0.5pt] ($(p2.north east)+(0.12,0)$) -- ($(p3.south east)+(0.12,0)$);
\coordinate (am) at ($(p2.north east)!0.5!(p3.south east)+(0.30,0)$);
\node[figtext, anchor=west, align=center, rounded corners=3pt, line width=0.5pt, inner xsep=5pt, inner ysep=3pt, fill=white, draw=figgreen!70!black] (et) at ($(am)+(0.40,0)$) {energy\\test};
\draw[figarrow] (am) -- (et.west);
\node[bline, anchor=north] (t1) at ($(12.525,0 |- p3.south)+(0,-0.14)$) {For this incoming path, use the largest cutoff\\that passes the energy test.};
\coordinate (b1b) at ($(t1.south)+(0,-0.11)$);
\begin{scope}[on background layer]
\node[bbox, fill=figgreen!10, draw=figgreen!70!black, fit=(w1)(h1)(p1)(p3)(et)(t1)(b1b)] (box1) {};
\end{scope}
\node[spacer] (w2) at ($(12.525,0 |- box1.south)+(0,-0.28)$) {};
\node[figstep, anchor=north west, align=left, inner sep=0pt, text depth=0pt] (h2) at ($(8.60,0 |- w2)+(0,-0.20)$) {Compare neighboring states\\after transport};
\node[tagtxt] at (16.45,0 |- h2.south) {final round};
\node[spill, anchor=north west] (su) at ($(h2.south west)+(0,-0.80)$) {state at $u$};
\node[spill, anchor=north east] (sv) at (16.45,0 |- su.north) {state at $v$};
\node[font=\normalsize, inner sep=0pt, anchor=east] (ap) at ($(sv.west)+(-0.12,0)$) {$\approx$};
\draw[figarrow] ($(su.east)+(0.10,0)$) -- ($(ap.west)+(-0.10,0)$) node[midway, above, align=center, font=\normalsize, inner sep=2.4pt] {transport at\\$(1-\epsilon_{\mathrm{thresh}})\gamma_{\mathrm{thresh}}$};
\node[bline, anchor=north] (t2) at ($(12.525,0 |- su.south)+(0,-0.15)$) {Accept the edge if the states agree.};
\coordinate (b2b) at ($(t2.south)+(0,-0.11)$);
\begin{scope}[on background layer]
\node[bbox, fill=figblue!10, draw=figblue, fit=(w2)(h2)(su)(sv)(t2)(b2b)] (box2) {};
\end{scope}
\node[spacer] (w3) at ($(12.525,0 |- box2.south)+(0,-0.54)$) {};
\node[figstep, anchor=north west, inner sep=0pt, text depth=0pt] (h3) at ($(8.60,0 |- w3)+(0,-0.20)$) {Certifying an endpoint link};
\coordinate (ch) at ($(12.525,0 |- h3.south)+(0,-0.46)$);
\foreach \i in {1,...,5} \coordinate (c\i) at ($(ch)+(1.0*\i-3.0,0)$);
\draw[black!70, line width=1pt] (c1) -- (c5);
\foreach \i/\j in {1/2,2/3,3/4,4/5} \pic at ($(c\i)!0.5!(c\j)+(0,0.13)$) {tick};
\foreach \i in {2,3,4} \node[cvert, fill=black!22, draw=black!40, line width=0.4pt] at (c\i) {};
\foreach \i in {1,5} \node[cvert, fill=figblue!12, draw=figblue, line width=1pt] at (c\i) {};
\node[stmt] (s1) at ($(9.15,0 |- ch)+(0,-0.58)$) {Chain of accepted edges, total length $\le L_*$};
\node[stmt] (s2) at ($(s1.west)+(0,-0.45)$) {Accurate ground states at both endpoints};
\node[stmt] (s3) at ($(s2.west)+(0,-0.45)$) {Intermediate vertices may be unresolved};
\draw[black!70, line width=1pt] (8.60,0 |- s1.center) -- (9.00,0 |- s1.center);
\pic at ($(8.80,0 |- s1.center)+(0,0.12)$) {tick};
\node[ckey, fill=figblue!12, draw=figblue, line width=0.8pt] at (8.80,0 |- s2.center) {};
\node[ckey, fill=black!22, draw=black!40, line width=0.4pt] at (8.80,0 |- s3.center) {};
\coordinate (bandtop) at ($(s3.south)+(0,-0.11)$);
\node[bline, font=\normalsize\bfseries, anchor=north] (t3d) at ($(12.525,0 |- bandtop)+(0,-0.11)$) {The endpoint ground states are\\not in different phases.};
\coordinate (b3b) at ($(t3d.south)+(0,-0.11)$);
\begin{scope}[on background layer]
\node[bbox, outer sep=0pt, fill=figpurple!10, fit=(w3)(h3)(s1)(s3)(t3d)(b3b)] (box3) {};
\begin{scope}
\clip[rounded corners=4pt] (box3.south west) rectangle (box3.north east);
\fill[figpurple!25] (box3.south west) rectangle (box3.east |- bandtop);
\end{scope}
\draw[figpurple, line width=0.6pt, rounded corners=4pt] (box3.south west) rectangle (box3.north east);
\end{scope}
\coordinate (ax) at (11.23, 0);
\draw[figarrow] (box2.south -| ax) -- (box3.north -| ax);
\node[figtext, anchor=west, inner sep=0pt] at ($(box2.south -| ax)!0.5!(box3.north -| ax)+(0.12,0)$) {accepted edges};
\node[figtitle, anchor=north west, align=left, inner sep=0pt] at (-0.72, 2.0) {(a) Exploration\\\phantom{(a) }round 7};
\begin{scope}[gridxf]
\foreach \r/\c/\n in {1/1/2,1/2/3,1/3/4,1/4/5,1/5/6, 2/1/1,2/2/2,2/3/3,2/4/4,2/5/5,2/6/6, 3/2/1, 4/1/1,4/2/2, 5/1/2,5/2/3,5/3/4,5/4/5,5/5/6}
  \node[cellnum] at (\c-0.5, -\r+0.5) {\n};
\node[cell, fill=figblue, text=white, font=\bfseries] at (0.5,-2.5) {0};
\foreach \r/\c in {1/6, 5/6}
  \node[cellnew] at (\c-0.5, -\r+0.5) {7};
\foreach \i in {0,...,8} \draw[gridline] (\i,0) -- (\i,-5);
\foreach \j in {0,...,5} \draw[gridline] (0,-\j) -- (8,-\j);
\fill[black!22] (2,-2) rectangle (5,-4);
\draw[gridline] (2,-2) rectangle (5,-4);
\draw[figorange, dash pattern=on 3pt off 2pt, line width=0.7pt] (4.0,-2) .. controls (4.15,-2.55) and (5.15,-2.95) .. (4.85,-4);
\node[bline] at (3.17,-2.435) {$\gamma<\gamma_{\mathrm{lb}}$\\unresolved};
\node[font=\normalsize, text=figorange, align=center, inner sep=0pt] at (3.17,-3.391) {possible\\discontinuous\\transition};
\draw[figblue, line width=1.2pt] (0,0) rectangle (6,-5);
\draw[figblue, line width=1.2pt] (2,-2) rectangle (5,-4);
\node[cell, fill=white, draw=figorange, line width=1.2pt, text=figorange, font=\bfseries] (c36) at (5.5,-2.5) {7};
\draw[figarrow, figgreen, line width=0.9pt] (5.2,-1.36) -- (5.2,-1.98);
\node[ulab, text=figgreen!70!black] at (5.2,-1.20) {$y$};
\node[ulab, text=figgreen!70!black] at (5.2,-2.16) {$x$};
\draw[figarrow, black!70] (4.45,-2.5) -- (5,-2.5);
\draw[figarrow, black!70] (6.6,-2.5) -- (6,-2.5);
\draw[figarrow, black!70] (5.2,-3.55) -- (5.2,-3);
\node[figtext, anchor=east, inner sep=0pt] at (-0.12, -2.5) {$x_{\mathrm{ref}}$};
\node[figtext, text=figblue, anchor=west, inner sep=0pt] at (0.60, 0.37) {connected component: $\gamma\ge\gamma_{\mathrm{lb}}$};
\draw[figarrow, figblue] (0.50, 0.37) to[out=180, in=90] (0.11, 0.045);
\end{scope}
\draw[zoom] (c36.north east) -- (box1.north west);
\draw[zoom] (c36.south east) -- (box1.south west);
\node[lsq, fill=figblue, text=white, font=\bfseries] at ($(0,0 |- box3.south)+(0,1.525)$) {0};
\node[figtext, anchor=west] at ($(0.95,0 |- box3.south)+(0,1.525)$) {reference point};
\node[lsq, fill=figamber!25, text=figorange, font=\bfseries] at ($(4.5,0 |- box3.south)+(0,1.525)$) {7};
\node[figtext, anchor=west] at ($(5.45,0 |- box3.south)+(0,1.525)$) {reached in\\round 7};
\node[lsq, fill=figblue!12] at ($(0,0 |- box3.south)+(0,0.375)$) {$k$};
\node[figtext, anchor=west] at ($(0.95,0 |- box3.south)+(0,0.375)$) {reached by\\round $k$};
\node[lsq, fill=black!22] at ($(4.5,0 |- box3.south)+(0,0.375)$) {};
\node[figtext, anchor=west] at ($(5.45,0 |- box3.south)+(0,0.375)$) {$\gamma<\gamma_{\mathrm{lb}}$\\(unresolved)};
\end{tikzpicture}
\caption{\textbf{Dynamic-programming exploration and phase connectivity.}
(a) Starting from a successful reference point $x_{\mathrm{ref}}$,
accurate candidate paths advance one edge per round along gapped grid paths.
Numbers indicate the earliest round in which an accurate candidate path
arrives along a gapped grid path. The gray region is unresolved.
(b) The path stored at a neighbor $y$ of a vertex $x$ is extended
to $x$ by quasi-adiabatic transport with each of two cutoffs, $\gamma_{\mathrm{lb}}/2$ and the
larger $(1-\epsilon_{\mathrm{thresh}})\gamma_{\mathrm{thresh}}$.
For this path, the largest cutoff that passes the energy test is kept.
In the final round, neighboring states stored in the preceding round
are compared after transport at
$(1-\epsilon_{\mathrm{thresh}})\gamma_{\mathrm{thresh}}$.
A chain of accepted edges whose total geometric length is at most $L_*$ certifies that
the endpoint ground states, if accurately represented, are not in different
phases, even when intermediate vertices are unresolved.}
\label{fig:dp-exploration}
\end{figure}

Dynamic programming avoids trying grid paths one by one. As in a shortest-path computation, each vertex stores a single candidate path rather than every path that reaches it.
One candidate path suffices because an accurate state stays nearly as accurate after transport across a gapped edge, whatever path produced it.
In each round, every vertex uses the energy test (Subsubsection~\ref{app:exploration-certify}) to compare its candidate path with those extended from its neighbors. A reference point is admitted unless its reference state fails the purity test (Subsection~\ref{subsec:energy-rank-estimation}). At an admitted reference point, the comparison also includes the reference state, as the candidate state of the empty path.
At a gapped vertex, the energy test keeps an accurate candidate path whenever one is present (Lemma~\ref{lem:selection}).
Accurate candidate paths thus advance one edge per round along every gapped grid path from a successful reference point (Figure~\ref{fig:dp-exploration}(a)); after as many rounds as there are grid points, every vertex on such a path holds one (Corollary~\ref{cor:large-gap}).

Our quantum algorithm (Algorithm~\ref{alg:overall-dp}) is not told the ground-state degeneracy $r(x)$, so it runs the dynamic program of Subsection~\ref{subsec:prepare-ground-states} once for each rank $r=1,\ldots,r_{\max}$.
At every grid point $x$, the rank-$r$ run stores a path $V_r[x]$ and an energy threshold $\mu_r[x]$ that characterizes the upper end of the lowest $r$ levels.
The selected rank of an input point, which estimates its ground-state degeneracy, is the largest $r$ whose energy threshold at that point is close to the smallest one there (Figure~\ref{fig:ranks-links}(a)).

In its final round, each run also tests the grid edges (Subsection~\ref{subsec:phase-guarantee}). An edge is accepted if the states stored at its ends in the preceding round agree after transport (Figure~\ref{fig:dp-exploration}(b)).
Chains of accepted edges between input points of the same selected rank give the links (Figure~\ref{fig:ranks-links}(b)).
The algorithm returns the links and their clusters (Figure~\ref{fig:ranks-links}(c)), which satisfy the full guarantee (Subsection~\ref{app:main-algorithm-proof}).

Transport across an edge uses the quasi-adiabatic propagator $U_\Delta$ of Definition~\ref{def:quasi-adiabatic-evolution} with one of two cutoffs. The band gap of the lowest $r$ levels is $\gamma_r=E_r-E_{r-1}$. A continuous path with band gap at least $\gamma_{\mathrm{lb}}$ has a grid approximation with band gap at least $3\gamma_{\mathrm{lb}}/4$ (Lemma~\ref{lem:gapped-grid-path}), and the factor of $1/2$ keeps the cutoff $\gamma_{\mathrm{lb}}/2$ below this gap. State preparation needs only the cutoff $\gamma_{\mathrm{lb}}/2$, while testing phase connectivity uses the constant cutoff $(1-\epsilon_{\mathrm{thresh}})\gamma_{\mathrm{thresh}}$. Both are tried on every edge, because a larger cutoff can give a cheaper transport. A passing candidate path of least cost is kept (Eq.~\eqref{eq:path-cost}). Ordinary adiabatic sweeps can replace these propagators (Subsection~\ref{subsec:dp-sweeps}).

\subsection{Efficient state preparation using dynamic programming}
\label{subsec:prepare-ground-states}

Fix $r\le r_{\max}$.
A rank-$r$ reference point $x_{\mathrm{ref}}$ is a point in $S$ of ground-state degeneracy $r$ whose preparation succeeds (Assumption~\ref{assumption-reference-state}).
A grid point is reachable at rank~$r$ if a grid path from a rank-$r$ reference point to it has band gap at least $3\gamma_{\mathrm{lb}}/4$ throughout (Eq.~\eqref{eq:reachable}).
For every reachable grid point, we aim to store a path whose state approximates the sector state $\rho_r(x)=P_r(x)/r$ of Definition~\ref{def:spectral-sector} with $d=r$, which is defined when $E_r(x)>E_{r-1}(x)$.
The connected component of the resolved region containing $x_{\mathrm{ref}}$ has ground-state degeneracy $r$ (Appendix~\ref{sec:appendix-A}).
Hence, on this component,
\begin{equation}
    P_r(x)=P_{\mathcal G}(x),
    \qquad
    \rho_r(x)=\rho_{\mathcal G}(x).
    \label{eq:sector-equals-gsm}
\end{equation}

Given a reference point $x_s\in S$ with reference state $C[x_s]$, a path $V$ from $x_s$ to $x$ is called a \emph{candidate path} at $x$.
We also write $V$ for its transport circuit. The candidate state is $\sigma_V=V C[x_s]V^\dagger$, as in Eq.~\eqref{eq:response-candidate-state}.
At every vertex where $E_r(x)>E_{r-1}(x)$, define
\begin{equation}
    \lambda_x(\sigma):=1-\operatorname{Tr}[P_r(x)\sigma],
    \qquad
    \lambda_x(V):=\lambda_x(\sigma_V)
    \label{eq:candidate-leakage}
\end{equation}
as the population outside the lowest $r$ levels.
We call $\lambda_x(V)$ its \emph{leakage}.
To control how leakage accumulates over rounds, the energy test compares candidate states through their overlaps with a sharp energy filter of the whole Hamiltonian $H(x)$.
The test applies the filter coherently, through a block encoding of $H(x)$ (see the proof of Lemma~\ref{lem:implementation-complexity}), so the dynamic program requires a fault-tolerant quantum computer.

Only the classical description of $V$ is stored.
Each observable $O$ is measured as $V^\dagger OV$ in $C[x_s]$.
We write $V[x]$ and $\mu[x]$ for the stored path and energy threshold at $x$, suppressing rank.

\subsubsection{Dynamic programming and its preparation guarantee}
\label{app:exploration-exact}

Let $V^{(k)}[x]$ denote the stored path at $x$ after $k$ rounds.
A reference point is admitted to the rank-$r$ run if the purity test of Subsection~\ref{subsec:energy-rank-estimation} admits it at rank $r$.
Initially, each admitted reference point stores the empty path, and all other stored paths are \textsc{Void}.
Each round simultaneously updates all vertices.

Let $\mathcal D$ be the set of cutoffs tested on each edge. The candidate pool is
\begin{equation}
\begin{aligned}
\mathcal C^{(k)}(x):={}&
\left\{V^{(k-1)}[x]:
V^{(k-1)}[x]\ne\textsc{Void}\right\}\\
&\cup\left\{\text{the empty path at }x:
C[x]\text{ is used by the run}\right\}\\
&\cup\left\{U_\Delta(y\to x)V^{(k-1)}[y]:
y\sim x,\ V^{(k-1)}[y]\ne\textsc{Void},\
\Delta\in\mathcal D\right\},
\end{aligned}
\label{eq:pool}
\end{equation}
where $y\sim x$ denotes adjacency.
If $x$ is an admitted reference point, $C[x]$ is used by the run.
Every $U_\Delta(y\to x)$ in the pool is implemented by an edge circuit.
Each vertex with a non-empty pool then stores one candidate path from its pool, using the energy test of Subsubsection~\ref{app:exploration-certify}.

We set
\begin{equation}
\mathcal D=\left\{
\frac{\gamma_{\mathrm{lb}}}{2},
(1-\epsilon_{\mathrm{thresh}})\gamma_{\mathrm{thresh}}
\right\}.
\label{eq:main-cutoffs}
\end{equation}
The larger cutoff $(1-\epsilon_{\mathrm{thresh}})\gamma_{\mathrm{thresh}}$ can give more efficient circuits (Subsubsection~\ref{subsubsec:coarse-cutoff}).

Theorem~\ref{thm:finite-precision} below derives the preparation guarantee from the following properties of these updates.

\emph{Transport across one edge.} An edge $e:u\to v$ has this property if some cutoff $\Delta\in\mathcal D$ gives an edge circuit $\widetilde U_\Delta(e)$ with
\[
    \lambda_v(\widetilde U_\Delta(e)\sigma\widetilde U_\Delta(e)^\dagger)
    \le\lambda_u(\sigma)+2\epsilon_U
\]
for every state $\sigma$.
Here $\epsilon_U$ is a small error per edge, tied to the accuracy of the edge circuits in Eq.~\eqref{eq:unitary-error}.
The bound holds for imperfect incoming states, so a candidate state produced in earlier rounds can be transported further.
The implementation of $\widetilde U_\Delta(e)$ is left unspecified here.

\emph{Selection at a vertex.}
A vertex $x$ has this property in a round with a known budget $\epsilon'$ if, whenever the pool at $x$ contains a candidate path with leakage at most $\epsilon'$, the stored path has leakage at most $\epsilon'+2\epsilon_p$.
Here $\epsilon_p$ sets the accuracy of the overlap and purity estimates.
Lemma~\ref{lem:selection} below gives conditions under which selection adds at most $2\epsilon_p$ leakage.

Starting from a reference state with leakage at most $\epsilon_{\mathrm{ref}}$, the \emph{leakage budget} $\theta_k$ of round $k$ is
\begin{equation}
    \theta_0:=\epsilon_{\mathrm{ref}},
    \qquad
    \theta_k:=\theta_{k-1}+2\epsilon_U+2\epsilon_p
             =\epsilon_{\mathrm{ref}}
              +2k(\epsilon_U+\epsilon_p).
    \label{eq:round-budget}
\end{equation}
In round $k$, an accurate incoming candidate path has leakage at most $\epsilon'=\theta_{k-1}+2\epsilon_U$.
After $N_{\mathrm{grid}}$ rounds, the total budget is
\begin{equation}
    \epsilon_{\mathrm{DP}}:=\theta_{N_{\mathrm{grid}}}
    =\epsilon_{\mathrm{ref}}
     +2N_{\mathrm{grid}}(\epsilon_U+\epsilon_p).
    \label{eq:dp-budget}
\end{equation}

\begin{theorem}[State preparation by dynamic programming]
\label{thm:finite-precision}
Fix a rank $r$, form the candidate pools by Eq.~\eqref{eq:pool}, and let round $k$ use the budget $\epsilon'=\theta_{k-1}+2\epsilon_U$.
Let $y_0,y_1,\ldots,y_\ell=x$ be a grid path with $E_r>E_{r-1}$ at its vertices. Suppose that the run admits $y_0$ and that the reference state at $y_0$ has leakage at most $\epsilon_{\mathrm{ref}}$. Suppose also that transport across one edge holds on each edge $y_{j-1}\to y_j$, and that selection at a vertex holds at each $y_j$ in every round.
Then, after $k$ rounds, $\lambda_{y_j}(V^{(k)}[y_j])\le\theta_k$ for every $j\le\min\{k,\ell\}$.
In particular, if $\ell<N_{\mathrm{grid}}$, then $\lambda_x(V^{(N_{\mathrm{grid}}-1)}[x])\le\theta_{N_{\mathrm{grid}}-1}$ and $\lambda_x(V[x])\le\epsilon_{\mathrm{DP}}$.
The dynamic program need not know the path or which preparations succeeded.
\end{theorem}

\begin{proof}
By induction on $k$, for every $j\le\min\{k,\ell\}$,
\begin{equation}
    \lambda_{y_j}\bigl(V^{(k)}[y_j]\bigr)\le\theta_k.
    \label{eq:induction}
\end{equation}
For $k=0$, the empty path stored at $y_0$ has leakage at most $\epsilon_{\mathrm{ref}}=\theta_0$.

Assume Eq.~\eqref{eq:induction} after $k-1$ rounds, and fix $j\le\min\{k,\ell\}$.
If $j<k$, the pool at $y_j$ still contains the earlier stored path with leakage at most $\theta_{k-1}$.
If $j=k$, consider the next edge $e_k:y_{k-1}\to y_k$ and a cutoff $\Delta_k\in\mathcal D$ for which transport across $e_k$ holds. The transported candidate state has leakage at most $\theta_{k-1}+2\epsilon_U$.
In either case, the pool at $y_j$ contains a candidate path with leakage at most $\epsilon'=\theta_{k-1}+2\epsilon_U$.
Selection at $y_j$ then implies Eq.~\eqref{eq:induction}, since
\[
    \theta_{k-1}+2\epsilon_U+2\epsilon_p
    =\theta_k.
\]
Taking $j=\ell$ with $k=N_{\mathrm{grid}}-1$ and $k=N_{\mathrm{grid}}$ gives the last claims.
\end{proof}

The stored path $V[x]$ need not follow the grid path of Theorem~\ref{thm:finite-precision}, because selection may keep a different candidate path, possibly from another reference point.
The theorem bounds the leakage of $V[x]$, and the edge tests of Subsection~\ref{subsec:phase-guarantee} use its bound on $V^{(N_{\mathrm{grid}}-1)}[x]$.
Corollary~\ref{cor:large-gap} gives conditions under which the hypotheses of the theorem hold along gapped grid paths.

\subsubsection{Selecting an accurate candidate path}
\label{app:exploration-certify}

To implement the selection at a vertex used in Theorem~\ref{thm:finite-precision}, we define, for an energy threshold $\nu$, a smooth step function $f_\nu$ that is approximately one below $\nu$ and approximately zero above $\nu$. We can measure $\operatorname{Tr}[f_\nu(H(x))\sigma_V]$ without diagonalizing $H(x)$. If the transition in $f_\nu$ lies between $E_{r-1}(x)$ and $E_r(x)$, then $f_\nu(H(x))$ is approximately $P_r(x)$. Since the energies are unknown, the \emph{threshold scan} increases $\nu$ until some candidate state has overlap close to one.

The scan works only if no candidate state has overlap close to one before the lowest $r$ levels are included, which requires every candidate state in a rank-$r$ run to have purity consistent with $1/r$.
Every candidate state in the pool is a unitary transport of an admitted reference state, so it satisfies the purity condition for rank $r$ (Subsection~\ref{subsec:energy-rank-estimation}).
Below, the purity condition always means this one.
If $2\epsilon_{\mathrm{ref}}+\epsilon_p\le1/(2r^2)$, the counting bound of Eq.~\eqref{eq:purity-counting} shows that no $r-1$ levels can have more than $1-1/(4r)$ of its weight.
Set
\begin{equation}
h_E:=\frac{\gamma_{\mathrm{lb}}}{16},
\label{eq:threshold-spacing}
\end{equation}
and consider energy thresholds $\nu$ given by the multiples of $h_E$ in $[-B,B]$.
Here $B=O(n)$ is a known bound from the local terms of $H$ that satisfies $B\ge\sup_x\|H(x)\|+5h_E$.
For each $\nu$, we choose a smooth step function $f_\nu$ that satisfies
\begin{equation}
\begin{aligned}
&0\le f_\nu(E)\le1,\\
&|f_\nu(E)-1|\le\epsilon_p/2
&&\text{for }E\le\nu-h_E,\\
&|f_\nu(E)|\le\epsilon_p/2
&&\text{for }E\ge\nu+h_E
\end{aligned}
\label{eq:smooth-step}
\end{equation}
on $[-B,B]$.
For each $V$ and $\nu$, we measure the filtered overlap
\begin{equation}
\operatorname{Tr}[f_\nu(H(x))\sigma_V]
=\operatorname{Tr}\left[C[x_s]V^\dagger f_\nu(H(x))V\right],
\label{eq:overlap-pullback}
\end{equation}
to obtain an estimate $\widehat p_\nu(V)$ with
\begin{equation}
\left|\widehat p_\nu(V)
-\operatorname{Tr}[f_\nu(H(x))\sigma_V]\right|
\le\frac{\epsilon_p}{2}.
\label{eq:overlap-estimate}
\end{equation}

\begin{corollary}[Overlap below an energy threshold]
\label{cor:smooth-step-overlap}
For every state $\sigma$ and smooth step satisfying
Eq.~\eqref{eq:smooth-step},
\begin{enumerate}
\item[(a)] if $\nu\le E_{r-1}-h_E$, then
$\operatorname{Tr}[f_\nu(H)\sigma]
\le\sqrt{(r-1)\operatorname{Tr}[\sigma^2]}+\epsilon_p/2$.
\end{enumerate}
When $E_r>E_{r-1}$, write
$\lambda=1-\operatorname{Tr}[P_r\sigma]$. Then also
\begin{enumerate}
\item[(b)] if $\nu\ge E_{r-1}+h_E$, then
$\operatorname{Tr}[f_\nu(H)\sigma]\ge1-\lambda-\epsilon_p/2$;
\item[(c)] if $\nu\le E_r-h_E$, then
$\operatorname{Tr}[f_\nu(H)\sigma]\le1-\lambda+\epsilon_p/2$.
\end{enumerate}
\end{corollary}

\begin{proof}
(a) Let $\Pi$ project onto some $r-1$ lowest eigenstates; then $f_\nu(H)\le\Pi+(\epsilon_p/2)I$. The result follows from Eq.~\eqref{eq:purity-counting}.
(b) $f_\nu(H)\ge(1-\epsilon_p/2)P_r$.
(c) $f_\nu(H)\le P_r+(\epsilon_p/2)(I-P_r)$.
Taking expectation values gives the inequalities.
\end{proof}

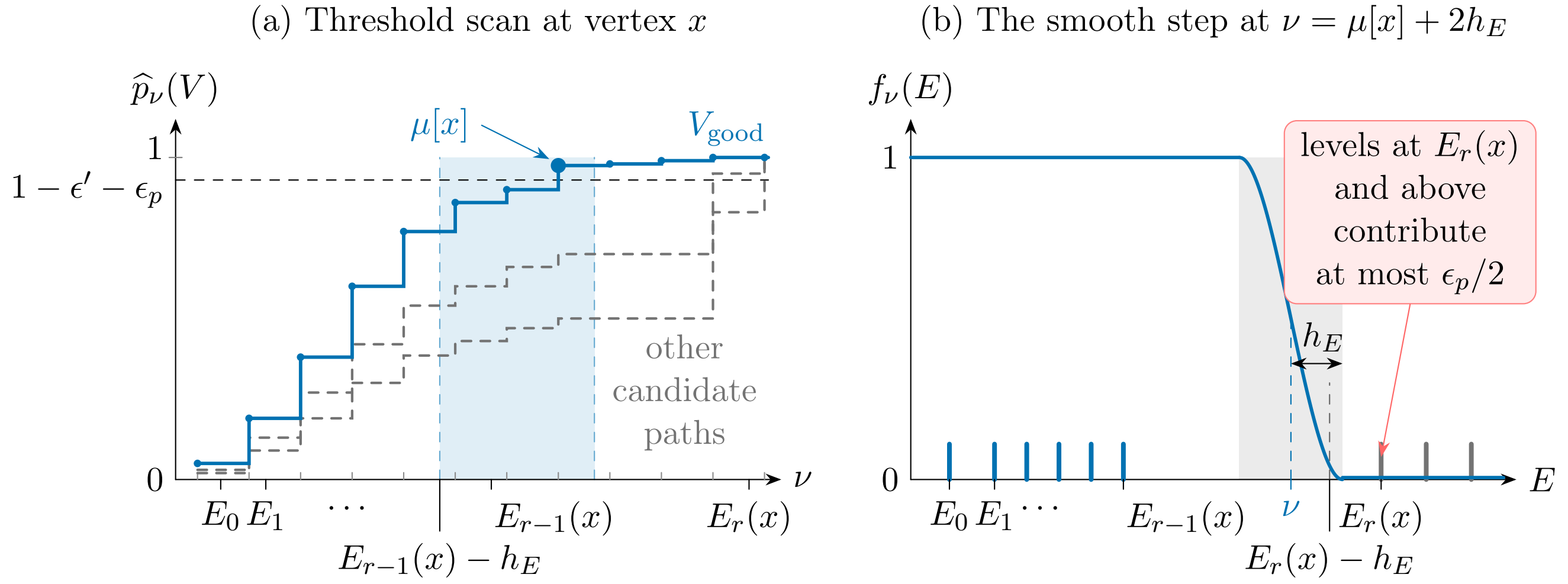
\begin{figure}[!t]
\centering
\begin{tikzpicture}[
    x=0.72cm,
    y=3.6cm,
    >={Stealth[length=5pt]},
    line cap=round
]
\definecolor{selblue}{HTML}{0072B2}
\node at (4.7,1.42) {(a) Threshold scan at vertex $x$};
\fill[selblue!12] (4.1,0) rectangle (6.5,1.0);
\draw[selblue!60, dashed] (4.1,0) -- (4.1,1.0);
\draw[selblue!60, dashed] (6.5,0) -- (6.5,1.0);
\draw[->] (0,0) -- (9.4,0) node[right] {$\nu$};
\draw[->] (0,0) -- (0,1.12)
    node[above] {$\widehat p_\nu(V)$};
\node[left, yshift=3pt] at (0,1) {$1$};
\node[left] at (0,0) {$0$};
\draw[black!50] (-0.1,1) -- (0.1,1);
\foreach \x in {
    0.34,1.14,1.94,2.74,3.54,4.34,
    5.14,5.94,6.74,7.54,8.34,9.14
}
  \draw[black!45] (\x,0) -- ++(0,2.2pt);
\draw[dashed] (0,0.93) -- (9.2,0.93);
\node[left, yshift=-3pt] at (0,0.93)
    {$1-\epsilon'-\epsilon_p$};
\draw[black!55, dashed, line width=0.8pt]
  (0.34,0.03) -- (1.14,0.03)
  -- (1.14,0.13) -- (1.94,0.13)
  -- (1.94,0.27) -- (2.74,0.27)
  -- (2.74,0.42) -- (3.54,0.42)
  -- (3.54,0.54) -- (4.34,0.54)
  -- (4.34,0.60) -- (5.14,0.60)
  -- (5.14,0.66) -- (5.94,0.66)
  -- (5.94,0.70) -- (7.54,0.70)
  -- (8.34,0.70)
  -- (8.34,0.95) -- (9.14,0.95)
  -- (9.14,1.0) -- (9.2,1.0);
\draw[black!55, dashed, line width=0.8pt]
  (0.34,0.02) -- (1.14,0.02)
  -- (1.14,0.09) -- (1.94,0.09)
  -- (1.94,0.19) -- (2.74,0.19)
  -- (2.74,0.30) -- (3.54,0.30)
  -- (3.54,0.385) -- (4.34,0.385)
  -- (4.34,0.43) -- (5.14,0.43)
  -- (5.14,0.47) -- (5.94,0.47)
  -- (5.94,0.50) -- (7.54,0.50)
  -- (8.34,0.50)
  -- (8.34,0.83) -- (9.14,0.83)
  -- (9.14,1.0) -- (9.2,1.0);
\node[black!55, align=center] at (7.9,0.27)
    {other\\candidate\\paths};
\draw[selblue, line width=1.1pt]
  (0.34,0.05) -- (1.14,0.05)
  -- (1.14,0.19) -- (1.94,0.19)
  -- (1.94,0.38) -- (2.74,0.38)
  -- (2.74,0.60) -- (3.54,0.60)
  -- (3.54,0.77) -- (4.34,0.77)
  -- (4.34,0.86) -- (5.14,0.86)
  -- (5.14,0.90) -- (5.94,0.90)
  -- (5.94,0.975) -- (6.74,0.975)
  -- (6.74,0.98) -- (7.54,0.98)
  -- (7.54,0.99) -- (8.34,0.99)
  -- (8.34,1.0) -- (9.2,1.0);
\foreach \x/\y in {
    0.34/0.05,1.14/0.19,1.94/0.38,2.74/0.60,
    3.54/0.77,4.34/0.86,5.14/0.90,6.74/0.98,
    7.54/0.99,8.34/1.0,9.14/1.0
}
  \fill[selblue] (\x,\y) circle (1.2pt);
\fill[selblue] (5.94,0.975) circle (2.4pt);
\draw[selblue, ->] (4.75,1.1)
    node[left] {$\mu[x]$} -- (5.8,0.995);
\node[selblue, above] at (8.55,1.0)
    {$V_{\mathrm{good}}$};
\foreach \x/\l in {
    0.7/{$E_0$},1.4/{$E_1$},8.9/{$E_r(x)$}
}
  \draw (\x,0) -- (\x,-0.03) node[below] {\l};
\draw (4.9,0) -- (4.9,-0.03)
    node[anchor=north west, xshift=-3pt] {$E_{r-1}(x)$};
\node[below] at (2.7,-0.03) {$\cdots$};
\draw (4.1,0) -- (4.1,-0.16)
    node[below] {$E_{r-1}(x)-h_E$};
\end{tikzpicture}\hfill
\begin{tikzpicture}[
    x=0.72cm,
    y=3.6cm,
    >={Stealth[length=5pt]},
    line cap=round
]
\definecolor{selblue}{HTML}{0072B2}
\node at (4.7,1.42)
    {(b) The smooth step at $\nu=\mu[x]+2h_E$};
\fill[black!8] (5.1,0) rectangle (6.7,1.0);
\draw[->] (0,0) -- (9.4,0) node[right] {$E$};
\draw[->] (0,0) -- (0,1.12)
    node[above] {$f_{\nu}(E)$};
\node[left] at (0,1) {$1$};
\node[left] at (0,0) {$0$};
\draw[black!50, dashed] (0,1) -- (5.1,1);
\foreach \x in {0.6,1.3,1.8,2.3,2.8,3.3}
    \draw[selblue, line width=1.5pt]
        (\x,0) -- (\x,0.11);
\foreach \x in {7.3,8.0,8.7}
    \draw[black!55, line width=1.5pt]
        (\x,0) -- (\x,0.11);
\draw[selblue, line width=1.1pt]
    (0,1) -- (5.1,1);
\draw[
    selblue,
    line width=1.1pt,
    domain=5.1:6.7,
    samples=40,
    smooth,
    variable=\x
]
  plot (\x, {
      1 - 3*((\x-5.1)/1.6)^2
        + 2*((\x-5.1)/1.6)^3
  });
\draw[selblue, line width=1.1pt]
    (6.7,0.006) -- (9.2,0.006);
\draw[<->] (5.9,0.36) -- (6.7,0.36);
\node[above] at (6.4,0.36) {$h_E$};
\draw[selblue, dashed] (5.9,0) -- (5.9,0.5);
\draw[black!60, dashed] (6.5,0) -- (6.5,0.3);
\node[
    draw=red!50,
    fill=red!8,
    rounded corners,
    align=center,
    text width=2.6cm,
    inner sep=3pt
] (call) at (7.75,0.83)
  {levels at $E_r(x)$ and above contribute at most $\epsilon_p/2$};
\draw[red!60, ->] (call.south) -- (7.3,0.07);
\foreach \x/\l in {
    0.6/{$E_0$},1.3/{$E_1$},7.3/{$E_r(x)$}
}
  \draw (\x,0) -- (\x,-0.03) node[below] {\l};
\draw (3.3,0) -- (3.3,-0.03)
    node[anchor=north west, xshift=-3pt] {$E_{r-1}(x)$};
\node[below] at (2.05,-0.03) {$\cdots$};
\draw[selblue] (5.9,0) -- (5.9,-0.03)
    node[below] {$\nu$};
\draw (6.5,0) -- (6.5,-0.16)
    node[below] {$E_r(x)-h_E$};
\end{tikzpicture}
\caption{Selecting an accurate candidate path
(Lemma~\ref{lem:selection}).
(a)~Estimated overlaps $\widehat p_\nu(V)$ of the candidate states at a
vertex $x$ for energy thresholds $\nu$ spaced by $h_E$ (small ticks),
examined in increasing order. By the purity condition and the counting
bound, all estimates remain below the target
$1-\epsilon'-\epsilon_p$ for $\nu\le E_{r-1}(x)-h_E$, while an
accurate candidate path $V_{\mathrm{good}}$ reaches the target for some
$\nu<E_{r-1}(x)+2h_E$. Hence the first energy threshold $\mu[x]$ at which the target is reached
lies in the shaded window of Eq.~\eqref{eq:sector-top-bracket}.
(b)~The smooth step $f_\nu$ at $\nu=\mu[x]+2h_E$, where the candidate states
are compared, drawn over the spectrum of $H(x)$ (blue bars: the lowest
$r$ levels; gray bars: higher levels). By Eq.~\eqref{eq:sector-top-bracket}
and the gap condition of Lemma~\ref{lem:selection},
$E_{r-1}(x)+h_E<\nu\le E_r(x)-h_E$. The filter is therefore within
$\epsilon_p/2$ of one on the lowest $r$ levels and at most $\epsilon_p/2$
on every level from $E_r(x)$ upward.}
\label{fig:selection}
\end{figure}

Suppose an unknown candidate path $V_{\mathrm{good}}$ satisfies $\lambda_x(V_{\mathrm{good}})\le\epsilon'$ for a known $\epsilon'$.
Let $\mu[x]$ be the smallest energy threshold $\nu$ at which some candidate path $V$ satisfies
\begin{equation}
\widehat p_\nu(V)\ge1-\epsilon'-\epsilon_p.
\label{eq:target-overlap}
\end{equation}
At this $\mu[x]$, however, the transition region of $f_{\mu[x]}$ might still overlap with $E_{r-1}(x)$, so we compare candidate states at $\mu[x]+2h_E$.
Under the gap condition of Lemma~\ref{lem:selection} below, the transition region then lies within the gap, so the filtered overlap controls the leakage (Figure~\ref{fig:selection}).
The scan terminates for any non-empty pool: for large $\nu$ every estimated overlap is at least $1-\epsilon_p$, and $[-B,B]$ contains $\mu[x]+2h_E$.
The threshold scan and the comparison at $\mu[x]+2h_E$ form the \emph{energy test}.
A candidate path \emph{passes} if its estimate at $\mu[x]+2h_E$ satisfies Eq.~\eqref{eq:target-overlap}; if some candidate path passes, the vertex stores one of least cost (Eq.~\eqref{eq:path-cost}).

\begin{lemma}[Selecting an accurate candidate path and estimating $E_{r-1}(x)$]
\label{lem:selection}
Suppose that $E_r(x)-E_{r-1}(x)\ge5h_E$.
Suppose that every candidate state in the pool satisfies the purity condition, and that the pool contains a candidate path $V_{\mathrm{good}}$ with $\lambda_x(V_{\mathrm{good}})\le\epsilon'$.
Suppose that $\epsilon'+2\epsilon_p<1/(4r)$ and $2\epsilon_{\mathrm{ref}}+\epsilon_p\le1/(2r^2)$, and that the estimates satisfy Eq.~\eqref{eq:overlap-estimate}.
Let $\mu[x]$ be the first energy threshold for which some candidate path satisfies Eq.~\eqref{eq:target-overlap}.
Then $\mu[x]$ satisfies
\begin{equation}
E_{r-1}(x)-h_E<\mu[x]<E_{r-1}(x)+2h_E.
\label{eq:sector-top-bracket}
\end{equation}
At $\nu=\mu[x]+2h_E$, every candidate path satisfies
\begin{equation}
\lambda_x(V)\le\epsilon'
\quad\Longrightarrow\quad
\widehat p_{\mu[x]+2h_E}(V)\ge1-\epsilon'-\epsilon_p
\quad\Longrightarrow\quad
\lambda_x(V)\le\epsilon'+2\epsilon_p.
\label{eq:shifted-threshold-selection}
\end{equation}
The lower bound of Eq.~\eqref{eq:sector-top-bracket} does not require an accurate candidate path or a positive gap above the lowest $r$ levels.
\end{lemma}

\begin{proof}
For the upper bound, let $\bar\nu$ denote the least energy threshold with $\bar\nu\ge E_{r-1}(x)+h_E$. Such $\bar\nu$ exists and satisfies $\bar\nu<E_{r-1}(x)+2h_E$. Corollary~\ref{cor:smooth-step-overlap}(b) and the estimation bound imply $\widehat p_{\bar\nu}(V_{\mathrm{good}}) \ge1-\epsilon'-\epsilon_p$, so $\mu[x]\le\bar\nu$.

For the lower bound, if $\nu\le E_{r-1}(x)-h_E$, then for every candidate path, Corollary~\ref{cor:smooth-step-overlap}(a) and the purity condition yield
\[
    \widehat p_\nu(V)
    \le1-\frac1{4r}+\epsilon_p
    <1-\epsilon'-\epsilon_p.
\]

Finally,
\[
    E_{r-1}(x)+h_E<\mu[x]+2h_E
    <E_{r-1}(x)+4h_E\le E_r(x)-h_E.
\]
Parts (b) and (c), with Eq.~\eqref{eq:overlap-estimate}, therefore imply that for all candidate paths $|\widehat p_{\mu[x]+2h_E}(V)-(1-\lambda_x(V))| \le\epsilon_p$, which gives both implications.
\end{proof}

Since $V_{\mathrm{good}}$ passes, the stored path $V[x]$ is a passing one, though not necessarily of least leakage. Eq.~\eqref{eq:shifted-threshold-selection} bounds the leakage of every passing candidate path, which gives the selection at a vertex used in Theorem~\ref{thm:finite-precision}:
\[
    \lambda_x(V_{\mathrm{good}})\le\epsilon'
    \quad\Longrightarrow\quad
    \lambda_x(V[x])\le\epsilon'+2\epsilon_p.
\]
In runs at incorrect ranks, the rank argument of Lemma~\ref{lem:rank-estimate} uses the lower bound on $\mu[x]$.

\subsubsection{Transport across one edge}
\label{subsubsec:coarse-cutoff}

We verify the transport across one edge assumed in Theorem~\ref{thm:finite-precision}, uniformly over imperfect incoming candidate states.
A cutoff $\Delta$ above the band gap $\gamma_r$ can give a more efficient circuit for transporting the state, at the cost of leakage. We bound this leakage to ensure that the candidate pool contains an accurate candidate path. For a grid edge $e:u\to v$, parameterized by $s\in[0,1]$ from $u$ to $v$, we abbreviate its quasi-adiabatic propagator $U_\Delta(u\to v)$ as $U_\Delta(e)$, with generator $D_\Delta(s)$.
If $\gamma_r>0$ throughout $e$, we define $a_\Delta(e)$ by
\begin{equation}
    a_\Delta(e):=\int_0^1
    \left\|(I-P_r(s))
    \bigl(\partial_sP_r(s)-i[D_\Delta(s),P_r(s)]\bigr)P_r(s)
    \right\|\,ds.
    \label{eq:edge-defect}
\end{equation}
The integrand is the norm of the diabatic coupling in Lemma~\ref{lemma:sparse-coupling-detailed}(a).
Only transitions out of the lowest $r$ levels with frequency in the window $[\gamma_r,\Delta)$ contribute to it (Eq.~\eqref{eq:Deff-bound-via-coupling}).
If $\gamma_r>\Delta$ throughout the edge, this window is always empty, so $a_\Delta(e)=0$.

\begin{lemma}[Transport of an imperfect candidate state]
\label{lem:imperfect-transport}
For every edge $e:u\to v$ with $\gamma_r>0$ throughout,
for every cutoff $\Delta>0$,
for every density matrix $\sigma$,
\begin{align}
    \|U_\Delta(e)^\dagger P_r(v)U_\Delta(e)-P_r(u)\|
    &\le a_\Delta(e),
    \label{eq:projector-transport-error}\\
    \lambda_v(U_\Delta(e)\sigma U_\Delta(e)^\dagger)
    &\le\lambda_u(\sigma)+a_\Delta(e).
    \label{eq:imperfect-transport}
\end{align}
\end{lemma}

\begin{proof}
For $U(s)=U_\Delta(s)$, we have
\[
    \frac{d}{ds}(U^\dagger P_rU)
    =U^\dagger\bigl(\partial_sP_r-i[D_\Delta,P_r]\bigr)U.
\]
The operator in parentheses is Hermitian and has only off-diagonal blocks with respect to $P_r$. Thus, its norm equals the integrand in Eq.~\eqref{eq:edge-defect}. Integration gives Eq.~\eqref{eq:projector-transport-error}. Taking the expectation value with respect to $\sigma$ gives Eq.~\eqref{eq:imperfect-transport}.
\end{proof}

Each quasi-adiabatic propagator $U_\Delta(e)$ is executed as an edge circuit $\widetilde U_\Delta(e)$ with accuracy
\begin{equation}
    \|\widetilde U_\Delta(e)-U_\Delta(e)\|
    \le\frac{\epsilon_U}{2}.
    \label{eq:unitary-error}
\end{equation}
Hence, a transported state changes by at most
\begin{equation}
    \|\widetilde U_\Delta\sigma\widetilde U_\Delta^\dagger
          -U_\Delta\sigma U_\Delta^\dagger\|_1
    \le2\|\widetilde U_\Delta-U_\Delta\|
    \le\epsilon_U.
    \label{eq:transport-error}
\end{equation}
Lemma~\ref{lem:imperfect-transport} then implies
\begin{equation}
    \lambda_v(\widetilde U_\Delta(e)\sigma
                    \widetilde U_\Delta(e)^\dagger)
    \le\lambda_u(\sigma)+a_\Delta(e)+\epsilon_U.
    \label{eq:implemented-leakage}
\end{equation}
Therefore, if $a_\Delta(e)\le\epsilon_U$, transport across a single edge introduces at most $2\epsilon_U$ leakage.
If another implementation of transport across an edge satisfies the same bound, it can replace the quasi-adiabatic one for state preparation. The adiabatic sweeps of Subsection~\ref{subsec:dp-sweeps} are one example.

We define the \emph{cost} of a path $V$ as
\begin{equation}
    c(V):=\sum_{e\in V}\frac1{\Delta_e},
    \label{eq:path-cost}
\end{equation}
where $\Delta_e$ is the cutoff used on edge $e$ of $V$.
The cost favors larger cutoffs.

\subsubsection{State preparation along gapped paths}
\label{subsubsec:finite-dp}
\label{app:exploration-algorithm}

To apply Theorem~\ref{thm:finite-precision} along gapped grid paths, we call a grid point $x$ \emph{reachable at rank $r$} if there exists a grid path from a rank-$r$ reference point $x_{\mathrm{ref}}$ to $x$ satisfying
\begin{equation}
    E_r(z)-E_{r-1}(z)
    \geq
    \frac{3\gamma_{\mathrm{lb}}}{4}
    \qquad
    \text{at every point $z$ of every edge and at $z=x_{\mathrm{ref}}$.}
    \label{eq:reachable}
\end{equation}

\begin{lemma}[Gapped grid path at fixed rank]
\label{lem:gapped-grid-path}
Let $x_{\mathrm{ref}}$ be a rank-$r$ reference point and let $x$ be a
grid point. If some continuous path from $x_{\mathrm{ref}}$ to $x$
satisfies $\gamma_r\geq\gamma_{\mathrm{lb}}$ at every point, then $x$ is
reachable at rank $r$ through a simple grid path of length at most
$N_{\mathrm{grid}}-1$. Here $N_{\mathrm{grid}}$ is the number of grid
points in Eq.~\eqref{eq:number-grid-points}. In particular, every grid point of the connected component
of the resolved region containing $x_{\mathrm{ref}}$ is reachable at rank $r$.
\end{lemma}

\begin{proof}
By uniform continuity, there exists a grid path within distance $h_{\mathrm{grid}}$ of the continuous path (Eq.~\eqref{eq:grid-approximation}). The band gap is Lipschitz with constant $L_\gamma$ (Eq.~\eqref{eq:grid-gap-lipschitz}). Along the grid path, the grid condition of Eq.~\eqref{eq:grid-large-gap} then gives $\gamma_r\geq\gamma_{\mathrm{lb}}-L_\gamma h_{\mathrm{grid}} \geq3\gamma_{\mathrm{lb}}/4$. Removing loops yields a simple path of length at most $N_{\mathrm{grid}}-1$. Since the connected component of the resolved region containing $x_{\mathrm{ref}}$ is path-connected and has ground-state degeneracy $r$ with $\gamma=\gamma_r\geq\gamma_{\mathrm{lb}}$, the first statement applies throughout this component.
\end{proof}

In round $k$, Eq.~\eqref{eq:target-overlap} gives the required overlap $p_k:=1-\theta_{k-1}-2\epsilon_U-\epsilon_p =1-\theta_k+\epsilon_p$.
Since $\epsilon'\le\epsilon_{\mathrm{DP}}$ and $r\le r_{\max}$, the conditions of Lemma~\ref{lem:selection} on $\epsilon'$, $\epsilon_p$ and $\epsilon_{\mathrm{ref}}$ hold in each round of every run if we require
\begin{equation}
    \epsilon_{\mathrm{DP}}+2\epsilon_p<\frac1{4r_{\max}},
    \qquad
    2\epsilon_{\mathrm{ref}}+\epsilon_p
    \le\frac1{2r_{\max}^2}.
    \label{eq:counting-condition}
\end{equation}

\emph{The update.}
At every admitted reference point $x$, the rank-$r$ run initializes $V^{(0)}[x]$ to the empty path and $\mu^{(0)}[x]$ to the first energy threshold with $\widehat p_\nu\ge1-\epsilon_{\mathrm{ref}}-\epsilon_p$. Elsewhere it sets $V^{(0)}[x]$ to \textsc{Void}.
In round $k=1,\ldots,N_{\mathrm{grid}}$, every vertex $x$ forms the pool in Eq.~\eqref{eq:pool}, using only the paths stored in the preceding round, the admitted reference states, and all cutoffs $\Delta\in\mathcal D$.
If the pool is empty, $V^{(k)}[x]$ is \textsc{Void}.
Otherwise, the run estimates the filtered overlaps and lets $\mu^{(k)}[x]$ be the first energy threshold at which a candidate path in the pool reaches $p_k$.
Among the candidate paths reaching $p_k$ at $\mu^{(k)}[x]+2h_E$, it sets $V^{(k)}[x]$ to one of smallest $c(V)$, breaking ties by the number of edges. If none passes, it retains a shortest candidate path reaching $p_k$ at $\mu^{(k)}[x]$. Every vertex with a non-empty pool thus stores a path.
In round $N_{\mathrm{grid}}$, the run also performs the edge tests of Subsubsection~\ref{subsubsec:phase-connectivity}, using the fixed paths $V^{(N_{\mathrm{grid}}-1)}$.
The run returns the final paths, the energy thresholds at vertices with a stored path, and the accepted graph $G_r$ of the edge tests.
Algorithm~\ref{alg:overall-dp} summarizes these steps. Figure~\ref{fig:dp-exploration} shows one exploration round and the final edge tests.

The edge tests in Lemma~\ref{lem:accepted-edge} require a stored state with small leakage at each reachable vertex. Determining the rank in Lemma~\ref{lem:rank-estimate} requires an energy threshold close to the top of the lowest $r$ levels.
The following corollary provides both.

\begin{corollary}[Exploration at reachable vertices]
\label{cor:large-gap}
Assume that the grid is fine enough (Eq.~\eqref{eq:grid-large-gap}),
that the edge circuits are accurate (Eq.~\eqref{eq:unitary-error}),
and that the precisions are small enough (Eq.~\eqref{eq:counting-condition}).
Assume also that all overlap and purity estimates have accuracy $\epsilon_p/2$. After
$N_{\mathrm{grid}}$ rounds, every vertex $x$ reachable at rank $r$
satisfies
\begin{align}
    \lambda_x(V[x])
        &\le\epsilon_{\mathrm{DP}},
        \label{eq:final-leakage}\\*
E_{r-1}(x)-h_E
        &<\mu[x]<E_{r-1}(x)+2h_E,
        \label{eq:final-bracket}\\*
    \|\sigma_{V[x]}-\rho_r(x)\|_1
        &\le2\sqrt{\epsilon_{\mathrm{DP}}}
        +\sqrt{2\epsilon_{\mathrm{DP}}+r(2\epsilon_{\mathrm{ref}}+\epsilon_p)},
        \nonumber
\end{align}
and $\lambda_x(V^{(N_{\mathrm{grid}}-1)}[x])\le\theta_{N_{\mathrm{grid}}-1}$.
Every vertex $x$ in the connected component of the resolved region
containing a rank-$r$ reference point is reachable at rank $r$ and satisfies
$\rho_r(x)=\rho_{\mathcal G}(x)$.
\end{corollary}

\begin{proof}
Consider any grid path $x_{\mathrm{ref}}=y_0,y_1,\ldots,y_\ell=x$ satisfying Eq.~\eqref{eq:reachable} with $\ell$ edges.
With loops removed, the path is simple, hence $\ell\le N_{\mathrm{grid}}-1$.
We verify the hypotheses of Theorem~\ref{thm:finite-precision} along it.
The run admits the rank-$r$ reference point $y_0$, whose reference state has leakage at most $\epsilon_{\mathrm{ref}}=\theta_0$.

By Eq.~\eqref{eq:reachable}, the band gap on each edge $e_j:y_{j-1}\to y_j$ is strictly above $\gamma_{\mathrm{lb}}/2$.
The cutoff $\gamma_{\mathrm{lb}}/2\in\mathcal D$ therefore satisfies $a_{\gamma_{\mathrm{lb}}/2}(e_j)=0$.
Eq.~\eqref{eq:implemented-leakage} then gives transport across $e_j$.

At each vertex $y_j$, Eq.~\eqref{eq:reachable} gives $E_r-E_{r-1}\ge3\gamma_{\mathrm{lb}}/4>5h_E$.
In each round $k$, all candidate states in the pool satisfy the purity condition.
Eq.~\eqref{eq:counting-condition} gives the conditions on $\epsilon'$, $\epsilon_p$ and $\epsilon_{\mathrm{ref}}$.
Lemma~\ref{lem:selection} therefore applies whenever the pool contains a candidate path with leakage at most $\epsilon'$.
Then, at the energy threshold $\mu^{(k)}[y_j]+2h_E$, every candidate path that reaches $p_k$ has leakage at most $\epsilon'+2\epsilon_p$.
In particular, the passing candidate path of least cost has leakage at most $\epsilon'+2\epsilon_p$, whichever cutoff it uses, so selection holds at $y_j$.

Since $\ell\le N_{\mathrm{grid}}-1$, Theorem~\ref{thm:finite-precision} gives Eq.~\eqref{eq:final-leakage} and the bound on $V^{(N_{\mathrm{grid}}-1)}[x]$.
In the final round, the pool at $x$ contains $V^{(N_{\mathrm{grid}}-1)}[x]$, whose leakage is at most $\theta_{N_{\mathrm{grid}}-1}\le\epsilon'$. Lemma~\ref{lem:selection} then gives Eq.~\eqref{eq:final-bracket}.

For the trace distance, write $P=P_r(x)$ and $\sigma=\sigma_{V[x]}$.
For a pure state $\psi$, write $\lambda=\lambda_x(\psi)$. The operator $\psi-P\psi P$ has rank at most two and eigenvalues $(\lambda\pm\sqrt{4\lambda-3\lambda^2})/2$. Hence $\|\psi-P\psi P\|_1\le2\sqrt{\lambda_x(\psi)}$.
Convexity of the trace norm and concavity of the square root extend this to $\|\sigma-P\sigma P\|_1\le2\sqrt{\lambda_x(\sigma)}$.
Since $P\sigma P-P/r$ acts on the $r$-dimensional range of $P$,
\[
    \|P\sigma P-P/r\|_1^2
    \le r\|P\sigma P-P/r\|_F^2
    =r\operatorname{Tr}[(P\sigma P)^2]-1+2\lambda_x(\sigma)
    \le2\lambda_x(\sigma)+r(2\epsilon_{\mathrm{ref}}+\epsilon_p),
\]
where the last step uses the purity condition and $\operatorname{Tr}[(P\sigma P)^2]\le\operatorname{Tr}[\sigma^2]$. This inequality holds because $P\sigma P$ is a diagonal block of $\sigma$. If $r=1$, $\operatorname{Tr}[\sigma^2]\le1$ replaces the purity condition.
The triangle inequality with $\rho_r(x)=P/r$ and $\lambda_x(\sigma)\le\epsilon_{\mathrm{DP}}$ gives the bound.

For the last statement, Lemma~\ref{lem:gapped-grid-path} implies reachability. Eq.~\eqref{eq:sector-equals-gsm} implies $\rho_r(x)=\rho_{\mathcal G}(x)$.
\end{proof}

\subsection{Testing phase connectivity during state preparation}
\label{subsec:phase-guarantee}

\subsubsection{Comparing neighboring stored states}
\label{subsubsec:phase-connectivity}

In its last round, the dynamic program tests each grid edge by transporting the stored state at one end across the edge and comparing the result with the stored state at the other end (Figure~\ref{fig:dp-exploration}(b)).
An edge that passes this edge test is accepted.
Informally, a chain is a short path of accepted edges.
A chain between two points with accurately stored ground states certifies that these ground states are not in different phases (Lemma~\ref{lem:linking}(i)).
Conversely, such a chain exists whenever a short grid path between the two points has spectral gap above $(1-\epsilon_{\mathrm{thresh}})\gamma_{\mathrm{thresh}}$ throughout and its vertices are reachable (Lemma~\ref{lem:linking}(ii)).

The energy tests certify stored states only at reachable vertices (Corollary~\ref{cor:large-gap}), yet a chain may pass through vertices that are not reachable.
Edge tests avoid this problem, since the discrepancies of accepted edges add up along a chain, whatever the stored states in between.

\emph{Accepted edges.} Fix a rank $r$ and the paths stored after round $N_{\mathrm{grid}}-1$. 
For each vertex $v$ whose stored path starts at the reference point $x_{s_v}$, define
\begin{equation}
    \sigma_r(v)
    :=V_r^{(N_{\mathrm{grid}}-1)}[v]\,
       C[x_{s_v}]\,
       V_r^{(N_{\mathrm{grid}}-1)}[v]^\dagger.
    \label{eq:frozen-state}
\end{equation}
Orient each grid edge with stored paths at both endpoints, in either direction, as $u\to v$.
With the cutoff $(1-\epsilon_{\mathrm{thresh}})\gamma_{\mathrm{thresh}}$, estimate
\begin{equation}
\begin{aligned}
    z_r(u,v)
    &:=\operatorname{Tr}\left[
        \sigma_r(v)\widetilde U_{uv}\sigma_r(u)
        \widetilde U_{uv}^\dagger
    \right],
    \qquad
    \widetilde U_{uv}
    :=\widetilde U_{(1-\epsilon_{\mathrm{thresh}})\gamma_{\mathrm{thresh}}}(u\to v),\\
    |\widehat z_r(u,v)-z_r(u,v)|&\le\frac{\epsilon_p}{2}.
\end{aligned}
\label{eq:measured-sector-transport}
\end{equation}

A simple path in the grid has at most $N_{\mathrm{grid}}$ edges.
The acceptance margin $\epsilon_{\mathrm{link}}$ below is chosen so that the overlap lost along that many accepted edges stays below $\eta_P$ of Definition~\ref{def:finite-size-phase} (Lemma~\ref{lem:linking}).
We accept $(u,v)$ if
\begin{equation}
    r\widehat z_r(u,v)\ge 1-\epsilon_{\mathrm{link}},
    \qquad
    \epsilon_{\mathrm{link}}
    :=\frac{\eta_P}{4(N_{\mathrm{grid}}+1)^2}.
    \label{eq:link-target}
\end{equation}
The \emph{accepted graph} $G_r$ contains all vertices in the grid and all accepted edges.
The edges are undirected: traversing one backwards with $\widetilde U_{uv}^\dagger$ leaves the overlap in Eq.~\eqref{eq:measured-sector-transport} unchanged.

\emph{Precision and agreement of neighboring states.}
Besides Eq.~\eqref{eq:counting-condition}, the error of the stored states, the implementation error and the estimation error must each be below the acceptance margin:
\begin{equation}
\begin{aligned}
    \epsilon_{\mathrm{DP}}
       +\frac{r_{\max}}{2}(2\epsilon_{\mathrm{ref}}+\epsilon_p)
        &\le\frac{\epsilon_{\mathrm{link}}}{8},\\
    [1+r_{\max}(2\epsilon_{\mathrm{ref}}+\epsilon_p)]\epsilon_U
        &\le\frac{\epsilon_{\mathrm{link}}}{4},\\
    \frac{r_{\max}\epsilon_p}{2}
        &\le\frac{\epsilon_{\mathrm{link}}}{8}.
\end{aligned}
    \label{eq:edge-precision}
\end{equation}
The lemmas below need the stored states close to their sector states in Hilbert--Schmidt norm.
At a reachable vertex, the path stored before the final round has leakage at most $\theta_{N_{\mathrm{grid}}-1}\le\epsilon_{\mathrm{DP}}$ (Corollary~\ref{cor:large-gap}). Hence,
\begin{equation}
\begin{aligned}
    \frac r2\|\sigma_r(v)-\rho_r(v)\|_F^2
    &=\frac r2\operatorname{Tr}[\sigma_r(v)^2]
       +\frac12-\operatorname{Tr}[P_r(v)\sigma_r(v)]\\
    &\le\epsilon_{\mathrm{DP}}
       +\frac r2(2\epsilon_{\mathrm{ref}}+\epsilon_p)
     \le\frac{\epsilon_{\mathrm{link}}}{8}.
\end{aligned}
\label{eq:hs-to-transported}
\end{equation}

\begin{lemma}[Accepted edges preserve the stored states]
\label{lem:accepted-edge}
Assume the hypotheses of Corollary~\ref{cor:large-gap},
the precision conditions in Eq.~\eqref{eq:edge-precision}, and the estimation accuracy in
Eq.~\eqref{eq:measured-sector-transport}. Every accepted edge satisfies
\begin{equation}
    \frac r2\left\|
        \sigma_r(v)-\widetilde U_{uv}\sigma_r(u)
                         \widetilde U_{uv}^\dagger
    \right\|_F^2
    \le\epsilon_{\mathrm{edge}},
    \label{eq:accepted-edge}
\end{equation}
where
\begin{equation}
    \epsilon_{\mathrm{edge}}
    :=r_{\max}(2\epsilon_{\mathrm{ref}}+\epsilon_p)
                         +\epsilon_{\mathrm{link}}
                         +\frac{r_{\max}\epsilon_p}{2}
    \le\frac{11\epsilon_{\mathrm{link}}}{8}.
    \label{eq:edge-budget}
\end{equation}
Conversely, an edge is
accepted if its endpoints are reachable at rank $r$ and
$\gamma_r>(1-\epsilon_{\mathrm{thresh}})\gamma_{\mathrm{thresh}}$ throughout the edge.
\end{lemma}

\begin{proof}
Write $\sigma_u$ and $\sigma_v$ for $\sigma_r(u)$ and $\sigma_r(v)$.
For an accepted edge, expanding the square and using the purity condition gives
\[
    \frac r2\|\sigma_v-\widetilde U_{uv}\sigma_u
                         \widetilde U_{uv}^\dagger\|_F^2
    \le 1+r(2\epsilon_{\mathrm{ref}}+\epsilon_p)-rz_r(u,v).
\]
Acceptance and the estimation error imply $rz_r(u,v)\ge1-\epsilon_{\mathrm{link}}-r\epsilon_p/2$.
Combining the two bounds gives Eq.~\eqref{eq:accepted-edge} with $\epsilon_{\mathrm{edge}}$ defined in Eq.~\eqref{eq:edge-budget}.
The bound $\epsilon_{\mathrm{edge}}\le11\epsilon_{\mathrm{link}}/8$ follows from the conditions on the stored states and on the estimation in Eq.~\eqref{eq:edge-precision}.

For the converse, we bound the overlap under exact transport and then add the implementation and estimation errors.
Let $U=U_{(1-\epsilon_{\mathrm{thresh}})\gamma_{\mathrm{thresh}}}(u\to v)$.
Since $\gamma_r>(1-\epsilon_{\mathrm{thresh}})\gamma_{\mathrm{thresh}}$ on the edge, Corollary~\ref{cor:exact-transport} with $d=r$ implies $U\rho_r(u)U^\dagger=\rho_r(v)$. Set $A_u=\sigma_u-\rho_r(u)$ and $A_v=\sigma_v-\rho_r(v)$. Expand $r\operatorname{Tr}[\sigma_vU\sigma_uU^\dagger]$ with $\sigma_u=\rho_r(u)+A_u$ and $\sigma_v=\rho_r(v)+A_v$. The two linear terms equal $-\lambda_u(\sigma_u)$ and $-\lambda_v(\sigma_v)$, with magnitude at most $\epsilon_{\mathrm{DP}}$. From Eq.~\eqref{eq:hs-to-transported},
\[
    \left|r\operatorname{Tr}[A_vUA_uU^\dagger]\right|
    \le r\|A_v\|_F\|A_u\|_F
    \le\frac{\epsilon_{\mathrm{link}}}{4}.
\]
Therefore,
\[
    r\operatorname{Tr}[\sigma_vU\sigma_uU^\dagger]
    \ge 1-2\epsilon_{\mathrm{DP}}
           -\frac{\epsilon_{\mathrm{link}}}{4}
    \ge 1-\frac{\epsilon_{\mathrm{link}}}{2}.
\]
We control implementation error using the Cauchy--Schwarz inequality in Hilbert--Schmidt norm,
\[
\begin{aligned}
    r\left|
      \operatorname{Tr}[\sigma_v\widetilde U_{uv}\sigma_u
                                      \widetilde U_{uv}^\dagger]
      -\operatorname{Tr}[\sigma_vU\sigma_uU^\dagger]
    \right|
    &\le 2r\|\widetilde U_{uv}-U\|
                     \|\sigma_v\|_F\|\sigma_u\|_F\\
    &\le[1+r(2\epsilon_{\mathrm{ref}}+\epsilon_p)]\epsilon_U.
\end{aligned}
\]
The estimation error and Eq.~\eqref{eq:edge-precision} then imply
\[
    r\widehat z_r(u,v)
    \ge 1-\frac{\epsilon_{\mathrm{link}}}{2}
           -\frac{\epsilon_{\mathrm{link}}}{4}
                   -\frac{r_{\max}\epsilon_p}{2}
    \ge 1-\frac{7\epsilon_{\mathrm{link}}}{8}
    >1-\epsilon_{\mathrm{link}}.
\]
\end{proof}

\subsubsection{From accepted edges to phase connectivity}

A \emph{chain} at rank $r$ is a simple path in $G_r$ of geometric length at most $L_*$.
Here $L_*$ is the $O(1)$ bound on the length of grid approximations from Subsection~\ref{subsec:local-higher-dimensional}.
We approximate the transport along a chain by an admissible circuit (Definition~\ref{def:admissible-circuits}).
The gates of such a circuit have diameter at most the radius function $R(n)$.
We take $R(n)$ at least the gate diameter that Proposition~\ref{prop:fdlu-from-gapped-path} gives for paths of length at most $L_*$.
Let $k_{\mathrm{circ}}$ be the exponent in the error $n^{-k_{\mathrm{circ}}}$ of this approximation.
We require
\begin{equation}
    N_{\mathrm{grid}}\epsilon_U+2n^{-k_{\mathrm{circ}}}<\frac{\eta_P}{2}.
    \label{eq:link-condition}
\end{equation}

\begin{lemma}[Phase guarantees for chains]
\label{lem:linking}
Assume the hypotheses of Lemma~\ref{lem:accepted-edge} and
Eq.~\eqref{eq:link-condition}. Let $x,z$ have spectral gap at least
$\gamma_{\mathrm{lb}}$ and the same ground-state degeneracy $r$.
Suppose that their connected components of the resolved region each contain a
rank-$r$ reference point.
\begin{enumerate}
\item[(i)] Every rank-$r$ chain from $x$ to $z$ gives an admissible
circuit $C$ with
\begin{equation}
\begin{aligned}
    \operatorname{Tr}\left[
        P_{\mathcal G}(z)C\rho_{\mathcal G}(x)C^\dagger
    \right]
    &\ge 1-\left(\sqrt{\epsilon_{\mathrm{link}}/2}
                    +N_{\mathrm{grid}}\sqrt{\epsilon_{\mathrm{edge}}}\right)^2
          -N_{\mathrm{grid}}\epsilon_U-2n^{-k_{\mathrm{circ}}}\\
    &>1-\eta_P.
\end{aligned}
\label{eq:link-certificate}
\end{equation}
Thus the endpoint states are not in different phases.

\item[(ii)] Every simple grid path from $x$ to $z$ of length at most
$L_*$ is a chain if its vertices are reachable at rank $r$ and $\gamma_r>(1-\epsilon_{\mathrm{thresh}})\gamma_{\mathrm{thresh}}$ throughout.
\end{enumerate}
\end{lemma}

\begin{proof}
For (i), write the chain as $x=y_0,\ldots,y_\ell=z$, with implemented edge unitaries $\widetilde U_1,\ldots,\widetilde U_\ell$, and set $\widetilde W=\widetilde U_\ell\cdots\widetilde U_1$.
Eq.~\eqref{eq:accepted-edge}, unitary invariance, and the triangle inequality give
\[
    \sqrt{\frac r2}
    \|\sigma_r(z)-\widetilde W\sigma_r(x)\widetilde W^\dagger\|_F
    \le\ell\sqrt{\epsilon_{\mathrm{edge}}}.
\]
By Lemma~\ref{lem:gapped-grid-path}, $x$ and $z$ are both reachable at rank $r$, because their connected components of the resolved region contain rank-$r$ reference points.
Hence Eq.~\eqref{eq:hs-to-transported} and the triangle inequality give
\begin{equation}
\begin{aligned}
    1-\operatorname{Tr}\left[
       P_r(z)\widetilde W\rho_r(x)\widetilde W^\dagger\right]
    &=\frac r2\left\|
       \rho_r(z)-\widetilde W\rho_r(x)\widetilde W^\dagger
       \right\|_F^2\\
    &\le\left(\sqrt{\epsilon_{\mathrm{link}}/2}
                 +\ell\sqrt{\epsilon_{\mathrm{edge}}}\right)^2.
\end{aligned}
\label{eq:chain-population}
\end{equation}
Let $W=U_\ell\cdots U_1$ be the product of the exact quasi-adiabatic propagators of the edges.
Eq.~\eqref{eq:unitary-error} gives $\|\widetilde W-W\|\le\ell\epsilon_U/2$.
Hence swapping $\widetilde W$ for $W$ moves the endpoint overlap by at most $\ell\epsilon_U$.
Since the chain has geometric length at most $L_*$ and constant cutoff, Proposition~\ref{prop:fdlu-from-gapped-path} provides an admissible circuit $C$ approximating $W$, with $\|W-C\|\le n^{-k_{\mathrm{circ}}}$.
Swapping $W$ for $C$ moves the endpoint overlap by at most $2n^{-k_{\mathrm{circ}}}$.
At the endpoints, $\rho_r=\rho_{\mathcal G}$ and $P_r=P_{\mathcal G}$.
This proves the first inequality in Eq.~\eqref{eq:link-certificate}.
Finally, since $\epsilon_{\mathrm{edge}}\le11\epsilon_{\mathrm{link}}/8$, we get
\[
    \left(\sqrt{\epsilon_{\mathrm{link}}/2}
             +N_{\mathrm{grid}}\sqrt{\epsilon_{\mathrm{edge}}}\right)^2
    \le\frac{11}{8}(N_{\mathrm{grid}}+1)^2\epsilon_{\mathrm{link}}
    =\frac{11\eta_P}{32}.
\]
Together with Eq.~\eqref{eq:link-condition}, this yields the strict bound.

For (ii), the vertices of the path are reachable at rank $r$ and $\gamma_r>(1-\epsilon_{\mathrm{thresh}})\gamma_{\mathrm{thresh}}$ on every edge, so the converse of Lemma~\ref{lem:accepted-edge} applies to each edge.
\end{proof}

\subsubsection{Determining the ranks and constructing the phase diagram}
\label{subsec:combining-rank-runs}

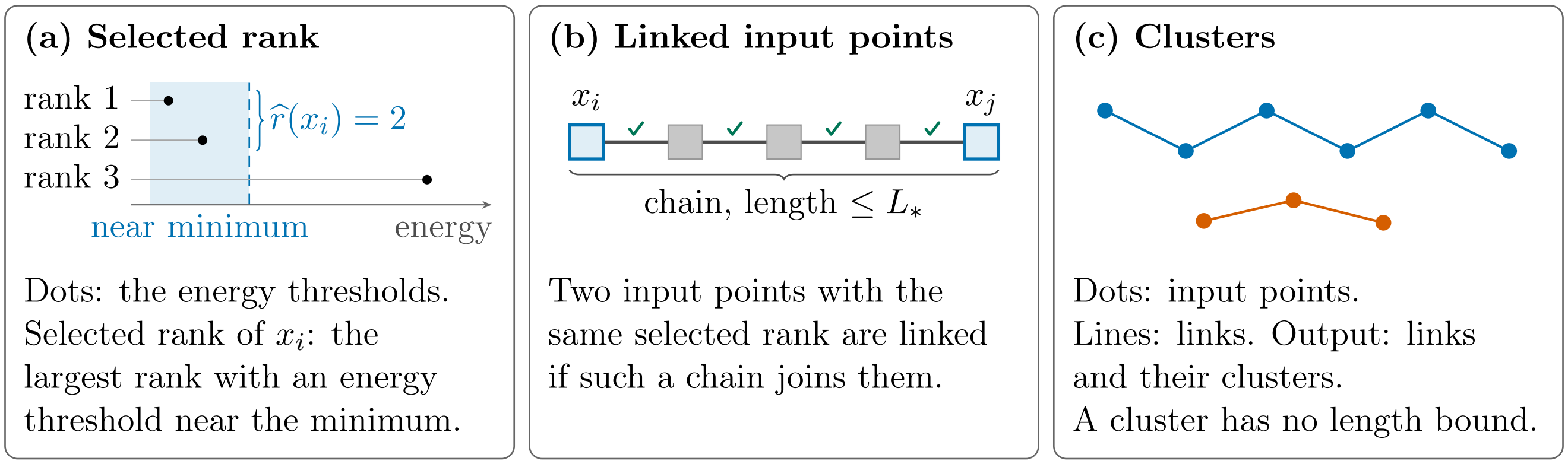
\begin{figure}[t]
\centering
\tikzset{
  txt/.style={figtext, anchor=base west, inner sep=0pt},
  st/.style={figstep, anchor=base west, inner sep=0pt},
  cvert/.style={minimum size=0.38cm, inner sep=0pt, outer sep=0pt},
  inp/.style={circle, inner sep=0pt, minimum size=5pt},
  tick/.pic={\draw[figgreen!75!black, line width=0.9pt, line cap=round, line join=round] (-0.07,0.015) -- (-0.025,-0.045) -- (0.075,0.085);},
}%
\begin{tikzpicture}[x=1cm, y=1cm]
\draw[figpanel] (0,0)     rectangle (5.67,-5.05);
\draw[figpanel] (5.84,0)  rectangle (11.51,-5.05);
\draw[figpanel] (11.68,0) rectangle (17.35,-5.05);
\node[st] at (0.2,-0.47) {(a) Selected rank};
\fill[figblue!12] (1.62,-2.22) rectangle (2.72,-0.86);
\draw[figblue, line width=0.5pt, dash pattern=on 3.5pt off 2.5pt] (2.72,-2.22) -- (2.72,-0.86);
\node[txt] at (0.2,-1.15) {rank 1};
\node[txt] at (0.2,-1.59) {rank 2};
\node[txt] at (0.2,-2.03) {rank 3};
\draw[black!35, line width=0.4pt] (1.40,-1.06) -- (1.82,-1.06); \fill (1.82,-1.06) circle (1.5pt);
\draw[black!35, line width=0.4pt] (1.40,-1.50) -- (2.20,-1.50); \fill (2.20,-1.50) circle (1.5pt);
\draw[black!35, line width=0.4pt] (1.40,-1.94) -- (4.70,-1.94); \fill (4.70,-1.94) circle (1.5pt);
\draw[figblue, line width=0.5pt, decorate, decoration={brace, amplitude=2pt}] (2.80,-0.94) -- (2.80,-1.62);
\node[anchor=west, inner sep=0pt, text=figblue, font=\normalsize] at (2.94,-1.28) {$\widehat r(x_i)=2$};
\draw[-{Stealth[length=3pt, width=2.5pt]}, black!60, line width=0.4pt] (1.40,-2.22) -- (5.42,-2.22);
\node[anchor=base, inner sep=0pt, font=\normalsize, text=figblue] at (2.17,-2.58) {near minimum};
\node[anchor=base east, inner sep=0pt, font=\normalsize, text=black!70] at (5.44,-2.58) {energy};
\node[txt] at (0.2,-3.30) {Dots: the energy thresholds.};
\node[txt] at (0.2,-3.78) {Selected rank of $x_i$: the};
\node[txt] at (0.2,-4.26) {largest rank with an energy};
\node[txt] at (0.2,-4.74) {threshold near the minimum.};
\node[st] at (6.04,-0.47) {(b) Linked input points};
\foreach \i in {1,...,5} \coordinate (c\i) at (5.375+1.1*\i,-1.52);
\draw[black!70, line width=1pt] (c1) -- (c5);
\foreach \i/\j in {1/2,2/3,3/4,4/5} \pic at ($(c\i)!0.5!(c\j)+(0,0.13)$) {tick};
\foreach \i in {2,3,4} \node[cvert, fill=black!22, draw=black!40, line width=0.4pt] at (c\i) {};
\foreach \i in {1,5} \node[cvert, fill=figblue!12, draw=figblue, line width=1pt] at (c\i) {};
\node[anchor=base, inner sep=0pt, font=\normalsize] at ($(c1)+(0,0.41)$) {$x_i$};
\node[anchor=base, inner sep=0pt, font=\normalsize] at ($(c5)+(0,0.41)$) {$x_j$};
\draw[black!70, line width=0.5pt, decorate, decoration={brace, amplitude=3pt, mirror}] ($(c1)+(-0.19,-0.31)$) -- ($(c5)+(0.19,-0.31)$);
\node[anchor=north, inner sep=0pt, font=\normalsize] at ($(c3)+(0,-0.52)$) {chain, length $\le L_*$};
\node[txt] at (6.04,-3.30) {Two input points with the};
\node[txt] at (6.04,-3.78) {same selected rank are linked};
\node[txt] at (6.04,-4.26) {if such a chain joins them.};
\node[st] at (11.88,-0.47) {(c) Clusters};
\draw[figblue, line width=0.8pt] (12.25,-1.17) -- (13.15,-1.62) -- (14.05,-1.17) -- (14.95,-1.62) -- (15.85,-1.17) -- (16.75,-1.62);
\foreach \p in {(12.25,-1.17),(13.15,-1.62),(14.05,-1.17),(14.95,-1.62),(15.85,-1.17),(16.75,-1.62)} \node[inp, fill=figblue] at \p {};
\draw[figorange, line width=0.8pt] (13.35,-2.40) -- (14.35,-2.17) -- (15.35,-2.42);
\foreach \p in {(13.35,-2.40),(14.35,-2.17),(15.35,-2.42)} \node[inp, fill=figorange] at \p {};
\node[txt] at (11.88,-3.30) {Dots: input points.};
\node[txt] at (11.88,-3.78) {Lines: links. Output: links};
\node[txt] at (11.88,-4.26) {and their clusters.};
\node[txt] at (11.88,-4.74) {A cluster has no length bound.};
\end{tikzpicture}
\caption{\textbf{Ranks, links, and clusters} (Algorithm~\ref{alg:overall-dp}).
(a) At an input point $x_i$, the dynamic-programming run of each rank $q$
stores an energy threshold $\mu_q[x_i]$. The selected rank
$\widehat r(x_i)$ estimates the ground-state degeneracy at $x_i$.
It is the largest rank whose energy threshold lies within $4h_E$ of the
smallest one (Eq.~\eqref{eq:rank-estimate}). Here
$\widehat r(x_i)=2$.
(b) Two input points with the same selected rank $r$ are linked if a chain joins
them. A chain is a path of edges accepted in the rank-$r$ run, of geometric
length at most $L_*$. Its intermediate vertices may be unresolved.
A link between input points whose ground states are accurately represented
certifies that these ground states are not in different phases
(Figure~\ref{fig:dp-exploration}(b)).
(c) The algorithm returns the links and the clusters.
A cluster joins input points through sequences of links, so it
has no length bound. The certificate holds for each link
separately. Input points in different clusters are not asserted to be in
different phases.}
\label{fig:ranks-links}
\end{figure}

\emph{Selecting the ranks.}
We select the rank of each input point from the energy thresholds $\mu_r$ (Figure~\ref{fig:ranks-links}(a)).
Let $x$ be a grid point with $\gamma(x)\ge\gamma_{\mathrm{lb}}$, set $r=r(x)$, and suppose its connected component of the resolved region contains a rank-$r$ reference point.
By Eq.~\eqref{eq:final-bracket}, the rank-$r$ run stores an energy threshold close to $E_{r-1}(x)$.
The lowest $r$ levels form the ground-state manifold, whose width $E_{r-1}(x)-E_0(x)$ is at most $\delta_{\mathrm{gsm}}$ (Definition~\ref{definition: ground-state manifold}), so this energy threshold is also close to the ground-state energy.
Since ranks below $r$ may also have energy thresholds close to the ground-state energy, we select the largest rank $q$ whose energy threshold lies within $4h_E$ of the minimum:
\begin{equation}
    \widehat r(x):=
    \max\left\{
        q:V_q[x]\ne\textsc{Void},\quad
        \mu_q[x]\le
        \min_{j:\,V_j[x]\ne\textsc{Void}}\mu_j[x]+4h_E
    \right\}.
    \label{eq:rank-estimate}
\end{equation}
If no rank has a stored path, the selected rank $\widehat{r}(x)$ is undefined.

Lemma~\ref{lem:linking} applies to a chain in $G_r$ only when both endpoints have ground-state degeneracy $r$, so the selected rank must equal the ground-state degeneracy; the next lemma shows that it does.

\begin{lemma}[Determining the ground-state degeneracy]
\label{lem:rank-estimate}
Assume the hypotheses of Corollary~\ref{cor:large-gap} for all
runs and $\delta_{\mathrm{gsm}}\le h_E$. If
$\gamma(x)\ge\gamma_{\mathrm{lb}}$, $r(x)\le r_{\max}$, and the
connected component of the resolved region containing $x$ also contains a
rank-$r(x)$ reference point, then $\widehat r(x)=r(x)$.
\end{lemma}

\begin{proof}
Fix any $x$ and let $r=r(x)$.
For every rank $q$, all candidate states in the rank-$q$ run satisfy the purity condition for rank $q$.
By Lemma~\ref{lem:selection} and Eq.~\eqref{eq:counting-condition}, $\mu_q[x]>E_{q-1}(x)-h_E\ge E_0(x)-h_E$, even if no candidate state is accurate.
In particular, the minimum stored energy threshold is greater than $E_0(x)-h_E$.
By Lemma~\ref{lem:gapped-grid-path}, $x$ is reachable at rank $r$, so Eq.~\eqref{eq:final-bracket} implies
\[
    \mu_r[x]<E_{r-1}(x)+2h_E
    \le E_0(x)+3h_E
    <\min_j\mu_j[x]+4h_E.
\]
Thus, rank $r$ satisfies the condition in Eq.~\eqref{eq:rank-estimate}.
For any $q>r$ with a stored path, using $\gamma_{\mathrm{lb}}=16h_E$,
\[
\begin{aligned}
    \mu_q[x]&>E_r(x)-h_E
             \ge E_0(x)+15h_E,\\
    \min_j\mu_j[x]+4h_E&\le\mu_r[x]+4h_E
                         <E_0(x)+7h_E,
\end{aligned}
\]
so no larger rank satisfies this condition.
Smaller ranks do not affect the maximum.
Hence, $\widehat r(x)=r$.
\end{proof}

\emph{Linking input points of equal rank.}
Two input points with the same selected rank $r$ are \emph{linked} if a chain in $G_r$ connects them (Figure~\ref{fig:ranks-links}(b)).
From each input point of selected rank $r$, we find shortest paths in $G_r$ under geometric edge lengths.
We link it to every other such input point at distance at most $L_*$ and keep the path with the link.
If several input points lie at the same location, these searches use only one of them.
We then restore the repeated input points whose selected rank is defined, using zero-length links.
Input points whose selected rank is undefined stay isolated.
Algorithm~\ref{alg:overall-dp} returns the links and the clusters (Figure~\ref{fig:ranks-links}(c)).

\begin{algorithm}[!htbp]
\caption{Learning the phase diagram with dynamic programming}
\label{alg:overall-dp}
\scriptsize
\begin{algorithmic}[1]
\Require $H(x)$, $x\in[-1,1]^m$; input points; reference states
$C[x_s]$, $x_s\in S$; set of cutoffs $\mathcal D$; length bound $L_*$.
\Ensure The linked relation and its clusters, with the guarantee of
Theorem~\ref{thm:detection}.
\State Form the refined grid of Subsection~\ref{subsec:grid-choice}.
\State If ranks $r>1$ are considered, admit a reference point to the
rank-$r$ run only if the estimated purity of its reference state is consistent with $1/r$.
\For{each $r\le r_{\max}$ admitting at least one reference point}
    \State Store the empty path at every admitted reference point
    and \textsc{Void} elsewhere.
    \State In round $k=1,\ldots,N_{\mathrm{grid}}$, at every vertex
    $x$: form the pool in Eq.~\eqref{eq:pool} from the paths stored
    in the preceding round, let $\mu[x]$ be the first energy threshold at
    which a candidate state reaches the required overlap $p_k$, and store a
    candidate path of least cost reaching $p_k$ at $\mu[x]+2h_E$.
    \State In the final round, run the edge tests on the states stored in
    the preceding round, with cutoff $(1-\epsilon_{\mathrm{thresh}})\gamma_{\mathrm{thresh}}$
    (Eq.~\eqref{eq:measured-sector-transport}), and let $G_r$ contain
    the accepted edges (Eq.~\eqref{eq:link-target}).
\EndFor
\State Select the rank of each input point using Eq.~\eqref{eq:rank-estimate}.
\State Link two input points of the same selected rank $r$ if they are connected by
a path of length at most $L_*$ in $G_r$; keep a shortest such path
with the link.
\State \Return The links and the clusters.
\end{algorithmic}
\end{algorithm}

\subsection{Correctness and resources}
\label{app:main-algorithm-proof}

We prove Theorem~\ref{thm:detection} and Corollary~\ref{cor:components-with-reference-point} for any constant parameter-space dimension $m$, first assuming that each estimate achieves the specified accuracy.
Lemma~\ref{lem:implementation-complexity} and Corollary~\ref{cor:reference-copy-count} then bound the resource complexities and failure probability at the precisions we set.
We assume $r_{\max}=\operatorname{poly}(n)$ throughout.
Eq.~\eqref{eq:1d-gsm-tolerance} gives the tolerance parameter $\delta_{\mathrm{gsm}}\le\gamma_{\mathrm{lb}}/16=h_E$, as Lemma~\ref{lem:rank-estimate} requires.
The grid is fine enough to satisfy Eqs.~\eqref{eq:grid-threshold-condition} and \eqref{eq:grid-large-gap}.
The reference precision $\epsilon_{\mathrm{ref}}$ and the other precisions are small enough to satisfy Eqs.~\eqref{eq:counting-condition}, \eqref{eq:edge-precision}, and \eqref{eq:link-condition}.

\subsubsection{Correctness}

\begin{proof}[Proof of Theorem~\ref{thm:detection}, general constant dimension]
Fix an input point $x_i$ with $\gamma(x_i)\ge\gamma_{\mathrm{lb}}$.
Assumption~\ref{assumption-reference-state} ensures a successful reference point in the connected component of the resolved region containing $x_i$.
By Corollary~\ref{cor:large-gap}, the run whose rank is the ground-state degeneracy of $x_i$ stores the state accurately.
Lemma~\ref{lem:rank-estimate} identifies this ground-state degeneracy.

Clause~(i) of Definition~\ref{def:phase-clustering} requires that linked input points with spectral gap at least $\gamma_{\mathrm{lb}}$ are not in different phases.
Consider two such points $x_i,x_j$.
Algorithm~\ref{alg:overall-dp} links only input points with the same selected rank.
By Lemma~\ref{lem:rank-estimate}, these selected ranks equal the ground-state degeneracies.
Hence, both input points have the same ground-state degeneracy, say $r$.
If $x_i$ and $x_j$ are at the same location, then the link between them has length zero, witnessed by the identity circuit with overlap one.
Otherwise, the path kept with the link is a shortest path of length at most $L_*$ in the accepted graph $G_r$ (Subsubsection~\ref{subsubsec:phase-connectivity}), and hence a rank-$r$ chain.
Lemma~\ref{lem:linking}(i) implies that the states at both ends of this chain are not in different phases.

For clause~(ii), consider two input points joined by a path of length at most $L_{\max}$ with $\gamma(x(s))\ge\gamma_{\mathrm{thresh}}$ for every $s\in[0,1]$.
Here $L_{\max}$ is the $O(1)$ length bound of clause~(ii).
As in Subsection~\ref{subsec:grid-choice}, the ground-state degeneracy along the path is a constant $r$.
Its grid approximation $\widetilde x$ of Eq.~\eqref{eq:grid-approximation} keeps a large band gap.
Eq.~\eqref{eq:grid-threshold-gap} gives $\gamma_r(\widetilde x(s))\ge(1-\epsilon_{\mathrm{thresh}}/2)\gamma_{\mathrm{thresh}}>(1-\epsilon_{\mathrm{thresh}})\gamma_{\mathrm{thresh}}$.
Every vertex of $\widetilde x$ is also reachable at rank $r$.
Its first vertex is reachable by Lemma~\ref{lem:gapped-grid-path}.
Along its edges, Eqs.~\eqref{eq:grid-large-gap} and~\eqref{eq:grid-gap-lipschitz} give $\gamma_r(\widetilde x(s))\ge\gamma_{\mathrm{thresh}}-\gamma_{\mathrm{lb}}/4\ge3\gamma_{\mathrm{lb}}/4$, so Eq.~\eqref{eq:reachable} holds.
Removing loops from $\widetilde x$ yields a simple path of length at most $L_*$ and at most $N_{\mathrm{grid}}$ edges.
By Lemma~\ref{lem:linking}(ii), this path is a chain in $G_r$.
Since both input points have selected rank $r$, the search from the location of one reaches the other within distance $L_*$.
They therefore lie in the same cluster.
Thus, no such path can exist between input points in different clusters, proving (ii).
\end{proof}

\begin{proof}[Proof of Corollary~\ref{cor:components-with-reference-point}, general constant dimension]
Corollary~\ref{cor:large-gap} and Lemma~\ref{lem:rank-estimate} apply to each connected component of the resolved region with a successful reference point.
The other connected components need no successful reference point.
The argument for clause~(i) applies whenever both endpoints are in components with a successful reference point.
For clause~(ii), a path with $\gamma\ge\gamma_{\mathrm{thresh}}$ connecting two such endpoints extends rank-$r$ reachability, Eq.~\eqref{eq:reachable}, along its length, so the same argument applies.
The algorithm need not know which preparations succeeded or which components contain them.
\end{proof}

\subsubsection{Precision and resources}\label{subsubsec:reference-copies}

We can satisfy Eqs.~\eqref{eq:counting-condition}, \eqref{eq:edge-precision}, and~\eqref{eq:link-condition} with inverse-polynomial precisions. We choose
\begin{equation}
    \epsilon_U=\epsilon_p:=\min\left\{
       \frac{\eta_P}
       {128(N_{\mathrm{grid}}+1)^2
       (N_{\mathrm{grid}}+r_{\max}+1)},
       \frac{1}{32r_{\max}
       (N_{\mathrm{grid}}+r_{\max}+1)}
    \right\}.
    \label{eq:overall-precision}
\end{equation}
We require successful preparations to have accuracy $\epsilon_{\mathrm{ref}}=\epsilon_p$. Since
\[
    \epsilon_{\mathrm{DP}}
       \le (4N_{\mathrm{grid}}+1)\epsilon_p,
    \qquad
    2\epsilon_{\mathrm{ref}}+\epsilon_p\le3\epsilon_p,
\]
Eqs.~\eqref{eq:counting-condition} and~\eqref{eq:edge-precision} are satisfied, with $\epsilon_{\mathrm{DP}}$ the total budget of Eq.~\eqref{eq:dp-budget}.
We choose $k_{\mathrm{circ}}$ sufficiently large to ensure $2n^{-k_{\mathrm{circ}}}\le\eta_P/4$.
Then, $N_{\mathrm{grid}}\epsilon_U+2n^{-k_{\mathrm{circ}}}<\eta_P/2$, as required by Eq.~\eqref{eq:link-condition}.

We fix a total failure probability $\delta_{\mathrm{tot}}=1/\mathrm{poly}(n)$. We allot $\delta_{\mathrm{tot}}/2$ to the estimates of the filtered overlaps, the purity tests and the edge tests (or to batches of these estimates). Then, Lemma~\ref{lem:adaptive-confidence}(a) ensures that all these estimates are accurate with probability at least $1-\delta_{\mathrm{tot}}/2$. The remaining half of $\delta_{\mathrm{tot}}$ accounts for the reuse of unique ground states in Corollary~\ref{cor:reference-copy-count}(a).

\begin{lemma}[Implementation and resources]
\label{lem:implementation-complexity}
With the precisions chosen above, Algorithm~\ref{alg:overall-dp} takes $\mathrm{poly}(n)$ time and consumes $\mathrm{poly}(n)$ independent copies of the reference states.
\end{lemma}

\begin{proof}
\emph{Number of estimates.}
There are $O(n/\gamma_{\mathrm{lb}})$ energy thresholds, spaced as in Eq.~\eqref{eq:threshold-spacing}.
The shifted energy threshold $\mu[x]+2h_E$ of Eq.~\eqref{eq:shifted-threshold-selection} is one of them, so it needs no new measurement.
The grid graph has degree at most $2m$ and hence each pool has size at most $2+2m|\mathcal D|$ (Eq.~\eqref{eq:pool}).
Including the initialization of the dynamic program at the $|S|$ reference points, the total number of filtered overlaps for all $r$ is
\begin{equation}
    Q_{\mathrm{filter}}
    =O\!\left(
       \frac{r_{\max}n}{\gamma_{\mathrm{lb}}}\left[
           |S|+N_{\mathrm{grid}}^2
               (2+2m|\mathcal D|)
       \right]
    \right).
    \label{eq:filter-count}
\end{equation}
For the edge tests, we use the cutoff $(1-\epsilon_{\mathrm{thresh}})\gamma_{\mathrm{thresh}}$ and one orientation of each edge in the set $E$ of undirected grid edges.
This adds
\begin{equation}
    Q_{\mathrm{edge}}\le r_{\max}|E|
       =O(r_{\max}N_{\mathrm{grid}})
    \label{eq:edge-measurement-count}
\end{equation}
expectations.
These observables depend only on the paths stored after round $N_{\mathrm{grid}}-1$, Eq.~\eqref{eq:frozen-state}, so they are fixed before the last round and can be measured in that round together with the filtered overlaps.
Each purity test (Subsection~\ref{subsec:energy-rank-estimation}) runs once per reference state and is used for all ranks.
In total,
\[
    Q:=1+Q_{\mathrm{filter}}+Q_{\mathrm{edge}}+|S|
\]
bounds the number of estimates.
Here $Q_{\mathrm{filter}}$ and $Q_{\mathrm{edge}}$ come from Eqs.~\eqref{eq:filter-count} and~\eqref{eq:edge-measurement-count}, and the last term counts the purity tests.

\emph{Circuits for the estimates.}
Each stored path has at most $N_{\mathrm{grid}}$ edge circuits and each cutoff has size at least $\gamma_{\mathrm{lb}}/2$. Hence, each stored path can be implemented with polynomially many gates according to Lemma~\ref{lemma:generator-discretization}.
Each filtered overlap uses a block encoding of a smooth step that satisfies Eq.~\eqref{eq:smooth-step}, which we construct as follows.
We strengthen the condition on $B$ to $B\ge\sup_x\alpha_x+5h_E$.
Here $\alpha_x=O(n)$ is the sum of the absolute Pauli coefficients of $H(x)$.
Then a linear combination of unitaries block-encodes $(H(x)-\nu I)/(2B)$ with normalization one.
We follow the construction of \cite[Section~2]{LinTong2020}, with their threshold $\mu$ set to $\nu$.
Quantum singular value transformation~\cite[Theorem~1]{LinTong2020} block-encodes any odd real polynomial of this operator that is bounded by one on $[-1,1]$.
We use the odd polynomial $-S(\cdot;\delta,\epsilon_p)$ of \cite[Lemma~3]{LinTong2020} with $\delta=h_E/(2B)$.
The polynomial $S$ approximates the sign function.
The circuit of \cite[Eq.~(2)]{LinTong2020} then block-encodes, with normalization one,
\[
    f_\nu(H(x)):=\tfrac12\Bigl[I-S\Bigl(\tfrac{H(x)-\nu I}{2B};\delta,\epsilon_p\Bigr)\Bigr]
\]
using $O((B/h_E)\log(1/\epsilon_p))$ queries.
This $f_\nu$ satisfies Eq.~\eqref{eq:smooth-step}.
Indeed, the polynomial $S$ is bounded by one in absolute value on $[-1,1]$ and lies within $\epsilon_p$ of the sign function on $[-1,-\delta]\cup[\delta,1]$.
It is evaluated at $(E-\nu)/(2B)\in[-1,1]$ for $E,\nu\in[-B,B]$.
We estimate each filtered overlap to accuracy $\epsilon_p/2$ by conjugating with the stored path and running Hadamard tests on fresh copies.
The implementation and statistical errors are $\epsilon_p/4$ each.

To estimate the overlap $z_r(u,v)$ of Eq.~\eqref{eq:measured-sector-transport}, fix the stored paths $V_u,V_v$ at the endpoints. They start at the reference points with indices $s_u,s_v$. Consider the Hermitian observable
\begin{equation}
    (V_u^\dagger\widetilde U_{uv}^\dagger\otimes V_v^\dagger)
    \operatorname{SWAP}
    (\widetilde U_{uv}V_u\otimes V_v).
    \label{eq:swap-pullback}
\end{equation}
This observable has norm one and polynomial circuit complexity. The SWAP ensures its expectation value in $C[x_{s_u}]\otimes C[x_{s_v}]$ is exactly the desired overlap. When $s_u=s_v$, use two independent copies of that reference state.

\emph{Copies and time.}
The total number of copies of reference states needed to estimate all these bounded observables on fresh copies is
\begin{equation}
    O\!\left(
        Q\epsilon_p^{-2}
        \log\frac{Q}{\delta_{\mathrm{tot}}}
    \right).
    \label{eq:fresh-copy-count}
\end{equation}
The circuit complexity is the number of copies times the polynomial cost of performing one measurement.
This count remains the same if the observables are chosen adaptively, because each measurement uses fresh copies.
Selecting the ranks, the Dijkstra searches for shortest paths and union-find for the clusters run classically in polynomial time.
At the precision fixed in Eq.~\eqref{eq:overall-precision}, all counts are polynomial.
\end{proof}

Reference states that are unique ground states or approximate Gibbs states need far fewer copies than Eq.~\eqref{eq:fresh-copy-count} counts.

\begin{corollary}[Reference-state copy counts]
\label{cor:reference-copy-count}
Consider Algorithm~\ref{alg:overall-dp} with the precisions chosen above.
\begin{enumerate}
\item[(a)] (Unique ground states.) Under Assumption~\ref{assumption-reference-state}, suppose that the reference states are unique ground states, as in Theorem~\ref{thm:detection}.
Suppose also that controlled evolution under their known Hamiltonians is efficient, as in Subsection~\ref{subsec:resources}.
Then every resolved point is non-degenerate, so we may take $r_{\max}=1$ and skip the purity test.
A single copy of each reference state serves all estimates, even though the observables are chosen adaptively.
For any constant $c>0$, the algorithm can be run in $\mathrm{poly}(n)$ time with the following property.
Averaged over the measurement record, the final joint state of the copies is within $e^{-cn}$ of their initial joint state in trace norm.
In the same run, all estimates are accurate except with probability at most $\delta_{\mathrm{tot}}/2$.
\item[(b)] (Gibbs reference states.) Suppose every reference state is within $\epsilon_{\mathrm{Gibbs}}$ in trace norm of a Gibbs state of a known
Hamiltonian, with polynomially bounded inverse temperature and
efficient controlled evolution. Run the dynamic program for all ranks together, one round at a time.
The number of copies consumed from each reference state is
\begin{equation}
    O\!\left(
        N_{\mathrm{grid}}
        \epsilon_p^{-2}
        \log\frac{Q}{\delta_{\mathrm{tot}}}
    \right).
    \label{eq:gibbs-copy-count}
\end{equation}
If $\epsilon_{\mathrm{Gibbs}}\le\epsilon_p/16$ and $\epsilon_{\mathrm{Gibbs}}$ times the total number of copies is at most $\delta_{\mathrm{tot}}/2$, all estimates are accurate except with probability at most $\delta_{\mathrm{tot}}$.
\end{enumerate}
\end{corollary}

\begin{proof}
For (a), a successful preparation at $x_s$ outputs a pure state $\psi_s=|\psi_s\rangle\langle\psi_s|$.
Write $r=r(x_s)$, so that $\rho_{\mathcal G}(x_s)=P_{\mathcal G}(x_s)/r$.
The variational characterization of the trace distance, applied to the projector $\psi_s$, gives
\[
    \tfrac12\bigl\|\psi_s-\rho_{\mathcal G}(x_s)\bigr\|_1
    \ge1-\frac{\langle\psi_s|P_{\mathcal G}(x_s)|\psi_s\rangle}{r}
    \ge1-\frac1r .
\]
If $r\ge2$, then $\|\psi_s-\rho_{\mathcal G}(x_s)\|_1\ge1>\epsilon_{\mathrm{ref}}$, because $\epsilon_{\mathrm{ref}}=\epsilon_p<1$ (Eq.~\eqref{eq:overall-precision}).
This contradicts Eq.~\eqref{eq:successful-preparation}.
Thus the ground-state degeneracy of $x_s$ is $r(x_s)=1$.
The ground-state degeneracy is constant on each connected component of the resolved region, as shown after Definition~\ref{def:large-gap-component}.
Assumption~\ref{assumption-reference-state} places a successful reference point in every such component, so every resolved point is non-degenerate.
We may therefore take $r_{\max}=1$.
This removes the purity test, whose swap tests need pairs of copies (Subsection~\ref{subsec:energy-rank-estimation}).

Lemma~\ref{lemma:catalytic-tomography} estimates every filtered overlap from the single copy of its reference state.
The edge observable in Eq.~\eqref{eq:swap-pullback} instead acts on two registers and requires a product reference state.
For different reference points, this product state is the unique ground state of the sum of the two known Hamiltonians, each acting on its own register.
The gap of this sum is the smaller of their two gaps.
Controlled evolution under this sum is the two controlled evolutions applied in turn, so it is also efficient.
Lemma~\ref{lemma:catalytic-tomography} can be applied with this product reference state and one copy of each of its factors.

Write $|\psi_s\rangle$ for the reference state at $x_s$, successful or not.
When $s_u=s_v=s$, two copies are not required, since
\[
    z_1(u,v)
      =\left|
          \langle\psi_s|
          V_v^\dagger\widetilde U_{uv}V_u
          |\psi_s\rangle
        \right|^2.
\]
The real and imaginary parts of the unitary in this matrix element are Hermitian contractions with efficient block encodings.
Thus Lemma~\ref{lemma:catalytic-tomography} estimates their expectation values up to $\epsilon_p/16$.
Clipping to $[-1,1]$, squaring, and adding the two gives an $\epsilon_p/4$-accurate estimate for the overlap.

Let $N_{\mathrm{meas}}=\mathrm{poly}(n)$ bound the number of catalytic calls, counting two calls for each overlap with $s_u=s_v$.
Suppose first that every call is exact.
Each call then disturbs the pure joint reference state $\bigotimes_s|\psi_s\rangle$ by at most $\delta_{\mathrm{cat}}$ on average.
Lemma~\ref{lem:adaptive-confidence}(b), applied with marginal disturbance $d_c=\delta_{\mathrm{cat}}$ and at most $N_{\mathrm{meas}}$ calls, bounds the total effect of these disturbances, even though each call may depend on earlier outcomes.
We choose $\delta_{\mathrm{cat}}$ with equality in Eq.~\eqref{eq:catalytic-disturbance-choice}.
Then, as in Eq.~\eqref{eq:catalytic-correlator-error}, the estimates fail with total probability at most $3\delta_{\mathrm{tot}}/8$.
Averaged over the record, the final joint state of the copies is then within $e^{-cn}/2$ of the initial joint state $\bigotimes_s|\psi_s\rangle\langle\psi_s|$ in trace norm.

If controlled evolutions and elementary gates are implemented approximately, we can distribute a total channel error of at most $\min\{\delta_{\mathrm{tot}}/8,e^{-cn}/8\}$ in diamond norm~\cite{watrous2018book} to the catalytic protocols.
Replacing the exact protocols by their implementations one at a time changes the joint output of the run, the record together with the copies, by at most this error in trace norm.
Since the accuracies above already include the errors from implementing the observables approximately, all estimates are accurate except with probability at most $3\delta_{\mathrm{tot}}/8+\delta_{\mathrm{tot}}/8=\delta_{\mathrm{tot}}/2$.
The averaged final state is within $e^{-cn}/2+e^{-cn}/8<e^{-cn}$ of the initial one.

Simulating and synthesizing the catalytic protocols takes time polynomial in the evolution time and in the logarithm of the inverse channel error.
This logarithm is $O(n)$.
In Lemma~\ref{lemma:catalytic-tomography}, the controlled-evolution time and the number of block-encoding calls grow only polylogarithmically in $1/\delta_{\mathrm{cat}}$.
The choice of $\delta_{\mathrm{cat}}$ above gives $\log(1/\delta_{\mathrm{cat}})=O(n)$, so the total time stays polynomial.

For (b), we first suppose that every reference state equals its Gibbs state, and remove this assumption at the end.
Since each round's outcomes fix the next round's transport circuits and threshold observables, in each round we batch all ranks' filtered overlaps by reference state and apply Lemma~\ref{lem:thermal-shadow-tomography}.
The initialization and $N_{\mathrm{grid}}$ rounds require at most $N_{\mathrm{grid}}+1$ batches for each reference state.
The initial purity tests require an extra batch over two independent copies.

We can put all observables in Eq.~\eqref{eq:swap-pullback} into batches labeled by unordered pairs of reference points, with a fixed ordering of registers.
A product of two exact Gibbs states is again a Gibbs state, at inverse temperature one, of the sum of the two Hamiltonians each scaled by their respective inverse temperatures.
Controlled evolution under this sum takes polynomial time, so Lemma~\ref{lem:thermal-shadow-tomography} can handle all observables for one pair in a single batch.
For a pair $(s,s)$, each product copy uses two copies of the reference state at $x_s$, which changes the number of copies by only a constant factor.

Each batch contains at most $Q$ observables.
With $O(\epsilon_p^{-2}\log(Q/\delta_{\mathrm{tot}}))$ copies of a reference state or a product reference state, a batch achieves accuracy $\epsilon_p$ and failure probability $\delta_{\mathrm{tot}}/Q$ up to constant factors.
Each reference state appears in at most $|S|\le N_{\mathrm{grid}}$ pair batches, leading to Eq.~\eqref{eq:gibbs-copy-count}.
Every batch has polynomial cost by Lemma~\ref{lem:thermal-shadow-tomography}.

Finally, let each reference state be only within $\epsilon_{\mathrm{Gibbs}}$ of its Gibbs state.
We compare the run with the same run on exact Gibbs states, with each batch at accuracy $\epsilon_p/8$.
This accuracy changes the copy count by only a constant factor.
Replacing the copies one at a time changes the joint input by at most $\epsilon_{\mathrm{Gibbs}}$ per copy.
Since $\epsilon_{\mathrm{Gibbs}}$ times the number of copies is at most $\delta_{\mathrm{tot}}/2$, the joint inputs of the two runs differ by at most $\delta_{\mathrm{tot}}/2$ in trace norm.
The run, adaptive choices included, is a quantum channel from the copies to the classical record.
Since channels are contractive in trace norm~\cite{nielsen2002quantum}, the probability that some batch fails grows by at most $\delta_{\mathrm{tot}}/4$.
Implementing the controlled evolutions and elementary gates of all batches to a total error of $\delta_{\mathrm{tot}}/4$ in diamond norm adds at most $\delta_{\mathrm{tot}}/4$, as in part (a).
Since part (b) reuses no copy, it can spend the half of $\delta_{\mathrm{tot}}$ reserved for reuse in part (a) on these two increases.
Outside this failure event, every estimate is within $\epsilon_p/8$ of the expectation value of its observable in the exact Gibbs states.
Each observable has norm at most one and acts on one copy or on two.
Its expectation value in the actual copies is therefore within $2\epsilon_{\mathrm{Gibbs}}\le\epsilon_p/8$ of the one in the exact Gibbs states.
Together with the implementation error $\epsilon_p/4$, as for fresh copies, every estimate has error at most $\epsilon_p/2$.
\end{proof}

\subsection{Implementation by adiabatic sweeps}
\label{subsec:dp-sweeps}

Ordinary adiabatic sweeps, which need no quasi-adiabatic filter, can replace the quasi-adiabatic propagators of the dynamic program with the same guarantee (Corollary~\ref{cor:dp-sweeps}).
The propagators move candidate states across gapped edges, and their product along a chain is close to an admissible circuit (Definition~\ref{def:admissible-circuits}), which has constant depth and gates of diameter at most the radius function $R(n)$.
This circuit approximation shows that the endpoint states are not in different phases (Lemma~\ref{lem:linking}).
A sweep across a gapped edge moves the lowest $r$ levels accurately (Lemma~\ref{lem:sweep-transport}).
If $R(n)$ is large enough and the protecting symmetry, if any, is unitary, a single sweep of polylogarithmic duration is close to an admissible circuit, by the Lieb--Robinson bound.

\subsubsection{Transport across a gapped edge}

\emph{Sweeps of grid edges.}
We take the spacing of the grid of Subsection~\ref{subsec:grid-choice} small enough that also $r_{\max}Jh_{\mathrm{grid}}\le\gamma_{\mathrm{lb}}/64$, where $J$ counts the local terms.
The grid stays polynomial in size.
For an edge $e:u\to v$ and a duration $T$, the \emph{sweep} of $e$ is the Schr\"odinger evolution under
\begin{equation}
    H(u)+g(t/T)\bigl(H(v)-H(u)\bigr),
    \qquad
    0\le t\le T,
    \label{eq:edge-sweep}
\end{equation}
where $g:[0,1]\to[0,1]$ is a fixed increasing function with $g(0)=0$ and $g(1)=1$.
For accurate sweeps, we also require that all derivatives of $g$ vanish at $0$ and $1$ and satisfy $|g^{(k)}|\le A^{k+1}(k!)^2$ for a constant $A$.
Ref.~\cite{ge2016rapid} gives an example.
If $H(x)$ is affine in $x$, as for the models of Appendix~\ref{app:numerics}, this is the parameter sweep along $e$.

\begin{lemma}[Sweeping the lowest $r$ levels across a gapped edge]
\label{lem:sweep-transport}
There is a constant $c$, depending only on $g$, with the following property.
Let $e:u\to v$ be a grid edge with $\gamma_r(u)\ge\Delta\ge\gamma_{\mathrm{lb}}/2$, let $0<\epsilon<1$, and let $U$ be the sweep of $e$ with duration $T\ge c\,\Delta^{-1}\log^2(c/\epsilon)$.
Then
\begin{equation}
    \|U^\dagger P_r(v)U-P_r(u)\|\le\epsilon.
    \label{eq:sweep-transport}
\end{equation}
\end{lemma}

\begin{proof}
Write $\delta H=H(v)-H(u)$.
By Eq.~\eqref{eq:local-term-bounds}, $r\|\delta H\|\le r_{\max}Jh_{\mathrm{grid}}\le\Delta/32$.
By Weyl's inequality, the lowest $r$ levels of $H(u)+s\,\delta H$ have band gap at least $15\Delta/16$ for every $s\in[0,1]$.

Let each of $r$ copies of the system evolve under the Hamiltonian in Eq.~\eqref{eq:edge-sweep}.
The sweep then acts as $U^{\otimes r}$ on their antisymmetric subspace.
There, the ground state is the Slater determinant of the lowest $r$ levels.
The next level lies a band gap higher.

Theorem~1 in the Supplemental Material of Ref.~\cite{ge2016rapid} is an adiabatic theorem for a gapped ground state of zero energy, so we shift the Hamiltonian of the $r$ copies by its ground-state energy.
This energy is the sum of the lowest $r$ eigenvalues of $H(u)+s\,\delta H$ at $s=g(t/T)$.
The theorem also needs bounds on all derivatives of this Hamiltonian in $t/T$.
Since $r\|\delta H\|$ is small compared with the band gap, this eigenvalue sum is analytic in $s$ within distance $7$ of $[0,1]$~\cite[Section~II.3]{kato1995perturbation}, where it changes by $O(r\|\delta H\|)$ per unit of $s$.
The theorem's Gevrey condition bounds the $k$-th derivative by a prefactor times $(k!)^{1+\alpha}$, up to a factor exponential in $k$.
With Cauchy's estimate and the bound on $g$, the derivatives in $t/T$ vanish at $0$ and $1$ and satisfy this condition with $\alpha=1$.
The prefactor is a large enough multiple of $\Delta$, as the proof of the theorem requires.
The theorem bounds the distance between the swept and final Slater determinants, up to a phase, by $c\exp\bigl(-\sqrt{T\Delta/c}\bigr)$, which is at most $\epsilon$ for the duration in the lemma.

Let $\varphi_1,\ldots,\varphi_r$ be the principal angles between the ranges of $UP_r(u)U^\dagger$ and $P_r(v)$.
The two Slater determinants then have overlap of modulus $\prod_j\cos\varphi_j$.
A distance at most $\epsilon$ up to a phase makes this modulus at least $1-\epsilon^2/2$.
Hence $\max_j\sin\varphi_j\le\epsilon$, which is Eq.~\eqref{eq:sweep-transport}.
\end{proof}

As in Lemma~\ref{lem:imperfect-transport}, a sweep therefore adds leakage at most $\epsilon$, so sweeps with $\epsilon\le\epsilon_U$ give the transport across one edge that Theorem~\ref{thm:finite-precision} needs.
For a constant cutoff $\Delta$ and $\epsilon=1/\mathrm{poly}(n)$, the duration is $O(\log^2n)$.

\subsubsection{Sweeps used for phase certification}

Phase certification needs the product of the sweeps along a chain to be close to an admissible circuit (Lemma~\ref{lem:linking}).
But a chain may cross polynomially many short edges, so the total duration of its sweeps need not be polylogarithmic.
We therefore run every other sweep backward in time, which requires that the sweeps of two consecutive edges both end, or both start, at their shared vertex.
Their common dynamics cancel, and only the small differences of the Hamiltonians along the path accumulate.

Along each axis, number the grid values in increasing order.
The two endpoints of an edge then differ by one in exactly one of these coordinate indices.
We orient each grid edge from its endpoint whose coordinate indices have an even sum to its other endpoint.
Each edge $e:u\to v$ is traversed forward by its sweep $U$ and backward by the \emph{inverse sweep} $U^\dagger$, which is the evolution of duration $T$ under $-\bigl[H(u)+g(1-t/T)\bigl(H(v)-H(u)\bigr)\bigr]$.
Since the endpoints of every edge have opposite parity, sweeps and inverse sweeps alternate along any grid path.
Like the rest of the dynamic program, the inverse sweep runs on a fault-tolerant quantum computer.
For symmetry-protected phases, this subsection takes the protecting symmetry to be unitary, because Schr\"odinger evolution does not commute with an anti-unitary symmetry of the Hamiltonian.

\begin{proposition}[Sweeps along a short grid path are close to an admissible circuit for large enough $R(n)$]
\label{prop:sweep-circuit}
Let $y_0,\ldots,y_\ell$ be a grid path of geometric length at most $L_*$ and let $T=\mathrm{polylog}(n)$.
Let $W$ be the product, in order, of the sweeps of duration $T$ of its edges, with the inverse sweep for each edge traversed against its orientation.
For every $k>0$, there is a circuit $C$ of depth $2^{\dlat}$ and gate diameter $\mathrm{polylog}(n)$, independent of $\ell$ and of the spectrum along the path, with $\|W-C\|\le n^{-k}$.
The circuit $C$ is admissible provided that $R(n)$ is no smaller than its gate diameter.
\end{proposition}

\begin{proof}
We group the factors of $W$ into consecutive pairs, leaving at most one factor unpaired.
The two sweeps $U_a$ and $U_b$ of a pair share an endpoint $y$.
The pair is $U_b^\dagger U_a$ if both sweeps end at $y$, and $U_aU_b^\dagger$ if both start at $y$.
Consider $U_b^\dagger U_a$, with $U_a$ from $a$ to $y$ and $U_b$ from $b$ to $y$; the other case is the same with time reversed.
Write $U_b(t)$ for $U_b$ up to time $t$.
The pair is then the evolution of duration $T$ generated by
\[
    U_b(t)^\dagger\bigl(1-g(t/T)\bigr)\bigl(H(a)-H(b)\bigr)U_b(t),
\]
the difference of the two sweep Hamiltonians in the interaction picture of $U_b$.
By Eq.~\eqref{eq:local-term-bounds}, each local term of $H(a)-H(b)$ has norm at most $\|a-b\|$, which is at most the total length of the two edges.
By the Lieb--Robinson bound~\cite{haah2018quantum}, conjugation by $U_b(t)$ spreads such a term over a distance $O(T+\log n)$, up to an error $n^{-k-2}$ times its norm.
We truncate every conjugated term to that distance with the normalized partial trace of Eq.~\eqref{eq:partial-trace-normalized}, which projects onto the operators supported there.
Since the edge lengths add up to at most $L_*$, this changes $W$ by $O(JTL_*n^{-k-2})$, which is at most $n^{-k}/2$.
After the truncation, the terms generating each pair have range $\mathrm{polylog}(n)$ and norms bounded by the length of its two edges.
We rescale the time of each pair by the length of its two edges.
Then $W$ becomes an evolution of total duration at most $(1+L_*)T$ under terms of norm at most one and range $\mathrm{polylog}(n)$.
The unpaired factor contributes duration $T$.
Such an evolution has a Lieb--Robinson light cone of polylogarithmic radius.
The block decomposition used in Proposition~\ref{prop:fdlu-from-gapped-path}~\cite{coser2019classification} turns it into a circuit of depth $2^{\dlat}$ with gates of diameter $\mathrm{polylog}(n)$, up to error $n^{-k}/2$.
If every $h_j(x)$ respects a unitary on-site symmetry, so does every gate, because the sweeps, the truncation and the block decomposition all preserve it.
\end{proof}

\subsubsection{Implementation of the full algorithm}

\begin{corollary}[Dynamic programming with sweeps]
\label{cor:dp-sweeps}
On the refined grid whose spacing also satisfies $r_{\max}Jh_{\mathrm{grid}}\le\gamma_{\mathrm{lb}}/64$, choose the precisions by Eq.~\eqref{eq:overall-precision}.
In Algorithm~\ref{alg:overall-dp}, replace each quasi-adiabatic propagator $U_\Delta(u\to v)$, $\Delta\in\mathcal D$, by the sweep of the edge with duration
\[
    T_\Delta=c\,\Delta^{-1}\log^2(8c/\epsilon_U),
\]
or by its inverse sweep if $u\to v$ is against the orientation, implemented with accuracy $\epsilon_U/4$.
Here $c$ is the constant of Lemma~\ref{lem:sweep-transport}.
Suppose that $R(n)$ is at least the gate diameter of Proposition~\ref{prop:sweep-circuit} for $T=T_{(1-\epsilon_{\mathrm{thresh}})\gamma_{\mathrm{thresh}}}=O(\log^2n)$.
Then the resulting algorithm satisfies Theorem~\ref{thm:detection} and Corollary~\ref{cor:components-with-reference-point}.
\end{corollary}

\begin{proof}
Sweeps are unitary, so every candidate state still satisfies the purity condition.
The accuracy $\epsilon_U/4$ gives Eq.~\eqref{eq:unitary-error}.
The proofs above use the quasi-adiabatic propagators only in exploration, in accepting gapped edges, in the circuit approximation of chains, and in the resource count.

\emph{Exploration.}
An edge of a path satisfying Eq.~\eqref{eq:reachable} has $\gamma_r\ge3\gamma_{\mathrm{lb}}/4$.
With $\Delta=\gamma_{\mathrm{lb}}/2$ and $\epsilon=\epsilon_U/8$, Lemma~\ref{lem:sweep-transport} therefore applies to the sweep replacing $U_{\gamma_{\mathrm{lb}}/2}(u\to v)$.
By unitary invariance, it also applies when this is an inverse sweep.
As in Lemma~\ref{lem:imperfect-transport}, the leakage of the candidate state increases by at most $\epsilon_U/8$.
Thus Eq.~\eqref{eq:implemented-leakage} holds with $\epsilon_U/8\le\epsilon_U$ in place of $a_\Delta(e)$.
The proof of Corollary~\ref{cor:large-gap} then applies unchanged.

\emph{Acceptance of gapped edges.}
The converse in Lemma~\ref{lem:accepted-edge} uses a unitary $U$ with $U\rho_r(u)U^\dagger=\rho_r(v)$ and $\|\widetilde U_{uv}-U\|\le\epsilon_U/2$.
If $\gamma_r>(1-\epsilon_{\mathrm{thresh}})\gamma_{\mathrm{thresh}}$ throughout the edge, Lemma~\ref{lem:sweep-transport} gives $\|U'P_r(u)U'^\dagger-P_r(v)\|\le\epsilon_U/8$ for the sweep $U'$ replacing $U_{(1-\epsilon_{\mathrm{thresh}})\gamma_{\mathrm{thresh}}}(u\to v)$.
Composing $U'$ with the direct rotation~\cite{davis1970rotation} from $U'P_r(u)U'^\dagger$ to $P_r(v)$ gives such a $U$ with $\|U-U'\|\le\sqrt2\,\epsilon_U/8$.
Hence $\|\widetilde U_{uv}-U\|\le\epsilon_U/4+\sqrt2\,\epsilon_U/8<\epsilon_U/2$, so the converse holds, and with it Lemma~\ref{lem:linking}(ii).

\emph{Chains.}
With sweeps, the proof of Lemma~\ref{lem:linking}(i) approximates the product $W$ of the exact sweeps along the chain by an admissible circuit $C$ with $\|W-C\|\le n^{-k_{\mathrm{circ}}}$.
Proposition~\ref{prop:sweep-circuit} provides this circuit, in place of Proposition~\ref{prop:fdlu-from-gapped-path}.
The rest of the proof is unchanged.

\emph{Resources.}
Since each sweep is evolution under a time-dependent local Hamiltonian for time at most $T_{\gamma_{\mathrm{lb}}/2}=\mathrm{poly}(n)$, the sweep, its inverse sweep and its controlled version can be implemented with accuracy $\epsilon_U/4$ by polynomially many gates~\cite{haah2018quantum}.
Hence Lemma~\ref{lem:implementation-complexity} and Corollary~\ref{cor:reference-copy-count} hold on this grid.
Theorem~\ref{thm:detection} and Corollary~\ref{cor:components-with-reference-point} then follow as in Subsection~\ref{app:main-algorithm-proof}.
\end{proof}

\section{Learning the Phase Diagram with a Classical Algorithm} \label{sec:classical-algorithm}
\label{subsec:classical-main-results}
This appendix proves that a deterministic classical algorithm can learn the phase diagram to constant precision $\gamma_{\mathrm{thresh}} = \Theta(1)$ (Theorem~\ref{thm:classical-classification}), under two additional assumptions, \emph{local visibility} and \emph{rank stability}.
The algorithm computes the filtered response classically.
This is possible because the quasi-adiabatic generator $D_\Delta$ of Definition~\ref{def:quasi-adiabatic-evolution} is quasi-local when $\Delta$ is constant.
For a transport path of length polylogarithmic in $n$, exact diagonalization on a patch of radius polylogarithmic in $n$ suffices to simulate the evolved probe approximately (Lemma~\ref{lem:evolution-truncation}).
The algorithm caps the length of the grid paths it explores, so every transport path is short.

Local visibility is needed because the classical algorithm uses $U_\Delta$ only for a constant cutoff $\Delta = O(1)$.
For this reason, it cannot rely on faithful transport through inverse-polynomial gap closings.
Theorem~\ref{lem:path-gradient-spectral-weight} guarantees that a failed transport is detectable, but using it requires the correct state along the path.
Without faithful transport, the classical algorithm cannot use that theorem.
Instead, before the gap falls below the cutoff $\Delta$, the algorithm must stop at the set $\mathcal S_{\mathrm{thresh}}$ of points whose gap lies just below the detection threshold $(1-\epsilon_{\mathrm{thresh}})\gamma_{\mathrm{thresh}}$ (Eq.~\eqref{eq:classical-threshold-shell}).
There, the gap closing must already be visible to some local probe.

\begin{assumption}[Local visibility]
\label{assumption:local-visibility}
Fix a margin $0<\epsilon_{\mathrm{thresh}}<1$ and constants $a>0$, $1-\epsilon_{\mathrm{thresh}}<\eta<1-\epsilon_{\mathrm{thresh}}/2$.
Define the region
\begin{equation}
    \mathcal S_{\mathrm{thresh}}
    :=
    \left\{
        x\in\mathcal M:
        \frac{(1-\epsilon_{\mathrm{thresh}})^2\gamma_{\mathrm{thresh}}}{\eta}
        \le
        \gamma(x)
        \le
        (1-\epsilon_{\mathrm{thresh}})\gamma_{\mathrm{thresh}}
    \right\}.
    \label{eq:classical-threshold-shell}
\end{equation}
Let $\delta_r=1/\mathrm{poly}(n)$.
For every $x\in\mathcal S_{\mathrm{thresh}}$, we require that
\begin{equation}
    \max_{
        x'\in B(x,\delta_r)\cap\mathcal M
    }
    \;
    \max_{O\in\mathcal L}
    W^O_{
        [\gamma(x'),\,
        \eta\gamma_{\mathrm{thresh}})
    }(x')
    \ge
    n^{-a},
    \label{eq:local-visibility}
\end{equation}
where $\mathcal L$ is the class of normalized local observables in Eq.~\eqref{eq:local-class}.
We also require that the \emph{same} radius $\delta_r$ satisfies
\begin{equation}
    L_\gamma\delta_r
    \le
    \frac{(1-\epsilon_{\mathrm{thresh}})^2\gamma_{\mathrm{thresh}}}{4\eta},
    \qquad
    L_S\delta_r
    \le
    \frac{n^{-a}}{8},
    \label{eq:local-visibility-radius}
\end{equation}
with $L_\gamma$ the Lipschitz constant of the gap and $L_S$ that of the filtered response for the kernel chosen below.
Both inequalities admit an inverse-polynomial $\delta_r$, since $L_\gamma, L_S = \mathrm{poly}(n)$ for constant kernel parameters.
A $\delta_r$ that gives visibility cannot simply be shrunk to meet these bounds, because visibility somewhere in a larger ball does not imply visibility in every smaller ball.
\end{assumption}

\emph{Parameters.} The response kernel $\mathcal K_{\Lambda,\delta}$ of Eq.~\eqref{eq:quantum-kernel} keeps the frequencies between $\delta$ and $\Lambda-\delta$ and removes those above $\Lambda+\delta$.
Assumption~\ref{assumption:local-visibility} supplies spectral weight
up to the energy \(\eta\gamma_{\mathrm{thresh}}\). We therefore choose the kernel parameters as
\begin{equation}
    \delta_{\mathrm{cl}}
    :=
    \min
    \left\{
        \frac{(1-\epsilon_{\mathrm{thresh}})^2\gamma_{\mathrm{thresh}}}{2\eta},
        \frac{
            \left[
                1-\epsilon_{\mathrm{thresh}}/2-\eta
            \right]
            \gamma_{\mathrm{thresh}}
        }{2}
    \right\},
    \quad
    \Lambda_{\mathrm{cl}}
    :=
    \eta\gamma_{\mathrm{thresh}}
    +
    \delta_{\mathrm{cl}}.
    \label{eq:classical-kernel-parameters}
\end{equation}
These parameters satisfy
\begin{equation}
    \Lambda_{\mathrm{cl}}-\delta_{\mathrm{cl}}
    =
    \eta\gamma_{\mathrm{thresh}},
    \qquad
    \Lambda_{\mathrm{cl}}+\delta_{\mathrm{cl}}
    \le
    \left(
        1-\frac{\epsilon_{\mathrm{thresh}}}{2}
    \right)
    \gamma_{\mathrm{thresh}}.
    \label{eq:classical-kernel-edges}
\end{equation}
Here the filtered response is that of the $d$ lowest levels.
Let $\gamma_d(x)$ be the band gap above them.
Lemma~\ref{lem:quantum-kernel} applies wherever $\gamma_d(x)\ge\delta_{\mathrm{cl}}$.
Hence, whenever
\begin{equation}
    \gamma_d(x)
    \ge
    \frac{(1-\epsilon_{\mathrm{thresh}})^2\gamma_{\mathrm{thresh}}}{2\eta},
\end{equation}
the lemma gives
\begin{equation}
    W^O_{
        [\gamma_d,\,
        \eta\gamma_{\mathrm{thresh}})
    }(x)
    \le
    S^O_{
        \mathcal K_{
            \Lambda_{\mathrm{cl}},
            \delta_{\mathrm{cl}}
        }
    }(x)
    \le
    W^O_{
        [\gamma_d,\,
        (1-\epsilon_{\mathrm{thresh}}/2)
        \gamma_{\mathrm{thresh}})
    }(x).
    \label{eq:classical-response-sandwich}
\end{equation}
In particular, the filtered response vanishes whenever
\begin{equation}
    \gamma_d(x)
    \ge
    \left(
        1-\frac{\epsilon_{\mathrm{thresh}}}{2}
    \right)
    \gamma_{\mathrm{thresh}}.
    \label{eq:classical-response-null}
\end{equation}

For Theorem~\ref{thm:classical-classification}, the quantities $\gamma_{\mathrm{thresh}}$, $\epsilon_{\mathrm{thresh}}$, and $\eta$ are constants, with
\begin{equation}
    1-\frac{\epsilon_{\mathrm{thresh}}}{2}-\eta
    =
    \Theta(1).
\end{equation}
We define
\begin{equation}
    L_S
    :=
    L_{
        S,
        \mathcal K_{
            \Lambda_{\mathrm{cl}},
            \delta_{\mathrm{cl}}
        }
    }
\end{equation}
as the Lipschitz constant of the corresponding filtered response, uniform over $O\in\mathcal L$.
Lemma~\ref{lem:filtered-response-stability} gives it on segments where the band gap $\gamma_d$ is at least $g_*=(1-\epsilon_{\mathrm{thresh}})\gamma_{\mathrm{thresh}}/2$.

The ground-state degeneracy $r(x)$ counts the levels within $\delta_{\mathrm{gsm}}$ of the ground-state energy (Definition~\ref{definition: ground-state manifold}).
Where $r(x)=d$, we have $\gamma_d(x)=\gamma(x)$.
We call the largest filtered response over $O\in\mathcal L$ the \emph{signal}.
It is zero whenever Eq.~\eqref{eq:classical-response-null} holds.
If $r(x)=d$ near a point of $\mathcal S_{\mathrm{thresh}}$, local visibility and the lower bound in Eq.~\eqref{eq:classical-response-sandwich} make the signal at least inverse-polynomial near that point.
The following theorem exploits this contrast to explore from a single reference point $x_{\mathrm{ref}}$.

The theorem also assumes rank stability, because the algorithm transports the $d=r(x_{\mathrm{ref}})$ lowest levels and does not estimate the ground-state degeneracy.
A splitting among these levels can lower the ground-state degeneracy, and hence change the phase, without producing a signal.

\begin{theorem}[Efficient classical algorithm for certified discovery of quantum phase diagrams to constant precision]
\label{thm:classical-classification}
Let $H(x)$, $x \in [-1, 1]^m$, be a smooth family of geometrically local Hamiltonians (Definition~\ref{definition: geometrically local hamiltonians}) given by its classical description, and assume local visibility (Assumption~\ref{assumption:local-visibility}) holds near the detection threshold. Let $x_{\mathrm{ref}}$ be a point with $\gamma(x_{\mathrm{ref}}) > \gamma_{\mathrm{thresh}}$ whose ground state $\rho_{\mathcal G}(x_{\mathrm{ref}})$ has local expectation values that can be computed classically with the cost in Eq.~\eqref{eq:pointwise-local-evaluability}. Fix a path-length bound $L_{\max}$. Assume \emph{rank stability}: the ground-state degeneracy equals $r(x_{\mathrm{ref}})$ on the connected component of $\{x : \gamma_{r(x_{\mathrm{ref}})}(x) \ge (1-\epsilon_{\mathrm{thresh}})^2\gamma_{\mathrm{thresh}}/(2\eta)\}$ that contains $x_{\mathrm{ref}}$, with $\eta$ as in Assumption~\ref{assumption:local-visibility}. Then there is a deterministic classical algorithm (Algorithm~\ref{alg:classical}) that covers the connected component of $\{x : \gamma(x) \ge \gamma_{\mathrm{thresh}}\}$ containing $x_{\mathrm{ref}}$, within path length $L_{\max}$ of $x_{\mathrm{ref}}$. It returns a set $\mathcal V$ of vertices of the refined grid of Subsection~\ref{subsec:grid-choice} such that
\begin{enumerate}
\item[(a)] $\mathcal V$ contains every grid point joined to $x_{\mathrm{ref}}$ by a path of length at most $L_{\max}$ inside that component;
\item[(b)] every vertex of $\mathcal V$ is joined to $x_{\mathrm{ref}}$ by a grid path of length at most $3mL_{\max}$ along which $\gamma > (1-\epsilon_{\mathrm{thresh}})\gamma_{\mathrm{thresh}}$.
\end{enumerate}
Count $x_{\mathrm{ref}}$ as an input point. For $L_{\max}=O(1)$ and every pair that contains $x_{\mathrm{ref}}$, linking $x_{\mathrm{ref}}$ to the input points in $\mathcal V$ satisfies clauses~(i) and~(ii) of Definition~\ref{def:phase-clustering}. These clauses use the constant resolution $\gamma_{\mathrm{lb}}=(1-\epsilon_{\mathrm{thresh}})\gamma_{\mathrm{thresh}}$ and, in clause~(ii), the path-length bound $L_{\max}$. The algorithm runs in time
\begin{equation}\label{eq:Tclass}
T_{\mathrm{class}}
 = \exp\!\left[ O\!\left(
 \bigl((1+L_{\max})\,\mathrm{polylog}(n^{4 + 2a},L_{\max})\bigr)^{\dlat}
 \right) \right],
\end{equation}
where $a$ is the exponent in Assumption~\ref{assumption:local-visibility} and $\dlat$ is the spatial dimension. In particular, $T_{\mathrm{class}}$ is quasipolynomial whenever $L_{\max}=\mathrm{polylog}(n)$.
\end{theorem}

Property~(a) gives clause~(ii). For $L_{\max}=O(1)$, property~(b) gives clause~(i), because, with its cutoff set to $(1-\epsilon_{\mathrm{thresh}})\gamma_{\mathrm{thresh}}$, Corollary~\ref{cor:gapped-path-same-phase} puts both ends of each grid path in the same phase. It applies because the grid path has constant length and the ground-state degeneracy cannot change where $\gamma>(1-\epsilon_{\mathrm{thresh}})\gamma_{\mathrm{thresh}}>\delta_{\mathrm{gsm}}$. The radius function $R(n)$ (Definition~\ref{def:admissible-circuits}) is taken large enough for paths of this length. Now run the algorithm with $L_{\max}=O(1)$ from one reference point satisfying the hypotheses of the theorem in each component of $\{x : \gamma(x) \ge \gamma_{\mathrm{thresh}}\}$ that contains input points. Suppose every two points of each such component are joined by a path of length at most $L_{\max}$ inside it. Then the union of the resulting links satisfies clauses~(i) and~(ii) for all pairs of input points.

For $r(x_{\mathrm{ref}})=1$, rank stability holds whenever $\delta_{\mathrm{gsm}}<(1-\epsilon_{\mathrm{thresh}})^2\gamma_{\mathrm{thresh}}/(2\eta)$, since $\gamma_1>\delta_{\mathrm{gsm}}$ forces $r=1$.

\label{sec:virtual-probes-trivial-closings}
\emph{Trivial gap closings.} Our classical algorithm also detects visible trivial gap closings that involve no bulk phase transition.
Such closings can be visible because $O$ ranges over all of $\mathcal{L}$, including probes that perturb the Hamiltonian outside the physical parameter space.

A simple example is an accidental conserved charge. Consider
\begin{equation}\label{eq:accidental-charge-model}
H(b_1, w_1) = b_1 Z_1 + w_1 Z_1 V + H',
\end{equation}
where $V$ and $H'$ act only on the other sites or, more generally, commute with $Z_1$. Then
\[
[H(b_1,w_1), Z_1] = 0,
\]
so the Hilbert space splits into the $Z_1 = \pm 1$ symmetry sectors, on which the Hamiltonian becomes
\[
H_\pm(b_1, w_1) = \pm b_1 + H' \pm w_1 V.
\]

With $e_\pm(w_1)$ the ground-state energies of $H' \pm w_1 V$, the lowest branches $\pm b_1 + e_\pm(w_1)$ cross whenever
\begin{equation}\label{eq:sector-crossing-condition}
2 b_1 = e_-(w_1) - e_+(w_1).
\end{equation}

There the two lowest levels may meet, but this is a crossing of disconnected symmetry sectors, not a bulk critical mode. If $H' \pm w_1 V$ have unique ground states, each gapped by more than $\delta_{\mathrm{gsm}}$, the ground-state degeneracy is two where the branch splitting is at most $\delta_{\mathrm{gsm}}$ and one elsewhere. Where the branch splitting exceeds $\delta_{\mathrm{gsm}}$, both sides of the crossing have ground-state degeneracy one. Definition~\ref{def:finite-size-phase} therefore compares them by admissible circuits. Let $\ket{g_\pm}$ be the ground states of $H' \pm w_1 V$. The $X_1$ matrix element between the two branches is the overlap of their states on the remaining sites, $\bra{1}\!\bra{g_-}\,X_1\,\ket{0}\!\ket{g_+} = \langle g_-|g_+\rangle$. So with $O = X_1$, the crossing meets Assumption~\ref{assumption:local-visibility} if the squared overlap obeys $|\langle g_-|g_+\rangle|^2 \ge n^{-a}$ and the branch splitting falls inside the energy window of Eq.~\eqref{eq:local-visibility}.

In contrast, the local quantum algorithms in Appendix~\ref{sec:local-detection} probe only $\partial_s H(s)$. Hence, these algorithms are less likely to detect a trivial closing that only probes outside the physical parameter space reveal.

\subsection{Preliminaries on Locality and Lieb--Robinson Bounds}\label{sec:definitions-locality}

In this appendix, as in~\cite{haah2018quantum}, every term $h_j$ has diameter at most $1$: if some term is larger, we divide all lattice distances by the largest term diameter.
For geometrically local Hamiltonians, this shrinks distances by at most a constant factor.
Radii quoted from Appendix~\ref{sec:appendix-A}, such as the probe range $\ell$, remain valid upper bounds.

The patch radius rests on a Lieb--Robinson bound (Lemma~\ref{lem:LR-quasi-adiabatic}) for the quasi-adiabatic propagator $U_\Delta(s)$, and applying this bound requires splitting the generator $D_\Delta(s)$ into terms supported on balls.

\subsubsection{Decomposing the generator into terms supported on balls}\label{sec:E-setup}

As in BMNS~\cite[Sec.~4]{bachmann2011automorphic}, we split $D_\Delta(s)$ into pieces supported on balls of growing radius around each Hamiltonian term. For each Hamiltonian term $h_j$ and each integer $m \ge 0$, we set
\begin{equation}\label{eq:Psi-decomp}
D_\Delta(s) = \sum_{j}\sum_{m \ge 0}\Psi_{\Delta,j,m}(s),
\qquad
\Psi_{\Delta,j,m}(s) := i\!\int_{-\infty}^{\infty}\!\! F_\Delta(t)\,
\bigl(\mathcal E_{B_m(j)} - \mathcal E_{B_{m-1}(j)}\bigr)\!
\bigl(e^{itH(s)}\partial_s h_j(s)\,e^{-itH(s)}\bigr)\,dt,
\end{equation}
with the convention $\mathcal E_{B_{-1}(j)} := 0$. Here, $B_m(j)$ is the ball of radius $m$ around the support of $h_j$, and $d(j,\cdot)$ denotes the distance from that support. The map $\mathcal E_Y$ is the normalized partial trace of Eq.~\eqref{eq:partial-trace-normalized}. Each term $\Psi_{\Delta,j,m}$ has support in $B_m(j)$, and summing over $m$ recovers the exact local generator $D_{\Delta,j}(s)$ of Eq.~\eqref{eq:DDelta-j-exact}. For a region $Z$, let $\Psi_{\Delta,Z}(s)$ be the sum of terms with $B_m(j) = Z$.

\begin{remark}
For a region $Y \subseteq \Lambda$ of the lattice and an operator $A$ acting on the full Hilbert space, the normalized partial trace onto $Y$ is defined as
\begin{equation}\label{eq:partial-trace-normalized}
\mathcal E_Y(A) := \frac{1}{\dim(\mathcal H_{\Lambda\setminus Y})}\,
   \mathrm{Tr}_{\Lambda\setminus Y}(A) \otimes I_{\Lambda\setminus Y}.
\end{equation}
This operation is contractive ($\|\mathcal E_Y(A)\| \le \|A\|$) and
fixes the algebra $\mathcal A_Y$ of operators supported on $Y$.
\end{remark}

\subsubsection{Lieb--Robinson bounds}\label{sec:R-LR}

\begin{lemma}[Lieb--Robinson bound for local Hamiltonian evolution~\cite{haah2018quantum}]\label{lem:LR-strict}
Let $H = \sum_j h_j$ be a geometrically local Hamiltonian (Definition~\ref{definition: geometrically local hamiltonians}) on a lattice $\Lambda$. Let $O_X$ have support in a subregion $X \subseteq \Lambda$, and let $\Omega \subseteq \Lambda$ be a region containing $X$. Define the boundary distance $\ell = \lfloor d(X, \Lambda\setminus\Omega)\rfloor$. Write $U^H_t = e^{-itH}$ and $U^{H_\Omega}_t = e^{-itH_\Omega}$, where $H_\Omega$ is the sum of the terms $h_j$ supported in $\Omega$. Then
\begin{equation}\label{eq:LR-strict}
\bigl\|(U^H_t)^\dagger O_X U^H_t - (U^{H_\Omega}_t)^\dagger O_X U^{H_\Omega}_t\bigr\|
 \;\le\; |X|\,\|O_X\|\,\frac{(2\zeta_0|t|)^\ell}{\ell!},
\end{equation}
whenever $2\zeta_0|t| \le (\ell+1)/2$. Here $Z_j$ is the support of $h_j$. We set $\zeta_0 := \max_{p \in \Lambda}\sum_{j:\, Z_j \ni p}|Z_j|\,\|h_j\| = O(1)$. There exist constants $v_{\mathrm{LR}}, \mu_{\mathrm{LR}} > 0$ such that, whenever $\ell \ge v_{\mathrm{LR}}|t|$, the error in Eq.~\eqref{eq:LR-strict} is at most $O(|X|\,\|O_X\|\,e^{-\mu_{\mathrm{LR}}\ell})$.
\end{lemma}

\begin{proof}
As in the proof in~\cite{haah2018quantum}, the left side of Eq.~\eqref{eq:LR-strict} is at most $\tfrac12|X|\,\|O_X\|\sum_{k\ge\max\{\ell,1\}}(2\zeta_0|t|)^k/k!$. The factor $\tfrac12$ arises because terms crossing the boundary of $\Omega$ act on at least two sites.
When $2\zeta_0|t|\le(\ell+1)/2$, this series is at most twice its first term, which gives Eq.~\eqref{eq:LR-strict}; for $\ell=0$ this uses $2\zeta_0|t|\le1$.
With $v_{\mathrm{LR}}=2e^2\zeta_0$, the condition $\ell\ge v_{\mathrm{LR}}|t|$ implies $2\zeta_0|t|\le(\ell+1)/2$, and $\ell!\ge(\ell/e)^\ell$ gives the second claim with $\mu_{\mathrm{LR}}=1$.
\end{proof}

The propagator $U_\Delta(s)$ arises from a quasi-local $D_\Delta(s)$ rather than from a geometrically local Hamiltonian, so Lemma~\ref{lem:LR-strict} does not apply.
To prove Lemma~\ref{lem:patched-generator-error}, we still need a Lieb--Robinson bound for this propagator.

\begin{definition}[Lieb--Robinson decay functions and norms~\cite{bachmann2011automorphic}]\label{def:LR-decay-functions}
In this definition, $x$, $y$ and $z$ denote lattice sites.
Given a geometrically local Hamiltonian $H(s)$, its quasi-adiabatic generator $D_\Delta(s)$ (Definition~\ref{def:quasi-adiabatic-evolution}), and a BMNS filter (Lemma~\ref{lem:BMNS-filter}), we define the following decay parameters:
\begin{enumerate}
\item[\textnormal{(i)}] \emph{Spatial decay function.} Let $v_a := \max\{2e\zeta_0, \Delta/2\}$. We set
$F_\Psi(r) := \tilde u_\mu\!\bigl(\tfrac{\Delta}{8 v_a} r\bigr)\,
   h\!\bigl(\tfrac{\Delta}{8 v_a} r\bigr)$,
where $h(r) = (1 + r)^{-(\dlat+1)}$. The function $\tilde u_\mu(w) := u_\mu(\max\{w, e^2\})$ regularizes the sub-exponential function $u_\mu(w) = e^{-\mu w/\ln^2 w}$, for a fixed $\mu \in (0, 2/7)$. This $F_\Psi$ is the decay function of~\cite[Sec.~4.1]{bachmann2011automorphic} with the cutoff $\Delta$ in place of the gap. The constant $v_a$ is a Lieb--Robinson velocity of $H(s)$. Indeed, by $\ell! \ge (\ell/e)^\ell$, the bound of Lemma~\ref{lem:LR-strict} is at most $|X|\,\|O_X\|\,(v_a|t|/\ell)^\ell$ for $\ell\ge1$.

\item[\textnormal{(ii)}] \emph{Weighted interaction norm.}
\begin{equation}\label{eq:Psi-norm-def}
\|\Psi_\Delta\|_{F_\Psi} := \sup_{x, y}\,\frac{1}{F_\Psi(d(x, y))}
   \sum_{Z: x, y \in Z}\|\Psi_{\Delta, Z}\|.
\end{equation}
Here $\|\Psi_{\Delta,Z}\|$ stands for $\sup_{s\in[0,1]}\|\Psi_{\Delta,Z}(s)\|$, as in the norm of~\cite[Theorem~4.6]{bachmann2011automorphic}.
\item[\textnormal{(iii)}] \emph{Geometric constant.}
\begin{equation}\label{eq:CFPsi-def}
C_{F_\Psi} := \sup_x \sum_y F_\Psi(d(x, y)).
\end{equation}

\item[\textnormal{(iv)}] \emph{Convolution constant.}
\begin{equation}\label{eq:CFPsi-conv-def}
C'_{F_\Psi} := \sup_{x, y}\,\frac{1}{F_\Psi(d(x, y))}
   \sum_z F_\Psi(d(x, z))\,F_\Psi(d(z, y)).
\end{equation}
The function $F_\Psi$ has the convolution property of Lieb--Robinson decay functions, because $\tilde u_\mu$ is non-increasing and its logarithm is superadditive~\cite{bachmann2011automorphic}.
Hence the weight $\tilde u_\mu$ does not increase the convolution constant of the polynomial base function $h(\tfrac{\Delta}{8 v_a}\,\cdot)$.
That convolution constant is at most $2^{\dlat+1}\sup_x\sum_y h\bigl(\tfrac{\Delta}{8 v_a}\,d(x,y)\bigr)$~\cite{nachtergaele2019quasi}.

\item[\textnormal{(v)}] \emph{Time growth function.} A function $g(s)$ satisfying
\begin{equation}\label{eq:gs-def}
C'_{F_\Psi}\!\cdot\! g(s) =
\begin{cases}
e^{2\|\Psi_\Delta\|_{F_\Psi} C'_{F_\Psi}|s|} - 1, & d(X, Y) > 0,\\
e^{2\|\Psi_\Delta\|_{F_\Psi} C'_{F_\Psi}|s|}, & \text{otherwise.}
\end{cases}
\end{equation}
Here $X$ and $Y$ are the supports of $A$ and $B$ in Lemma~\ref{lem:LR-quasi-adiabatic}.
\end{enumerate}
\end{definition}

\begin{lemma}[Generalized Lieb--Robinson bound for quasi-adiabatic evolution~\cite{bachmann2011automorphic}]\label{lem:LR-quasi-adiabatic}
Let $H(s)$ be geometrically local, and let $D_\Delta(s)$ be its quasi-adiabatic generator. Let $\alpha_s(A) := U_\Delta(s)^\dagger A\,U_\Delta(s)$. Whenever $0 \le s \le 1$, for every $A \in \mathcal A_X$, $B \in \mathcal A_Y$,
\begin{equation}\label{eq:LR-quasi-adiabatic}
\|[\alpha_s(A), B]\|
 \;\le\; 2\|A\|\,\|B\|\,\min\!\left\{1,\;
   g(s)\!\!\sum_{x \in X,\,y \in Y}\! F_\Psi(d(x, y))\right\},
\end{equation}
with $F_\Psi$, $g(s)$, $\|\Psi_\Delta\|_{F_\Psi}$, $C'_{F_\Psi}$ as in Definition~\ref{def:LR-decay-functions}.
\end{lemma}

\begin{proof}
\tolerance=2000 \emergencystretch=1.5em
The bound follows from~\cite[Theorem~4.6]{bachmann2011automorphic}, which needs no spectral gap.
The propagator $U_\Delta(s)$ satisfies $\partial_s U_\Delta(s) = -i\bigl(-D_\Delta(s)\bigr)U_\Delta(s)$, so in the convention $\tfrac{d}{dt}U = -iH(t)U$ of~\cite[Sec.~4.2]{bachmann2011automorphic} it is generated by the time-dependent interaction $-\Psi_\Delta(s)$.
The terms of this interaction (Eq.~\eqref{eq:Psi-decomp}) are supported on balls, and they are self-adjoint because the filter $F_\Delta$ is purely imaginary.
We apply the theorem with $F = F_\Psi$. This $F$ is positive and non-increasing.
Lemma~\ref{lem:qa-locality-constant-scaling} below shows that its constants $C_{F_\Psi}$, $C'_{F_\Psi}$ and the norm in Eq.~\eqref{eq:Psi-norm-def} are finite.
The constant $C'_{F_\Psi}$ is the convolution constant of~\cite[Theorem~4.6]{bachmann2011automorphic}.
The theorem then gives Eq.~\eqref{eq:LR-quasi-adiabatic} with the time growth function $g$.
We do not use~\cite[Theorem~4.5]{bachmann2011automorphic}, whose hypotheses include a uniform spectral gap (their Assumption~4.3).
\end{proof}

\subsection{Two-Stage Truncation of Local Observables}
\label{subsec:two-stage-truncation}
To simulate the quasi-adiabatic evolution by exact diagonalization on a patch, we truncate its generator in two stages.
We split the global generator into local generators, one for each term $h_j(s)$ in the Hamiltonian,
\begin{equation}\label{eq:DDelta-j-exact}
D_{\Delta,j}(s)
 := i\!\int_{-\infty}^{\infty}\!\!
   F_\Delta(t)\,e^{itH(s)}\,\partial_s h_j(s)\,e^{-itH(s)}\,dt,
\qquad
D_\Delta(s) = \sum_j D_{\Delta,j}(s).
\end{equation}
Consider a target observable $O = \sum_{i=1}^n O_i$ with at most $n$ local terms, each $O_i$ supported on a finite region $Z_i$.
The bounds track $|Z_i|$ explicitly because the observables we evolve later, which enter the filtered response, are supported on polylogarithmic patches.
There are two sources of error, one controlled by each stage of the truncation:
\begin{enumerate}
\item
\textit{Local Hamiltonian truncation (finite balls):}
Each exact local generator $D_{\Delta,j}(s)$ is replaced by a finite-region approximation:
\begin{equation}\label{eq:Djfilter-def}
    D_{\Delta,j}^\Omega(s) := i\!\int_{-\infty}^{\infty}\!\!
       F_\Delta(t)\,e^{itH_j^\Omega(s)}\,\partial_s h_j(s)\,e^{-itH_j^\Omega(s)}\,dt,
    \end{equation}
where $H_j^\Omega(s)$ is the sum of the Hamiltonian terms supported in the ball $B_\Omega(j)$.
Thus, $D_{\Delta,j}^\Omega(s)$ can be computed efficiently by exact diagonalization on $B_\Omega(j)$.
Lemma~\ref{lem:local-truncation} bounds the resulting local truncation error.
\item
\textit{Restriction to the patch $\Lambda_R$:}
The normalized partial trace $\mathcal E_{\Lambda_R}$ of Eq.~\eqref{eq:partial-trace-normalized} restricts the generator to a patch $\Lambda_R := B_R(Z_i)$ around $Z_i$, the support of the current $O_i$.
This truncation discards quasi-adiabatic interactions that leave the patch:
\begin{equation}\label{eq:DDelta-truncated}
    D_\Delta^{\Omega, R}(s)
     := \mathcal E_{\Lambda_R}\!\!\left(\sum_{j:\, B_\Omega(j) \cap \Lambda_R \ne \emptyset}\!
       D_{\Delta,j}^\Omega(s)\right).
    \end{equation}
These interactions still affect the observable by propagating from the boundary of $\Lambda_R$ to $Z_i$.
Lemma~\ref{lem:patched-generator-error} bounds the resulting error.
\end{enumerate}
Let $U_\Delta(s, u)$ and $U_\Delta^{\Omega, R}(s, u)$ be the propagators from $u$ to $s$ generated by $D_\Delta$ and $D_\Delta^{\Omega, R}$, respectively.
The exact and truncated Heisenberg evolutions of a local term $O_i$ are
\begin{equation}\label{eq:Heisenberg-pair}
O_{i,\Delta}(s, u) := U_\Delta(s, u)^\dagger O_i\,U_\Delta(s, u),
\quad
O_{i,\Delta}^{\Omega, R}(s, u) := U_\Delta^{\Omega, R}(s, u)^\dagger O_i\,U_\Delta^{\Omega, R}(s, u).
\end{equation}

\subsubsection{Local truncation error}\label{sec:local-truncation}

Each $h_j$ has parameter derivatives of norm at most one (Eq.~\eqref{eq:local-term-bounds}).
For a path of length $L$ (Definition~\ref{def:path}), the chain rule therefore gives the uniform bound $\|\partial_s h_j(s)\| = O(L)$.

\begin{lemma}[Spatial truncation of local generator terms]\label{lem:local-truncation}
Let $D_{\Delta,j}^\Omega(s)$ be as in Eq.~\eqref{eq:Djfilter-def}. Suppose that every local term $h_j(s)$ has support contained in a ball around $j$ of radius $r_0 = O(1)$.
Then for every $j$,
\begin{equation}\label{eq:local-truncation-bound}
\|D_{\Delta,j}(s) - D_{\Delta,j}^\Omega(s)\|
 \;\le\; \epsilon^{\Omega}
 := \frac{\sup_{j,s}\|\partial_s h_j(s)\|}{\Delta}\!\left[
   c_F\,|X_j|\,2^{-(\Omega - r_0 - 1)}
   + 2C_{\mathrm{tail}}\,F_{\mathrm{ball}}\!\Bigl(\frac{\Delta(\Omega - r_0 - 1)}{4 e \zeta_0}\Bigr)\right],
\end{equation}
where $X_j := \mathrm{supp}(\partial_s h_j(s))$ and $\zeta_0$ is the $O(1)$ constant from Lemma~\ref{lem:LR-strict}. The constants $c_F$, $C_{\mathrm{tail}}$ and the sub-exponentially decaying envelope $F_{\mathrm{ball}}$ of the filter tail are those of Lemma~\ref{lem:filter-constants}. Thus, $\epsilon^\Omega$ is sub-exponentially small in $\Omega$.
\end{lemma}

\begin{proof}
We write $O_j := \partial_s h_j(s)$ and $\ell_j^\Omega := \lfloor d(X_j, \Lambda\setminus B_\Omega(j))\rfloor \ge \Omega - r_0 - 1$.
Subtracting Eq.~\eqref{eq:Djfilter-def} from Eq.~\eqref{eq:DDelta-j-exact} gives
\[
D_{\Delta,j}(s) - D_{\Delta,j}^\Omega(s) = i\!\int_{-\infty}^{\infty}\!F_\Delta(t)\bigl(e^{itH(s)}O_j\,e^{-itH(s)} - e^{itH_j^\Omega(s)}O_j\,e^{-itH_j^\Omega(s)}\bigr)\,dt,
\]
and this time integral is split at $T := \ell_j^\Omega/(4e\zeta_0)$.
Such a split is standard for localizing the quasi-adiabatic generator~\cite{hastings2005quasiadiabatic,bachmann2011automorphic}.
For short times, $|t| \le T$: Lemma~\ref{lem:LR-strict} with $\ell! \ge (\ell/e)^\ell$ bounds the norm of the difference in the brackets by $|X_j|\,\|O_j\|\,2^{-\ell_j^\Omega}$.
The filter mass $c_F/\Delta$ of Lemma~\ref{lem:filter-constants} yields the first term of Eq.~\eqref{eq:local-truncation-bound}.
For long times, $|t| > T$: that norm is at most $2\|O_j\|$.
The filter tail of Lemma~\ref{lem:filter-constants} at $\xi = \Delta T \ge \Delta(\Omega - r_0 - 1)/(4e\zeta_0)$ yields the second term, because $F_{\mathrm{ball}}$ is non-increasing.
\end{proof}

\subsubsection{Patch truncation}\label{sec:macroscopic-truncation}
The next lemma bounds the total error of both stages, including a far-field error from the local generators $D_{\Delta,j}$ with $B_\Omega(j) \cap \Lambda_R = \emptyset$, which Eq.~\eqref{eq:DDelta-truncated} drops.

\begin{lemma}[Patch truncation error]\label{lem:patched-generator-error}
Let $O_i$ be an observable supported on a finite region $Z_i$. Let $\Lambda_R := B_R(Z_i)$ be the simulation patch of radius $R$ around $Z_i$. Let $O_{i,\Delta}(s)$ and $O_{i,\Delta}^{\Omega,R}(s)$ be the exact and truncated Heisenberg evolutions along the path from $0$ to $s$, respectively. Then for all $s \in [0, 1]$,
\begin{equation}\label{eq:macroscopic-truncation-bound}
\|O_{i,\Delta}(s) - O_{i,\Delta}^{\Omega,R}(s)\|
 \;\le\;E_{LR}(s) + E_{\text{ball}}(s),
\end{equation}
where
\begin{enumerate}
\item[\textnormal{(i)}]
\emph{Error from restricting the generator to $\Lambda_R$:}
\begin{equation}\label{eq:LRR-def}
   E_{LR}(s) := 8\|O_i\|\,|Z_i|\,\|\Psi_\Delta\|_{F_\Psi} C_{F_\Psi}
       \!\left(\!\int_0^s\!\!
         \frac{e^{2\|\Psi_\Delta\|_{F_\Psi} C'_{F_\Psi}|s-u|}}{C'_{F_\Psi}}\,du\!\right)\!
       \sup_{x}\sum_{y:\, d(x,y) \ge R}
         F_\Psi\!\Bigl(\frac{d(x,y)}{2}\Bigr).
    \end{equation}
\item[\textnormal{(ii)}]
\emph{Finite-ball generator error:}
\begin{equation}\label{eq:eROmega-def}
    E_{\text{ball}}(s) := 2s\,\|O_i\| \Biggl(
    \underbrace{\sum_{j:\, B_\Omega(j) \cap \Lambda_R \ne \emptyset}\!\epsilon^\Omega}_{\textnormal{near-field error}}
       \;+\; \underbrace{\sum_{j:\, B_\Omega(j) \cap \Lambda_R = \emptyset}\!\epsilon^{\Omega_j(R)}}_{\textnormal{far-field error}}
    \Biggr),
    \qquad
    \Omega_j(R) := \lfloor d(j, \Lambda_R)\rfloor - 1,
    \end{equation}
where $\Omega > 0$ is the radius of the finite balls. Here $\epsilon^{(\cdot)}$ is the bound of Lemma~\ref{lem:local-truncation} on the local truncation error. Its superscript is the truncation radius.
\end{enumerate}
\end{lemma}

\begin{proof}
In the notation of Eq.~\eqref{eq:Heisenberg-pair}, $O_{i,\Delta}(s) = O_{i,\Delta}(s, 0)$ and $O_{i,\Delta}^{\Omega,R}(s) = O_{i,\Delta}^{\Omega,R}(s, 0)$.
Applying the Duhamel formula for two time-dependent evolutions~\cite{childs2019theory}, we have
\begin{equation}\label{eq:Oi-diff-integral}
\|O_{i,\Delta}(s) - O_{i,\Delta}^{\Omega,R}(s)\| \le \int_0^s\!\! \|[V(u),\, O_{i,\Delta}(s, u)]\|\,du,
\end{equation}
where $V(u) := D_\Delta(u) - D_\Delta^{\Omega,R}(u)$.

We split the generator mismatch as $V(u) = V_{\mathrm{LR}}(u) + V_{\mathrm{near}}(u) + V_{\mathrm{far}}(u)$.
First, restrict the exact generator to $\Lambda_R$ to obtain $D_{\Delta,R}(u) := \mathcal{E}_{\Lambda_R}(D_\Delta(u))$.
Terms of Eq.~\eqref{eq:Psi-decomp} with balls contained in the patch are fixed by $\mathcal E_{\Lambda_R}$ and cancel in the difference.
Define
\begin{align}
V_{\mathrm{LR}}(u) &:= D_\Delta(u) - D_{\Delta,R}(u) = \sum_{X \not\subseteq \Lambda_R}\bigl(\Psi_{\Delta,X}(u) - \mathcal E_{\Lambda_R}(\Psi_{\Delta,X}(u))\bigr), \label{eq:VLR-def}\\
V_{\mathrm{near}}(u) &:= \mathcal{E}_{\Lambda_R}\!\!\left(\sum_{j:\, B_\Omega(j) \cap \Lambda_R \ne \emptyset}\!\bigl(D_{\Delta,j}(u) - D_{\Delta,j}^\Omega(u)\bigr)\right), \label{eq:Vnear-def}\\
V_{\mathrm{far}}(u) &:= \mathcal{E}_{\Lambda_R}\!\!\left(\sum_{j:\, B_\Omega(j) \cap \Lambda_R = \emptyset}\!D_{\Delta,j}(u)\right). \label{eq:Vfar-def}
\end{align}

\emph{The $V_{\mathrm{LR}}$ term.} Write $B_X(u) := \Psi_{\Delta,X}(u) - \mathcal E_{\Lambda_R}(\Psi_{\Delta,X}(u))$. Each $B_X(u)$ is supported on the ball $X$, because the partial trace onto $\Lambda_R$ maps operators on $X$ to operators on $X \cap \Lambda_R \subseteq X$. By contractivity, $\|B_X(u)\| \le 2\|\Psi_{\Delta,X}(u)\|$. By the triangle inequality,
\begin{equation}\label{eq:VLR-triangle}
\|[V_{\mathrm{LR}}(u),\, O_{i,\Delta}(s, u)]\| \le \sum_{X \not\subseteq \Lambda_R} \|[B_X(u),\, O_{i,\Delta}(s, u)]\|.
\end{equation}
We apply the generalized Lieb--Robinson bound (Lemma~\ref{lem:LR-quasi-adiabatic}) with $A = O_i$ supported on $Z_i$ and $B = B_X(u)$ supported on $X$. Since $g(s-u) \le \exp\bigl(2\|\Psi_\Delta\|_{F_\Psi} C'_{F_\Psi} |s-u|\bigr) / C'_{F_\Psi}$ by Definition~\ref{def:LR-decay-functions}(v), we obtain
\begin{equation}\label{eq:LR-each-term}
\|[B_X(u),\, O_{i,\Delta}(s, u)]\| \le 4\|O_i\|\,\|\Psi_{\Delta,X}(u)\|\,\frac{e^{2\|\Psi_\Delta\|_{F_\Psi} C'_{F_\Psi} |s-u|}}{C'_{F_\Psi}} \sum_{x \in X,\, z \in Z_i} F_\Psi(d(x, z)).
\end{equation}
Because $X \not\subseteq \Lambda_R = B_R(Z_i)$, any such set $X$ must contain at least one site $y$ satisfying $d(y, Z_i) \ge R$. Using the indicator bound $\mathbf{1}_{\{X \not\subseteq \Lambda_R\}} \le \sum_{y \in X} \mathbf{1}_{\{d(y, Z_i) \ge R\}}$, we reorder the summation over $X$:
\begin{equation}\label{eq:reorder-sum}
\sum_{X \not\subseteq \Lambda_R} \|\Psi_{\Delta,X}(u)\| \sum_{\substack{x \in X \\ z \in Z_i}} F_\Psi(d(x,z)) 
\le \sum_{z \in Z_i} \sum_{y:\, d(y,Z_i) \ge R} \sum_x F_\Psi(d(x,z)) \sum_{X:\, x,y \in X} \|\Psi_{\Delta,X}(u)\|.
\end{equation}
By Eq.~\eqref{eq:Psi-norm-def}, the innermost sum over $X$ is bounded by $\|\Psi_\Delta\|_{F_\Psi} F_\Psi(d(x, y))$. For the remaining sum over $x$, we use the triangle inequality: any lattice site $x$ must satisfy either $d(x, z) \ge d(z, y)/2$ or $d(x, y) \ge d(z, y)/2$. We split the sum over $x$ into these two regions. In each region, we bound the factor whose distance is at least $d(z,y)/2$ by $F_\Psi(d(z,y)/2)$, because $F_\Psi$ is non-increasing, and pull this factor outside the sum. The remaining sum of $F_\Psi$ over all lattice sites is uniformly bounded by $C_{F_\Psi}$ of Eq.~\eqref{eq:CFPsi-def}, yielding:
\begin{equation}\label{eq:spatial-triangle}
\sum_x F_\Psi(d(x,z)) F_\Psi(d(x,y)) \le 2 C_{F_\Psi} F_\Psi\!\left(\frac{d(z,y)}{2}\right).
\end{equation}
Substituting Eq.~\eqref{eq:spatial-triangle} into Eq.~\eqref{eq:reorder-sum}, we find that the spatial sums are bounded by
\begin{equation}\label{eq:after-triangle}
2\|\Psi_\Delta\|_{F_\Psi} C_{F_\Psi} \sum_{z \in Z_i} \sum_{y:\, d(y,Z_i) \ge R} F_\Psi\!\left(\frac{d(z,y)}{2}\right).
\end{equation}
As $d(y,Z_i) \ge R$ implies $d(z,y) \ge R$ for all $z \in Z_i$, we can replace $z$ by a free site $x$ and take the supremum over $x$, which bounds the $y$-sum uniformly in $z$. Summing over $z \in Z_i$ then gives a factor $|Z_i|$:
\begin{equation}\label{eq:VLR-final-bound}
\|[V_{\mathrm{LR}}(u),\, O_{i,\Delta}(s, u)]\| \le 8\|O_i\|\,|Z_i|\,\|\Psi_\Delta\|_{F_\Psi} C_{F_\Psi}\, \frac{e^{2\|\Psi_\Delta\|_{F_\Psi} C'_{F_\Psi} |s-u|}}{C'_{F_\Psi}} \sup_{x}\sum_{y:\, d(x,y) \ge R} F_\Psi\!\left(\frac{d(x,y)}{2}\right).
\end{equation}

\emph{The $V_{\mathrm{near}}$ and $V_{\mathrm{far}}$ terms.} We bound both remaining pieces with the following generic estimate, in which $V(u)$ stands for either piece:
\begin{equation}
\|[V(u),\, O_{i,\Delta}(s,u)]\| \le 2\,\|O_{i,\Delta}(s,u)\|\,\|V(u)\| = 2\|O_i\|\,\|V(u)\|.
\end{equation}
The equality holds because unitary Heisenberg evolution preserves the operator norm. We also use that $\mathcal{E}_{\Lambda_R}$ is contractive, $\|\mathcal{E}_{\Lambda_R}(\cdot)\| \le \|\cdot\|$.
For the near piece, the triangle inequality applied to Eq.~\eqref{eq:Vnear-def} and the bound $\epsilon^\Omega$ of Lemma~\ref{lem:local-truncation} give
\begin{equation}\label{eq:Vnear-bound}
\|V_{\mathrm{near}}(u)\| \le \sum_{j:\, B_\Omega(j) \cap \Lambda_R \ne \emptyset} \epsilon^\Omega .
\end{equation}
For the far piece, we fix $j$ with $B_\Omega(j) \cap \Lambda_R = \emptyset$. The ball $B_{\Omega_j(R)}(j)$, with $\Omega_j(R) = \lfloor d(j,\Lambda_R)\rfloor - 1$, is disjoint from $\Lambda_R$.
Therefore, the finite-ball generator $D_{\Delta,j}^{\Omega_j(R)}(u)$ is supported outside the patch.
This generator is also traceless, because the condition $\widetilde{F}_\Delta(0)=0$ of Definition~\ref{definition:filter-requirements} means that the filter integrates to zero over time.
An operator that is traceless and supported outside $\Lambda_R$ is sent to zero by the normalized partial trace onto $\Lambda_R$; therefore
\begin{equation}\label{eq:Vfar-single-j}
\|\mathcal{E}_{\Lambda_R}(D_{\Delta,j}(u))\|
= \|\mathcal{E}_{\Lambda_R}\bigl(D_{\Delta,j}(u)-D_{\Delta,j}^{\Omega_j(R)}(u)\bigr)\|
\le \epsilon^{\Omega_j(R)},
\end{equation}
by contractivity and Lemma~\ref{lem:local-truncation}.
Summing over all such $j$ gives $\|V_{\mathrm{far}}(u)\| \le \sum_{j:\, B_\Omega(j) \cap \Lambda_R = \emptyset} \epsilon^{\Omega_j(R)}$.

We combine the three commutator bounds in Eq.~\eqref{eq:Oi-diff-integral} and integrate over $u \in [0,s]$. The $V_{\mathrm{LR}}$ bound in Eq.~\eqref{eq:VLR-final-bound} integrates to $E_{LR}(s)$. The $V_{\mathrm{near}}$ and $V_{\mathrm{far}}$ bounds do not depend on $u$, so they integrate to $E_{\mathrm{ball}}(s)$ of Eq.~\eqref{eq:eROmega-def}.
\end{proof}

The error $E_{LR}(s)$ contains $\|\Psi_\Delta\|_{F_\Psi}$ and $C'_{F_\Psi}$ in the exponent.
The patch radius in Lemma~\ref{lem:evolution-truncation} requires the following lemma to control their growth in $L$ and $\Delta$.

\begin{lemma}[Scaling of the quasi-adiabatic locality constants]\label{lem:qa-locality-constant-scaling}
Let $x : [0, 1] \to \mathcal{M}$ be a path of length $L$ (Definition~\ref{def:path}), parameterized such that $|\dot{x}(s)| \le \zeta_s L$, where $\zeta_s = O(1)$ is a constant.
Let $H(s) := H(x(s))$ be the corresponding path in the space of geometrically local Hamiltonians (Definition~\ref{definition: geometrically local hamiltonians}).
Let $D_\Delta(s)$ be its quasi-adiabatic generator, constructed with a BMNS filter with energy cutoff $\Delta$.
Let $\Psi_\Delta(s)$ be the spatial interaction decomposition of this generator, with decay function $F_\Psi$.
Then its norm and the constants of Definition~\ref{def:LR-decay-functions} satisfy the following scalings,
\begin{equation}\label{eq:Psi-norm-scaling}
\|\Psi_\Delta(s)\|_{F_\Psi} = O\!\left(\frac{v_a^{\dlat+2}\,L}{\Delta^{\dlat+3}}\right), \qquad
C_{F_\Psi},\; C'_{F_\Psi} = O\!\left(\frac{v_a^{\dlat}}{\Delta^{\dlat}}\right),
\end{equation}
where the hidden constants depend only on $\dlat$, the lattice constants, $\zeta_s$, $\mu$, and the filter.
The first bound holds uniformly in $n$, even for the norm in Definition~\ref{def:LR-decay-functions}(ii), whose sum includes the supremum over $s \in [0,1]$, and hence for every fixed $s$.
\end{lemma}

\begin{proof}
Throughout, $c := \Delta/(2v_a)$, which is at most $1$ because $v_a = \max\{2e\zeta_0, \Delta/2\}$ (Definition~\ref{def:LR-decay-functions}(i)). For $b > 0$ we write $\mathcal F_b(u) := \exp\bigl(-b\,u/\log^2(e+u)\bigr)$. This function is non-increasing in $u \ge 0$, because the derivative of $u/\log^2(e+u)$ has the sign of $\log(e+u) - 2u/(e+u)$, whose minimum over $u \ge 0$ is $\ln 2$ at $u = e$. Lemma~\ref{lem:filter-constants} proves Eq.~\eqref{eq:filter-tail-bound} with $\mu_1 = 1/7$, so the bound holds with $\mathcal F_{1/7}$ in place of $F_{\mathrm{ball}}$. We name the $O(1)$ constants of Definition~\ref{definition: geometrically local hamiltonians}: each $h_j$ acts on at most $\bar s$ sites of a fixed set $S_j \subseteq B_{r_0}(p_j)$ around a centre site $p_j$. We also use the standard meaning of a lattice in $\dlat$ dimensions: every ball of radius $r$ contains at most $c_{\mathrm{vol}}(1+r)^{\dlat}$ sites. Since each site lies in the supports of $O(1)$ terms, such a ball also contains the centres of at most $c_{\mathrm{vol}}(1+r)^{\dlat}$ terms $h_j$. Lemma~\ref{lem:LR-strict} relies on this lattice structure for $\zeta_0 = O(1)$, and Lemma~\ref{lem:evolution-truncation} relies on it for its shell counts.

\emph{One term.} Fix $j$ and $s$ and let $A_j := \partial_s h_j(s)$, which is supported on $X_j \subseteq S_j$ with $\|A_j\| \le \zeta_s L$ by Eqs.~\eqref{eq:local-term-bounds} and~\eqref{eq:path-speed}.
The normalized partial traces are linear and bounded, so they commute with the time integral. Thus, $\Psi_{\Delta,j,m} = (\mathcal E_{B_m(j)} - \mathcal E_{B_{m-1}(j)})(D_{\Delta,j})$.
For $m = 0$, this implies $\|\Psi_{\Delta,j,0}\| \le \|D_{\Delta,j}\| \le c_F\|A_j\|/\Delta$.
For $m \ge 1$, let $\Omega := B_{m-1}(j)$ and let $D'$ be defined as $D_{\Delta,j}(s)$ in Eq.~\eqref{eq:DDelta-j-exact} where $H(s)$ is replaced by its restriction $H_\Omega(s)$ to the terms supported in $\Omega$.
Then $D' \in \mathcal A_\Omega$ is fixed by both partial traces. Hence,
\[
\Psi_{\Delta,j,m} = \bigl(\mathcal E_{B_m(j)} - \mathcal E_{B_{m-1}(j)}\bigr)\bigl(D_{\Delta,j} - D'\bigr),
\qquad
\|\Psi_{\Delta,j,m}\| \le 2\,\|D_{\Delta,j} - D'\|.
\]
Every site outside $\Omega$ is farther than $m-1$ from $X_j$, so $\ell := \lfloor  d(X_j, \Lambda\setminus\Omega)\rfloor \ge m-1$.
Splitting the time integral at $T := \ell/(4e\zeta_0)$ as in the proof of Lemma~\ref{lem:local-truncation}, we obtain
\[
\|D_{\Delta,j} - D'\| \le \frac{\|A_j\|}{\Delta}\Bigl[c_F\,\bar s\,2^{-\ell} + 2C_{\mathrm{tail}}\,\mathcal F_{1/7}(\Delta T)\Bigr].
\]
Because $4e\zeta_0 \le 2v_a$, we have $\Delta T \ge c\ell$, and because $c \le 1$, we have $2^{-\ell} \le e^{-c\ell/7} \le \mathcal F_{1/7}(c\ell)$.
With $\ell \ge m-1$ and the case $m = 0$, every term obeys
\begin{equation}\label{eq:Psi-term-bound}
\|\Psi_{\Delta,j,m}(s)\| \le C_1\,\frac{L}{\Delta}\,\mathcal F_{1/7}\bigl(c\,(m-1)_+\bigr),
\qquad
C_1 := 2\zeta_s\bigl(c_F\,\bar s + 2C_{\mathrm{tail}}\bigr),
\end{equation}
uniformly in $j$ and $s$.

\emph{Balls through two sites.}
Fix sites $x, y$. The triangle inequality gives
\[
\sum_{Z \ni x, y}\|\Psi_{\Delta,Z}\| \le \sum_{(j,m):\ x,y \in B_m(j)} \|\Psi_{\Delta,j,m}\|.
\]
If $x, y \in B_m(j)$, then $d(x,y) \le 2m + 2r_0$ and $d(x, p_j) \le m + r_0$.
Only $m \ge m_0 := \max\{0, \lceil d(x,y)/2 - r_0\rceil\}$ contribute. Each $m$ has at most $c_{\mathrm{vol}}(1+r_0+m)^{\dlat}$ indices $j$.
Set $k := (m-1)_+$. Then $1 + r_0 + m \le (2+r_0)(1+k) \le (2+r_0)(1+ck)/c$. Each $k$ is obtained for at most two values of $m$, and $k \ge k_0 := (m_0-1)_+$.
With $f(u) := (1+u)^{\dlat}\mathcal F_{1/7}(u)$, Eq.~\eqref{eq:Psi-term-bound} gives
\[
\sum_{Z \ni x,y}\|\Psi_{\Delta,Z}\| \le 2c_{\mathrm{vol}}(2+r_0)^{\dlat}C_1\,\frac{L}{\Delta}\,c^{-\dlat}\sum_{k \ge k_0} f(ck).
\]
The majorant $\hat f(u) := \sup_{u' \ge u} f(u')$ is non-increasing. Compare the sum to an integral and use $c \le 1$:
\[
\sum_{k \ge k_0} f(ck) \le \hat f(ck_0) + \frac1c\int_{ck_0}^\infty \hat f(u)\,du \le \frac{\widehat G(ck_0)}{c},
\qquad
\widehat G(u) := \hat f(u_+) + \int_{u_+}^\infty \hat f(u')\,du'.
\]
Since $\widehat G$ is non-increasing and $ck_0 \ge c\,d(x,y)/2 - r_0 - 1$, we have
\begin{equation}\label{eq:Psi-pair-sum}
\sum_{Z \ni x,y}\|\Psi_{\Delta,Z}\| \le 2c_{\mathrm{vol}}(2+r_0)^{\dlat}C_1\,\frac{L}{\Delta}\Bigl(\frac{2v_a}{\Delta}\Bigr)^{\dlat+1}\widehat G\Bigl(\frac{c\,d(x,y)}{2} - r_0 - 1\Bigr).
\end{equation}

\emph{Comparison with $F_\Psi$.} With $w := \Delta\,d(x,y)/(8v_a) = c\,d(x,y)/4$, Definition~\ref{def:LR-decay-functions}(i) gives $F_\Psi(d(x,y)) = \tilde u_\mu(w)(1+w)^{-(\dlat+1)}$ and the argument of $\widehat G$ in Eq.~\eqref{eq:Psi-pair-sum} is $2w - r_0 - 1$.
Dividing by $F_\Psi(d(x,y))$ and taking the supremum over $x, y$, we have
\[
\|\Psi_\Delta(s)\|_{F_\Psi} \le 2^{\dlat+2}c_{\mathrm{vol}}(2+r_0)^{\dlat}C_1\,q_\mu\,\frac{v_a^{\dlat+1}L}{\Delta^{\dlat+2}},
\qquad
q_\mu := \sup_{w \ge 0}\frac{\widehat G(2w - r_0 - 1)\,(1+w)^{\dlat+1}}{\tilde u_\mu(w)},
\]
and the same bound holds with the supremum over $s$ inside the sum of Eq.~\eqref{eq:Psi-norm-def}, because Eq.~\eqref{eq:Psi-term-bound} is uniform in $s$. The constant $q_\mu$ is finite because $\mu < 2/7$. To see this, fix $b \in (\mu/2, 1/7)$. Then $f(u) \le C_2(1+u)^{-2}\mathcal F_b(u)$ with $C_2 := \sup_{u \ge 0}(1+u)^{\dlat+2}\mathcal F_{1/7-b}(u) < \infty$. The right side is non-increasing, so $\widehat G(u) \le 2C_2\,\mathcal F_b(u)$ for $u \ge 0$. For large $w$, minus the logarithm of the ratio in $q_\mu$ is therefore at least $(2b - \mu - o(1))\,w/\log^2 w - (\dlat+1)\log(1+w) - \log(2C_2)$, which tends to infinity. On bounded ranges of $w$, the ratio is bounded.
Finally, $\Delta \le 2v_a$ gives $v_a^{\dlat+1}/\Delta^{\dlat+2} \le 2v_a^{\dlat+2}/\Delta^{\dlat+3}$, which is the first bound of Eq.~\eqref{eq:Psi-norm-scaling}.

\emph{The geometric constants.} Since $\tilde u_\mu \le 1$, writing $h(c\,d(x,y)/4)$ as $\int_{c\,d(x,y)/4}^\infty(-h'(u))\,du$ and exchanging the sum over $y$ with the integral gives
\[
C_{F_\Psi} \le \sup_x \int_0^\infty (\dlat+1)(1+u)^{-(\dlat+2)}\,\bigl|\{y : c\,d(x,y)/4 < u\}\bigr|\,du \le c_{\mathrm{vol}}(\dlat+1)\Bigl(\frac{8v_a}{\Delta}\Bigr)^{\dlat},
\]
using $|\{y: c\,d(x,y)/4 < u\}| \le c_{\mathrm{vol}}(1+4u/c)^{\dlat} \le c_{\mathrm{vol}}\,(4/c)^{\dlat}(1+u)^{\dlat}$ (as $c \le 1$), $\int_0^\infty (1+u)^{-2}\,du = 1$, and $4/c = 8v_a/\Delta$. By Definition~\ref{def:LR-decay-functions}(iv), $C'_{F_\Psi} \le 2^{\dlat+1}c_{\mathrm{vol}}(\dlat+1)(8v_a/\Delta)^{\dlat}$.
\end{proof}

\subsubsection{Patch radius for quasi-adiabatic evolution}\label{sec:evolution-radius}

\begin{lemma}[Patch radius]\label{lem:evolution-truncation}
Suppose the hypotheses of Lemma~\ref{lem:patched-generator-error} hold, where each term $O_i$ is supported on a region $Z_i$ of radius at most $r_Z$.
Fix a path $H(s) = H(x(s))$, $s \in [0, 1]$, of length $L$ in parameter space.
Let $O_\Delta(1) := \sum_{i=1}^n O_{i,\Delta}(1)$ be the exact evolution of the target observable. Let $O_\Delta^{\Omega, R}(1) := \sum_{i=1}^n O_{i,\Delta}^{\Omega, R}(1)$ be its fully truncated evolution.

For any target precision $\epsilon > 0$, choose the finite-ball radius $\Omega = \Theta(R)$ and a patch radius $R \ge r_Z$ with
\begin{equation}\label{eq:path-radius}
R \;=\; O\!\left(r_Z +
   \left(\frac{1}{\Delta} + \frac{L}{\Delta^{2\dlat+4}}\right)
   \mathrm{polylog}\!\left(\frac{n\,\Xi}{\epsilon},\,\frac{L}{\Delta^{2\dlat+3}}\right)\right),
\qquad
\Xi := \max_i \|O_i\|\,|Z_i|,
\end{equation}
to ensure a total simulation error of $\|O_\Delta(1) - O_\Delta^{\Omega, R}(1)\| \le \epsilon$.

Each truncated observable $O_{i,\Delta}^{\Omega, R}$ is supported on a patch of $\Theta((r_Z+R)^{\dlat})$ sites.

\end{lemma}

\begin{proof}
We apply the triangle inequality over the $n$ local terms,
\begin{equation}\label{eq:union-bound}
\|O_\Delta(1) - O_\Delta^{\Omega, R}(1)\| \le \sum_{i=1}^n \|O_{i,\Delta}(1) - O_{i,\Delta}^{\Omega, R}(1)\|
\le \sum_{i=1}^n \bigl( E_{LR}(1) + E_{\mathrm{ball}}(1) \bigr),
\end{equation}
where $E_{LR}$ and $E_{\mathrm{ball}}$ are the two sources of error in Lemma~\ref{lem:patched-generator-error}. Throughout this proof, $\Omega = \kappa R$ for some constant $\kappa > 0$, and $v_a$ and $\kappa$ are considered as $O(1)$. Here, $C, C', c, c'$ are positive constants that may change between lines and depend only on $\dlat$, $\kappa$ and the filter. They differ from $C'_{F_\Psi}$ and from $c = \Delta/(2v_a)$.

\emph{The error $E_{LR}(1)$.} We first control the spatial tail,
\begin{equation}\label{eq:S-R-def}
S(R) := \sup_x\sum_{y:\,d(x,y)\ge R}F_\Psi\!\left(\frac{d(x,y)}{2}\right).
\end{equation}

On a $\dlat$-dimensional lattice, the number of sites $y$ at distance $r$ from a fixed site $x$ is $O(r^{\dlat-1})$, so we group the terms into shells of radius $r \ge R$.
Inserting Definition~\ref{def:LR-decay-functions}(i) and comparing the sum to an integral with the change of variables $w=\frac{\Delta r}{16v_a}$ gives
\begin{equation}\label{eq:S-R-integral}
S(R) \le \frac{C}{\Delta^{\dlat}}\int_{w_0}^{\infty}w^{\dlat-1}\,\tilde{u}_\mu(w)\,(1+w)^{-(\dlat+1)}\,dw,
\qquad w_0 = \frac{\Delta(R-1)}{16v_a}.
\end{equation}

We use $w^{\dlat-1}(1+w)^{-(\dlat+1)}\le (1+w)^{-2}$ and the monotonicity of $\tilde{u}_\mu$. We also absorb the $O(1)$ shift between $w_0$ and $\frac{\Delta R}{16 v_a}$ into constants. Then
\begin{equation}\label{eq:S-R-final}
S(R) \le \frac{C}{\Delta^{\dlat}}\,\frac{\tilde{u}_\mu(c\,\Delta R)}{1 + c\,\Delta R}.
\end{equation}

Setting $z=x$ in Eq.~\eqref{eq:CFPsi-conv-def} gives $C'_{F_\Psi} \ge F_\Psi(0) = \Theta(1)$, so $1/C'_{F_\Psi} = O(1)$. We write $E_{LR}(1)$ of Eq.~\eqref{eq:LRR-def} in terms of $S(R)$ and use $\int_0^1 e^{A(1-u)}\,du \le e^A$ to obtain
\begin{equation}\label{eq:ELR-assembled}
E_{LR}(1)
\le C\,\|O_i\|\,|Z_i|\,\|\Psi_\Delta\|_{F_\Psi} C_{F_\Psi}\,
e^{2\|\Psi_\Delta\|_{F_\Psi} C'_{F_\Psi}}\, S(R),
\qquad
\|\Psi_\Delta\|_{F_\Psi} C_{F_\Psi},\;
\|\Psi_\Delta\|_{F_\Psi} C'_{F_\Psi}
= O\!\left(\frac{L}{\Delta^{2\dlat+3}}\right),
\end{equation}
where Lemma~\ref{lem:qa-locality-constant-scaling} ensures the scaling of the constants.

\emph{The finite-ball error $E_{\mathrm{ball}}(1)$.}
By Lemma~\ref{lem:local-truncation} and the bound~$\sup_{j,s}\|\partial_s h_j(s)\| = O(L)$, for sufficiently large $r$, the local truncation error at radius $r$ satisfies
\begin{equation}\label{eq:eps-r-envelope}
\epsilon^{r} \le C \frac{L}{\Delta}
\left[
2^{-(r - r_0 - 1)}
+
F_{\mathrm{ball}}\!\left(\frac{\Delta (r-r_0-1)}{4e\zeta_0}\right)
\right]
\le C' \frac{L}{\Delta}\, F_{\mathrm{ball}}(c\,\Delta r),
\end{equation}
where the second inequality uses a change of constants that lets the envelope $F_{\mathrm{ball}}$ of Lemma~\ref{lem:filter-constants} absorb the shift by $r_0+1$. For $\Delta = O(1)$, the envelope also absorbs the exponential term, which decays faster.

The near-field sum in Eq.~\eqref{eq:eROmega-def} has at most $O((r_Z+R+\Omega)^{\dlat}) = O((r_Z+R)^{\dlat})$ terms. Therefore,
\begin{equation}\label{eq-near-field}
\sum_{j:\,B_\Omega(j)\cap \Lambda_R\neq \emptyset}
\epsilon^\Omega
\le
C \frac{L}{\Delta}\, (r_Z+R)^{\dlat}\, F_{\mathrm{ball}}(c\,\Delta R).
\end{equation}
For the far-field sum, we group the indices $j$ according to their integer distance $m=\lfloor d(j,\Lambda_R)\rfloor$ to the patch.
There are $O((r_Z+R+m)^{\dlat-1})$ such indices at distance $m$, and the far-field condition requires $m\ge \Omega = \kappa R$.
Using Eq.~\eqref{eq:eps-r-envelope} with $\Omega_j(R) = m - 1$ and comparing the sum to an integral as in Eq.~\eqref{eq:S-R-integral}, we obtain
\begin{equation}\label{eq-far-field}
\begin{aligned}
\sum_{j:\,B_\Omega(j)\cap \Lambda_R=\emptyset}
\epsilon^{\Omega_j(R)}
&\le
C \frac{L}{\Delta}
\sum_{m\ge \kappa R}
(r_Z+R+m)^{\dlat-1}\,F_{\mathrm{ball}}(c\,\Delta(m-1))
\\
&\le
\frac{C'}{\Delta^{\dlat}}\,\frac{L}{\Delta}\,\bigl(1+\Delta (r_Z+R)\bigr)^{\dlat}\, F_{\mathrm{ball}}(c'\,\Delta R).
\end{aligned}
\end{equation}
Combining Eqs.~\eqref{eq-near-field} and~\eqref{eq-far-field} with the prefactor in Eq.~\eqref{eq:eROmega-def}, we obtain
\begin{equation}\label{eq:Eball-assembled}
E_{\mathrm{ball}}(1)
\le
C\,\|O_i\|\,\frac{L}{\Delta^{\dlat+1}}\,
\bigl(1+\Delta (r_Z+R)\bigr)^{\dlat}\, F_{\mathrm{ball}}(c\,\Delta R).
\end{equation}

\emph{Choice of the radius.} We insert the bounds on $E_{LR}(1)$ and $E_{\mathrm{ball}}(1)$ into the sum over the $n$ local terms in Eq.~\eqref{eq:union-bound}. We keep the support factor $\Xi = \max_i \|O_i\|\,|Z_i|$.
To put both error terms under a single envelope, we fix $\mu_1 \le \mu$ (without loss of generality) in the envelope $F_{\mathrm{ball}}$ defined in Lemma~\ref{lem:filter-constants}. Then $\tilde u_\mu(w) \le C\, F_{\mathrm{ball}}(w)$ for all $w \ge 0$. Hence, the total truncation error obeys
\begin{equation}\label{eq:total-truncation-bound}
\|O_\Delta(1) - O_\Delta^{\Omega, R}(1)\|
\;\le\;
M\, F_{\mathrm{ball}}(c\, \Delta R),
\qquad
M := C\, \frac{n L\,\Xi}{\Delta^{3\dlat+3}}
\exp\!\left(C'\,\frac{L}{\Delta^{2\dlat+3}}\right),
\end{equation}
where we used $R \ge r_Z$ to bound the polynomial factor $(1+\Delta(r_Z+R))^{\dlat}$ of Eq.~\eqref{eq:Eball-assembled} by $2^{\dlat}(1+\Delta R)^{\dlat}$. We then absorbed this factor into the sub-exponential envelope by lowering its constant $\mu_1$.
Thus, it suffices to choose $R$ such that $F_{\mathrm{ball}}(c\,\Delta R) \le \epsilon/M$.

To invert the envelope, set $\ell := \log(M/\epsilon)$ and $w := \frac{4}{\mu_1}\,\ell\,(\log \ell)^2$, where $\mu_1$ is the constant in $F_{\mathrm{ball}}$ after the lowering above.
For sufficiently large $\ell$, we have $\log(e+w) \le 2\log\ell$, so
\begin{equation}\label{eq:verify-inversion}
\mu_1\,\frac{w}{\log^2(e+w)}
\;\ge\;
\mu_1\,\frac{\tfrac{4}{\mu_1}\,\ell\,(\log\ell)^2}{4(\log\ell)^2}
= \ell,
\qquad\text{hence}\qquad
F_{\mathrm{ball}}(w) \le e^{-\ell} = \frac{\epsilon}{M}.
\end{equation}
Substituting $w = c\,\Delta R$ gives
\begin{equation}\label{eq:R-precise}
R \;\ge\; \frac{C}{\Delta}\,
\log\frac{M}{\epsilon}\,\Bigl(\log\log\frac{M}{\epsilon}\Bigr)^{2},
\qquad
\log\frac{M}{\epsilon}
= O\!\left(\frac{L}{\Delta^{2\dlat+3}} + \log\frac{n\,\Xi}{\epsilon} + \log\frac{1}{\Delta}\right),
\end{equation}
which holds by the scaling of $R$ in Eq.~\eqref{eq:path-radius}.
Finally, each truncated observable $O_{i,\Delta}^{\Omega,R}$ acts on $\Theta((r_Z+R)^{\dlat})$ sites.
\end{proof}

\subsubsection{Classical simulation of constant-cutoff transport}\label{sec:macroscopic-path}
The algorithm of Theorem~\ref{thm:classical-classification} transports observables along paths on the grid of Subsection~\ref{subsec:grid-choice}, at one constant cutoff $\Delta=\Theta(1)$ given in Eq.~\eqref{eq:classical-transport-cutoff} below.
Let $\pi$ be a piecewise-smooth grid path of total geometric length $L_\pi$, let $U_\Delta[\pi]$ be the quasi-adiabatic propagator along $\pi$ at cutoff $\Delta$, and let $A$ be an observable supported on $\Omega_A$ with radius $r_Z$.
Lemma~\ref{lem:evolution-truncation} truncates the pullback $U_\Delta[\pi]^\dagger A\, U_\Delta[\pi]$. Applying it to $A$ with initial support radius $r_Z$ and path length $L = L_\pi$ yields a patch radius $\widetilde R \ge r_Z$ with
\begin{equation}\label{eq:transport-truncation-radius}
\widetilde R \;=\;
O\!\left(r_Z +
   \left(\frac{1}{\Delta} + \frac{L_\pi}{\Delta^{2\dlat+4}}\right)
   \mathrm{polylog}\!\left(\frac{n\,\|A\|\,|\Omega_A|}{\epsilon},\,
   \frac{L_\pi}{\Delta^{2\dlat+3}}\right)\right),
\end{equation}
at which the truncated pullback $A^{\widetilde R}_{\pi}$ is supported on $B_{\widetilde R}(\Omega_A)$ and obeys $\|U_\Delta[\pi]^\dagger A\,U_\Delta[\pi] - A^{\widetilde R}_{\pi}\|
\le \epsilon$.

The time-evolved observables of the filtered response are only approximately local. Suppose therefore that $A$ has a localization error: $\|A - A'\| \le \epsilon_1$ for an observable $A'$ supported on $\Omega_A$, as in Corollary~\ref{cor:classical-reference-evaluation} below.
By the unitary invariance of the operator norm,
\begin{equation}\label{eq:transport-error-separation}
\|U_\Delta[\pi]^\dagger A\,U_\Delta[\pi] - (A')^{\widetilde R}_{\pi}\|
\;\le\; \epsilon_1 \;+\;
\|U_\Delta[\pi]^\dagger A'\,U_\Delta[\pi] - (A')^{\widetilde R}_{\pi}\|,
\end{equation}
so the localization and transport errors add.

\subsubsection{Evaluation of transported observables on the reference ground state}\label{sec:classical-EstimateSignal}
The truncated pullback $(A')^{\widetilde R}_{\pi}$ is supported on the patch $B_{\widetilde R}(\Omega_A)$ containing $\Theta\bigl((r_Z + \widetilde R)^{\dlat}\bigr)$ sites.
Computing its expectation value only requires the marginal of $\rho_{\mathcal G}(x_{\mathrm{ref}})$ on this patch.
Exact diagonalization on the patch yields the truncated operator, and the local evaluation assumed in Theorem~\ref{thm:classical-classification} gives its expectation value.
Each evaluation therefore adds the error of that local evaluation to the localization and transport errors.
When the transport is exact, these evaluations give the entries of the response matrix of Eq.~\eqref{eq:classical-path-pullback}, whose largest eigenvalue is the signal. Corollary~\ref{cor:classical-reference-evaluation} splits the precision budget over these entries.

\subsection{Proof of
Theorem~\ref{thm:classical-classification}}
\label{sec:proof-classical-quantum-classification}

We prove Theorem~\ref{thm:classical-classification} in two steps. First, we show that the response test is reliable whenever the reference ground state is transported exactly: a large gap at a grid point gives an estimated signal below $n^{-a}/2$ there, while a transport path that crosses the detection threshold $(1-\epsilon_{\mathrm{thresh}})\gamma_{\mathrm{thresh}}$ gives a grid point whose estimated signal exceeds $n^{-a}/2$. Second, we show that the search never accepts an edge that crosses the detection threshold, so every transport it performs is exact.

By Lemma~\ref{lem:optimize-filtered-response}, the signal at a grid point $x$ is the largest eigenvalue of the response matrix $M_{\mathcal K}(x)$.
The algorithm computes an approximation $\widehat M_{\mathcal K}(x)$ of this matrix, and we call its largest eigenvalue the estimated signal.
Algorithm~\ref{alg:classical} runs Dijkstra's shortest-path search on the grid from $x_{\mathrm{ref}}$, with edges weighted by their geometric length and ties broken by a fixed order of the grid edges.
To test an edge $(v,v')$ from an accepted vertex $v$, the search transports the reference ground state along the stored path $\pi_v$ followed by $(v,v')$ and evaluates the signal at $v'$. The edge is accepted if the estimated signal is below $n^{-a}/2$.
Every path is capped at length $3mL_{\max}$, and by Eq.~\eqref{eq:grid-approximation} the grid approximation of any path of length at most $L_{\max}$ fits within this cap.
Since edge lengths are positive, vertices are accepted in nondecreasing order of the length of their stored paths.
The kernel $\mathcal K$ has the parameters of Eq.~\eqref{eq:classical-kernel-parameters}.

\begin{algorithm}[!htbp]
\caption{Learning the phase diagram with a classical algorithm}
\label{alg:classical}
\scriptsize
\begin{algorithmic}[1]
\Require Classical description of $H(x)$, $x\in[-1,1]^m$; reference
point $x_{\mathrm{ref}}$ whose ground state has local expectation
values that can be computed classically; path-length bound $L_{\max}$.
\Ensure A set $\mathcal V$ of grid points, with the guarantee of
Theorem~\ref{thm:classical-classification}.
\State Form the grid of Subsection~\ref{subsec:grid-choice}.
\State Evaluate the estimated signal
$\lambda_{\max}\bigl(\widehat M_{\mathcal K}(x_{\mathrm{ref}})\bigr)$
and accept $x_{\mathrm{ref}}$ with the empty stored path.
\While{a new vertex $v$ is accepted with stored path $\pi_v$}
    \State Test each edge $(v,v')$ to a vertex $v'$ not yet accepted,
    provided that $\pi_v$ followed by $(v,v')$ has length at most
    $3mL_{\max}$: transport the reference ground state along this path with
    $U_\Delta$ (Eq.~\eqref{eq:classical-transport-cutoff}) and
    evaluate the signal at $v'$.
    \State Accept each tested edge if the estimated signal is below
    $n^{-a}/2$ (Eq.~\eqref{eq:classical-signal-threshold}); otherwise
    reject it.
    \State Among all accepted edges $(u,u')$ whose endpoint $u'$ is
    not yet accepted, if any, pick one that minimizes the length of
    $\pi_u$ followed by $(u,u')$; accept $u'$ and store this path as
    $\pi_{u'}$.
\EndWhile
\State \Return The set $\mathcal V$ of accepted vertices.
\end{algorithmic}
\end{algorithm}

\emph{Grid approximation.} The grid approximation $\widetilde x(s)$ of Subsection~\ref{subsec:grid-choice} to a path $x(s)$ stays within distance $h_{\mathrm{grid}}$ of $x(s)$.
The grid spacing is chosen to satisfy three bounds: $L_\gamma h_{\mathrm{grid}}\le\epsilon_{\mathrm{thresh}}\gamma_{\mathrm{thresh}}/2$ from Eq.~\eqref{eq:grid-threshold-condition}, a bound on the variation of the filtered response along an edge
\begin{equation}
    L_Sh_{\mathrm{grid}}
    \leq
    \frac{n^{-a}}{8},
    \label{eq:grid-local-visibility-signal}
\end{equation}
and a bound on the gap variation along an edge
\begin{equation}
    L_\gamma h_{\mathrm{grid}}
    \leq
    \frac{(1-\epsilon_{\mathrm{thresh}})^2\gamma_{\mathrm{thresh}}}{4\eta}.
    \label{eq:grid-local-visibility-gap}
\end{equation}

\emph{Transport at a constant cutoff.} The algorithm transports the reference ground state using the single cutoff
\begin{equation}
    \Delta
    :=
    \frac{(1-\epsilon_{\mathrm{thresh}})^2\gamma_{\mathrm{thresh}}}{4\eta}.
    \label{eq:classical-transport-cutoff}
\end{equation}
When the signal is evaluated at a grid point \(x\),
let \(\pi_x\) be the grid path from \(x_{\mathrm{ref}}\) to \(x\) along which this evaluation transports the reference ground state.
This path need not be the one that the search stores for \(x\).
We write
\begin{equation}
    U_{\mathrm{past}}(x)
    :=
    U_{\Delta}[\pi_x].
    \label{eq:classical-past-propagator}
\end{equation}
The
patch radius of Eq.~\eqref{eq:transport-truncation-radius} and the
runtime bounds below depend on the total geometric length
\(L_\pi\) of \(\pi_x\).
The length cap gives \(L_\pi \le 3mL_{\max}\) at every evaluation.

Whenever the band gap $\gamma_{r(x_{\mathrm{ref}})}$ along $\pi_x$ is at least $\Delta$ and $r(x)=r(x_{\mathrm{ref}})$, exact quasi-adiabatic transport yields
\begin{equation}
    U_{\mathrm{past}}(x)
    \rho_{\mathcal G}(x_{\mathrm{ref}})
    U_{\mathrm{past}}(x)^\dagger
    =
    \rho_{\mathcal G}(x).
    \label{eq:classical-exact-past-transport}
\end{equation}

Let $\mathcal P_\ell$ be the set of Pauli strings that act nontrivially on at most $\ell$ sites in a ball of radius $\ell$.
For $P,Q\in\mathcal P_\ell$, we define
\begin{equation}
    A_{P,Q}(t;x)
    :=
    \frac{i}{2}
    \left(
        [P_x(t),Q]+[Q_x(t),P]
    \right),
    \qquad
    P_x(t):=
    e^{iH(x)t}Pe^{-iH(x)t}.
    \label{eq:classical-response-entry-observable}
\end{equation}
By Eq.~\eqref{eq:filtered-response-matrix-time}, the response matrix integrates against the kernel the expectation values of these observables in the sector state $\rho_d(x)$ with $d=r(x_{\mathrm{ref}})$.
When Eq.~\eqref{eq:classical-exact-past-transport} holds, its entries pull back to the reference ground state:
\begin{align}
    [M_{\mathcal K}(x)]_{P,Q}
    &=
    \int_{-\infty}^{\infty}
    \mathcal K_{
        \Lambda_{\mathrm{cl}},
        \delta_{\mathrm{cl}}
    }(t)
    \operatorname{Tr}
    \left[
        \rho_{\mathcal G}(x_{\mathrm{ref}})
        U_{\mathrm{past}}(x)^\dagger
        A_{P,Q}(t;x)
        U_{\mathrm{past}}(x)
    \right]
    dt.
    \label{eq:classical-path-pullback}
\end{align}
Using Corollary~\ref{cor:classical-reference-evaluation}, we compute a matrix \(\widehat M_{\mathcal K}(x)\) satisfying
\begin{equation}
    \left\|
        \widehat M_{\mathcal K}(x)
        -
        M_{\mathcal K}(x)
    \right\|
    \le
    \frac{n^{-a}}{8},
    \label{eq:classical-matrix-precision}
\end{equation}
and reject the tested edge whenever
\begin{equation}
    \lambda_{\max}
    \left(
        \widehat M_{\mathcal K}(x)
    \right)
    \ge
    \frac{n^{-a}}{2}.
    \label{eq:classical-signal-threshold}
\end{equation}

\emph{Rank stability.} Every point reached below is joined to $x_{\mathrm{ref}}$ by a path on which $\gamma_{r(x_{\mathrm{ref}})}\ge(1-\epsilon_{\mathrm{thresh}})^2\gamma_{\mathrm{thresh}}/(2\eta)$.
Each gap bound below checks this with a Lipschitz estimate (Lemma~\ref{cor:gap-lipschitz}(iii)) and $\eta>1-\epsilon_{\mathrm{thresh}}$.
Rank stability therefore gives $r=r(x_{\mathrm{ref}})$ on such paths, and hence $\gamma=\gamma_{r(x_{\mathrm{ref}})}$ there.

\emph{Step 1: the test under exact transport.} In this step, we assume that every evaluation we consider satisfies Eq.~\eqref{eq:classical-exact-past-transport}; Step~2 shows that every evaluation of the search does.

\emph{The large-gap case.} Suppose that a path $x(s)$ from $x_{\mathrm{ref}}$ satisfies
\begin{equation}
    \gamma(x(s))
    \ge
    \gamma_{\mathrm{thresh}}
    \qquad
    \text{for every }s\in[0,1].
    \label{eq:classical-connected-promise}
\end{equation}
Along such a path, $\gamma\ge\gamma_{\mathrm{thresh}}>\delta_{\mathrm{gsm}}$, so the ground-state degeneracy stays $r(x_{\mathrm{ref}})$.
From the Lipschitz continuity of the gap and Eq.~\eqref{eq:grid-threshold-condition}, we have
\begin{equation}
    \gamma(\widetilde x(s))
    \ge
    \left(
        1-\frac{\epsilon_{\mathrm{thresh}}}{2}
    \right)
    \gamma_{\mathrm{thresh}}
    \qquad
    \text{for every }s\in[0,1].
    \label{eq:classical-connected-grid-gap}
\end{equation}
From Eq.~\eqref{eq:classical-response-null}, this implies
\begin{equation}
    S^O_{
        \mathcal K_{
            \Lambda_{\mathrm{cl}},
            \delta_{\mathrm{cl}}
        }
    }(x)
    =
    0
\end{equation}
at every vertex of the grid path, for every $O\in\mathcal L$. Hence,
\begin{equation}
    \lambda_{\max}
    \left(
        \widehat M_{\mathcal K}(x)
    \right)
    \le
    \frac{n^{-a}}{8}
    <
    \frac{n^{-a}}{2},
\end{equation}
so every evaluation at a vertex of the grid path is below $n^{-a}/2$.

\emph{Detection in the region $\mathcal S_{\mathrm{thresh}}$.} Suppose that the path of an evaluation reaches the region $\gamma\le(1-\epsilon_{\mathrm{thresh}})\gamma_{\mathrm{thresh}}$. By continuity, the path contains a point $x\in\mathcal S_{\mathrm{thresh}}$ with $\gamma(x)=(1-\epsilon_{\mathrm{thresh}})\gamma_{\mathrm{thresh}}$.
According to Assumption~\ref{assumption:local-visibility}, there exists a point $x'\in B(x,\delta_r)$ and an observable $O\in\mathcal L$ such that
\begin{equation}
    W^O_{
        [\gamma(x'),\,
        \eta\gamma_{\mathrm{thresh}})
    }(x')
    \ge
    n^{-a}.
    \label{eq:classical-visible-weight}
\end{equation}
By the Lipschitz continuity of the gap and Eq.~\eqref{eq:local-visibility-radius},
\begin{equation}
    \gamma(x')
    \ge
    \frac{(1-\epsilon_{\mathrm{thresh}})^2\gamma_{\mathrm{thresh}}}{\eta}
    -
    L_\gamma\delta_r
    \ge
    \frac{3(1-\epsilon_{\mathrm{thresh}})^2\gamma_{\mathrm{thresh}}}{4\eta}
    \ge
    \delta_{\mathrm{cl}}.
    \label{eq:classical-witness-gap}
\end{equation}
In particular, $\gamma(x')$ exceeds $(1-\epsilon_{\mathrm{thresh}})^2\gamma_{\mathrm{thresh}}/(2\eta)$.
The lower bound in Eq.~\eqref{eq:classical-response-sandwich} therefore applies at $x'$ and yields
\begin{equation}
    S^O_{
        \mathcal K_{
            \Lambda_{\mathrm{cl}},
            \delta_{\mathrm{cl}}
        }
    }(x')
    \ge
    n^{-a}.
    \label{eq:classical-witness-signal}
\end{equation}

Let $v$ be a grid point within distance $h_{\mathrm{grid}}$ of $x$, such as the closest one. Then,
\begin{equation}
    \|v-x'\|
    \le
    h_{\mathrm{grid}}+\delta_r.
\end{equation}
Every point of the segment from $x'$ to $v$ lies within $\max\{\delta_r,h_{\mathrm{grid}}\}$ of $x$. There, Eqs.~\eqref{eq:local-visibility-radius} and~\eqref{eq:grid-local-visibility-gap} give $\gamma_{r(x_{\mathrm{ref}})}\ge(1-\epsilon_{\mathrm{thresh}})\gamma_{\mathrm{thresh}}-(1-\epsilon_{\mathrm{thresh}})^2\gamma_{\mathrm{thresh}}/(4\eta)>(1-\epsilon_{\mathrm{thresh}})\gamma_{\mathrm{thresh}}/2$.
Since $L_S$ is defined with $g_*=(1-\epsilon_{\mathrm{thresh}})\gamma_{\mathrm{thresh}}/2$, the filtered response is $L_S$-Lipschitz on this segment.
By this Lipschitz bound and Eqs.~\eqref{eq:local-visibility-radius} and~\eqref{eq:grid-local-visibility-signal}, the filtered response at $v$ satisfies
\begin{align}
    S^O_{
        \mathcal K_{
            \Lambda_{\mathrm{cl}},
            \delta_{\mathrm{cl}}
        }
    }(v)
    &\ge
    S^O_{
        \mathcal K_{
            \Lambda_{\mathrm{cl}},
            \delta_{\mathrm{cl}}
        }
    }(x')
    -
    L_S(\delta_r+h_{\mathrm{grid}})
    \nonumber\\
    &\ge
    \frac{3}{4}n^{-a}.
    \label{eq:classical-grid-witness}
\end{align}
The signal at $v$ is at least the filtered response of this observable.
With the precision $n^{-a}/8$ of $\widehat M_{\mathcal K}(v)$, Eq.~\eqref{eq:classical-grid-witness} therefore gives
\begin{equation}
    \lambda_{\max}
    \left(
        \widehat M_{\mathcal K}(v)
    \right)
    \ge
    \frac{3}{4}n^{-a}
    -
    \frac{1}{8}n^{-a}
    =
    \frac{5}{8}n^{-a}
    >
    \frac{1}{2}n^{-a}.
\end{equation}

\emph{Step 2: every accepted edge stays above the detection threshold.} Step~1 guarantees accurate estimated signals only for exactly transported states, so we now show that every transport the search performs is exact. Recall that $d=r(x_{\mathrm{ref}})$, and call an edge \emph{good} if $\gamma_d>(1-\epsilon_{\mathrm{thresh}})\gamma_{\mathrm{thresh}}$ at every point of the closed edge. Suppose that some accepted edge is not good, and let $(v,v')$ be the first such edge in the order in which the search tests edges.

Every edge that the search accepts before it tests $(v,v')$ is good. Each edge of the stored path $\pi_v$ was tested and accepted before $v$ was accepted, and hence before $(v,v')$ was tested, so $\pi_v$ consists of good edges, or is empty if $v=x_{\mathrm{ref}}$. Thus $\gamma_d>(1-\epsilon_{\mathrm{thresh}})\gamma_{\mathrm{thresh}}$ along $\pi_v$ and at $v$, where for $v=x_{\mathrm{ref}}$ we use $\gamma_d(x_{\mathrm{ref}})=\gamma(x_{\mathrm{ref}})>\gamma_{\mathrm{thresh}}$. Along $(v,v')$, which has length at most $\delta x\le h_{\mathrm{grid}}$, Lemma~\ref{cor:gap-lipschitz}(iii) and Eq.~\eqref{eq:grid-local-visibility-gap} bound the variation of $\gamma_d$ by $L_\gamma h_{\mathrm{grid}}\le\Delta$, so every point of the closed edge satisfies
\begin{equation}
    \gamma_d
    \;\ge\;
    (1-\epsilon_{\mathrm{thresh}})\gamma_{\mathrm{thresh}}
    -
    \Delta
    \;>\;
    2\Delta,
    \label{eq:classical-gap-before-detection}
\end{equation}
where the last inequality is equivalent to $4\eta>3(1-\epsilon_{\mathrm{thresh}})$, which follows from $\eta>1-\epsilon_{\mathrm{thresh}}$.
Since also $(1-\epsilon_{\mathrm{thresh}})\gamma_{\mathrm{thresh}}>2\Delta$, the path $\pi_v$ followed by $(v,v')$ lies in the connected component of $\{x:\gamma_d(x)\ge2\Delta\}$ that contains $x_{\mathrm{ref}}$. This is the component in the rank-stability assumption, because $2\Delta=(1-\epsilon_{\mathrm{thresh}})^2\gamma_{\mathrm{thresh}}/(2\eta)$.
Rank stability gives $r=d$ on this path, and $\gamma_d>\Delta$ along it, so the quasi-adiabatic propagator of Eq.~\eqref{eq:classical-past-propagator} transports the ground-state manifold exactly.
Hence the evaluation at $v'$ that tests $(v,v')$ satisfies Eq.~\eqref{eq:classical-exact-past-transport}.

Since $(v,v')$ is not good and $\gamma_d(v)>(1-\epsilon_{\mathrm{thresh}})\gamma_{\mathrm{thresh}}$, continuity of $\gamma_d$ gives a point $x$ of the closed edge with $\gamma_d(x)=(1-\epsilon_{\mathrm{thresh}})\gamma_{\mathrm{thresh}}$. As $r=d$ on the edge, $\gamma(x)=\gamma_d(x)$, so $x\in\mathcal S_{\mathrm{thresh}}$.
Because $x$ lies on the closed edge, $\|v'-x\|\le h_{\mathrm{grid}}$, so by Step~1, applied to the grid point $v'$, this evaluation has estimated signal at least $\tfrac58n^{-a}>\tfrac12n^{-a}$, and $(v,v')$ is rejected, a contradiction.
Hence every accepted edge is good, and by rank stability, applied as above, $\gamma=\gamma_d>(1-\epsilon_{\mathrm{thresh}})\gamma_{\mathrm{thresh}}$ on it.
It follows that every evaluation of the search satisfies Eq.~\eqref{eq:classical-exact-past-transport}: the first one trivially, because it transports along the empty path, and every other one because it transports along a stored path of good edges followed by one tested edge, which satisfies Eq.~\eqref{eq:classical-gap-before-detection} by the argument above.

\emph{Properties~(a) and~(b).} Each accepted vertex $v$ is joined to $x_{\mathrm{ref}}$ by its stored path $\pi_v$, which has length at most $3mL_{\max}$ and consists of accepted edges, so Step~2 gives property~(b).
For property~(a), let a grid point $x$ be joined to $x_{\mathrm{ref}}$ by a path of length $L\le L_{\max}$ inside the component of $\{y:\gamma(y)\ge\gamma_{\mathrm{thresh}}\}$ containing $x_{\mathrm{ref}}$.
By Eq.~\eqref{eq:grid-approximation}, the grid approximation of this path has length at most $3mL\le3mL_{\max}$.
By Step~2, every evaluation satisfies Eq.~\eqref{eq:classical-exact-past-transport}, so the large-gap case puts every evaluation at its vertices below $n^{-a}/2$.
Walking along the grid approximation shows that each of its vertices is accepted with a stored path no longer than the grid approximation up to that vertex.
If a vertex is not yet accepted when its predecessor is accepted, the edge between them is tested within the cap and accepted.
The vertex is then accepted later, with a stored path no longer than the predecessor's stored path plus this edge.
Otherwise the vertex was accepted earlier, so its stored path is no longer than the predecessor's.
In particular, $x\in\mathcal V$.
Grid points in the region $(1-\epsilon_{\mathrm{thresh}})\gamma_{\mathrm{thresh}}<\gamma<\gamma_{\mathrm{thresh}}$, and large-gap vertices reached from $x_{\mathrm{ref}}$ only through this region or only by paths longer than $L_{\max}$, may or may not be returned.

\emph{Classical runtime.} For Theorem~\ref{thm:classical-classification}, the quantities $\gamma_{\mathrm{thresh}}$, $\Delta$, $\delta_{\mathrm{cl}}$, and $\Lambda_{\mathrm{cl}}$ are $\Theta(1)$.
Hence, Corollary~\ref{cor:classical-reference-evaluation} computes $\widehat M_{\mathcal K}(x)$ to precision $\epsilon=n^{-a}/8$ in time $\operatorname{poly}(n)\exp[O((R+\widetilde R)^{\dlat})]$.
The corollary uses the response radius $R$, the patch radius $\widetilde R$, the number $N_{\mathcal L}=|\mathcal P_\ell|=O(n)$ of Pauli strings, and the total mass $K_0=O(1)$ of $|\mathcal K|$.
The localizations $A'_{P,Q}(t;x)$ of that corollary have norm $O(1)$ and support on $O(R^{\dlat})$ sites.
Thus, for large $n$, every logarithm in the formulas for $R$ and $\widetilde R$ has argument $O(L_\pi)$ or at most $n^{4+2a}$.
The largest argument of the second kind, $O(n^{2+a}R^{\dlat})$, comes from the precision $\epsilon/(8N_{\mathcal L}K_0)$.
With $\Delta=\Theta(1)$ and $L_\pi\le3mL_{\max}$, this gives $R=\mathrm{polylog}(n^{4+2a})$ and
\[
    \widetilde R=O\bigl((1+L_{\max})\,\mathrm{polylog}(n^{4+2a},L_{\max})\bigr).
\]
Each accepted vertex tests at most $2m$ edges, and there are polynomially many grid points (Eq.~\eqref{eq:number-grid-points}), so the signal is evaluated polynomially many times.
Hence the total runtime obeys Eq.~\eqref{eq:Tclass}.

\subsection{Classical Simulation of the Filtered Response}
\label{sec:optimization-filtered-weight}

Evaluating the signal at a grid point $x$ requires $\widehat M_{\mathcal K}(x)$ to the precision of Eq.~\eqref{eq:classical-matrix-precision}.
We compute it classically, reusing the kernel $\mathcal{K}=\mathcal{K}_{\Lambda,\delta}$ of Eq.~\eqref{eq:quantum-kernel}.
We first treat the case where the local expectation values of the state at $x$ itself can be computed classically (Proposition~\ref{prop:classical-computation-filtered-response}).
We then evaluate the matrix from the reference ground state alone (Corollary~\ref{cor:classical-reference-evaluation}).

Fix a parameter point $x$.
The response matrix of Eq.~\eqref{eq:filtered-response-matrix-time} uses the maximally mixed sector state $\rho_d(x)$, and $\rho_d(x)=\rho_{\mathcal G}(x)$ in the phase learning application.
Suppose that the local expectation values of $\rho_d(x)$ can be computed classically.
That is, for every precision $\epsilon'>0$, every ball $B$ of radius $r$, and every operator $A$ supported in $B$, the expectation value $\operatorname{Tr}[\rho_d(x)A]$ can be computed up to additive error $\epsilon'$ in time
\begin{equation}
    \operatorname{poly}
    \left(
        n,\frac{1}{\epsilon'}
    \right)
    \exp
    \left[
        O\left(r^{\dlat}\right)
    \right].
    \label{eq:pointwise-local-evaluability}
\end{equation}

We truncate the time integrals that define $M_{\mathcal K}(x)$ at a time $T$.
Define the kernel masses:
\begin{equation}
    K_0
    :=
    \int_{-\infty}^{\infty}|\mathcal{K}(t)|\,dt,
    \qquad
    K_{\mathrm{tail}}(T)
    :=
    \int_{|t|>T}|\mathcal{K}(t)|\,dt.
    \label{eq:classical-kernel-masses}
\end{equation}

\begin{proposition}[Classical computation of \(M_{\mathcal K}(x)\)]
\label{prop:classical-computation-filtered-response}
Let $N_{\mathcal L}=|\mathcal P_\ell|=O(n)$. Fix $\epsilon\in(0,1)$. Choose $T>0$ such that
\begin{equation}
    K_{\mathrm{tail}}(T)
    \le
    \frac{\epsilon}{16N_{\mathcal L}}.
    \label{eq:classical-time-cutoff}
\end{equation}
There exists a constant $C_{\mathrm{loc}}>0$ such that it suffices to simulate on regions of radius
\begin{equation}
    R
    =
    C_{\mathrm{loc}}
    \left(
        v_{\mathrm{LR}}T
        +
        \log\frac{eN_{\mathcal L}K_0}{\epsilon}
    \right).
    \label{eq:classical-response-radius}
\end{equation}
The constant $C_{\mathrm{loc}}$ depends only on the locality parameters of the Hamiltonian family.
Under the assumption in Eq.~\eqref{eq:pointwise-local-evaluability}, there exists a deterministic classical algorithm that outputs a real symmetric matrix $\widehat M_{\mathcal K}(x)$ that satisfies
\begin{equation}
    \left\|
        \widehat M_{\mathcal K}(x)-M_{\mathcal K}(x)
    \right\|
    \le
    \epsilon
    \label{eq:classical-response-matrix-error}
\end{equation}
in time
\begin{equation}
    \operatorname{poly}
    \left(
        n,\frac{1}{\epsilon}
    \right)
    \exp
    \left[
        O\left(R^{\dlat}\right)
    \right].
    \label{eq:classical-response-runtime}
\end{equation}
The tail decay of the kernel $\mathcal{K}_{\Lambda,\delta}$ in Eq.~\eqref{eq:quantum-kernel-tail} allows us to choose
\begin{equation}
    T
    =
    \widetilde O
    \left(
        \frac{1}{\delta}
        \log\frac{N_{\mathcal L}}{\epsilon}
    \right),
    \qquad
    R
    =
    \widetilde O
    \left(
        \frac{1}{\delta}
        \log\frac{n}{\epsilon}
    \right).
    \label{eq:classical-response-radius-scaling}
\end{equation}
\end{proposition}

\begin{proof}
We bound the error entry by entry, truncating in time and in space.
Set
\begin{equation}
    \epsilon_{\mathrm{ent}}
    :=
    \frac{\epsilon}{N_{\mathcal L}}.
\end{equation}
Additive error $\epsilon_{\mathrm{ent}}$ in every entry of $M_{\mathcal K}(x)$ is sufficient, because
\begin{equation}
    \left\|
        \widehat M_{\mathcal K}(x)-M_{\mathcal K}(x)
    \right\|
    \le
    N_{\mathcal L}
    \max_{P,Q}
    \left|
        [\widehat M_{\mathcal K}(x)]_{P,Q}
        -
        [M_{\mathcal K}(x)]_{P,Q}
    \right|.
    \label{eq:entrywise-to-operator-error}
\end{equation}
Since $P$ and $Q$ are Pauli operators,
\begin{equation}
    \|[P_x(t),Q]\|\le 2,
    \qquad
    \|[Q_x(t),P]\|\le 2.
\end{equation}
Therefore, the part of the time integral in Eq.~\eqref{eq:filtered-response-matrix-time} with $|t|>T$ is bounded in magnitude by
\begin{equation}
    2K_{\mathrm{tail}}(T)
    \le
    \frac{\epsilon_{\mathrm{ent}}}{8}.
    \label{eq:classical-response-time-error}
\end{equation}
For $P,Q\in\mathcal P_\ell$, define their support distance
\begin{equation}
    d(P,Q)
    :=
     d
    \left(
        \operatorname{supp}P,
        \operatorname{supp}Q
    \right).
\end{equation}
The Lieb--Robinson bound (Lemma~\ref{lem:LR-strict}) gives constants $C_{\mathrm{LR}},\mu_{\mathrm{LR}},v_{\mathrm{LR}}>0$, independent of $n$, such that
\begin{align}
    \|[P_x(t),Q]\|
    +
    \|[Q_x(t),P]\|
    \le
    C_{\mathrm{LR}}
    \exp
    \left[
        -\mu_{\mathrm{LR}}
        \left(
            d(P,Q)-v_{\mathrm{LR}}|t|
        \right)
    \right]
    \label{eq:response-lieb-robinson}
\end{align}
whenever $d(P,Q)\ge v_{\mathrm{LR}}|t|$.
Combining this with the time tail, we have
\begin{align}
    \left|
        [M_{\mathcal K}(x)]_{P,Q}
    \right|
    \le
    C_{\mathrm{LR}}K_0
    \exp
    \left[
        -\mu_{\mathrm{LR}}
        \left(
            d(P,Q)-v_{\mathrm{LR}}T
        \right)
    \right]
    +
    2K_{\mathrm{tail}}(T).
    \label{eq:distant-response-entry}
\end{align}
Choose $C_{\mathrm{loc}}$ large enough in Eq.~\eqref{eq:classical-response-radius}. Then every pair with $d(P,Q)\ge 2R$ satisfies
\begin{equation}
    \left|
        [M_{\mathcal K}(x)]_{P,Q}
    \right|
    \le
    \frac{\epsilon_{\mathrm{ent}}}{2}.
\end{equation}
These entries are set to zero.

The remaining entries are those with $d(P,Q)<2R$. Let $\Omega_{P,Q}$ be the $R$-neighborhood of $\operatorname{supp}P\cup\operatorname{supp}Q$.
Let $H_{\Omega_{P,Q}}(x)$ be the part of $H(x)$ consisting of the Hamiltonian terms supported in this region. The restricted Heisenberg evolutions are
\begin{equation}
    P_{\Omega,x}(t)
    :=
    e^{iH_{\Omega_{P,Q}}(x)t}
    P
    e^{-iH_{\Omega_{P,Q}}(x)t},
    \qquad
    Q_{\Omega,x}(t)
    :=
    e^{iH_{\Omega_{P,Q}}(x)t}
    Q
    e^{-iH_{\Omega_{P,Q}}(x)t}.
\end{equation}
The reduced state is
\begin{equation}
    \rho_{d,\Omega_{P,Q}}(x)
    :=
    \operatorname{Tr}_{\Lambda\setminus\Omega_{P,Q}}
    \rho_d(x).
\end{equation}
We approximate the matrix entry by
\begin{align}
    [\widetilde M_{\mathcal K}(x)]_{P,Q}
    &:=
    \frac{1}{2}
    \int_{-T}^{T}
    \mathcal{K}(t)\,
    i\operatorname{Tr}
    \left[
        \rho_{d,\Omega_{P,Q}}(x)
        \left(
            [P_{\Omega,x}(t),Q]
            +
            [Q_{\Omega,x}(t),P]
        \right)
    \right]
    dt.
    \label{eq:truncated-response-entry}
\end{align}
The Lieb--Robinson bound yields, uniformly for $|t|\le T$,
\begin{align}
    \|P_x(t)-P_{\Omega,x}(t)\|
    +
    \|Q_x(t)-Q_{\Omega,x}(t)\|
    \le
    C_{\mathrm{LR}}
    \exp
    \left[
        -\mu_{\mathrm{LR}}
        \left(
            R-v_{\mathrm{LR}}T
        \right)
    \right].
    \label{eq:restricted-hamiltonian-error}
\end{align}
Consequently,
\begin{align}
    \left|
        [M_{\mathcal K}(x)]_{P,Q}
        -
        [\widetilde M_{\mathcal K}(x)]_{P,Q}
    \right|
    \le
    2K_{\mathrm{tail}}(T)
    +
    C_{\mathrm{LR}}K_0
    \exp
    \left[
        -\mu_{\mathrm{LR}}
        \left(
            R-v_{\mathrm{LR}}T
        \right)
    \right].
    \label{eq:response-entry-truncation-error}
\end{align}
Eqs.~\eqref{eq:classical-time-cutoff} and~\eqref{eq:classical-response-radius} ensure that this error is at most a fixed fraction of $\epsilon_{\mathrm{ent}}$.

We diagonalize $H_{\Omega_{P,Q}}(x)$, evaluate the finite-time integral in Eq.~\eqref{eq:truncated-response-entry} by numerical quadrature, and take the expectation values of the resulting local operators in $\rho_d(x)$.
Choosing the quadrature and local expectation value errors such that their sum is within the remaining fraction of $\epsilon_{\mathrm{ent}}$, we obtain
\begin{equation}
    \max_{P,Q}
    \left|
        [\widehat M_{\mathcal K}(x)]_{P,Q}
        -
        [M_{\mathcal K}(x)]_{P,Q}
    \right|
    \le
    \epsilon_{\mathrm{ent}}.
\end{equation}
Symmetrize the numerical matrix if necessary.
Eq.~\eqref{eq:entrywise-to-operator-error} implies Eq.~\eqref{eq:classical-response-matrix-error}.

The number of entries is at most $ N_{\mathcal L}^2=O(n^2) $.
Each retained entry is computed on $ O(R^{\dlat}) $ sites, so exact diagonalization costs $ \exp[O(R^{\dlat})] $.
All entry counts, quadrature costs, and matrix diagonalization costs contribute only polynomial factors.
This establishes Eq.~\eqref{eq:classical-response-runtime}.

For suitable constants $c, C > 0$, the tail bound of Eq.~\eqref{eq:quantum-kernel-tail} gives
\begin{equation}
    K_{\mathrm{tail}}(T)
    \le
    C
    \exp
    \left[
        -c
        \frac{
            \delta T
        }{
            \log^2(e+\delta T)
        }
    \right].
\end{equation}
Together with
\begin{equation}
    K_0
    \le
    \frac{2}{\pi}
    \log\frac{\Lambda}{\delta}
    +
    C
\end{equation}
from Lemma~\ref{lem:quantum-kernel-moments}, this gives the choices in Eq.~\eqref{eq:classical-response-radius-scaling}.
\end{proof}

The algorithm of Theorem~\ref{thm:classical-classification} evaluates local expectations only in the reference ground state, so Proposition~\ref{prop:classical-computation-filtered-response} does not apply directly.

\begin{corollary}[Classical evaluation of the pulled-back response
matrix on the reference ground state]
\label{cor:classical-reference-evaluation}
Fix a precision $\epsilon\in(0,1)$ and the cutoff $\Delta$
of Eq.~\eqref{eq:classical-transport-cutoff}. Let $\pi_x$ be a
grid path from $x_{\mathrm{ref}}$ to $x$ of total geometric
length $L_\pi$. Let
$U_{\mathrm{past}}(x) = U_\Delta[\pi_x]$. In this corollary, $M_{\mathcal K}(x)$ denotes the pulled-back response matrix of
Eq.~\eqref{eq:classical-path-pullback}. Assume that the reference
ground state $\rho_{\mathcal G}(x_{\mathrm{ref}})$ has local
expectation values that can be computed classically with the cost in
Eq.~\eqref{eq:pointwise-local-evaluability}. Choose $T$ as in
Eq.~\eqref{eq:classical-time-cutoff}. Choose the response radius $R$ as in
Eq.~\eqref{eq:classical-response-radius}. Choose the patch radius
$\widetilde R$ as in Eq.~\eqref{eq:transport-truncation-radius} for
the path $\pi_x$, the initial support radius $r_Z = 3(R + \ell) \le \widetilde R$, and
the precision $\epsilon/(8N_{\mathcal L}K_0)$. Then a deterministic classical
algorithm outputs a real symmetric matrix
$\widehat M_{\mathcal K}(x)$ satisfying
\begin{equation}
\bigl\|\widehat M_{\mathcal K}(x) - M_{\mathcal K}(x)\bigr\|
\le \epsilon,
\label{eq:classical-pullback-matrix-error}
\end{equation}
in time
\begin{equation}
\operatorname{poly}\!\left(n,\frac{1}{\epsilon}\right)
\exp\!\left[O\bigl((R+\widetilde R)^{\dlat}\bigr)\right],
\label{eq:classical-pullback-runtime}
\end{equation}
using expectation values only in the reference ground state. For
$\Delta, \gamma_{\mathrm{thresh}} = \Theta(1)$ and paths of
polylogarithmic geometric length, both radii are
$\mathrm{polylog}(n/\epsilon)$. The runtime is then
$\exp[\mathrm{polylog}(n/\epsilon)]$.
\end{corollary}

\begin{proof}
Fix an entry $(P,Q)$.

\emph{Time truncation and spatial truncation of the Hamiltonian evolution.}
We truncate the time integral to $|t|\le T$, which by Eq.~\eqref{eq:classical-time-cutoff} costs at most $2K_{\mathrm{tail}}(T) \le \epsilon/(8N_{\mathcal L})$ per entry.
We replace the observable $A_{P,Q}(t;x)$ in Eq.~\eqref{eq:classical-response-entry-observable} with its localization $A'_{P,Q}(t;x)$.
The localization uses the restricted Heisenberg evolutions $P_{\Omega,x}(t)$ and $Q_{\Omega,x}(t)$ of the proof of Proposition~\ref{prop:classical-computation-filtered-response} in place of $P_x(t)$ and $Q_x(t)$, and it is supported on the patch $\Omega_{P,Q}$.
By the Lieb--Robinson bound, $\|A_{P,Q}(t;x) - A'_{P,Q}(t;x)\| \le \epsilon/(8N_{\mathcal L}K_0)$ uniformly for $|t|\le T$.
The localization has norm $O(1)$, and its support lies within distance $r_Z = 3(R+\ell)$ of the center of the ball containing $\operatorname{supp}P$.
Indeed, both Pauli strings lie in balls of radius $\ell$, retained pairs lie within distance $2R$, and $\Omega_{P,Q}$ extends their supports by $R$.

\emph{Transport pullback with spatial truncation.}
We pull $A'_{P,Q}(t;x)$ back along $\pi_x$ as in Subsubsection~\ref{sec:macroscopic-path}, applying Lemma~\ref{lem:evolution-truncation} with this $r_Z$ and $\Xi = O(\|A'\|\,|\Omega_{P,Q}|)$.
For every $|t|\le T$, the lemma yields a truncated pullback $(A'_{P,Q})^{\widetilde R}_{\pi_x}(t)$ on $B_{\widetilde R}(\Omega_{P,Q})$ satisfying
\[
\bigl\|
U_{\mathrm{past}}(x)^\dagger A'_{P,Q}(t;x)\,U_{\mathrm{past}}(x)
- (A'_{P,Q})^{\widetilde R}_{\pi_x}(t)
\bigr\| \le \frac{\epsilon}{8N_{\mathcal L}K_0}.
\]
Computing this pullback involves a numerical integration along $\pi_x$.
By Eq.~\eqref{eq:transport-error-separation}, the localization and transport error budgets add.

\emph{Evaluation in the reference ground state.}
Each expectation value $\operatorname{Tr}[\rho_{\mathcal G}(x_{\mathrm{ref}})\,
(A'_{P,Q})^{\widetilde R}_{\pi_x}(t)]$ involves an operator supported on a ball of radius $r_Z + \widetilde R$.
Exact diagonalization on the patch yields the truncated operator.
One local evaluation as in Eq.~\eqref{eq:pointwise-local-evaluability}, at precision $\epsilon/(8N_{\mathcal L}K_0)$, suffices.

\emph{Error budget and matrix accuracy.}
Per retained entry, the time tail, the spatial truncation of the Hamiltonian evolution, the transport truncation, the evaluation, the numerical integration in $t$, and the numerical integration along $\pi_x$ each contribute an error of at most $\epsilon/(8N_{\mathcal L})$ after integrating against $|\mathcal K|$, whose total mass is $K_0$.
These six errors sum to at most $3\epsilon/(4N_{\mathcal L})$.
The error from setting entries with $d(P,Q)\ge 2R$ to zero is at most $\epsilon/(2N_{\mathcal L})$, as in Proposition~\ref{prop:classical-computation-filtered-response}.
Every entry of $\widehat M_{\mathcal K}(x)$ is therefore accurate to $\epsilon/N_{\mathcal L}$, and by Eq.~\eqref{eq:entrywise-to-operator-error} the operator-norm error is at most $\epsilon$.
There are $O(N_{\mathcal L}^2)$ entries and polynomially many quadrature times.
Each entry at each quadrature time requires one local evaluation and polynomially many diagonalizations on $\Theta((r_Z+\widetilde R)^{\dlat})$ sites, which gives the stated runtime.
\end{proof}

\section{Details of Numerical Simulations}
\label{app:numerics}

This appendix describes the models and numerical parameters of the three benchmarks of Section~\ref{sec:numerics-main}.
The algorithm receives only two inputs.
The first is the Hamiltonian family $H(x)$, $x \in [-1,1]^m$, of Definition~\ref{definition: geometrically local hamiltonians}.
The second is a single reference ground state $\rho_{\mathrm{ref}}$ at a gapped, exactly solvable reference point $x_{\mathrm{ref}}$, together with its gap $\gamma(x_{\mathrm{ref}})$.
The local expectation values of $\rho_{\mathrm{ref}}$ must be computable classically with the cost in Eq.~\eqref{eq:pointwise-local-evaluability}.
The algorithm receives no other physical information, such as an order parameter, a symmetry-breaking pattern, or an expected universality class.

At every grid point $x$, the filtered-response test of Algorithm~\ref{alg:gap_cert} is applied to detect whether the spectral gap $\gamma(x)$ falls below the threshold $\gamma_{\mathrm{thresh}}$.
When the ground-state manifold contains several levels, the spectral gap of Definition~\ref{definition: spectral gap} is the gap above the whole manifold, not the splitting within it.
With the margin $\epsilon_{\mathrm{thresh}}$, the resolution error band is the region where $(1-\epsilon_{\mathrm{thresh}})\gamma_{\mathrm{thresh}} \le \gamma(x) \le \gamma_{\mathrm{thresh}}$.
In this band, the filtered-response test may return either verdict.
Each benchmark figure shows this band.
All three benchmarks use $\epsilon_{\mathrm{thresh}} = 1/11$.
The threshold and grid spacing of each model are given below.

\subsection{Initialization from Solvable Points}
\label{subsec:numerics-solvable-points}

In practice, the most useful reference points are the solvable points of a phase, where the ground state is known explicitly.
From such a point, the algorithm explores the surrounding parameter space as far as gapped paths reach.

Many strongly correlated phases contain frustration-free points, where the ground states are exactly the states that minimize every local term of the Hamiltonian at once.
Examples with commuting terms are topological stabilizer codes such as the toric code (Subsection~\ref{subsec:toric-code-benchmark}) and symmetry-protected topological phases at their fixed points, such as the 1D cluster state.
Further frustration-free examples are the parent Hamiltonians of injective tensor networks, such as the AKLT model.
At $t=t'=0$, our spinless Fermi--Hubbard benchmark (Subsection~\ref{subsec:fermion-hubbard-benchmark}) is frustration-free and commuting.

In some models, a phase contains integrable points, where all eigenstates are known exactly, not only the ground states.
Examples are free boson and fermion models, the $XXZ$ model, and the Kitaev honeycomb model at zero magnetic field (Subsection~\ref{subsec:kitaev-honeycomb-benchmark}).

\subsection{Spinless Fermi--Hubbard Model on the Square Lattice}
\label{subsec:fermion-hubbard-benchmark}

The spinless Fermi--Hubbard model with next-nearest-neighbor hopping is an interacting fermion model with a fermion sign problem in quantum Monte Carlo (QMC).
Our algorithm treats interacting fermions directly, so the sign problem does not arise.
We expect the detected boundary to be the spontaneous symmetry-breaking transition between the charge-density-wave (CDW) phase and the translation-symmetric gapless phase.
The rest of this subsection supports this expectation.
We also expect this transition to be second order.

\emph{Model.}
We consider spinless fermions on an $N_x\times N_y$ square lattice with periodic boundary conditions (PBC), with annihilation and creation operators $c_i$ and $c_i^\dagger$ at site $i$.
The Hamiltonian in Eq.~\eqref{eq:fermion-hamiltonian-main} has nearest-neighbor (NN) hopping $t$, next-nearest-neighbor (NNN) hopping $t'$, and NN repulsion $V>0$.
We work at half filling, with $N/2$ fermions on $N=N_xN_y$ sites.
With PBC, the Hamiltonian is invariant under the lattice translations $T_x,T_y$.
We choose $N_x$ and $N_y$ even, so the lattice is bipartite, with two sublattices $A$ and $B$ of $N/2$ sites each.
Site $(i_x, i_y)$ belongs to $A$ when $i_x+i_y$ is even and to $B$ when it is odd.
The NN hopping $t$ connects opposite sublattices, whereas the NNN hopping $t'$ stays within one sublattice.
With NN hopping only ($t'=0$), the hopping is bipartite and admits sign-problem-free QMC formulations~\cite{huffman2014solution,li2015solving}.
Once $t'\neq 0$, the hopping is no longer bipartite and the fermion sign problem appears.

\emph{Exactly solvable reference point.}
At $t=t'=0$, the ground-state manifold is spanned by the two checkerboard CDW states:
\begin{equation}\label{eq:cdw-states}
|\psi_A\rangle = \prod_{i\in A} c_i^\dagger|0\rangle,\qquad
|\psi_B\rangle = \prod_{i\in B} c_i^\dagger|0\rangle.
\end{equation}
Each is a classical product state with one sublattice completely filled and the other empty, and both have zero energy.
They are ground states because the Hamiltonian is non-negative at $t=t'=0$.
No other classical product state at half filling has zero energy.
At fixed filling, the lowest excitation moves one particle to a neighboring site, creating three occupied nearest-neighbor bonds on the square lattice and costing energy $3V$.
We take this point as the reference point, with $\gamma(x_{\mathrm{ref}}) = 3V$.

\emph{Translation-symmetry breaking (CDW order).}
The reference point lies in the CDW phase, which spontaneously breaks translation symmetry.
A one-step translation in either lattice direction exchanges the two sublattices, so
\begin{equation}\label{eq:translation-cdw}
T_x|\psi_A\rangle = s_x|\psi_B\rangle,\quad
T_y|\psi_A\rangle = s_y|\psi_B\rangle,\qquad
T_x|\psi_B\rangle = s_x|\psi_A\rangle,\quad
T_y|\psi_B\rangle = s_y|\psi_A\rangle,
\end{equation}
with signs $s_x,s_y\in\{\pm 1\}$ fixed by the ordering of the fermionic creation operators in Eq.~\eqref{eq:cdw-states}.
The signs are the same for both states because $N_x$ and $N_y$ are even.
Neither CDW product state is translation invariant.

\emph{Phase diagram.}
At $t'=0$, there is a CDW phase for all $t$~\cite{gubernatis1985two,ma2021interacting}, due to Fermi-surface nesting.
At $t=0$, both CDW product states are exact eigenstates for any $t'$, because NNN hopping stays within a sublattice and each sublattice is either full or empty in both states.
Along the $t'$ axis, they remain the ground states up to a critical $|t'|$, beyond which the translation-symmetric gapless phase begins.

The filtered-response test is run with the two coordinate probes $\partial_t H$ and $\partial_{t'} H$.
On the $t'$ axis inside the CDW phase, only the transverse probe $\partial_t H$ can respond.
The probe $\partial_{t'} H$ moves particles only within a sublattice, so it annihilates both CDW states:
\begin{equation}\label{eq:pauli-blocking}
\partial_{t'} H|\psi_A\rangle = 0,\qquad
\partial_{t'} H|\psi_B\rangle = 0 .
\end{equation}
The band-restricted spectral weight of a probe (Definition~\ref{def:band-restricted-local-spectral-weight}) is built from its matrix elements between the ground states and the higher levels less than $\Lambda$ above them.
Hence, on the $t'$ axis inside the CDW phase, the weight of $\partial_{t'} H$ vanishes for every $\Lambda$:
\begin{equation}\label{eq:vanishing-tp-coupling}
W_{[\gamma,\Lambda)}^{\partial_{t'} H}(t=0,t') = 0 .
\end{equation}
By contrast, $\partial_t H$ moves a particle from one sublattice to the other, and at the solvable point each such hop costs an excitation energy of exactly $3V$, so
\begin{equation}\label{eq:nonvanishing-t-coupling}
W_{[\gamma,\Lambda)}^{\partial_t H}(t=0,t'=0) \neq 0
\end{equation}
for every $\Lambda > 3V$.

\emph{Finite-size ground-state manifold.}
For $t\neq 0$ inside the CDW phase, the exact ground states at finite size are two states $|\psi_\pm\rangle$.
As $t\to 0$, they tend to the symmetric and antisymmetric superpositions
\begin{equation}\label{eq:cdw-eigenstates}
|\psi_\pm\rangle \to
\frac{1}{\sqrt{2}}\left(|\psi_A\rangle \pm |\psi_B\rangle\right)
\end{equation}
of the CDW states.
The states $|\psi_\pm\rangle$ are eigenstates of translation, with the signs $s_x,s_y$ of Eq.~\eqref{eq:translation-cdw}:
\begin{equation}\label{eq:translation-eigenstates}
T_x|\psi_\pm\rangle = \pm s_x|\psi_\pm\rangle,\qquad
T_y|\psi_\pm\rangle = \pm s_y|\psi_\pm\rangle .
\end{equation}
Their energy splitting decays exponentially in system size.
Because the filtered-response test checks the gap above the twofold ground-state manifold spanned by $|\psi_+\rangle$ and $|\psi_-\rangle$, this exponentially small splitting does not register as a gap closing.

\emph{Numerical parameters.}
Figure~\ref{fig:hubbard-main} shows the certified region in the $(t/V, t'/V)$ plane, with threshold $\gamma_{\mathrm{thresh}} = 1.1$ and grid spacing $0.0015$, both in units of $V$.

\subsection{Kitaev Honeycomb Model (Topological Phase Boundary)}
\label{subsec:kitaev-honeycomb-benchmark}

\emph{Model.}
The Kitaev honeycomb model~\cite{kitaev2006anyons} is an exactly solvable spin-$1/2$ model with bond-dependent Ising interactions on the honeycomb lattice.
In its Hamiltonian, Eq.~\eqref{eq:kitaev-hamiltonian-main}, the sums run over the three inequivalent link directions of the lattice, $\sigma_j^{x,y,z}$ are the Pauli operators on site $j$, and $J_x,J_y,J_z$ are the coupling constants.
Kitaev's exact solution reduces this spin model to a problem of free Majorana fermions.

\emph{Phase diagram.}
To obtain a two-dimensional parameter space, we restrict to the standard simplex $J_x + J_y + J_z = 1$, $J_x,J_y,J_z \ge 0$.
On the simplex, the phase diagram consists of three disconnected gapped regions $\mathcal{A}_x,\mathcal{A}_y,\mathcal{A}_z$ near the corners, each with the same topological order as the toric code, and a gapless spin-liquid region $\mathcal{B}$ in the center.
The transition from $\mathcal{A}$ to $\mathcal{B}$ is therefore a topological quantum phase transition, rather than a conventional Landau transition driven by a local order parameter.
The phase boundaries are the lines $J_\alpha = 1/2$ for $\alpha \in \{x,y,z\}$, on each of which one coupling equals the sum of the other two.

\emph{Reference ground states.}
We initialize the algorithm at one gapped reference point in each of the three $\mathcal{A}$ regions, marked by stars in Figure~\ref{fig:kitaev-main}.
We consider a periodic honeycomb lattice of $n = 144$ spins.
On the torus, the ground-state manifold of the $\mathcal{A}$ phase is fourfold degenerate, as for the toric code, and lies in the vortex-free sector~\cite{kitaev2006anyons}.
At each reference point, $\rho_{\mathrm{ref}}$ is the equal mixture of the four vortex-free Wilson-loop ground states.

\emph{Numerical parameters.}
Figure~\ref{fig:kitaev-main} shows the certified regions on the simplex, with threshold $\gamma_{\mathrm{thresh}} = 0.11$ (in units where $J_x + J_y + J_z = 1$) and grid spacing $10^{-4}$ in $(J_x, J_y)$.
The detected boundary undulates, and the same undulations appear in the resolution error band, which we compute independently of the filtered-response test.
The agreement between the boundary and the band supports reading the undulations as finite-size spectral effects.
To compute the band, we diagonalize the quadratic Majorana Hamiltonians in the relevant flux and Wilson-loop sectors.
From their spectra we reconstruct the low-energy spectrum of the spin model, and this spectrum gives the spectral gap of Definition~\ref{definition: spectral gap}.
At finite system size and nonzero $\gamma_{\mathrm{thresh}}$, the detected boundary and the band need not coincide with the straight thermodynamic phase boundaries $J_\alpha = 1/2$, drawn for comparison.
The algorithm certifies that each grid point of the certified regions is connected to a reference point by a path along which the spectral gap of the finite system remains above $(1-\epsilon_{\mathrm{thresh}})\gamma_{\mathrm{thresh}}$.

\subsection{Toric Code in Uniform Fields (Topological Phase Boundary)}
\label{subsec:toric-code-benchmark}

\emph{Model.}
We finally consider the toric code in uniform $x$- and $z$-fields, Eq.~\eqref{eq:toric-field-hamiltonian-main}, with qubits on the edges of a periodic square lattice.
Here, $A_s$ and $B_p$ are the star and plaquette operators.
The reference ground state is a stabilizer ground state at the exactly solvable point $h_x = h_z = 0$, which lies in a gapped $\mathbb{Z}_2$ topologically ordered phase.
The stabilizer terms have coefficient $J = 1$, and at this point the Hamiltonian is a sum of commuting stabilizers.
On the torus, the model has four topologically degenerate ground states, and its elementary excitations are $\mathbb{Z}_2$ electric charges and magnetic vortices.
A violated stabilizer costs energy $2J$, and on the torus violated stabilizers come in pairs, so the gap at the reference point is $\gamma(x_{\mathrm{ref}}) = 4J$.

\emph{Phase diagram.}
The field strengths $(h_x,h_z)$ define a two-dimensional parameter space.
The $h_z$ perturbation creates and moves charge excitations, and the $h_x$ perturbation creates and moves vortex excitations.
Increasing either field eventually destroys the topological phase through quasiparticle proliferation, so the phase diagram contains a topological phase surrounding the origin, separated from the trivial phase at larger fields.

\emph{Numerical parameters.}
Figure~\ref{fig:toric-main} shows the certified region for $n = 8$ qubits in the $(h_x/J, h_z/J)$ plane, with threshold $\gamma_{\mathrm{thresh}} = 2.3485$ and grid spacing $0.002$, both in units of $J$.
The certified region surrounds the origin and approximates the $\mathbb{Z}_2$ topologically ordered phase.
For the thermodynamic phase diagram of the same quantum Hamiltonian, see Ref.~\cite{WuDengProkofev2012}.

\bibliography{refs}

\begin{thebibliography}{10}

\bibitem{ebadi2021quantum}
Sepehr Ebadi, Tout~T Wang, Harry Levine, Alexander Keesling, Giulia Semeghini, Ahmed Omran, Dolev Bluvstein, Rhine Samajdar, Hannes Pichler, Wen~Wei Ho, et~al.
\newblock Quantum phases of matter on a 256-atom programmable quantum simulator.
\newblock {\em Nature}, 595(7866):227--232, 2021.

\bibitem{scholl2021quantum}
Pascal Scholl, Michael Schuler, Hannah~J. Williams, Alexander~A. Eberharter, Daniel Barredo, Kai-Niklas Schymik, Vincent Lienhard, Louis-Paul Henry, Thomas~C. Lang, Thierry Lahaye, Andreas~M. L{\"a}uchli, and Antoine Browaeys.
\newblock Quantum simulation of {2D} antiferromagnets with hundreds of {R}ydberg atoms.
\newblock {\em Nature}, 595(7866):233--238, 2021.

\bibitem{semeghini2021probing}
Giulia Semeghini, Harry Levine, Alexander Keesling, Sepehr Ebadi, Tout~T Wang, Dolev Bluvstein, Ruben Verresen, Hannes Pichler, Marcin Kalinowski, Rhine Samajdar, et~al.
\newblock Probing topological spin liquids on a programmable quantum simulator.
\newblock {\em Science}, 374(6572):1242--1247, 2021.

\bibitem{shao2024antiferromagnetic}
Hou-Ji Shao, Yu-Xuan Wang, De-Zhi Zhu, Yan-Song Zhu, Hao-Nan Sun, Si-Yuan Chen, Chi Zhang, Zhi-Jie Fan, Youjin Deng, Xing-Can Yao, Yu-Ao Chen, and Jian-Wei Pan.
\newblock Antiferromagnetic phase transition in a {3D} fermionic {H}ubbard model.
\newblock {\em Nature}, 632(8024):267--272, 2024.

\bibitem{yang2020observation}
Bing Yang, Hui Sun, Robert Ott, Han-Yi Wang, Torsten~V. Zache, Jad~C. Halimeh, Zhen-Sheng Yuan, Philipp Hauke, and Jian-Wei Pan.
\newblock Observation of gauge invariance in a 71-site {B}ose--{H}ubbard quantum simulator.
\newblock {\em Nature}, 587(7834):392--396, 2020.

\bibitem{zhou2022thermalization}
Zhao-Yu Zhou, Guo-Xian Su, Jad~C. Halimeh, Robert Ott, Hui Sun, Philipp Hauke, Bing Yang, Zhen-Sheng Yuan, Jürgen Berges, and Jian-Wei Pan.
\newblock Thermalization dynamics of a gauge theory on a quantum simulator.
\newblock {\em Science}, 377(6603):311--314, 2022.

\bibitem{ambainis2014physical}
Andris Ambainis.
\newblock On physical problems that are slightly more difficult than {QMA}.
\newblock In {\em 2014 IEEE 29th Conference on Computational Complexity (CCC)}, pages 32--43. IEEE, 2014.

\bibitem{gharibian2019complexity}
Sevag Gharibian and Justin Yirka.
\newblock The complexity of simulating local measurements on quantum systems.
\newblock {\em Quantum}, 3:189, 2019.

\bibitem{xu2026provable}
Jue Xu, Xiao Yuan, and Qi~Zhao.
\newblock Provable quantum advantage for dynamical phase transition, 2026.
\newblock arXiv:2606.30396.

\bibitem{cubitt2015undecidability}
Toby~S. Cubitt, David P{\'e}rez-Garc{\'i}a, and Michael~M. Wolf.
\newblock Undecidability of the spectral gap.
\newblock {\em Nature}, 528(7581):207--211, 2015.

\bibitem{bausch2020undecidability1d}
Johannes Bausch, Toby~S. Cubitt, Angelo Lucia, and David P{\'e}rez-Garc{\'i}a.
\newblock Undecidability of the spectral gap in one dimension.
\newblock {\em Physical Review X}, 10(3):031038, 2020.

\bibitem{bausch2021uncomputability}
Johannes Bausch, Toby~S. Cubitt, and James~D. Watson.
\newblock Uncomputability of phase diagrams.
\newblock {\em Nature Communications}, 12:452, 2021.

\bibitem{cugini2025spectral}
Davide Cugini, Francesco Ghisoni, Angela~Rosy Morgillo, and Francesco Scala.
\newblock Spectral gap estimation via adiabatic preparation.
\newblock {\em Physica Scripta}, 101(24):245109, 2026.

\bibitem{pages2026lowresource}
Hugo Pages, Chusei Kiumi, Yuto Morohoshi, B{\'a}lint Koczor, and Kosuke Mitarai.
\newblock Low-resource quantum energy gap estimation via randomization.
\newblock {\em Physical Review Research}, 2026.
\newblock arXiv:2601.13881.

\bibitem{carrasquilla2017machine}
Juan Carrasquilla and Roger~G Melko.
\newblock Machine learning phases of matter.
\newblock {\em Nat. Phys.}, 13(5):431--434, 2017.

\bibitem{vannieuwenburg2017learning}
Evert P.~L. van Nieuwenburg, Ye-Hua Liu, and Sebastian~D. Huber.
\newblock Learning phase transitions by confusion.
\newblock {\em Nature Physics}, 13(5):435--439, 2017.

\bibitem{wang2016discovering}
Lei Wang.
\newblock Discovering phase transitions with unsupervised learning.
\newblock {\em Physical Review B}, 94(19):195105, 2016.

\bibitem{wetzel2017unsupervised}
Sebastian~J. Wetzel.
\newblock Unsupervised learning of phase transitions: From principal component analysis to variational autoencoders.
\newblock {\em Physical Review E}, 96(2):022140, 2017.

\bibitem{broecker2017machine}
Peter Broecker, Juan Carrasquilla, Roger~G. Melko, and Simon Trebst.
\newblock Machine learning quantum phases of matter beyond the fermion sign problem.
\newblock {\em Scientific Reports}, 7(1):8823, 2017.

\bibitem{chng2017machine}
Kelvin Ch'ng, Juan Carrasquilla, Roger~G. Melko, and Ehsan Khatami.
\newblock Machine learning phases of strongly correlated fermions.
\newblock {\em Physical Review X}, 7(3):031038, 2017.

\bibitem{zhang2017quantum}
Yi~Zhang and Eun-Ah Kim.
\newblock Quantum loop topography for machine learning.
\newblock {\em Phys. Rev. Lett.}, 118(21):216401, 2017.

\bibitem{rodrigueznieva2019identifying}
Joaquin~F. Rodriguez-Nieva and Mathias~S. Scheurer.
\newblock Identifying topological order through unsupervised machine learning.
\newblock {\em Nature Physics}, 15(8):790--795, 2019.

\bibitem{greitemann2019probing}
Jonas Greitemann, Ke~Liu, and Lode Pollet.
\newblock Probing hidden spin order with interpretable machine learning.
\newblock {\em Physical Review B}, 99(6):060404, 2019.

\bibitem{carleo2019machine}
Giuseppe Carleo, Ignacio Cirac, Kyle Cranmer, Laurent Daudet, Maria Schuld, Naftali Tishby, Leslie Vogt-Maranto, and Lenka Zdeborov{\'a}.
\newblock Machine learning and the physical sciences.
\newblock {\em Reviews of Modern Physics}, 91(4):045002, 2019.

\bibitem{che2025circuit}
Yanming Che, Clemens Gneiting, Xiaoguang Wang, and Franco Nori.
\newblock Quantum circuit complexity and unsupervised machine learning of topological order.
\newblock {\em Nature Communications}, 17(1):5179, 2026.

\bibitem{grossifonseca2025gradient}
Andr{\'e} Grossi~Fonseca, Eric Wang, Sachin Vaidya, Patrick~J. Ledwith, Ashvin Vishwanath, and Marin Solja{\v{c}}i{\'c}.
\newblock Gradient-based search of quantum phases: discovering unconventional fractional {C}hern insulators, 2025.
\newblock arXiv:2509.10438.

\bibitem{khosrojerdi2025unsupervised}
Mehran Khosrojerdi, Alessandro Cuccoli, Paola Verrucchi, and Leonardo Banchi.
\newblock Unsupervised learning to recognize quantum phases of matter.
\newblock {\em AVS Quantum Science}, 8(3):033801, 2026.

\bibitem{malyshev2026distance}
Oleksii Malyshev, Simon~M. Linsel, Fabian Grusdt, Annabelle Bohrdt, Eugene Demler, and Ivan Morera.
\newblock Distance learning from projective measurements as an information-geometric probe of many-body physics, 2026.
\newblock arXiv:2603.13485.

\bibitem{huang2026detecting}
Ching-Yu Huang and Min-Fong Yang.
\newblock Detecting quantum phase transitions via shallow variational quantum circuits.
\newblock {\em Physical Review B}, 2026.
\newblock arXiv:2607.25345.

\bibitem{zhou2026autonomous}
Shiyu Zhou, Yuxuan Zhang, Sebastian Wetzel, Roger Melko, and Xiu-Zhe Luo.
\newblock Autonomous phase discovery, 2026.
\newblock arXiv:2609.33802.

\bibitem{nys2026quantummatterexplorer}
Jannes Nys, Andrew Jreissaty, and Juan Carrasquilla.
\newblock {QuantumMatterExplorer}: An autonomous agent for the discovery of quantum many-body phases of matter, 2026.
\newblock arXiv:2610.32001.

\bibitem{knabe1988energy}
Stefan Knabe.
\newblock Energy gaps and elementary excitations for certain {VBS}-quantum antiferromagnets.
\newblock {\em Journal of Statistical Physics}, 52(3):627--638, 1988.

\bibitem{rai2026hierarchy}
Kshiti~Sneh Rai, Ilya Kull, Patrick Emonts, Jordi Tura, Norbert Schuch, and Flavio Baccari.
\newblock A hierarchy of spectral gap certificates for frustration-free spin systems.
\newblock {\em Quantum}, 10:2065, 2026.

\bibitem{lake2025exact}
Ethan Lake, Shankar Balasubramanian, and Soonwon Choi.
\newblock Exact quantum algorithms for quantum phase recognition: renormalization group and error correction.
\newblock {\em PRX Quantum}, 6:010350, 2025.

\bibitem{chaduteau2026agnostic}
Alexandre Chaduteau, Hao Chen, Juraj Hasik, and Titus Neupert.
\newblock Agnostic and efficient detection of $\mathbb{Z}_2$ symmetry-protected topological phases in projected entangled-pair states, 2026.
\newblock arXiv:2611.07030.

\bibitem{huang2022provably}
Hsin-Yuan Huang, Richard Kueng, Giacomo Torlai, Victor~V Albert, and John Preskill.
\newblock Provably efficient machine learning for quantum many-body problems.
\newblock {\em Science}, 377(6613):eabk3333, 2022.

\bibitem{wu2023quantum}
Yusen Wu, Bujiao Wu, Jingbo Wang, and Xiao Yuan.
\newblock Quantum phase recognition via quantum kernel methods.
\newblock {\em Quantum}, 7:981, 2023.

\bibitem{jager2026provable}
Jonas J{\"a}ger, Paolo Braccia, Pablo Bermejo, Manuel~G. Algaba, Diego Garc{\'\i}a-Mart{\'\i}n, and M.~Cerezo.
\newblock Provable and scalable quantum {G}aussian processes for quantum learning, 2026.
\newblock arXiv:2605.00099.

\bibitem{chen2010local}
Xie Chen, Zheng-Cheng Gu, and Xiao-Gang Wen.
\newblock Local unitary transformation, long-range quantum entanglement, wave function renormalization, and topological order.
\newblock {\em Physical Review B}, 82(15):155138, 2010.

\bibitem{watson2021complexityapproximatingcriticalpoints}
James~D. Watson and Johannes Bausch.
\newblock The complexity of approximating critical points of quantum phase transitions, 2021.
\newblock arXiv:2105.13350.

\bibitem{laumann2012quantum}
C.~R. Laumann, R.~Moessner, A.~Scardicchio, and S.~L. Sondhi.
\newblock Quantum adiabatic algorithm and scaling of gaps at first-order quantum phase transitions.
\newblock {\em Physical Review Letters}, 109:030502, 2012.

\bibitem{campostrini2014finite}
Massimo Campostrini, Jacopo Nespolo, Andrea Pelissetto, and Ettore Vicari.
\newblock Finite-size scaling at first-order quantum transitions.
\newblock {\em Physical Review Letters}, 113:070402, 2014.

\bibitem{keesling2019quantum}
Alexander Keesling, Ahmed Omran, Harry Levine, Hannes Bernien, Hannes Pichler, Soonwon Choi, Rhine Samajdar, Sylvain Schwartz, Pietro Silvi, Subir Sachdev, Peter Zoller, Manuel Endres, Markus Greiner, Vladan Vuleti{\'c}, and Mikhail~D. Lukin.
\newblock Quantum {K}ibble--{Z}urek mechanism and critical dynamics on a programmable {R}ydberg simulator.
\newblock {\em Nature}, 568(7751):207--211, 2019.

\bibitem{mazurenko2017cold}
Anton Mazurenko, Christie~S. Chiu, Geoffrey Ji, Maxwell~F. Parsons, M{\'a}rton Kan{\'a}sz-Nagy, Richard Schmidt, Fabian Grusdt, Eugene Demler, Daniel Greif, and Markus Greiner.
\newblock A cold-atom {F}ermi--{H}ubbard antiferromagnet.
\newblock {\em Nature}, 545(7655):462--466, 2017.

\bibitem{hastings2005quasiadiabatic}
Matthew~B Hastings and Xiao-Gang Wen.
\newblock Quasiadiabatic continuation of quantum states: The stability of topological ground-state degeneracy and emergent gauge invariance.
\newblock {\em Phys. Rev. B}, 72(4):045141, 2005.

\bibitem{hastings2010locality}
Matthew~B Hastings.
\newblock Locality in quantum systems.
\newblock In {\em Quantum Theory from Small to Large Scales}, pages 171--212. Oxford University Press, 2012.

\bibitem{bachmann2011automorphic}
Sven Bachmann, Spyridon Michalakis, Bruno Nachtergaele, and Robert Sims.
\newblock Automorphic equivalence within gapped phases of quantum lattice systems.
\newblock {\em Communications in Mathematical Physics}, 309(3):835--871, 2012.

\bibitem{jones2026quasiphases}
Nick~G. Jones, Paul Fendley, Nathanan Tantivasadakarn, and Ruben Verresen.
\newblock Quasi-phases of quantum matter, 2026.
\newblock arXiv:2610.40526.

\bibitem{uhrich2017noninvasive}
Philipp Uhrich, Salvatore Castrignano, Hermann Uys, and Michael Kastner.
\newblock Noninvasive measurement of dynamic correlation functions.
\newblock {\em Physical Review A}, 96:022127, 2017.

\bibitem{chen2024catalytictomography}
Chi-Fang Chen and Robbie King.
\newblock Catalytic tomography of ground states, 2025.
\newblock arXiv:2512.10247.

\bibitem{chen_gilyen_thermal_shadows}
Chi-Fang Chen and Andr\'as Gily\'en.
\newblock Efficiently estimating quantum thermal properties from exponentially fewer samples, 2026.
\newblock arXiv:2603.16845.

\bibitem{li2026stability}
Kangle Li and Liujun Zou.
\newblock Stability of energy gap of macroscopic quantum matter, 2026.
\newblock arXiv:2611.06528.

\bibitem{bernien2017probing}
Hannes Bernien, Sylvain Schwartz, Alexander Keesling, Harry Levine, Ahmed Omran, Hannes Pichler, Soonwon Choi, Alexander~S Zibrov, Manuel Endres, Markus Greiner, et~al.
\newblock Probing many-body dynamics on a 51-atom quantum simulator.
\newblock {\em Nature}, 551(7682):579--584, 2017.

\bibitem{haah2016invariant}
Jeongwan Haah.
\newblock An invariant of topologically ordered states under local unitary transformations.
\newblock {\em Communications in Mathematical Physics}, 342(3):771--801, 2016.

\bibitem{bravyi2006lieb}
Sergey Bravyi, Matthew~B Hastings, and Frank Verstraete.
\newblock {L}ieb-{R}obinson bounds and the generation of correlations and topological quantum order.
\newblock {\em Physical Review Letters}, 97(5):050401, 2006.

\bibitem{Chen_2024}
Xie Chen, Arpit Dua, Michael Hermele, David~T. Stephen, Nathanan Tantivasadakarn, Robijn Vanhove, and Jing-Yu Zhao.
\newblock Sequential quantum circuits as maps between gapped phases.
\newblock {\em Physical Review B}, 109(7):075116, February 2024.

\bibitem{huffman2014solution}
Emilie~Fulton Huffman and Shailesh Chandrasekharan.
\newblock Solution to sign problems in half-filled spin-polarized electronic systems.
\newblock {\em Phys. Rev. B}, 89:111101, 2014.

\bibitem{li2015solving}
Zi-Xiang Li, Yi-Fan Jiang, and Hong Yao.
\newblock Solving the fermion sign problem in quantum {Monte Carlo} simulations by {Majorana} representation.
\newblock {\em Phys. Rev. B}, 91:241117, 2015.

\bibitem{kitaev2006anyons}
Alexei Kitaev.
\newblock Anyons in an exactly solved model and beyond.
\newblock {\em Annals of Physics}, 321(1):2--111, 2006.

\bibitem{Tupitsyn_2010}
I.~S. Tupitsyn, A.~Kitaev, N.~V. Prokof’ev, and P.~C.~E. Stamp.
\newblock Topological multicritical point in the phase diagram of the toric code model and three-dimensional lattice gauge {H}iggs model.
\newblock {\em Physical Review B}, 82(8):085114, August 2010.

\bibitem{WuDengProkofev2012}
Fengcheng Wu, Youjin Deng, and Nikolay Prokof'ev.
\newblock Phase diagram of the toric code model in a parallel magnetic field.
\newblock {\em Physical Review B}, 85:195104, 2012.

\bibitem{LinTong2020}
Lin Lin and Yu~Tong.
\newblock Near-optimal ground state preparation.
\newblock {\em Quantum}, 4:372, 2020.

\bibitem{DingChenLin2024}
Zhiyan Ding, Chi-Fang Chen, and Lin Lin.
\newblock Single-ancilla ground state preparation via {L}indbladians.
\newblock {\em Physical Review Research}, 6:033147, 2024.

\bibitem{LiYangLin2026}
Hao-En Li, Yilun Yang, and Lin Lin.
\newblock Dissipative phase decision without ground-state preparation, 2026.
\newblock arXiv:2606.28750.

\bibitem{schuster2025hardness}
Thomas Schuster, Dominik Kufel, Norman~Y. Yao, and Hsin-Yuan Huang.
\newblock Hardness of recognizing phases of matter, 2025.
\newblock arXiv:2510.08503.

\bibitem{coser2019classification}
Andrea Coser and David P{\'e}rez-Garc{\'\i}a.
\newblock Classification of phases for mixed states via fast dissipative evolution.
\newblock {\em Quantum}, 3:174, 2019.

\bibitem{rakovszky2024defining}
Tibor Rakovszky, Sarang Gopalakrishnan, and Curt von Keyserlingk.
\newblock Defining stable phases of open quantum systems.
\newblock {\em Physical Review X}, 14:041031, 2024.

\bibitem{sang2024mixed}
Shengqi Sang, Yijian Zou, and Timothy~H. Hsieh.
\newblock Mixed-state quantum phases: Renormalization and quantum error correction.
\newblock {\em Physical Review X}, 14:031044, 2024.

\bibitem{sang2025stability}
Shengqi Sang and Timothy~H. Hsieh.
\newblock Stability of mixed-state quantum phases via finite {M}arkov length.
\newblock {\em Physical Review Letters}, 134:070403, 2025.

\bibitem{ellison2025towards}
Tyler~D. Ellison and Meng Cheng.
\newblock Toward a classification of mixed-state topological orders in two dimensions.
\newblock {\em PRX Quantum}, 6:010315, 2025.

\bibitem{bao2026mixed}
Yimu Bao, Ruihua Fan, Ashvin Vishwanath, and Ehud Altman.
\newblock Mixed-state topological order and the errorfield double formulation of decoherence-induced transitions.
\newblock {\em Physical Review Letters}, 136:220402, 2026.

\bibitem{fan2024diagnostics}
Ruihua Fan, Yimu Bao, Ehud Altman, and Ashvin Vishwanath.
\newblock Diagnostics of mixed-state topological order and breakdown of quantum memory.
\newblock {\em PRX Quantum}, 5:020343, 2024.

\bibitem{ma2025topological}
Ruochen Ma, Jian-Hao Zhang, Zhen Bi, Meng Cheng, and Chong Wang.
\newblock Topological phases with average symmetries: The decohered, the disordered, and the intrinsic.
\newblock {\em Physical Review X}, 15:021062, 2025.

\bibitem{lessa2025strong}
Leonardo~A. Lessa, Ruochen Ma, Jian-Hao Zhang, Zhen Bi, Meng Cheng, and Chong Wang.
\newblock Strong-to-weak spontaneous symmetry breaking in mixed quantum states.
\newblock {\em PRX Quantum}, 6:010344, 2025.

\bibitem{Bha97:Matrix-Analysis}
Rajendra Bhatia.
\newblock {\em Matrix analysis}, volume 169 of {\em Graduate Texts in Mathematics}.
\newblock Springer-Verlag, New York, 1997.

\bibitem{nachtergaele2019quasi}
Bruno Nachtergaele, Robert Sims, and Amanda Young.
\newblock Quasi-locality bounds for quantum lattice systems. i. {L}ieb-{R}obinson bounds, quasi-local maps, and spectral flow automorphisms.
\newblock {\em J. Math. Phys.}, 60(6):061101, 2019.

\bibitem{nielsen2002quantum}
Michael~A. Nielsen and Isaac~L. Chuang.
\newblock {\em {Quantum Computation and Quantum Information}}.
\newblock Cambridge University Press, 10th anniversary edition, 2010.

\bibitem{essin2013classifying}
Andrew~M. Essin and Michael Hermele.
\newblock Classifying fractionalization: Symmetry classification of gapped $\mathbb{Z}_2$ spin liquids in two dimensions.
\newblock {\em Physical Review B}, 87(10):104406, 2013.

\bibitem{cheng2016translational}
Meng Cheng, Michael Zaletel, Maissam Barkeshli, Ashvin Vishwanath, and Parsa Bonderson.
\newblock Translational symmetry and microscopic constraints on symmetry-enriched topological phases: A view from the surface.
\newblock {\em Physical Review X}, 6(4):041068, 2016.

\bibitem{barkeshli2019symmetry}
Maissam Barkeshli, Parsa Bonderson, Meng Cheng, and Zhenghan Wang.
\newblock Symmetry fractionalization, defects, and gauging of topological phases.
\newblock {\em Physical Review B}, 100(11):115147, 2019.

\bibitem{kato1950adiabatic}
Tosio Kato.
\newblock On the adiabatic theorem of quantum mechanics.
\newblock {\em Journal of the Physical Society of Japan}, 5(6):435--439, 1950.

\bibitem{jansen2007bounds}
Sabine Jansen, Mary-Beth Ruskai, and Ruedi Seiler.
\newblock Bounds for the adiabatic approximation with applications to quantum computation.
\newblock {\em Journal of Mathematical Physics}, 48(10):102111, 2007.

\bibitem{pedernales2014efficient}
J.~S. Pedernales, R.~Di~Candia, I.~L. Egusquiza, J.~Casanova, and E.~Solano.
\newblock Efficient quantum algorithm for computing $n$-time correlation functions.
\newblock {\em Physical Review Letters}, 113(2):020505, 2014.

\bibitem{childs2019theory}
Andrew~M Childs, Yuan Su, Minh~C Tran, Nathan Wiebe, and Shuchen Zhu.
\newblock Theory of {T}rotter error with commutator scaling.
\newblock {\em Physical Review X}, 11(1):011020, 2021.

\bibitem{berry2015simulating}
Dominic~W. Berry, Andrew~M. Childs, Richard Cleve, Robin Kothari, and Rolando~D. Somma.
\newblock Simulating {H}amiltonian dynamics with a truncated {T}aylor series.
\newblock {\em Physical Review Letters}, 114(9):090502, 2015.

\bibitem{berry2020timedependent}
Dominic~W. Berry, Andrew~M. Childs, Yuan Su, Xin Wang, and Nathan Wiebe.
\newblock Time-dependent {H}amiltonian simulation with {$L^1$}-norm scaling.
\newblock {\em Quantum}, 4:254, 2020.
\newblock arXiv:1906.07115.

\bibitem{watrous2018book}
John Watrous.
\newblock {\em The Theory of Quantum Information}.
\newblock Cambridge University Press, 2018.

\bibitem{vershynin2018high}
Roman Vershynin.
\newblock {\em High-dimensional probability}, volume~47 of {\em Cambridge Series in Statistical and Probabilistic Mathematics}.
\newblock Cambridge University Press, Cambridge, 2018.
\newblock An introduction with applications in data science, With a foreword by Sara van de Geer.

\bibitem{ge2016rapid}
Yimin Ge, Andr{\'a}s Moln{\'a}r, and J.~Ignacio Cirac.
\newblock Rapid adiabatic preparation of injective projected entangled pair states and {G}ibbs states.
\newblock {\em Physical Review Letters}, 116(8):080503, 2016.

\bibitem{kato1995perturbation}
Tosio Kato.
\newblock {\em Perturbation Theory for Linear Operators}.
\newblock Classics in Mathematics. Springer, Berlin, 1995.

\bibitem{haah2018quantum}
Jeongwan Haah, Matthew~B Hastings, Robin Kothari, and Guang~Hao Low.
\newblock Quantum algorithm for simulating real time evolution of lattice {H}amiltonians.
\newblock {\em SIAM Journal on Computing}, 52(6):FOCS18--250, 2023.

\bibitem{davis1970rotation}
Chandler Davis and W.~M. Kahan.
\newblock The rotation of eigenvectors by a perturbation. {III}.
\newblock {\em SIAM Journal on Numerical Analysis}, 7(1):1--46, 1970.

\bibitem{gubernatis1985two}
J.~E. Gubernatis, D.~J. Scalapino, R.~L. Sugar, and W.~D. Toussaint.
\newblock Two-dimensional spin-polarized fermion lattice gases.
\newblock {\em Phys. Rev. B}, 32:103, 1985.

\bibitem{ma2021interacting}
Kou-Han Ma and Ning-Hua Tong.
\newblock Interacting spinless fermions on the square lattice: Charge order, phase separation, and superconductivity.
\newblock {\em Phys. Rev. B}, 104:155116, 2021.

\end{thebibliography}
\bibliographystyle{unsrt}

\end{document}